\documentstyle[12pt,graphicx,epsfig,epsf,feynmp,rotating]{article}

\catcode`@=11
\def\citer{\@ifnextchar
[{\@tempswatrue\@citexr}{\@tempswafalse\@citexr[]}}

\def\@citexr[#1]#2{\if@filesw\immediate\write\@auxout{\string\citation{#2}}\fi
  \def\@citea{}\@cite{\@for\@citeb:=#2\do
    {\@citea\def\@citea{--\penalty\@m}\@ifundefined
       {b@\@citeb}{{\bf ?}\@warning
       {Citation `\@citeb' on page \thepage \space undefined}}%
\hbox{\csname b@\@citeb\endcsname}}}{#1}}
\catcode`@=12

\newcommand{\nn}{\noindent}

\newcommand{\s}{\\ \vspace*{-3mm}}
\newcommand{\lsim}{\raisebox{-0.13cm}{~\shortstack{$<$ \\[-0.07cm] $\sim$}}~}
\newcommand{\gsim}{\raisebox{-0.13cm}{~\shortstack{$>$ \\[-0.07cm] $\sim$}}~}

\newcommand{\beq}{\begin{eqnarray}}
\newcommand{\eeq}{\end{eqnarray}}
\newcommand{\bq}{\begin{equation}}
\newcommand{\eq}{\end{equation}}
\newcommand{\be}{\begin{equation}}
\newcommand{\ee}{\end{equation}}

\begin{document}

\begin{center}

\vspace*{0.4cm}

{\Large\sc \bf THE QCD AND STANDARD MODEL WORKING GROUP: }

\vspace*{0.4cm}

{\Large\sc \bf Summary Report} 

\vspace*{0.4cm} 

Conveners: \\[0.2cm] 
{\sc S. Catani$^1$, M. Dittmar$^2$, J. Huston$^3$, D. Soper$^4$, S. Tapprogge$^5$ }

\vspace*{0.6cm}

 Working Group: \\[0.2cm]
{\sc 

P.~Aurenche$^6$,
C.~Bal\'azs$^7$,
R.~D.~Ball$^8$,
T.~Binoth$^6$,
E.~Boos$^9$,
J.~Collins$^{10}$,
V.~del Duca$^{1}$,
M.~Fontannaz$^{11}$,
S.~Frixione$^1$,
J.P.~Guillet$^6$,
G.~Heinrich$^{11}$,
V.~Ilyin$^9$,
Y.~Kato$^{12}$,
K.~Odagiri$^{13}$,
F.~Paige$^{14}$,
E.~Pilon$^{6}$,
A.~Pukhov$^{9}$,
I.~Puljak$^{15,16}$,
A.~Semenov$^{9}$,
A.~Skatchkova$^{9}$,
V.~Tano$^{17}$,
W.K.~Tung$^{3}$,
W.~Vogelsang$^{18}$,
M.~Werlen$^{6}$,
D.~Zeppenfeld$^{1,19}$.
}

\vspace*{0.6cm}

{\small

$^1$ CERN, Theory Division, CH-1211 Geneva 23, Switzerland\\
$^2$ Institute for Particle Physics (IPP), ETH Zurich, CH-8093 Zurich, Switzerland\\
$^3$ Physics Dept., Michigan State University, East Lansing, MI 48824 USA \\
$^4$ Institute of Theoretical Science, University of Oregon, Eugene, Oregon 97403 USA \\
$^5$ CERN, EP Division, CH-1211 Geneva 23, Switzerland\\
$^6$ LAPTH, BP 110, F--74941 Annecy le Vieux Cedex, France. \\
$^7$ Department of Physics and Astronomy, University of Hawaii, Honolulu, 
HI, 96822 USA \\
$^8$ Department of Physics and Astronomy, University of Edinburgh, EH9 3JZ, Scotland \\
$^9$ Institute of Nuclear Physics, MSU, 119899 Moscow, Russia\\
$^{10}$ Department of Physics, Penn State University, 
University Park, PA 16802, USA \\ 
$^{11}$ Laboratoire de Physique Th\'eorique, Universit\'e de Paris XI,
Batiment 210, F-91405, Orsay Cedex, France  \\
$^{12}$ Kogakuin University, Shinjuku, Tokyo 163-8677, Japan \\
$^{13}$ Rutherford Appleton Laboratory, Chilton, Didcot, Oxon OX11 0QX, UK \\
$^{14}$ Physics Department, Brookhaven National Laboratory, Upton, NY 11973, USA \\
$^{15}$ Laboratoire de Physique Nucl\'eaire et des Houtes Energies, Ecole Polytechnique, 91128 Palaiseau, France\\
$^{16}$ University of Split, 21000 Split, Croatia\\
$^{17}$ Technische Hochschule Aachen, III Physikalisches Institut, Sommerfeldestrasse 26-28, D-52-56 Aachen, Germany; MPI fuer Physik, Foehringer Ring , 80805 Muenchen\\
$^{18}$ C.N. Yang Institute for Theoretical Physics, SUNY Stony Brook, Stony Brook, New York 11794, USA\\
$^{19}$ Department of Physics, University of Wisconsin, Madison, WI 53706, USA\\
}

\vspace*{0.5cm}

{\it Report of the QCD/SM working group for the Workshop \\[0.1cm]
``Physics at TeV Colliders", Les Houches, France 8--18 June 1999.}

\end{center} 

\newpage


\begin{center}
{\bf \large CONTENTS} 
\end{center} 


\nn {\bf 0. Introduction} \hfill 3 \\

\nn {\bf 1. Aspects of QCD, from the Tevatron to the LHC } 
\hfill 4 \\[0.2cm] \hspace*{0.5cm}
S.~Catani. \\

\nn {\bf 2. Partons for the LHC} 
\hfill 34 \\[0.2cm] \hspace*{0.5cm}
R.~.D.~Ball and J. Huston. \\

\nn {\bf 3. Generalized Factorization and Resummation }
\hfill 62 \\[0.2cm]  \hspace*{0.5cm}
C.~Bal\'azs, J.~Collins, D.~Soper. \\

\nn {\bf 4. A Comparison of the Predictions from Monte Carlo Programs and 
Transverse Momentum Resummation }
\hfill 82 \\[0.2cm]  \hspace*{0.5cm}
C.~Bal\'azs, J.~Huston, I.~Puljak. \\

\nn {\bf 5. Automatic Computation of LHC Processes}
\hfill 108 \\[0.2cm]  \hspace*{0.5cm}
E.~Boos, V.~Ilyin, K.~Kato, A.~Pukhov, A.~Semenov, A.~Skatchkova. \\

\nn {\bf 6. Monte Carlo Event Generators at NLO}
\hfill 117 \\[0.2cm]  \hspace*{0.5cm}
J. Collins. \\

\nn {\bf 7. NLO and NNLO Calculations}
\hfill 122 \\[0.2cm] \hspace*{0.5cm}
V. del Duca, G. Heinrich. \\

\nn {\bf 8. Jet Algorithms } 
\hfill 132 \\[0.2cm] \hspace*{0.5cm}
S.~Catani and D.~Zeppenfeld. \\

\nn {\bf 9. Underlying Event in Jet Events } 
\hfill 141 \\[0.2cm] \hspace*{0.5cm}
J.~Huston and V.~Tano. \\

\nn {\bf 10. Isolated Photon Production } 
\hfill 152 \\[0.2cm] \hspace*{0.5cm}
S. Frixione, W. Vogelsang. \\

\nn {\bf 11. Direct Photon Pair Production at Colliders}
\hfill 161 \\[0.2cm] \hspace*{0.5cm}
T. Binoth, J.P.~Guillet, V. Ilyin, E. Pilon, M. Werlen. \\

\newpage


\begin{center}
{\large\sc {\bf Introduction}} 
\end{center}

\vspace*{0.1cm}

The  Les Houches  Workshop on Physics at TeV Colliders  took place  from
June 8-18, 1999. One of the three working groups at  Les Houches  concentrated on
QCD issues, both at the Tevatron Collider  and at the LHC. Besides the interest  in
QCD in its own right, QCD dynamics  plays an important  role in the production
mechanisms  for any new physics processes that might be observed at either
collider, as well as any processes  that 
may form backgrounds to the new physics.  As might  be expected,
there was a great deal of overlap with the other two working groups, and especially
with the Higgs working group. 

To provide a more specific focus, each day at
Les Houches  was devoted to a specific topic. The topic and speakers  are listed
below: 

\begin{itemize}

\item Thursday:	parton distibutions (W.K. Tung, R.  Ball)
\item	Friday:	photons (E. Pilon, M. Fontannaz, S. Frixione); jet definitions (J. Huston, D. Zeppenfeld)
\item	Saturday:	Monte Carlos (F. Paige, J.Collins, K. Odagiri, Y. Kato, E. Boos, A. Skatchkova, V. Ilyin)
\item	Monday:	joint meeting with Higgs group (C. Balazs)
\item	Tuesday:	(re)summation (C. Balazs)
\item	Wednesday:	direct photons and pions (P. Aurenche), heavy flavor (S. Frixione)
\item	Thursday:	N$^n$LO (V. del Duca, G. Henrich)

\end{itemize}

	This writeup for the QCD working group is not intended to be a 
comprehensive summary of all of the QCD  issues currently  important  in high
energy  physics, or expected to be important  for the LHC. Rather, we have
chosen to concentrate  in detail on a few selected topics  and to summarize in a
pedagogical  manner  the current  status and the progress expected in the near
future.  The expertise of the people attending the workshop, and the timeliness
of the issues, resulted in a great deal of concentration on resummation
calculations, and their relation to Monte Carlos. This is  also reflected in the
writeup. 
The writeup is organized into  chapters  that  roughly follow the organization
shown above. Preceding the discussion of the individual  topics  is a general 
introduction  to QCD phenomenology relevant for the LHC, by S. Catani. 

\bigskip

Joey Huston (for the Working Group). \bigskip

\noindent {\bf Acknowledgements}: \smallskip

\noindent We would like to thank the organizers of the workshop for the
pleasant and stimulating atmosphere. Les Houches is a beautiful place
to do physics; we should do this every two years. 

\newpage




\def\naive{na\"{\i}ve}
\def\ltap{\raisebox{-.4ex}{\rlap{$\,\sim\,$}} \raisebox{.4ex}{$\,<\,$}}
\def\gtap{\raisebox{-.4ex}{\rlap{$\,\sim\,$}} \raisebox{.4ex}{$\,>\,$}}
\newcommand\as{\alpha_{\mathrm{S}}}
\def\beq{\begin{equation}}
\def\eeq{\end{equation}}
\def\beeq{\begin{eqnarray}}
\def\eeeq{\end{eqnarray}}
\def\bom#1{{\mbox{\boldmath $#1$}}}
\def\to{\rightarrow}
\def\kper{k_{\perp}}
\def\nn{\nonumber}



\begin{center}
\vspace*{1.2cm}
{\Large\sc \bf Aspects of QCD, from the Tevatron to the LHC } \\
\vspace*{1.cm} 
{\sc S. Catani}
\vspace*{1.cm}
\end{center}







\par \vspace{2mm}
\begin{center} {\large \bf Abstract} \end{center}
\begin{quote}
\pretolerance 10000

This contribution presents a selection of the topics 
(parton densities, fixed-order calculations, parton showers, soft-gluon
resummation) discussed in my introductory lectures at the Workshop and 
includes a pedagogical overview of the corresponding theoretical
tools.

\end{quote}



\section{Introduction}
\label{intro}

The production cross sections for all the processes at hadron-collider
experiments are controlled by strong interaction physics and, hence,
by its underlying field theory, QCD (see recent overviews in 
Refs.~[\ref{eps99}, \ref{lp99webber}, \ref{ichep98}, \ref{lp97}]).
Studies of QCD at the Tevatron and the LHC
have two main purposes [\ref{proceeding}, \ref{proctev}, \ref{proclhc}]. 
First, they are important to test the predictions of
QCD, to measure its fundamental parameters (e.g. the strong coupling $\as$) and
to extract quantitative information on its non-perturbative dynamics (e.g. the
distribution of partons in the proton). Second, they are relevant to a precise 
estimate of the background to other Standard Model processes and to signals of
new physics.

This contribution is not a comprehensive review of QCD at high-energy hadron
colliders. It is based on a selection of the topics presented in my
introductory lectures at this Workshop. The selection highlights
the QCD subjects that were most discussed during the Workshop 
and includes a pedagogical overview of some of the corresponding theoretical
tools. 

After the introduction of the general theoretical framework, 
I~summarize in Sect.~\ref{secpdf} 
the present knowledge on the parton densities and its impact on
QCD predictions for hard-scattering processes at the Tevatron and the LHC.
In Sect.~\ref{secgdensity}, I then discuss some issues related to processes
that are sensitive to the gluon density and, hence, to its 
determination. Section~\ref{secpxs} presents a dictionary of 
different approaches (fixed-order expansions, resummed calculations, 
parton showers) to perturbative QCD calculations. 
The dictionary continues in Sect.~\ref{secsoftg}, where I review soft-gluon
resummation and discuss some recent phenomenological applications of threshold
resummation to hadron collisions.

The QCD framework to describe any inclusive hard-scattering process,
\beq
\label{hardpro}
h_1(p_1) + h_2(p_2) \to H(Q,\{ \dots \}) + X \;\;,
\eeq
in hadron--hadron collisions is based on perturbation theory and on the 
factorization theorem of mass singularities. The corresponding cross section
is computed by using the factorization formula [\ref{factform}]
\beeq
\sigma(p_1,p_2;Q, \{ \dots \} ) \!\!\!&=&\!\!\! 
\sum_{a,b} \int_{x_{\rm min}}^1 dx_1 \, dx_2 \,f_{a/h_1}(x_1, \mu_F^2)
\, f_{b/h_2}(x_2, \mu_F^2) \; 
{\hat \sigma}_{ab}(x_1p_1,x_2p_2;Q, \{ \dots \}; \mu_F^2)
\nn \\
\label{factfor}
\!\!\!&+&\!\!\! {\cal O}\left( (\Lambda_{QCD}/Q )^p \right) \;\;.
\eeeq

The colliding hadrons $h_1$ and $h_2$ have momenta $p_1$ and $p_2$, $H$ denotes
the triggered hard probe (vector bosons, jets, heavy quarks, Higgs bosons,
SUSY particles and so on) and $X$ stands for 
any unobserved particle produced by the collision. The typical scale $Q$
of the scattering process is set by the invariant mass or the transverse
momentum of the hard probe, and the notation $\{ \dots \}$ stands for any other
relevant scale and kinematic variable of the process. For instance,
in the case of $W$ production we have $Q=M_W$ and $\{ \dots \}= \{ Q_{\perp},
y, \dots \}$, where $M_W, Q_{\perp}$ and $y$ are the mass of the vector boson, 
its transverse momentum and its rapidity, respectively.

The factorization formula (\ref{factfor}) involves the convolution
of the partonic cross sections ${\hat \sigma}_{ab}$ (where $a,b=q,{\bar q},g)$ 
and the parton distributions $f_{a/h}(x, \mu_F^2)$ 
of the colliding hadrons. If the hard probe $H$ is a hadron or a photon,
the factorization formula has to include an additional convolution with the
corresponding parton fragmentation function $d_{a/H}(z, \mu_F^2)$.

The term ${\cal O}\left( (\Lambda_{QCD}/Q )^p \right)$ on the right-hand side 
of Eq.~(\ref{factfor}) generically denotes non-perturbative contributions
(hadronization effects, multiparton interactions, contributions of the soft
underlying event, and so on). Provided the hard-scattering process 
(\ref{hardpro}) is sufficiently inclusive\footnote{ 
More precisely, it has to be defined
in an infrared- and collinear-safe manner.}, 
${\hat \sigma}_{ab}$ is computable as a power series expansion in $\as(Q^2)$
and the non-perturbative contributions are (small) power-suppressed corrections
(i.e. the power $p$ is positive) as long as the hard-scattering scale
$Q$ is larger than few hundred MeV, the typical size of the QCD scale 
$\Lambda_{QCD}$.

The parton densities $f_{a/h}(x, \mu_F^2)$ are phenomenological distributions 
that describe how partons are bounded in the colliding hadrons. Although they 
are not calculable in QCD perturbation theory, the parton densities
are universal (process-independent) quantities. The scale $\mu_F$ is a
factorization scale introduced in Eq.~(\ref{factfor}) to separate the
bound-state effects from the perturbative interactions of the partons.
The physical cross section $\sigma(p_1,p_2;Q, \{ \dots \} )$ does not depend on
this arbitrary scale, but parton densities and partonic cross sections 
separately depend on $\mu_F$. In particular, higher-order contributions to 
${\hat \sigma}_{ab}(x_1p_1,x_2p_2;Q, \{ \dots \}; \mu_F^2)$ contain corrections
of relative order $(\as(Q^2) \ln Q^2/\mu_F^2)^n$. If $\mu_F$ is very different
from $Q$, these corrections become large and spoil the reliability of the
perturbative expansion. Thus, in practical applications of the factorization
formula (\ref{factfor}), the scale $\mu_F$ is set approximately equal to
the hard scale $Q$ and variations of $\mu_F$ around this central value
are used to estimate the uncertainty of the perturbative expansion.

\begin{figure}
  \centerline{
    \setlength{\unitlength}{1cm}
    \begin{picture}(0,6)
       \put(0,0){\includegraphics{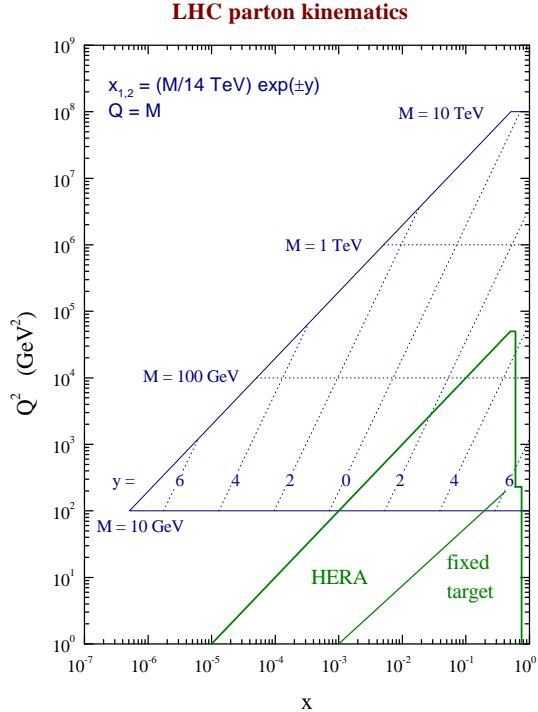}}
    \end{picture}}
\vspace*{3.0cm}    
\caption{\sf The $(x,Q^2)$ plane of the parton kinematics for the production of
a heavy system of invariant mass $M$ and rapidity $y$ at LHC, HERA and
fixed-target experiments.
\label{lhckin}}
\end{figure}

The lower limit $x_{\rm min}$ of the integrations over the parton momentum
fractions $x_1$ and $x_2$, as well as the values of $x_1$ and $x_2$ that
dominate the convolution integral in Eq.~(\ref{factfor}), are controlled
by the kinematics of the hard-scattering process. Typically we have
$x_{\rm min} \gtap Q^2/S$, where $S=(p_1+p_2)^2$ is the square of the
centre-of-mass energy of the collision. If the hard probe is a state of
invariant mass $M$ and rapidity $y$, the dominant values of the momentum
fractions are $x_{1,2} \sim (M e^{\pm y})/{\sqrt S}$
(see Fig.~\ref{lhckin}). 
Thus varying $M$ and
$y$ at fixed ${\sqrt S}$, we are sensitive to partons with different
momentum fractions. Increasing ${\sqrt S}$ the parton densities are probed in
a kinematic range that extends towards larger values of $Q$ and smaller values
of $x_{1,2}$.

\section{Parton densities}
\label{secpdf}

The parton densities are an essential ingredient to study hard-scattering
collisions. Once the partonic cross sections have been perturbatively
computed, cross section measurements can be used to determine the parton
densities. Then, they can in turn be used to predict cross sections
for other hard-scattering processes.

The dependence of the parton densities\footnote{In the following the parton
densities of the proton $f_{a/p}$ are simply denoted by $f_{a}$ and
those of the antiproton are obtained by using charge-conjugation invariance,
i.e. $f_{a/{\bar p}}= f_{{\bar a}/p} = f_{{\bar a}}$.}
$f_{a}(x, \mu^2)$ on the momentum fraction $x$ and their absolute value
at any fixed scale $\mu$
are not computable in perturbation theory. However, the scale dependence is
perturbatively controlled by the DGLAP evolution equation~[\ref{DGLAP}] 
\beq
\label{evequa}
\frac{d \,f_{a}(x, \mu^2)}{d \ln \mu^2} = 
\sum_{b} \int_{x}^1 \frac{dz}{z} \, P_{ab}(\as(\mu^2), z) \,f_{a}(x/z, \mu^2)
\;\;.
\eeq
The kernels $P_{ab}(\as, z)$ are the Altarelli--Parisi (AP) splitting functions.
As the partonic cross sections in Eq.~(\ref{factfor}), the AP splitting 
functions can be computed as a power series expansion in $\as$:
\beq
\label{apexp}
P_{ab}(\as, z) = \as P_{ab}^{(LO)}(z) + \as^2 P_{ab}^{(NLO)}(z)
+ \as^3 P_{ab}^{(NNLO)}(z) + {\cal O}(\as^4) \;\;.
\eeq
The leading order (LO) and next-to-leading order (NLO) terms 
$P_{ab}^{(LO)}(z)$ and $P_{ab}^{(NLO)}(z)$ in the expansion 
are known [\ref{NLOAP}]. These first two terms are used
in most of the QCD studies.
Having determined $f_{a}(x, Q_0^2)$ at a given input scale $\mu = Q_0$, the
evolution equation (\ref{evequa}) can be used to compute the parton densities
at different perturbative scales $\mu$ and larger values of $x$.

The parton densities are determined by performing global fits 
[\ref{pdffit}] to data from deep-inelastic scattering (DIS), Drell--Yan (DY),
prompt-photon and jet production. The method consists in parametrizing
the parton densities at some input scale $Q_0$ and then adjusting the 
parameters to fit the data. The parameters are usually constrained by imposing
the positivity of the parton densities $(f_{a}(x, \mu^2) \geq 0)$ and
the momentum sum rule $(\sum_a \int_0^1 dx \,x \,f_{a}(x, \mu^2) =1)$.

\begin{figure}
  \centerline{
    \setlength{\unitlength}{1cm}
    \begin{picture}(0,6)
       \put(0,0){\includegraphics{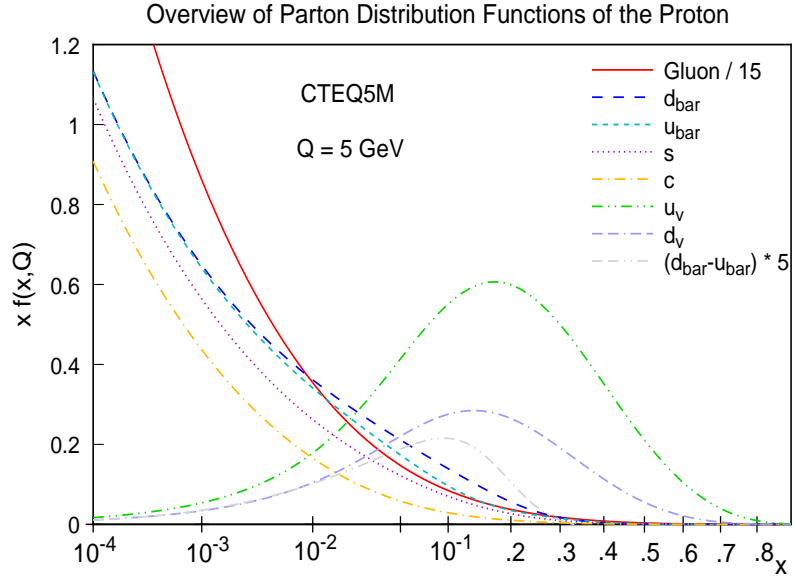}}
    \end{picture}}
\vspace*{1.3cm}    
\caption{\sf Typical $x$-shape of the parton densities (set CTEQ5M at $Q=5$~GeV).
\label{pdfplot}}
\end{figure}

The present knowledge on the parton densities of the proton is reviewed in
Refs.~[\ref{proclhc}, \ref{lhcpdf}, \ref{pdfrball}].
Their typical behaviour is shown in 
Fig.~\ref{pdfplot}. 
All densities decrease at large $x$. At small $x$ the valence quark densities
vanish and the gluon density dominates. The sea-quark densities also increase
at small $x$ because they are driven by the strong rise of the gluon density
and the splitting of gluons in $q {\bar q}$ pairs. Note that the quark
densities are not flavour-symmetric either in the valence sector
$(u_v \neq d_v)$ or in the sea sector $({\bar u} \neq {\bar d})$. 

In addition to having the best estimate of the parton densities, it is 
important to quantify the corresponding uncertainty. This is a difficult issue.
The uncertainty depends on the kinematic range in $x$ and $Q^2$. Moreover,
it cannot be reliably estimated by simply comparing the parton densities
obtained by different global fits. In fact, a lot of common systematic and
common assumptions affect the global-fit procedures.
Recent attempts to
obtain parton densities with error bands that take into account correlations
between experimental errors are described in Refs.~[\ref{proclhc}, 
\ref{pdfrball}].
Some important theoretical uncertainties that are still to be understood
are also discussed in Ref.~[\ref{proclhc}]. 

The overall conclusion is that
the quark densities\footnote{Uncertainties on the determination of the quark
densities at very high $x$ are discussed in Refs.~[\ref{bodek}, 
\ref{Melnitchouk}, \ref{highxcteq}].}
are reasonably well constrained and determined by
DIS and DY processes, while the gluon density is certainly more 
uncertain [\ref{pdffit}, \ref{gluoncteq}].
At small $x$ $(x \ltap 10^{-3})$, the gluon density $f_g$ is at present 
constrained by a {\em single} process, namely DIS at HERA. Thus, 
large higher-order corrections of the type $(\as \ln 1/x)^n$ could possibly 
affect the extraction of $f_g$.
{\em Assuming}
that $f_g$ is well determined at small $x$, the momentum sum rule reasonably
constrains $f_g$ at intermediate values of $x$ $(x \sim 10^{-2})$.
Jet production at the Tevatron at low to moderate values of the
jet transverse energy $E_T$ can also be useful
in constraining the gluon distribution in the range $0.05 \ltap x \ltap 0.2$.
At large $x$ $(x \gtap 10^{-1})$, the most sensitive process to $f_g$
is prompt-photon production. Since, at present, prompt-photon data are not well
described/predicted by perturbative QCD calculations, they cannot be used for a
precise determination of $f_g$. 
Further discussion on these points is given
in Sect.~\ref{secgdensity}.
 
\begin{figure}
  \centerline{
    \setlength{\unitlength}{1cm}
    \begin{picture}(0,6)
       \put(0,0){\includegraphics{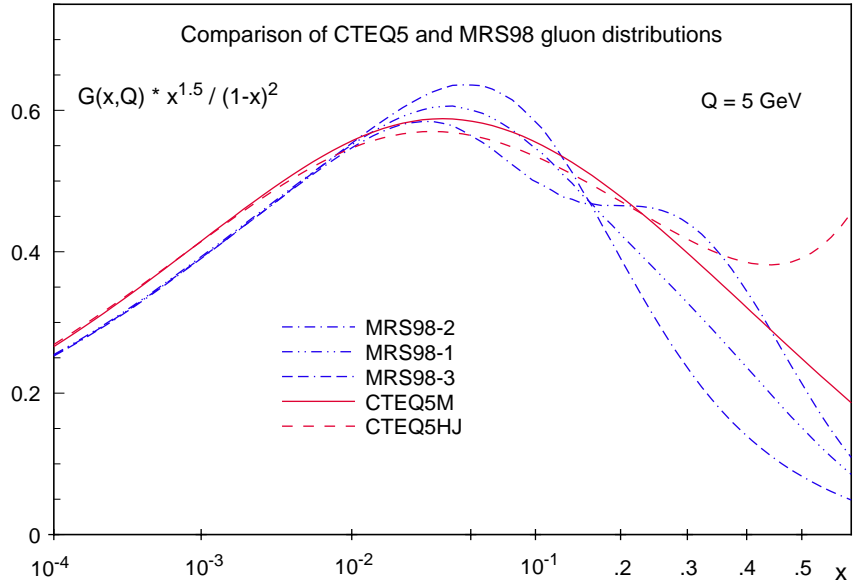}}
    \end{picture}}
\vspace*{1.5cm}    
\caption{\sf Comparison between the gluon densities of the CTEQ and MRST groups.
\label{differentfg}}
\end{figure}

The conclusion that the gluon density is not well known can also be drawn
by inspection (see Fig.~\ref{differentfg})
of the differences between the most updated analyses performed
by the CTEQ Collaboration and the MRST group.

The differences between the MRST gluons and the CTEQ ones are due to the fact
that the two groups used different data sets.
The various gluon densities are very
similar at small $x$, because in this region both groups used the HERA data.
The MRST group includes prompt-photon data in the global fit: these data 
constrain the gluon directly at $x \gtap 10^{-1}$ and indirectly (by the momentum
sum rule) at $x \sim 10^{-2}$. The CTEQ group does not use prompt-photon data,
but it includes Tevatron data on the one-jet inclusive cross section.
These data give a good constraint on $f_g$ in the region 
$0.05 \ltap x \ltap 0.2$.

There are also differences within the MRST and CTEQ sets. The various gluon
densities of the MRST set correspond to different values of the non-perturbative
transverse-momentum smearing that can be introduced to describe the differences
among the prompt-photon data that are available at several 
centre-of-mass energies. The CTEQ5M and CTEQ5HJ gluons correspond to different
assumptions on the parametrization of the functional form of $f_g(x,Q_0^2)$ at
large $x$;
the CTEQ5M set corresponds to the minimum-$\chi^2$ solution of the fit while
the CTEQ5HJ set (with a slightly higher $\chi^2$) provides the best fit to the
high-$E_T$ tail of the CDF {\em and} D0 jet cross sections. 

\begin{figure}
  \centerline{
    \setlength{\unitlength}{1cm}
    \begin{picture}(0,6)
       \put(0,0){\includegraphics{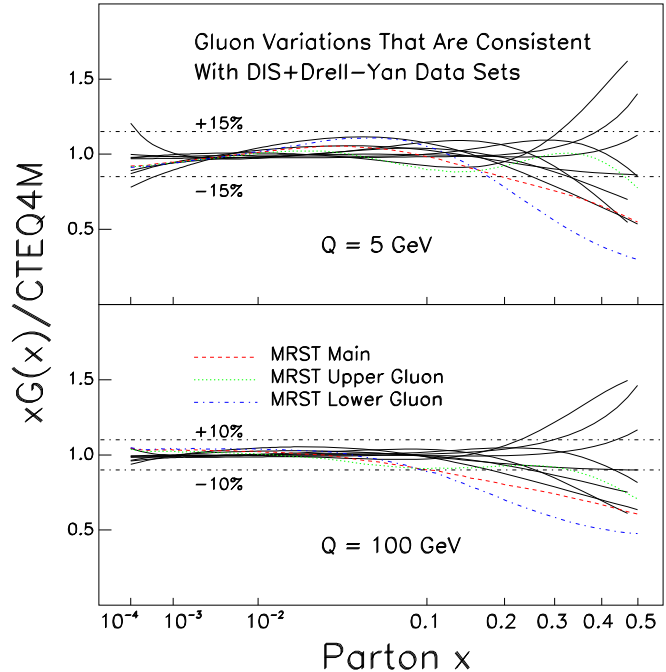}}
    \end{picture}}
\vspace*{2.8cm}    
\caption{\sf A picture of the gluon density uncertainty. The continuous (black)
lines refer to gluon densities that are constrained only by DIS and DY data.
The dashed (coloured) lines refer to gluon densities of the MRST group, which
uses also prompt-photon data.
\label{fguncertainty}}
\end{figure}

This brief illustration shows that the differences in the most recent
parton densities are mainly due to either inconsistencies between data sets
and/or poor theoretical understanding of them.
A more quantitative picture of the dependence on $x$ and $Q^2$
of the gluon density uncertainty is presented in Fig.~\ref{fguncertainty}.

We can see that the DIS and DY data sets weakly constrain $f_g$ for
$x \gtap 10^{-1}$. Since the AP splitting functions lead to negative scaling
violation at large $x$, when $f_g(x,Q^2)$ is evolved at larger scales $Q$
according to Eq.~(\ref{evequa}) the gluon uncertainty 
is diluted: it propagates at smaller values of $x$ and its size 
is reduced at fixed $x$.

\begin{figure}
  \centerline{
    \setlength{\unitlength}{1cm}
    \begin{picture}(0,7)
       \put(0,0){\includegraphics{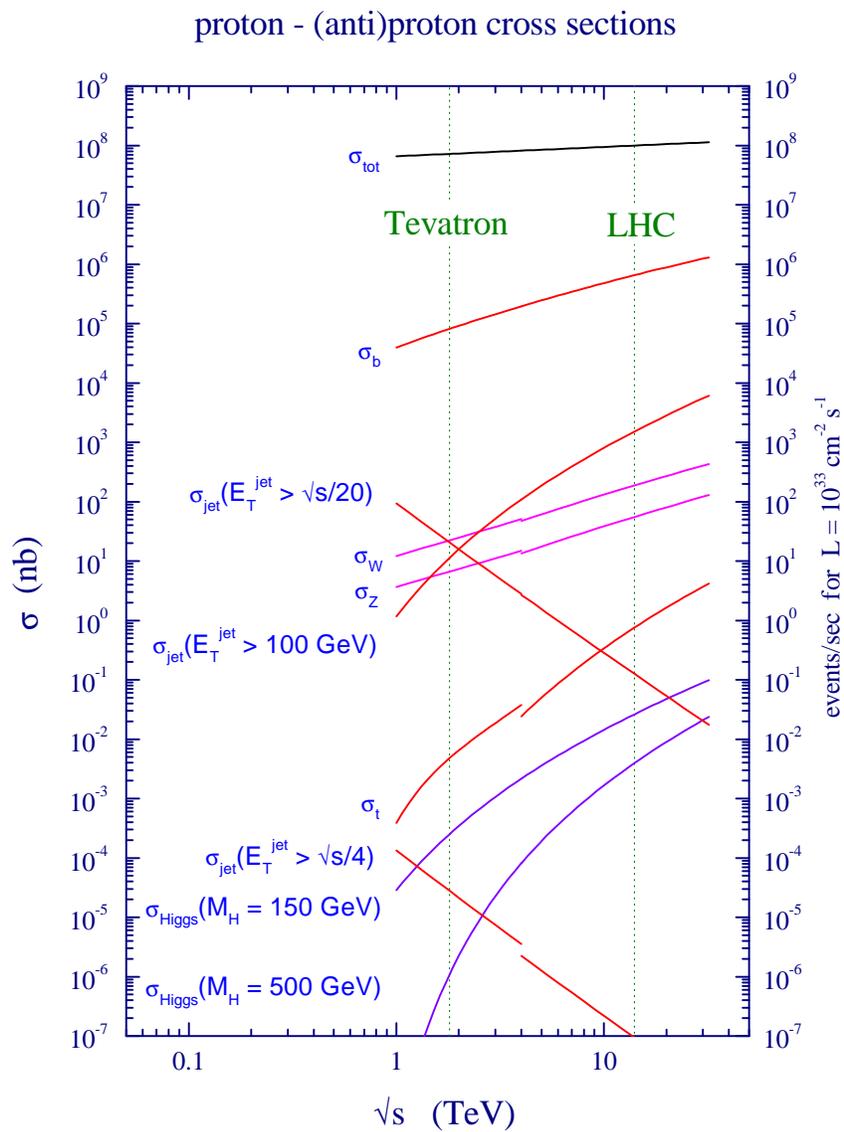}}
    \end{picture}}
\vspace*{4.7cm}    
\caption{\sf QCD predictions for hard-scattering cross sections at the Tevatron 
and the LHC.
\label{figxs}}
\end{figure}

Figure~\ref{figxs}
shows the typical predictions for hard-scattering cross sections at the 
Tevatron and the LHC, as obtained by using the parton densities of the MRST 
set. These predictions have to be supplemented with the corresponding
uncertainties [\ref{proclhc}]
coming from the determination of the parton 
densities and from perturbative corrections beyond the NLO.
Owing to the
increased centre-of-mass energy and to QCD scaling violation 
(see Fig.~\ref{fguncertainty}), the kinematic region with small uncertainties
is larger at the LHC than at the Tevatron. 

For most of the QCD processes at the LHC, the uncertainty from
the parton densities is smaller than $\pm 10\%$ and, in particular,
it is smaller than the uncertainty from higher-order corrections. Some 
relevant exceptions are the single-jet, $W/Z$ and top quark cross sections.
In the case of the single-jet inclusive cross section at high $E_T$
$(E_T \gtap 2$~TeV), the uncertainty from the poorly known gluon density at 
high $x$ is larger than that $(\sim \pm 10\%)$ from higher-order corrections.
The $W$ and $Z$ production cross sections are dominated by $q{\bar q}$
annihilation. Since the quark densities are well known, the ensuing uncertainty
on the $W/Z$ cross section is small $(\sim \pm 5\%)$. Nonetheless, in this case
the uncertainty from higher-order corrections is even smaller, since the
partonic cross sections for the DY process are known [\ref{NNLODY}] at the
next-to-next-to-leading order (NNLO) in perturbation theory. In the case
of top-quark production at the LHC, the gluon channel dominates and leads to
an uncertainty of $\pm 10\%$ on the total cross section. Also for this process,
however, the perturbative component is known beyond the NLO. Including
all-order resummation of soft-gluon contributions [\ref{txs}], 
the estimated uncertainty from unknown higher-order corrections is 
approximately $\pm 5\%$ [\ref{txs}, \ref{proclhc}].

\section{The gluon density issue}
\label{secgdensity}

At present, the processes\footnote{The r\^ole of jet production at the Tevatron
has briefly been recalled in Sect.~\ref{secpdf}, and it is discussed in detail
in Ref.~[\ref{lhcpdf}].} that are, in principle,
most sensitive to the gluon density are
DIS at HERA, $b$-quark production at the Tevatron, and prompt-photon production
at fixed-target experiments.
These processes constrain $f_g$
for $x \ltap 10^{-3}$, $x \sim 10^{-3}$--$10^{-2}$ and $x \gtap 10^{-1}$,
respectively.
Nonetheless, the gluon density is, in practice, not well determined. 
The issue (or, perhaps, the puzzle)
is that from a phenomenological viewpoint the standard theory, namely
perturbative QCD at NLO, works pretty well for $x \ltap 10^{-3}$ but not
so well at larger values of $x$, while from theoretical arguments we should
expect just the opposite to happen. This issue is discussed below mainly in its
perturbative aspects. We should however keep it in mind that all these processes 
are dominated by hard-scattering scales $Q$ of the order of few GeV. Different
types of non-perturbative contributions can thus be important.

From the study of DIS at HERA we can extract information on the gluon and
sea-quark densities of the proton. The main steps in the QCD analysis of the
structure functions at small values of the Bjorken variable $x$
are the following.
The measurement of the proton structure function
$F_2(x,Q^2) \sim q_S(x,Q^2)$ directly determines the sea-quark density
$q_S= x (f_q + f_{\bar q})$. Then, the DGLAP evolution equation (\ref{evequa})
or, more precisely, the following equations
(the symbol $\otimes$ denotes the convolution integral with respect to $x$):
\beeq
\label{df2}
dF_2(x,Q^2) / d\ln Q^2 
&\sim& P_{qq} \otimes q_S +  P_{qg} \otimes g \;\;, \\
\label{dfg}
d g(x,Q^2)/ d\ln Q^2 
&\sim& P_{gq} \otimes q_S +  P_{gg} \otimes g \;\;, 
\eeeq
are used to extract a gluon density $g(x,Q^2)= x f_g(x,Q^2)$ that agrees with 
the measured scaling violation in $dF_2(x,Q^2) / d\ln Q^2$ 
(according to Eq.~(\ref{df2}))
and fulfils the self-consistency equation (\ref{dfg}). 

The perturbative-QCD 
ingredients in this analysis are the AP splitting functions 
$P_{ab}(\as,x)$.
Once they are known (and only then), the non-perturbative gluon density
can be determined. 

The standard perturbative-QCD framework to extract $g(x,Q^2)$ consists 
in using the truncation of the AP splitting functions at the NLO. This
approach has been extensively compared with structure function
data over the last few years and it gives a good description of the HERA data,
down to low values of $Q^2 \sim 2 \, {\rm GeV^2}$. The NLO QCD fits simply 
require a slightly steep input gluon density at these low momentum scales.
Typically [\ref{pdffit}], we have $g(x,Q_0^2) \sim x^{-\lambda}$, 
with $\lambda \sim 0.2$ at $Q_0^2 \sim 2 \, {\rm GeV^2}$, and the data
constrain $g(x,Q_0^2)$ with an uncertainty of approximately $\pm 20\%$.

Although it is phenomenologically successful, the NLO approach is
not fully satisfactory from a theoretical viewpoint.
The truncation of the splitting functions
at a fixed perturbative order is equivalent to assuming that the dominant
dynamical mechanism leading to scaling violations is the evolution of
parton cascades with strongly-ordered transverse momenta. However, at high 
energy this evolution takes place over large rapidity intervals $(\Delta  y \sim
\ln1/x)$ and diffusion in transverse momentum becomes relevant. Formally, this
implies that higher-order corrections to $P_{ab}(\as,x)$ are logarithmically
enhanced:
\beq
\label{plog}
P_{ab}(\as,x) \sim \frac{\as}{x} + \frac{\as}{x} \;( \as \ln x ) + \dots
+ \frac{\as}{x} \;( \as \ln x )^n + \dots \;\;.
\eeq
At asymptotically small values of $x$, resummation of these corrections is 
mandatory to obtain reliable predictions.

Small-$x$ resummation is, in general, accomplished by the BFKL 
equation~[\ref{BFKL}]. In the context of structure-function
calculations, the BFKL equation provides us with improved expressions
of the AP splitting functions $P_{ab}(\as,x)$, in which the leading logarithmic
(LL) terms $(\as \ln x )^n$, the next-to-leading logarithmic (NLL) terms 
$\as (\as \ln x )^n$, and so forth, are systematically summed to all orders $n$
in $\as$. The present theoretical status of small-$x$ resummation is discussed
in Ref.~[\ref{proclhc}].
Since in the small-$x$ region the gluon channel 
dominates, only the gluon splitting functions $P_{gg}$ and $P_{gq}$ contain
LL contributions. These are known [\ref{BFKL}, \ref{Jar}] to be positive
but numerically smaller than naively expected
(the approach to the asymptotic regime is
much delayed by cancellations of logarithmic
corrections that occur at the first perturbative orders in $P_{gg}$ and 
$P_{gq}$). The NLL terms in the quark splitting functions $P_{qg}$ and 
$P_{qq}$ are known [\ref{CH}] and turn out to be positive and large.
 A very important progress is the recent calculation
[\ref{fadin}, \ref{CC}] of the NLL terms in $P_{gg}$, which are found to
be negative and large. The complete NLL terms in $P_{gq}$ are still unknown.

The results of Refs.~[\ref{fadin}, \ref{CC}], the large size of the NLL terms
and the alternating sign (from the LL to the NLL order and from
the gluon to the quark channel) of the resummed
small-$x$ contributions have 
prompted a lot of activity (see the list of references in Ref.~[\ref{proclhc}]) 
on the conceptual basis and the phenomenological implications of small-$x$ 
resummation. This activity is still in progress and definite quantitative
conclusions on the impact of small-$x$ resummation at HERA cannot be drawn yet.

\begin{figure}
  \centerline{
    \setlength{\unitlength}{1cm}
    \begin{picture}(0,6)
       \put(0,0){\includegraphics{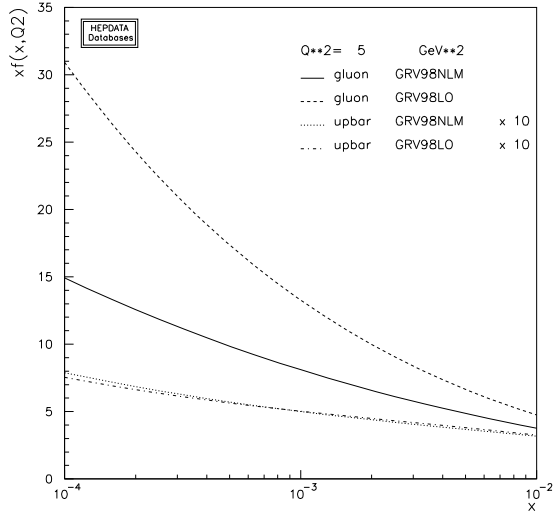}}
    \end{picture}}
\vspace*{0.7cm}    
\caption{\sf Comparison between the LO (GRV98LO) and NLO (GRV98NLM) GRV 
parametrizations of the gluon and sea-quark densities at $Q^2=5$~GeV$^2$.
\label{figgrv}}
\end{figure}
 
At the same time, the capability of the fixed-order approach to produce a good 
description of the proton
structure function $F_2(x,Q^2)$ at HERA cannot be used to conclude
that the small-$x$ behaviour of the gluon density is certainly well determined.
In fact, by comparing LO and NLO results, we could argue that the ensuing
theoretical uncertainty on $f_g$ is sizeable [\ref{lp97}].
Going from LO to NLO, we can obtain stable predictions for $F_2$,
but we have to vary the gluon density a lot.
As shown in Fig.~\ref{figgrv}, the NLO gluon density sizeably differs from its
LO parametrization, not only in absolute normalization but also in $x$-shape.
For instance, at $x=10^{-4}$ and $Q^2=5~{\rm GeV}^2$ the NLO gluon is a factor
of 2 smaller than the LO gluon.
This can be understood~[\ref{sdis}] from the fact that the scaling violation of 
$F_2$ is produced by the convolution $P_{qg} \otimes g$ (see the right-hand 
side of Eq.~(\ref{df2})). The quark splitting function $P_{qg}$ behaves as
\beq
\label{pqg}
P_{qg}(\as,x) \simeq \as P_{qg}^{(LO)}(x) \left[ 1 + 2.2 \frac{C_A \as}{\pi}
\frac{1}{x} + \dots \right] \;\;,
\eeq
where the LO term $P_{qg}^{(LO)}(x)$ is flat at small $x$, whereas the NLO
correction is steep. To obtain a stable evolution of $F_2$, the NLO steepness 
of $P_{qg}$ has to be compensated by a gluon density that is less steep 
at NLO than at LO. This has to be kept in mind when
concluding on the importance of small-$x$ resummation because
the NLO steepness of $P_{qg}$ is the lowest-order manifestation of BFKL 
dynamics in the quark channel. 

In the large-$x$ region, there is a well-known correlation between
$\as$ and $f_g$. At small $x$, there is an analogous strong correlation
between the $x$-shapes of $P_{qg}$ and $f_g$. In the fixed-order QCD analysis
of $F_2$, large NLO perturbative corrections at small $x$ can be balanced by
the extreme flexibility of parton density parametrizations. It is difficult
to disentangle this correlation between process-dependent
perturbative contributions and non-perturbative parton densities from the study 
of a single quantity, as in the case of $F_2$ at HERA. The uncertainty on the
gluon density at small $x$, as estimated from the NLO QCD fits of the
HERA data, is evidently only a lower limit on the actual uncertainty on $f_g$.

\begin{figure}
  \centerline{
    \setlength{\unitlength}{1cm}
    \begin{picture}(0,6)
       \put(0,0){\includegraphics{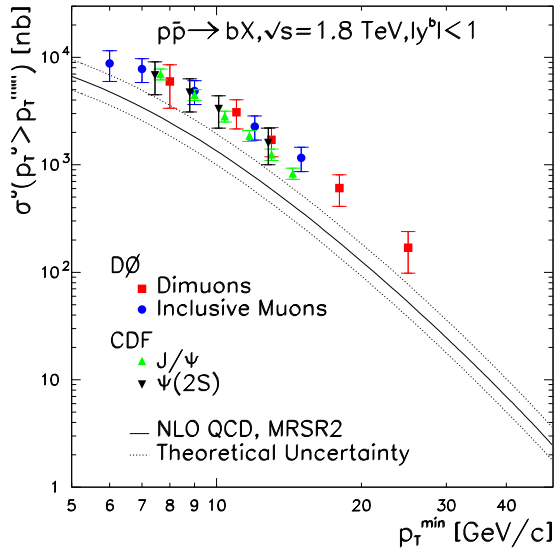}}
    \end{picture}}
\vspace*{1.7cm}    
\caption{\sf Comparison between Tevatron data and NLO QCD for $b$-quark production
[\ref{baarmand}]. The band is obtained by varying factorization
and renormalization scales in the NLO calculation.
\label{figbtev}}
\end{figure}

The production of $b$ quarks at the Tevatron is also sensitive
to the gluon density at relatively small values of $x$. The comparison
between Tevatron data and perturbative-QCD predictions at NLO [\ref{hqnlo}] 
is shown in Fig.~\ref{figbtev}. 
Using standard sets of parton densities, the theoretical predictions 
typically underestimate the measured cross section by a factor of 2.
This certainly is disappointing, although justifiable by the large theoretical
uncertainty of the perturbative calculation [\ref{bquarkrev}]. A lower limit
on this uncertainty can be estimated by studying the scale dependence and the
convergence of the perturbative expansion.
Varying the factorization
and renormalization scales by a factor of four around the $b$-quark mass $m_b$,
the NLO cross section varies by a factor of almost 2 at the Tevatron
and by a factor of 4--5 at the LHC [\ref{proclhc}]. 
Similar factors are obtained by considering
the ratio of the NLO and LO cross sections.

The present theoretical predictions for $b$-quark production at hadron 
colliders
certainly need to be improved [\ref{proclhc}]. Since the hard scale 
$Q \sim m_b$ is not very large, a possible improvement regards estimates of 
non-perturbative contributions (for instance, effects of the fragmentation
of the $b$-quark and of the intrinsic transverse momentum of the colliding 
partons). As for the evaluation of perturbative contributions at higher 
orders, 
the resummation of logarithmic terms of the type $\as^n \ln^n(p_t/m_b)$ is 
important [\ref{ptbres}] when the transverse momentum $p_t$ of the $b$ quark 
is much larger than $m_b$. The resummation of small-$x$ logarithmic 
contributions $\as^n \ln^n x$ can also be relevant, because 
$x \sim 2m_b/{\sqrt S}$ is as small as $\sim 10^{-3}$ at the Tevatron and
as $\sim 10^{-4}$ at the LHC. The theoretical tool to perform this resummation,
namely the $k_{\perp}$-factorization approach [\ref{ktfac}], is available.
Updated phenomenological studies based on this tool and on the information
from small-$x$ DIS at HERA would be interesting.

\begin{figure}
  \centerline{
    \setlength{\unitlength}{1cm}
    \begin{picture}(0,6)
       \put(0,0){\includegraphics{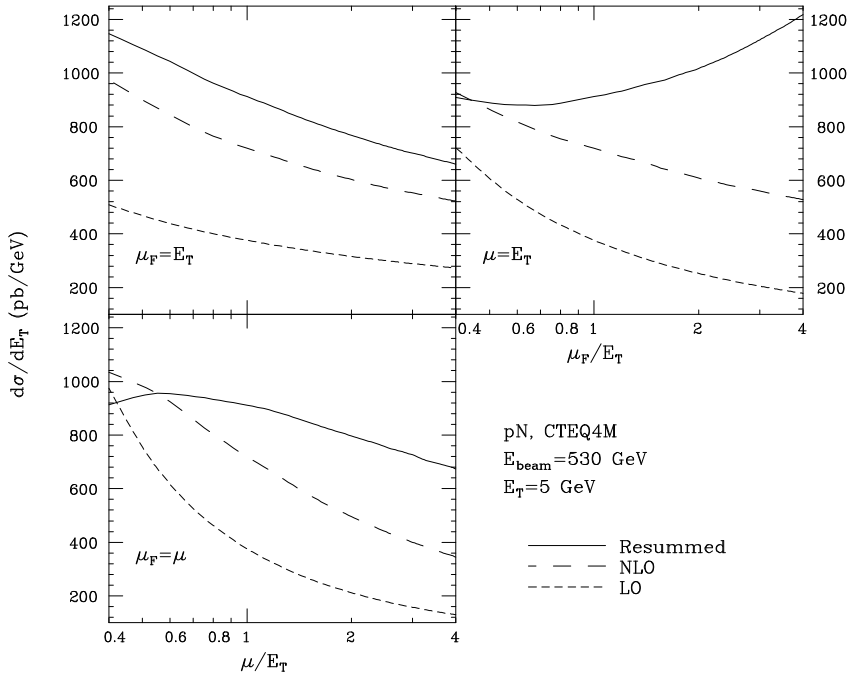}}
    \end{picture}}
\vspace*{2.5cm}    
\caption{\sf The dependence on the factorization $(\mu_F)$ and renormalization
$(\mu_R=\mu)$ scale of the LO and NLO prompt-photon cross section 
$d\sigma/dE_T$ in $pN$ collisions at $E_T=5$~GeV and ${\sqrt S}=31.6$~GeV. 
The resummed calculation is discussed in Sect.~\ref{secsoftg}.
\label{figgamma5}}
\end{figure}

Prompt-photon production at fixed-target experiments is sensitive to the
behaviour of the gluon density at large $x$ $(x \gtap 0.1)$. The theoretical
predictions for this process, however, are not very accurate. 
Figure~\ref{figgamma5} shows the 
factorization- and renormalization-scale dependence of the perturbative cross
section for the case of the E706 kinematics. If the scale is varied by a factor
of 4 around the transverse energy $E_T$ of the prompt photon, the LO
cross section varies by a factor of almost 4. Going to NLO 
[\ref{promptgnlo}] the situation
improves, but not very much, because the NLO cross section still varies by 
a factor of about 2.

\begin{figure}
  \centerline{
    \setlength{\unitlength}{1cm}
    \begin{picture}(0,6)
       \put(0,0){\includegraphics{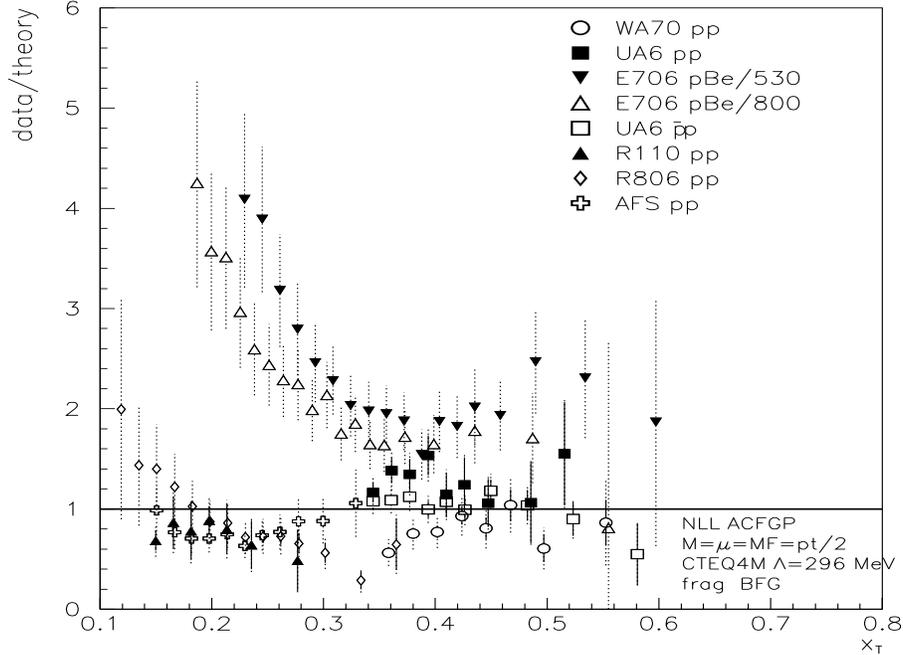}}
    \end{picture}}
\vspace*{2.5cm}    
\caption{\sf A comparison between NLO QCD calculations and data from the ISR and 
fixed-target experiments for the prompt-photon distribution $d\sigma/dE_T$ 
$(x_T=2E_T/{\sqrt S})$.
\label{figannecy}}
\end{figure}

A detailed comparison between NLO QCD calculations and data from the ISR and
fixed-target experiments has recently been performed in
Ref.~[\ref{annecygamma}]. As shown in Fig.~\ref{figannecy}, the overall
agreement with the theory is not satisfactory, even taking into account the
uncertainty coming from scale variations in the theoretical predictions.
Modifications of the gluon density can improve the agreement with some data
sets only at the expense of having larger disagreement with other data sets.
The differences between experiments at similar centre-of-mass energies
(see, for instance, E706 pBe/530 at ${\sqrt S}=31.6~{\rm GeV}$ and
WA70 pp at ${\sqrt S}=23~{\rm GeV})$ are much larger than expected from 
perturbative scaling violations. This can possibly suggest [\ref{annecygamma}]
inconsistencies of experimental origin. 

Another (not necessarily alternative)
origin of the differences between data and theory could be the presence of
non-perturbative effects that are not included in the NLO perturbative
calculation. This explanation has been put forward in 
Refs.~[\ref{E706}, \ref{ktcteq}] by introducing some amount of 
intrinsic\footnote{To be precise, in Ref.~[\ref{ktcteq}]
the $\langle k_{\perp} \rangle$ of the colliding partons is not called 
`intrinsic', but it is more
generically called the $\langle k_{\perp} \rangle$ `from initial-state
soft-gluon radiation'.} 
transverse momentum $\langle k_{\perp} \rangle$
of the colliding partons. Owing to the steeply
falling $E_T$ distribution $(d\sigma/dE_T \propto 1/E_T^7)$ of the 
prompt photon, even a small transverse-momentum 
kick\footnote{The $E_T$ distribution of the single-photon is not calculable
down to $E_T=0$ or, in other words, $d\sigma/dE_T$ is not integrable in the
entire kinematic range of $E_T$. Thus, the intrinsic $\langle k_{\perp} \rangle$
of the incoming partons does not simply produce a shift of events from
the low-$E_T$ to the high-$E_T$ region. For this reason, the terminology 
`$\langle k_{\perp} \rangle$ kick' seems to be more appropriate than
`$\langle k_{\perp} \rangle$ smearing'.}
can indeed produce
a large effect on the cross section, in particular, 
at small values of $E_T$. Phenomenological investigations [\ref{ktcteq}]
show that this additional $\langle k_{\perp} \rangle$ kick can lead to a
better agreement between calculations and data. The E706 data suggest the
value $\langle k_{\perp} \rangle \sim 1.2~{\rm GeV}$, the WA70 data prefer
no $\langle k_{\perp} \rangle$, and the UA6 data in the intermediate range of
centre-of-mass energy $({\sqrt S}=24.3~{\rm GeV}$) may prefer an intermediate
value of $\langle k_{\perp} \rangle$. Similar conclusions are obtained in the 
analysis by the MRST group [\ref{pdffit}].

A precise physical understanding of $\langle k_{\perp} \rangle$ 
effects is still missing. On one side,
since the amount of $\langle k_{\perp} \rangle$ 
suggested by prompt-photon data varies with ${\sqrt S}$, it is difficult to
argue that the transverse momentum is really `intrinsic' and has an
entirely non-perturbative origin. On the other side, in the case of the
inclusive production of a single photon, 
a similar effect cannot
be justified by higher-order {\em logarithmic}
corrections produced by perturbative
soft-gluon radiation (see Sect.~\ref{secsoftg}). A lot of model-dependent
assumptions (and ensuing uncertainties) certainly enter in the present 
implementations of the $\langle k_{\perp} \rangle$ kick. 
A general framework to {\em consistently} include non-perturbative 
transverse-momentum effects in perturbative calculations is not yet available. 
Recent proposals with this aim are presented in Refs.~[\ref{ktmartin}]
and [\ref{ktsterman}]. 

Further studies on the consistency between different
prompt-photon experiments and on the issue of
intrinsic-$\langle k_{\perp} \rangle$ effects in hadron--hadron collisions 
are necessary. Owing to the present theoretical
(and, possibly, experimental) uncertainties, it is difficult to use
prompt-photon data to accurately determine the gluon density at large $x$. 
Other recent theoretical improvements, such as soft-gluon resummation,
of the perturbative calculations for prompt-photon production at large 
$x_T=2E_T/{\sqrt S}$ are discussed in Sect.~\ref{secsoftg}. 

Studies of other single-particle inclusive cross sections, 
such as $\pi^0$ cross sections [\ref{ktcteq}, \ref{annecypi}, \ref{jh}], 
can be valuable to
constrain the parton densities and could possibly help to clarify some of
the experimental and theoretical issues arisen by prompt-photon production.

\section{Partonic cross sections: fixed-order expansions, \\
resummed calculations, parton showers}
\label{secpxs}

The calculation of hard-scattering cross sections according
to the factorization formula (\ref{factfor}) requires the knowledge of the 
partonic cross sections ${\hat \sigma}$, besides that of the parton densities.
The partonic cross sections are 
usually computed by truncating their perturbative expansion at a fixed order 
in $\as$:
\beeq
 \label{pertex}
\!\!\!\!\!\!
{\hat \sigma}(p_1,p_2;Q, \{Q_1, \dots \}; \mu_F^2) \!\!\!\!&=&\!\!\!\!
\as^k(\mu_R^2) \left\{
{\hat \sigma}^{(LO)}(p_1,p_2;Q, \{Q_1, \dots \}) \right.  \\
&~& \;\;\;\;\;\; \;\;\;\; + \,\as(\mu_R^2) \;
{\hat \sigma}^{(NLO)}(p_1,p_2;Q, \{Q_1, \dots \}; \mu_R^2; \mu_F^2) \nn \\
&~& \;\;\;\;\;\; \;\;\;\; +\left. \! \as^2(\mu_R^2) \;
{\hat \sigma}^{(NNLO)}(p_1,p_2;Q, \{Q_1, \dots \}; \mu_R^2; \mu_F^2) +
\dots \right\} \,.\nn
\eeeq
The scale $\mu_R$ is the arbitrary renormalization scale introduced to define
the perturbative expansion. Although the `exact' partonic cross section on 
the left-hand side of Eq.~(\ref{pertex}) does not depend on $\mu_R$, each term
on the right-hand side (and, hence, any fixed-order truncation) separately
depends on it.

The LO (or tree-level) term ${\hat \sigma}^{(LO)}$ gives only an estimate
of the order of magnitude of the partonic cross section, because
at this order $\as$ is not unambiguously defined. Equivalently, we can say
that since ${\hat \sigma}^{(LO)}$ does not 
depend on $\mu_R$, the size of its contribution can be varied quite 
arbitrarily by changing $\mu_R$ in its coefficient $\as^k(\mu_R^2)$. The strong
coupling $\as$ can be precisely defined only starting from NLO. A `reliable'
estimate of the central value of
${\hat \sigma}$ thus requires the knowledge of (at least)
the NLO term
${\hat \sigma}^{(NLO)}$. This term explicitly depends on $\mu_R$ and this
dependence begins to compensate that of $\as(\mu_R^2)$. 

In general, the
$n$-th term in the curly bracket of Eq.~(\ref{pertex}) contains contributions
of the type $( \as(\mu_R^2) \ln Q/\mu_R )^n$.
If $\mu_R$ is very different
from the hard scale $Q$, these contributions become large and spoil the 
reliability of the truncated expansion (\ref{pertex}).  
Thus, in practical applications the scale $\mu_R$ should be set approximately 
equal to the hard scale $Q$. As mentioned in Sect.~\ref{secgdensity}, 
variations of $\mu_R$ around this central value
are typically used to set a lower limit on the theoretical uncertainty
of the perturbative calculation.

A better estimate of the accuracy of any perturbative expansion is obtained
by considering the effect of removing the last perturbative term that has been
computed. Since $\as$ can be precisely defined only at NLO, this 
procedure can consistently be applied to Eq.~(\ref{pertex}) only as
from its NNLO term. A `reliable' estimate of the theoretical error on
${\hat \sigma}$ thus requires the knowledge of 
the NNLO term ${\hat \sigma}^{(NNLO)}$ in Eq.~(\ref{pertex}).

The LO and NLO approximations of ${\hat \sigma}$ are used at present
in (most of) the fixed-order QCD calculations. Prospects towards NNLO
calculations of partonic cross sections and AP splitting functions
are reviewed in Refs.~[\ref{proclhc}, \ref{delduca}].

The fixed-order expansion (\ref{pertex}) provides us with a well-defined and
systematic framework to compute the partonic cross section 
${\hat \sigma}(p_1,p_2;Q, \{Q_1, \dots \}; \mu_F^2)$ of any
hard-scattering process that is sufficiently inclusive or, more precisely,
that is defined in an infrared- and collinear-safe manner. However,
the fixed-order expansion is reliable only when all the kinematical scales
$Q, \{Q_1, \dots \}$ are of the same order of magnitude. 
When the hard-scattering process involves two (or several) very different 
scales, say $Q_1 \gg Q$, the ${\rm N}^n{\rm LO}$ term in Eq.~(\ref{pertex}) can 
contain double- and single-logarithmic contributions of the type
$(\as L^2)^n$ and $(\as L)^n$ with $L=\ln (Q_1/Q) \gg 1$. These terms spoil
the reliability of the fixed-order expansion and have to be summed to all orders
by systematically defining logarithmic expansions (resummed calculations).

Typical large logarithms, $L=\ln Q/Q_0$, are those related to the evolution of 
the parton densities from a low input scale $Q_0$ to the hard-scattering scale
$Q$. These logarithms are produced by collinear radiation from the colliding 
partons and give single-logarithmic contributions. They never explicitly appear
in the calculation of the partonic cross section, because they are 
systematically (LO, NLO and so forth) resummed in the evolved parton densities 
$f(x,Q^2)$ by using the DGLAP equation (\ref{evequa}).

Different large logarithms, $L=\ln Q/{\sqrt S}$, appear when the centre-of-mass
energy ${\sqrt S}$ of the collision is much larger than the hard scale $Q$. 
These small-$x$ $(x=Q/{\sqrt S})$ logarithms are produced by multiple radiation
over the wide rapidity range that is available at large energy. They usually
give single-logarithmic contributions that can be resummed by using the BFKL
equation. BFKL resummation is relevant
to DIS structure functions at small values of the Bjorken variable $x$ 
(see Sect.~\ref{secgdensity}) and it can also be important at the LHC for the 
production of $b$ quarks and of prompt photons at relatively low $E_T$.

Another class of large logarithms is associated to the bremsstrahlung
spectrum of soft gluons. Since soft gluons can be radiated collinearly,
they give rise to double-logarithmic contributions to the partonic cross
section:
\beq
\label{logxs}
{\hat \sigma} \sim \as^k {\hat \sigma}^{(LO)} 
\left\{ 1 + \sum_{n=1}^{\infty} \as^n \left( C_{2n}^{(n)} L^{2n} +  
C_{2n-1}^{(n)} L^{2n-1}
+ C_{2n-2}^{(n)} L^{2n-2} + \dots \right) \right\} \;\;.
\eeq
Soft-gluon resummation is discussed in Sect.~\ref{secsoftg}.

A related approach to evaluate higher-order contributions to the 
partonic cross sections is based on Monte Carlo parton showers 
(see [\ref{book}] and the updated list of references in 
[\ref{proctev}, \ref{proclhc}]). 
Rather than computing exactly ${\hat \sigma}^{(NLO)}$,
${\hat \sigma}^{(NNLO)}$ and so forth,
the parton shower gives an all-order approximation
of the partonic cross section in the soft and collinear regions.
In this respect, the computation of the partonic cross sections performed by
parton showers is somehow similar to that obtained by soft-gluon resummed 
calculations. There is, however, an important conceptual difference between
the two approaches. This difference and the limits of applicability of
the parton-shower method are briefly recalled below. Apart from these limits,
parton-shower calculations can give some advantages. Multiparton kinematics
can be treated exactly. The parton shower can be 
supplemented with models of non-perturbative effects
(hadronization, intrinsic $k_{\perp}$, soft underlying event)
to provide a complete description of the hard-scattering 
process at the hadron level. 

For a given cross section, resummed calculations can in 
principle be performed to any logarithmic accuracy. The logarithmic accuracy 
achievable by parton showers is instead intrinsically limited by quantum 
mechanics. The parton-shower algorithms are probabilistic.
Starting from the LO cross 
section, the parton shower generates multiparton final states according to
a probability distribution that approximates the {\em square} of the QCD 
matrix elements.
The approximation is based on 
the universal (process-independent) factorization properties of multiparton 
matrix elements in the soft and collinear limits. Although the matrix element
does factorize, its square contains quantum interferences, which are not 
positive-definite and, in general, cannot be used to define probability 
distributions. To leading infrared accuracy,
this problem is overcome by exploiting QCD coherence (see
Refs.~[\ref{book}, \ref{Dokbook}, \ref{BCM}] and referencees therein):
soft gluons radiated at
large angle from the partons involved in the LO subprocess destructively
interfere. This quantum mechanical effect can be simply implemented
by enforcing an angular-ordering constraint on the phase space available
for the parton shower evolution. Thus, angular-ordered parton showers can
{\em consistently} compute the first two dominant towers ($\as^n L^{2n}$ and 
$\as^n L^{2n-1}$) of logarithmic contributions in Eq.~(\ref{logxs}).
However, parton showers contain also some subleading logarithmic contributions.
For instance, they correctly compute the single-logarithmic terms $\as^n L^n$ 
of purely collinear origin that lead to the LO evolution of the parton 
densities. 
Moreover, as discussed in Ref.~[\ref{CMW}] by a comparison with 
resummed calculations,
in the case of hard-scattering processes whose LO subprocess involves 
two coloured partons (e.g. DIS or DY production), angular-ordered parton showers
have a higher logarithmic accuracy: they can
consistently evaluate the LL and NLL terms in Eq.~(\ref{resff}).
The extension of parton-shower algorithms to higher logarithmic 
accuracy is not necessarily feasible and is, in any case, challenging. 

Of course, because of
quantum interferences and quantum fluctuations, the probabilistic parton-shower
approach cannot be used to systematically perform exact calculations at NLO,
NNLO and so forth. Nonetheless, important progress has been made to include
matrix element corrections in parton shower algorithms 
[\ref{mecorsey}--\ref{mecorcol}]. The purpose is to consider the multiparton
configurations generated by parton showering from the LO matrix element and to
correct them in the hard (non-soft and non-collinear) region
by using the exact expressions of the higher-order matrix elements.
Hard matrix element corrections to parton showers have been implemented
for top quark decay [\ref{seytop}] and for production of $W,Z$ and 
DY lepton pairs [\ref{sjow}, \ref{mrenna}, \ref{corseyw}]. The same techniques
could be applied to other processes, as, for instance, production of Higgs 
boson [\ref{resvsps}] and vector-boson pairs [\ref{proclhc}].

Note also that, at present, angular-ordered parton showers 
cannot be considered as true `next-to-leading' tools, even where their
logarithmic accuracy is concerned. 
The consistent computation of the first two towers of
logarithmic contributions in Eq.~(\ref{logxs}) is not sufficient for this
purpose. For instance, to precisely introduce an NLO definition of
$\as$, we should control all the terms obtained by the replacement $\as \to
\as + c \;\as^2 + {\cal O}(\as^3)$. When it is introduced in the towers
of double-logarithmic terms $\as^n L^{2n}$ of Eq.~(\ref{logxs}), 
this replacement leads to contributions of the type $\as^{n+1} L^{2n}
\sim \as^{n} L^{2n-2}$. Since these contributions are not
fully computable at present, the parameter $\as$ used in the parton showers
corresponds to a simple LO parametrization of QCD running 
coupling.

\section{Soft-gluon resummation}
\label{secsoftg}

Double-logarithmic contributions due to soft gluons arise in all the 
kinematic configurations where radiation of real and virtual partons
is highly unbalanced (see Ref.~[\ref{softrev}] and references therein).
For instance, this happens in the case of transverse-momentum distributions
at low transverse momentum, in the case of hard-scattering production near
threshold or when the structure of the final state
is investigated with high resolution (internal jet structure, shape variables).

Soft-gluon resummation for jet shapes has been extensively
studied and applied to hadronic final states produced by $e^+e^-$ annihilation
[\ref{eps99}, \ref{lp97}, \ref{jetres}]. Applications to hadron--hadron
collisions have just begun to appear [\ref{seymour}]
and have a large, 
yet uncovered, potential (from $\as$ determinations to studies of
non-perturbative dynamics).

Transverse-momentum logarithms, $L = \ln Q^2/{\bom Q}_{\perp}^2$,
occur in the distribution of transverse momentum ${\bom Q}_{\perp}$
of systems with high mass $Q$ $(Q \gg Q_{\perp})$ that are produced with
a vanishing ${\bom Q}_{\perp}$ in the LO subprocess. Examples of such systems
are DY lepton pairs, lepton pairs produced by $W$ and $Z$ decay, heavy 
quark--antiquark pairs, photon pairs and Higgs bosons. In these processes the LO 
transverse-momentum distribution is sharply peaked around ${\bom Q}_{\perp}=0$
($d {\hat \sigma}/d^2{\bom Q}_{\perp} \propto \delta^{(2)}({\bom Q}_{\perp}$)).
If the heavy system is produced with ${\bom Q}_{\perp}^2 \ll Q^2$, the emission
of real radiation at higher orders is strongly suppressed and cannot balance 
the virtual contributions. The ensuing logarithms, 
$L = \ln Q^2/{\bom Q}_{\perp}^2$, diverge order by order when 
${\bom Q}_{\perp} \to 0$, but after all-order resummation they leads to a 
finite smearing of the LO distribution.

Threshold logarithms, $L = \ln (1-x)$, occur when the tagged final state
produced by the hard scattering is forced to carry a very large fraction
$x$ ($x \to 1$) of the available centre-of-mass energy $\sqrt S$. Also in
this case, the radiative tail of real emission is stronly suppressed
at higher perturbative orders. Oustanding examples of hard processes near
threshold are DIS at large $x$ (here $x$ is the Bjorken variable), production
of DY lepton pairs with large invariant mass $Q$ ($x = Q/{\sqrt S}$), 
production of heavy quark--antiquark pairs ($x = 2m_Q/{\sqrt S}$), production
of single jets and single photons at large transverse energy $E_T$ 
($x = 2E_T/{\sqrt S}$).

To emphasize the difference between transverse-momentum logarithms and
threshold logarithms generated by soft gluons, it can be instructive to 
consider prompt-photon production. In the case of production of a {\em
photon pair}\footnote{The same discussion applies to the production of a DY
lepton pair.} with invariant mass squared 
$Q^2= (p^{(\gamma)}_{1}+p^{(\gamma)}_{2})^2$
and {\em total} transverse momentum 
${\bom Q}_{\perp}= {\bom p}^{(\gamma)}_{1 \perp}+{\bom p}^{(\gamma)}_{2 \perp}$,
transverse-momentum logarithms and threshold logarithms
appear when ${\bom Q}_{\perp}^2 \ll Q^2$ and ${\bom Q}_{\perp}^2 \sim
(S/4 - Q^2)$, respectively. However, in the case of production of a {\em
single photon} with transverse energy (or, equivalently, transverse momentum)
$E_T$, soft gluons can produce logarithms
only in the threshold  region $x_T = 2E_T/{\sqrt S} \to 1$. If the prompt
photon has a transverse energy that is not close\footnote{Eventually,
when $x_T \ll 1$, higher-order corrections are single-logarithmically
enhanced. This small-$x$ logarithms, $(\as \ln x_T)^n$, have to be taken into
account by BFKL resummation.}
to its threshold value, 
the emission of accompanying radiation is not kinematically suppressed
and there are no soft logarithms analogous to those in the
transverse-momentum distribution of a photon pair. In particular, 
there are no double-logarithmic contributions of the type 
$(\as \ln^2 E_T^2/S)^n$, and perturbative soft gluons are not distinguishable
from perturbative hard gluons. 

Studies of soft-gluon resummation for transverse-momentum distributions
at low transverse momentum and for hard-scattering production near
threshold started two decades ago [\ref{BCM}, \ref{DDT}]. The physical bases
for a systematic all-order summation of the soft-gluon contributions are
dynamics and kinematics factorizations [\ref{softrev}]. The first factorization
follows from gauge invariance and unitarity: in the soft limit, multigluon
amplitudes fulfil factorization formulae given in terms of universal
(process-independent) soft contributions. The second factorization regards 
kinematics and strongly depends on the actual cross section to be evaluated.
{\em If}, in the appropriate soft limit, the multiparton phase space for
this cross section can be written in a factorized way, resummation
is analytically feasible in form of {\em generalized exponentiation}
of the universal soft contributions that appear in the factorization formulae
of the QCD amplitudes.

Note that the phase space depends in a 
non-trivial way on multigluon configurations and, in general, is not
factorizable in single-particle contributions\footnote{In the case of jet
cross sections, for instance, phase-space factorization depends on the
detailed definition of jets and it can easily be violated [\ref{BS}]. Some
jet algorithms, such as the $k_{\perp}$-algorithm 
[\ref{duralg}, \ref{ktalg}], have better
factorization properties.}.
Moreover, even when phase-space factorization is
achievable, it does not always occur in the space of the kinematic variables
where the cross section is defined. Usually, it is necessary to
introduce a conjugate space to overcome phase-space constraints. This is the
case for transverse-momentum distributions and hard-scattering production near
threshold. The relevant kinematical constraint for 
${\bom Q}_{\perp}$-distributions is (two-dimensional) 
transverse-momentum conservation and it
can be factorized by performing a Fourier transformation. Soft-gluon 
resummation for ${\bom Q}_{\perp}$-distributions is thus carried out
in ${\bom b}$-space [\ref{pp}, \ref{CSS}], where the impact parameter 
${\bom b}$ is the variable conjugate to ${\bom Q}_{\perp}$ via the Fourier
transformation. Analogously, the relevant kinematical constraint for
hard-scattering production near threshold is (one-dimensional) energy
conservation and it can be factorized by working in $N$-moment space 
[\ref{S}, \ref{CT}], 
$N$ being the variable conjugate to the threshold variable $x$ 
(energy fraction) via a Mellin or Laplace transformation.

Using a short-hand notation, 
the general structure of the partonic cross section ${\hat \sigma}$
after summation of soft-gluon contributions is
\beq
\label{resmatch}
{\hat \sigma} = {\hat \sigma}_{\rm res.} + {\hat \sigma}_{\rm rem.} \;.
\eeq
The term ${\hat \sigma}_{\rm res.}$ embodies the all-order resummation,
while the remainder ${\hat \sigma}_{\rm rem.}$ contains no large logarithmic
contributions. The latter has the form
\beq
{\hat \sigma}_{\rm rem.} = {\hat \sigma}^{({\rm f.o.})} -
\left[ \,{\hat \sigma}_{\rm res.} \,\right]^{({\rm f.o.})} \;\;,
\eeq
and it is obtained from ${\hat \sigma}^{({\rm f.o.})}$,
the truncation of the perturbative expansion for ${\hat \sigma}$ at a given
fixed order (LO, NLO, ...), by subtracting the corresponding truncation
$\left[ {\hat \sigma}_{\rm res.}\right]^{({\rm f.o.})}$ of the resummed part.
Thus, the expression on the right-hand side of Eq.~(\ref{resmatch}) includes
soft-gluon logarithms to all orders and it is {\it matched} to the exact (with
no logarithmic approximation) fixed-order calculation. It represents an
improved perturbative calculation that is everywhere as good as the fixed-order
result, and much better in the kinematics regions where the soft-gluon
logarithms become large ($\as L \sim 1$). Eventually, when $\as L \gtap 1$, the
resummed perturbative contributions are of the same size as the 
non-perturbative contributions and the effect of the latter has to be 
implemented in the resummed calculation.

The resummed cross section has the following typical form:
\beq
{\hat \sigma}_{\rm res.} = 
\as^k \int_{\rm inv.} {\hat \sigma}^{(LO)} \cdot C \cdot S \;,
\eeq
where the integral $\int_{\rm inv.}$ denotes the inverse tranformation from
the conjugate space where resummation is actually carried out.
Methods to perform the inverse transformation are discussed in 
Refs.~[\ref{qtback}] and [\ref{thback}] for $Q_{\perp}$-resummation and
threshold resummation, respectively. The $C$ term has the perturbative
expansion
\beq
C = 
1 + C_1 \,\as + C_2 \,\as^2 + \dots \,
\eeq
and contains all the constant contributions in the limit $L \to \infty$
(the coefficients $C_1, C_2, \dots$ do not depend on the conjugate variable).
The singular dependence on $L$ (more precisely, on the logarithm ${\tilde L}$
of the conjugate variable) is entirely {\it exponentiated} in the
factor $S$: 
\beq
\label{resff}
S = \exp \left\{ L \;g_1(\as L) + g_2(\as L) +
\as \;g_3(\as L) + \dots \right\} \;\;.
\eeq
In the exponent, the function $L \,g_1$
resums all the leading logarithmic (LL) contributions
$\as^n L^{n+1}$, while $g_2$
contains the next-to-leading logarithmic (NLL) terms $\as^n L^n$ and so 
forth\footnote{To compare this notation with that of Ref.~[\ref{qthere}], we 
can notice that our functions $g_i$ are obtained by the straightforward
integration over ${\overline \mu}$ of the functions $A(\as({\overline \mu}))$ 
and $B(\as({\overline \mu}))$ of Ref.~[\ref{qthere}]. In particular, our
terms $g_1, g_2, g_3$ are not to be confused with the non-perturbative
parameters of the same name used in Ref.~[\ref{qthere}].}
(all the functions $g_i$ are normalized as $g_i(\lambda=0) = 0$).
Note that the LL terms are formally suppressed by a power of $\as$ with respect
to the NLL terms, and so forth for the successive classes of logarithmic terms. 
Thus, this logarithmic expansion is as systematic
as the fixed-order expansion in Eq.~(\ref{pertex}).
In particular, using a matched NLL+NLO calculation, we can 
consistently $i)$ introduce a precise definition (say ${\overline {\rm MS}}$)
of $\as(\mu)$ and $ii)$ investigate the theoretical accuracy of the calculation
by studying its renormalization-scale dependence. 

The structure of the exponentiated resummed calculations discussed so far
has to be contrasted with that obtained by organizing the logarithmic expansion
on the right-hand side of Eq.~(\ref{logxs}) in terms of towers as 
\beq
\label{restowerex}
{\hat \sigma} \sim \as^k {\hat \sigma}^{(LO)} 
\left\{ t_1(\as L^2) + \as L \; t_2(\as L^2) + \as^2 L^2 \; t_3(\as L^2)
+ \dots  \right\} \;\;,
\eeq
where the double-logarithmic function $t_1(\as L^2)$ and the successive
functions are normalized as $t_i(0)= {\rm const.}$ While the ratio of two
successive terms in the exponent of Eq.~(\ref{resff}) is formally of the order
of $\as$, the ratio of two successive towers in Eq.~(\ref{restowerex}) is 
formally of the order of $\as L$. In other words, the tower expansion sums the
double-logarithmic terms $(\as L^2)^n$, then the terms $\as^n L^{2n-1} \sim 
\as L (\as L^2)^{n-1}$, and so forth; it thus 
assumes that the resummation procedure is carried out with respect to the 
large parameter $\as L^2$ ($\as L^2 \ltap 1$). On the contrary, in 
Eq.~(\ref{resff}) the large parameter is $\as L \ltap 1$. The tower expansion
allows us to formally extend the applicability of perturbative QCD to the
region $L \ltap 1/{\sqrt \as}$, and exponentiation extends it to the wider
region $L \ltap 1/\as$. This fact can also be argued by comparing the
amount of information on the logarithmic terms that is included in the
truncation of Eqs.~(\ref{resff}) and (\ref{restowerex}) at some logarithmic
accuracy. The reader can easily check that, after matching to the NLO (LO)
calculation as in Eq.~(\ref{resmatch}), the NLL (LL) result of Eq.~(\ref{resff})
contains all the logarithms of the first {\em four} ({\em two}) towers in
Eq.~(\ref{restowerex}) (and many more logarithmic terms).

In the case of $Q_{\perp}$-distributions, full NLL resummation has been 
performed for lepton pairs, $W$ and $Z$ bosons produced by the DY mechanism
[\ref{CSS}, \ref{dyqt}] and for Higgs bosons produced by gluon fusion
[\ref{Hqt}]. Corresponding resummed calculations are discussed in 
Refs.~[\ref{resvsps}, \ref{qthere}] and references therein.

Threshold logarithms in hadron collisions have been resummed to NLL
accuracy for DIS and DY production [\ref{S}, \ref{CT}, \ref{CMW}, \ref{CLS}]
and for Higgs boson production [\ref{Hth}].
Recent theoretical progress [\ref{KS}, \ref{KOS}, \ref{txs}]
regards the extension of NLL resummation
to processes produced by LO hard-scattering of more than two coloured partons,
such as heavy-quark hadroproduction [\ref{KS}, \ref{txs}] and leptoproduction
[\ref{LM}], as well as prompt-photon [\ref{LOS}--\ref{kow}],
quarkonium [\ref{cacciari}] and vector-boson [\ref{kdd}] production.

An important feature of threshold resummation
is that the resummed soft-gluon contributions regard the partonic cross
section rather than the hadronic cross section. This fact has two main
consequences: $i)$ soft-gluon contributions can be sizeable long before the 
threshold region in the hadronic cross section is actually approached, and 
$ii)$ the resummation effects typically enhance the fixed-order perturbative
calculations. 

The first consequence follows from the fact that the evolution of the parton
densities sizeably reduces the energy that is available in the 
partonic hard-scattering subprocess. Thus, the partonic cross section 
$\hat \sigma$ in the factorization formula
(\ref{factfor}) is typically evaluated much closer to threshold than the 
hadronic cross section. In other words, the
parton densities are strongly suppressed at large $x$ (typically, 
when $x\to 1$, $f(x,\mu^2) \sim (1-x)^\eta$ with $\eta \sim 3$ and 
$\eta \sim 6$ for valence quarks and sea-quarks or gluons, respectively); 
after integration over them, the dominant value of the square of the
partonic centre-of-mass energy $\langle {\hat s} \rangle = 
\langle x_1 x_2 \rangle S$ is therefore
substantially smaller than the corresponding hadronic value $S$.

The second consequence, which depends on the actual definition of the parton 
densities,
follows from the fact that the resummed contributions
are those soft-gluon effects that are left at the partonic level after 
factorization of the parton densities. After having absorbed part of the
full soft-gluon contributions in the customary definitions (for instance, those
in the ${\overline {\rm MS}}$ or DIS factorization schemes) of the parton 
densities, it turns out that the residual effect in the partonic cross section 
is positive and tends to enhance the perturbative predictions.

A quantitative illustration of these consequences is given below by discussing
top-quark and prompt-photon production. The discussion also shows another
relevant feature of NLO+NLL calculations, namely, their increased stability
with respect to scale variations.

\begin{figure}
  \centerline{
    \setlength{\unitlength}{1cm}
    \begin{picture}(0,6)
       \put(0,0){\includegraphics{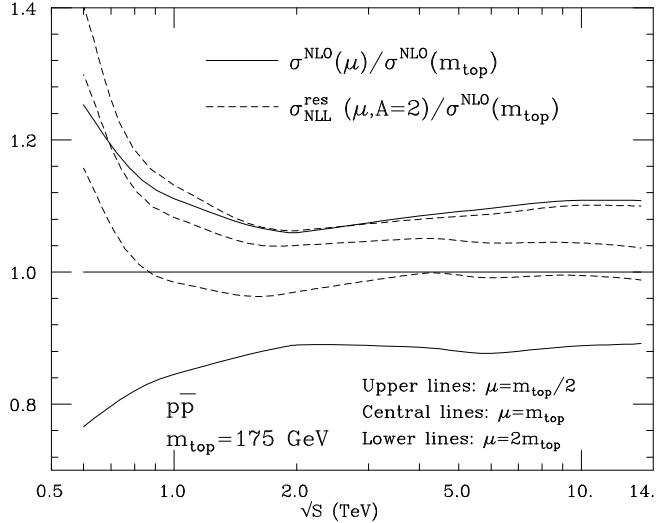}}
    \end{picture}}
\vspace*{1.3cm}    
\caption{\sf The $t{\bar t}$ production cross section in $p{\bar p}$ collisions as a
function of ${\sqrt S}$. The solid lines represent the NLO results for different
choices ($\mu=m_t/2$ and $\mu=2m_t$) of the renormalization/factorization 
scale $\mu=\mu_R=\mu_F$, normalized to the result with $\mu=m_t$. The dashed
lines represent the NLO+NLL results for different choices of $\mu$
($\mu=m_t/2, m_t$ and $2m_t$), normalized to the NLO result with 
$\mu=m_t$.
\label{figtop}}
\end{figure}

The effects of soft-gluon resummation on the top-quark production cross sections
at hadron colliders have been studied in Refs.~[\ref{txs}, 
\ref{tqll1}--\ref{tqll4}]. In the case of $p{\bar p}$ collisions, the 
comparison between QCD predictions at NLO and those after NLL resummation is shown in
Fig.~\ref{figtop} [\ref{txs}]. 
At the Tevatron the resummation effects are not very large 
and the NLO cross section is increased by about $4\%$. This had to be expected
because the top quark is not produced very close
to threshold ($x = 2m_t/{\sqrt S} \sim 0.2$, at the Tevatron). Note, however,
that the dependence on the factorization/renormalization scale of the
theoretical cross section is reduced by a factor of  almost 2 by including NLL
resummation. More precisely, the scale dependence $(\sim \pm 5\%)$
of the NLO+NLL calculation becomes comparable to that obtained by using 
different sets of parton densities [\ref{pdffit}]. Combining linearly scale
and parton density uncertainties, the NLO+NLL cross section is
$\sigma_{t{\bar t}}=5.0 \pm 0.6$, with $m_t=175~{\rm GeV}$ and 
${\sqrt S}=1.8~{\rm TeV}$ [\ref{txs}]. 

At the LHC $(x = 2m_t/{\sqrt S} \sim 0.03)$ the top quark is produced less close
to the hadronic threshold than at the Tevatron. However this is compensated
by the fact that the gluon channel\footnote{Since
$f_g$ is steeper than $f_q$ at large $x$, partonic cross sections in gluon
subprocesses are typically closer to threshold than in quark subprocesses.
Moreover, the intensity of soft-gluon radiation from gluons is larger than 
that from
quarks by a factor of $\sim C_A/C_F \sim 2$.} is more important at the LHC.
As a result, the effect of including
soft-gluon resummation to NLL accuracy is very similar: 
the NLO cross section is enhanced by $\sim 5\%$ and its scale dependence
is reduced from $\sim \pm 10\%$ to $\sim \pm 5\%$.
Note, however, that the uncertainty $(\sim \pm 10\%)$ coming from the parton
(gluon) densities is larger than at the Tevatron [\ref{proclhc}]. 

\begin{figure}
  \centerline{
    \setlength{\unitlength}{1cm}
    \begin{picture}(0,6)
       \put(0,0){\includegraphics{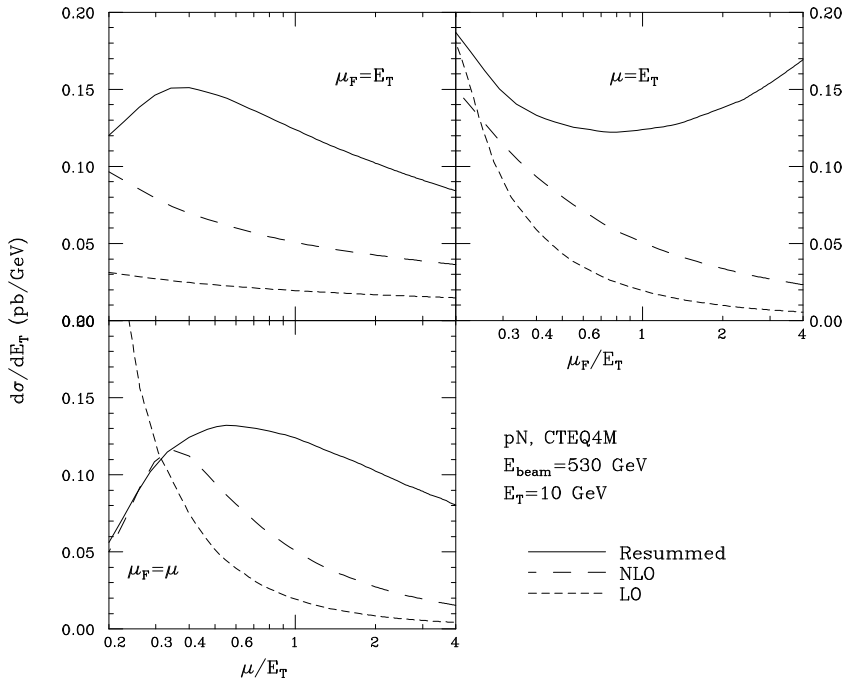}}
    \end{picture}}
\vspace*{3.0cm}    
\caption{\sf The dependence on the factorization $(\mu_F)$ and renormalization
$(\mu_R=\mu)$ scale of the prompt-photon cross section 
$d\sigma/dE_T$ in $pN$ collisions at $E_T=10$~GeV and ${\sqrt S}=31.6$~GeV. 
The short-dashed, long-dashed and solid lines are respectively the results at
LO, NLO and after NLO+NLL resummation. 
\label{figgamma10}}
\end{figure}

\begin{figure}
  \centerline{
    \setlength{\unitlength}{1cm}
    \begin{picture}(0,6)
       \put(0,0){\includegraphics{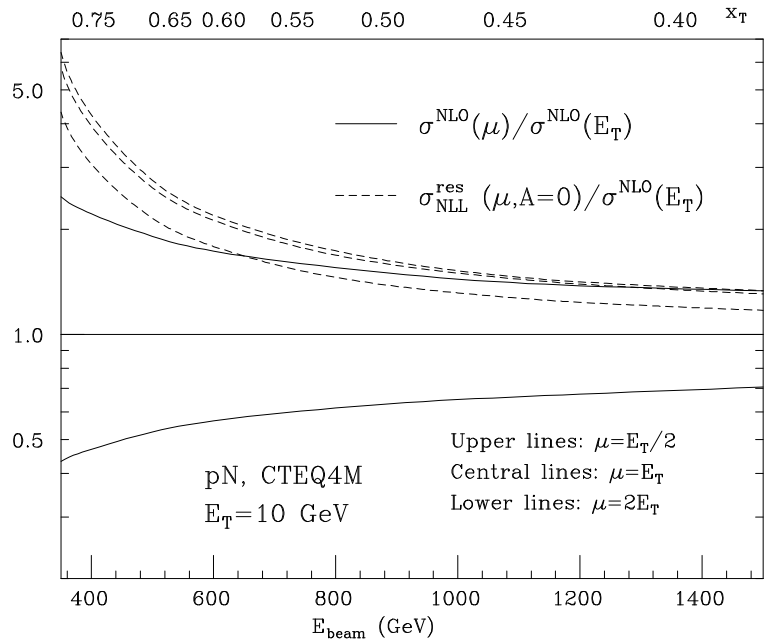}}
    \end{picture}}
\vspace*{2.5cm}    
\caption{\sf The prompt-photon cross section $d\sigma/dE_T$ in $pN$ collisions 
at $E_T=10$~GeV as a function of the energy $E_{{\rm beam}}$
of the proton beam. The solid lines represent the NLO results for 
different choices ($\mu=E_T/2$ and $\mu=2E_T$) of the 
renormalization/factorization 
scale $\mu=\mu_R=\mu_F$, normalized to the result with $\mu=E_T$. The dashed
lines represent the NLO+NLL results for different choices of $\mu$
($\mu=E_T/2, E_T$ and $2E_T$), normalized to the NLO result with $\mu=E_T$.
\label{figgammabeam}}
\end{figure}

Similar qualitative results are obtained [\ref{CMNOV}] when NLL resummation
is applied to prompt-photon production at fixed-target experiments. The scale
dependence of the theoretical calculation is highly reduced and the resummed
NLL contributions lead to large corrections at high $x_T=2E_T/{\sqrt S}$
(and smaller corrections at lower $x_T$). Of course, the impact of
soft-gluon resummation is quantitatively more sizeable in 
prompt-photon production than in top-quark production, because $x_T$
can be as large as 0.6, the hard scale $E_T$ is much smaller than $m_t$
(thus, $\as(E_T) > \as(m_t)$) and the gluon channel is always important. The
scale dependence of the theoretical cross section for the E706 kinematics
is shown in Fig.~\ref{figgamma10}. Fixing $\mu_R=\mu_F=\mu$ and varying $\mu$
in the range $E_T/2 < \mu < 2E_T$ with $E_T=10~{\rm GeV}$, the cross section
varies by a factor of $\sim 6$ at LO, by a factor of $\sim 4$ at NLO and
by a factor of $\sim 1.3$ after NLL resummation. The highly reduced scale
dependence of the NLO+NLL cross section is also visible in 
Fig.~\ref{figgammabeam}, which, in particular, shows that when 
$E_T=10~{\rm GeV}$ and $E_{\rm beam}=530~{\rm GeV}$ the central value (i.e. with
$\mu=E_T$) of the NLO cross section increases by a factor of $\sim 2.5$ after
NLL resummation. As expected, the size of these effects is reduced by increasing
${\sqrt S}$ at fixed $E_T$ (see Fig.~\ref{figgammabeam}) or by decreasing 
$E_T$ at fixed ${\sqrt S}$ (see Fig.~\ref{figgamma5}).

\begin{figure}
  \centerline{
    \setlength{\unitlength}{1cm}
    \begin{picture}(0,6)
       \put(0,0){\includegraphics{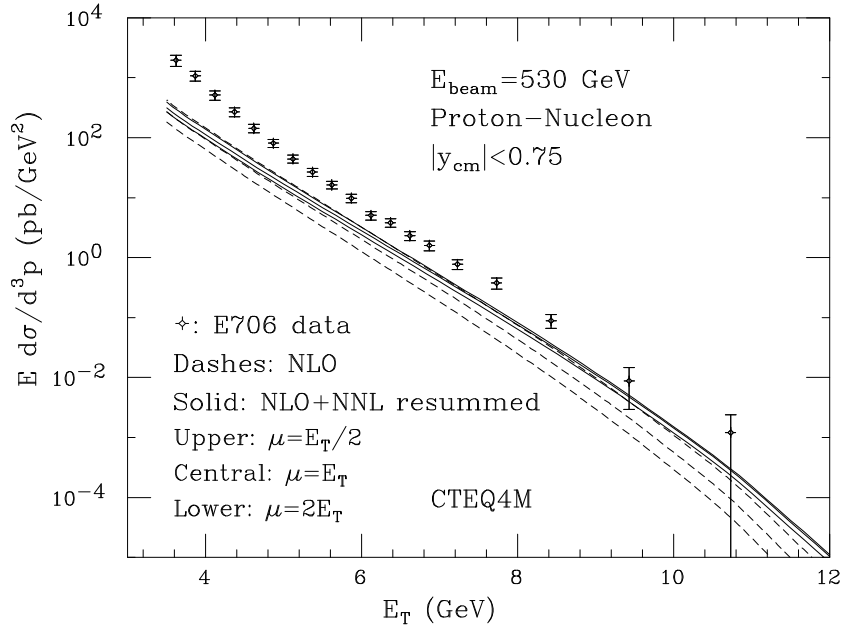}}
    \end{picture}}
\vspace*{2.7cm}    
\caption{\sf E706 prompt-photon data compared with theoretical calculations, 
which use the parton
densities of the set CTEQ4 and GRV photon fragmentation functions.
The solid and dashed lines correspond to the NLO+NLL and pure NLO calculations,
respectively.
\label{fige706vsres}}
\end{figure}

The comparison with the E706 data shown in Fig.~\ref{fige706vsres} suggests
that the NLO+NLL calculation can help to
better understand prompt-photon production at large $x_T$. Note, however, that
this comparison has to be regarded as preliminary in several respects 
[\ref{CMNOV}]. In particular, the parton densities used in 
Fig.~\ref{fige706vsres} are those extracted from NLO fits. Owing to the
soft-gluon enhancement at large $x_T$, refitting the parton densities may lead
to a smaller $f_g$ at large $x$ and, consequently (because of the momentum sum
rule), a larger $f_g$  at intermediate $x$. As a result, this procedure could
somehow increase the theoretical cross section also at smaller values of $x_T$.

Soft-gluon resummation at NLL accuracy is now available for all the processes
(namely, DIS, 
DY 
and prompt-photon 
production)
that are typically used to perform global fits to parton densities.
A detailed extraction/evolution of parton densities by consistently using
NLL resummed calculations is thus nowadays feasible.

\section{Other topics}

The activity of the QCD Working Group at this Workshop has also been devoted 
to other topics, such as
automatic computation of matrix elements and LO cross sections for
multiparticle processes at high-energy colliders, 
definition and properties of jets algorithms, 
definition of isolated photons and related NLO calculations. 
Corresponding contributions are included in these Proceedings. 

Other studies performed during this Workshop have a large overlap
with the activity of the related Workshops at FERMILAB and CERN 
and can be found in those Proceedings~[\ref{proctev},~\ref{proclhc}].

\noindent {\bf Acknowledgements}. This work was supported in part 
by the EU Fourth Framework Programme ``Training and Mobility of Researchers'', 
Network ``Quantum Chromodynamics and the Deep Structure of
Elementary Particles'', contract FMRX--CT98--0194 (DG 12 -- MIHT).
I would like to thank the members of the
Local Organizing Committee for the excellent Workshop. 

\section*{References}

\def\ac#1#2#3{Acta Phys.\ Polon.\ #1 (19#3) #2}
\def\ap#1#2#3{Ann.\ Phys.\ (NY) #1 (19#3) #2}
\def\ar#1#2#3{Annu.\ Rev.\ Nucl.\ Part.\ Sci.\ #1 (19#3) #2}
\def\cpc#1#2#3{Computer Phys.\ Comm.\ #1 (19#3) #2}
\def\ib#1#2#3{ibid.\ #1 (19#3) #2}
\def\np#1#2#3{Nucl.\ Phys.\ B#1 (19#3) #2}
\def\pl#1#2#3{Phys.\ Lett.\ #1B (19#3) #2}
\def\pr#1#2#3{Phys.\ Rev.\ D #1 (19#3) #2}
\def\prep#1#2#3{Phys.\ Rep.\ #1 (19#3) #2}
\def\prl#1#2#3{Phys.\ Rev.\ Lett.\ #1 (19#3) #2}
\def\rmp#1#2#3{Rev.\ Mod.\ Phys.\ #1 (19#3) #2}
\def\sj#1#2#3{Sov.\ J.\ Nucl.\ Phys.\ #1 (19#3) #2}
\def\zp#1#2#3{Z.\ Phys.\ C#1 (19#3) #2}
\def\epj#1#2#3{Eur.\ Phys.\ J. C#1 (19#3) #2}
\def\jhep#1#2#3{JHEP #1 (19#3) #2}

\begin{enumerate}

\item \label{eps99}
M.L.\ Mangano, plenary talk at the {\it 1999 International Europhysics 
Conference on High Energy Physics} Tampere, Finland, 15--21 July 1999, preprint
CERN-TH/99-337 (hep-ph/9911256), to appear in the Proceedings, 
and references therein.

\item \label{lp99webber}
B.R.\ Webber, plenary talk at the {\it 19th International Symposium on 
Lepton-Photon Interactions at High-Energies, LP 99}, Stanford,
California, 9--14 Aug 1999, preprint CERN-TH/99-387
(hep-ph/9912292), to appear in the Proceedings, 
and references therein. 

\item \label{ichep98}
J.\ Huston, in Proc. of the {\it 29th International Conference on High-Energy
Physics, ICHEP 98}, eds. A.~Astbury, D.~Axen and J. Robinson
(World Scientific, Singapore, 1999), vol.~1, p.~283, and references therein.

\item \label{lp97}
S.\ Catani, hep-ph/9712442, in Proc. of the {\it XVIII International Symposium
on Lepton-Photon Interactions, LP97}, eds. A.\ De Roeck and A.\ Wagner
(World Scientific, Singapore, 1998), p.~147,
and references therein.

\item \label{proceeding}
G.\ Jarlskog and D.\ Rein eds., Proc. of ECFA LHC Workshop, report CERN 90-10
(December 1990);
ATLAS Coll., ATLAS TDR 15, report CERN/LHCC/99-15 (May 1999).

\item \label{proctev}
Proceedings of the Workshop on {\it Physics at the Tevatron in Run II}, Fermilab,
2000 (to appear).
See: http://www-theory.fnal.gov/people/ellis/QCDWB/QCDWB.html

\item \label{proclhc}
Proceedings of the Workshop on {\it Standard Model Physics (and more) at the LHC},
CERN 1999 (to appear). 
See: http://home.cern.ch/\~{}mlm/lhc99/lhcworkshop.html

\item \label{factform}
See J.C.\ Collins, D.E.\ Soper and G.\ Sterman, in {\it Perturbative Quantum
Chromodynamics}, ed. A.H.\ Mueller (World Scientific, Singapore, 1989), p.~1,
and references therein.

\item \label{DGLAP}
V.N.\ Gribov and L.N.\ Lipatov, Sov. J. Nucl. Phys. 15 (1972) 438, 
675; G.\ Altarelli and G.\ Parisi,
\np{126}{298}{77}; Yu.L.\ Dokshitzer, Sov. Phys. JETP  46 (1977) 641.

\item \label{NLOAP}
E.~G.~Floratos, D.~A.~Ross and C.~T.~Sachrajda,
Nucl.\ Phys.\  B129 (1977) 66, E ibid. B139 (1978) 545,
Nucl.\ Phys.\  B152 (1979) 493;
A.~Gonzalez-Arroyo, C.~Lopez and F.~J.~Yndurain,
Nucl.\ Phys.\  B153 (1979) 161;
A.~Gonzalez-Arroyo and C.~Lopez, Nucl.\ Phys.\  B166 (1980) 429;
G.~Curci, W.~Furmanski and R.~Petronzio, Nucl.\ Phys.\  B175 (1980) 27;
W.~Furmanski and R.~Petronzio, Phys.\ Lett.\  B97 (1980) 437;
E.~G.~Floratos, C.~Kounnas and R.~Lacaze, Nucl.\ Phys.\  B192 (1981) 417.
 
\item \label{pdffit}
A.D.\ Martin, R.G.\ Roberts, W.J.\ Stirling and R.S. Thorne, \epj{4}{463}{98},
\pl{443}{301}{98},
preprint DTP-99-64 (hep-ph/9907231); 
M.\ Gl\"uck, E.\ Reya and A. Vogt, \epj{5}{461}{98}; 
H.L. Lai et al.,
Eur.\ Phys.\ J. C12 (2000) 375.

\item \label{lhcpdf}
LHC Guide to Parton Distribution Functions and Cross Sections, ATLAS note
ATL-PHYS-99-008, http://www.pa.msu.edu/\~{}huston/lhc/lhc$_{-}$pdfnote.ps .

\item \label{pdfrball}
R.D.~Ball and J. Huston, in these Proceedings, and references therein.

\item \label{bodek}
U.K.~Yang and A.~Bodek, \prl{82}{2467}{99},
preprint UR-1581 (hep-ex/9908058).

\item \label{Melnitchouk}
W.~Melnitchouk, I.R.~Afnan, F.~Bissey and A.W.~Thomas,
preprint ADP-99-47-T384 (hep-ex/9912001). 

\item \label{highxcteq}
S. Kuhlmann et al., hep-ph/9912283.

\item \label{gluoncteq}
J. Huston et al., \pr{58}{114034}{98}. 

\item \label{NNLODY}
R.\ Hamberg, W.L.\ van Neerven and T.\ Matsuura, \np{359}{343}{91};
W.L.~van Neerven and E.B.\ Zijlstra, \np{382}{11}{92}.

\item \label{txs}
R.\ Bonciani, S.\ Catani, M.L.\ Mangano and P.\ Nason, \np{529}{424}{98}.

\item \label{BFKL}
L.N.\ Lipatov, Sov. J. Nucl. Phys. 23 (1976) 338; E.A.\ Kuraev,
L.N.\ Lipatov and V.S.\ Fadin, Sov. Phys. JETP  45 (1977) 199; Ya.\
Balitskii and L.N.\ Lipatov, Sov. J. Nucl. Phys. 28 (1978) 822.

\item \label{Jar}
T.\ Jaroszewicz, \pl{116}{291}{82}.

\item \label{CH}
S.\ Catani and F.\ Hautmann, \pl{315}{157}{93}, \np{427}{475}{94}.

\item \label{fadin}
V.S.\ Fadin and L.N.\ Lipatov, \sj{50}{712}{89}, \pl{429}{127}{98}.
 
\item \label{CC}
M. Ciafaloni and G. Camici, 
\np{496}{305}{97}, \pl{430}{349}{98}. 

\item \label{sdis}
S.\ Catani, \zp{70}{263}{96}, \zp{75}{665}{97}.

\item \label{hqnlo}
P.\ Nason, S.\ Dawson and R.K.\ Ellis, \np{303}{607}{88};
W.\ Beenakker, H.~Kuijf, W.L.\ van Neerven and J.\ Smith, \pr{40}{54}{89};
M.L.\ Mangano, P.\ Nason and G.\ Ridolfi, \np{373}{295}{92}.

\item \label{baarmand}
M.~Baarmand, talk presented at the
Workshop on {\it Standard Model Physics (and more) at the LHC},
CERN, January 1999. \\
See: http://home.cern.ch/n/nason/www/lhc99/14-01-99/program-1-14-99

\item \label{bquarkrev}
S.~Frixione, M.L.~Mangano, P.~Nason and G.~Ridolfi, in {\it Heavy Flavours II},
eds. A.J.~Buras and M. Lindner (World Scientific, Singapore, 1998), p.~609, and
references therein.

\item \label{ptbres}
M.\ Cacciari, M.\ Greco and P. Nason, \jhep{05}{007}{98};
F.I.\ Olness, R.J.\ Scalise and Wu-Ki Tung, \pr{59}{014506}{99}.

\item \label{ktfac}
S. Catani, M. Ciafaloni and F. Hautmann, \pl{242}{97}{90}, \np{366}{135}{91};
J.C. Collins and R.K. Ellis, \np{360}{3}{91};
E.M.\ Levin, M.G.\ Ryskin, Yu.M.\ Shabel'skii and A.G.\ Shuvaev, Sov. J.
Nucl. Phys. 53 (1991) 657.

\item \label{promptgnlo}
P.\ Aurenche, R.\ Baier, M.\ Fontannaz and D.\ Schiff, \np{297}{661}{88};
H.~Baer, J. Ohnemus and J.F. Owens, \pr{42}{61}{90};
P.\ Aurenche, R.~Baier and M.\ Fontannaz, \pr{42}{1440}{90};
L.E.\ Gordon and W.~Vogelsang, \pr{48}{3136}{93}. 

\item \label{annecygamma}
P.\ Aurenche, M.\ Fontannaz, J.Ph.\ Guillet, B.\ Kniehl,
E.\ Pilon and M. Werlen, \epj{9}{107}{99}.

\item \label{E706}
E706 Coll., L. Apanasevich et al., \prl{81}{2642}{98}.

\item \label{ktcteq}
L. Apanasevich et al., \pr{59}{074007}{99}.

\item \label{ktmartin}
M.A.~Kimber, A.D.~Martin and M.G.~Ryskin, Eur. Phys. J. C12 (2000) 655. 

\item \label{ktsterman}
E.~Laenen, G.~Sterman and W.~Vogelsang, preprint YITP-99-69
(hep-ph/0002078).

\item \label{annecypi}
P.~Aurenche, M.~Fontannaz, J.P.~Guillet, B.A.~Kniehl and M.~Werlen, 
preprint LAPTH-751-99 (hep-ph/9910252).

\item \label{jh}
J.\ Huston et al., paper in preparation.

\item \label{delduca}
V.~Del Duca and G.~Heinrich, in these Proceedings, and references therein.

\item \label{book}
R.K.\ Ellis, W.J.\ Stirling and B.R.\ Webber, {\it QCD and Collider 
Physics} (Cambridge University Press, Cambridge, 1996) and references therein.

\item \label{Dokbook}
Yu.L.\ Dokshitzer, V.A.\ Khoze, A.H.\ Mueller and S.I.\ Troian, 
{\it Basics of Perturbative QCD} (Editions Fronti\`eres, Gif-sur-Yvette, 1991)
and references therein.

\item \label{BCM}
A.\ Bassetto, M.\ Ciafaloni and G.\ Marchesini, \prep{100}{201}{83}, and
references therein.

\item \label{CMW}
S.\ Catani, G.\ Marchesini and B.R. Webber, \np{349}{635}{91}.

\item \label{mecorsey}
M.H.~Seymour, Comput. Phys. Commun. 90 (1995) 95. 

\item \label{mecorsjo}
J.~Andre and T.~Sjostrand, Phys. Rev. D57 (1998) 5767.

\item \label{mecorfri}
C.~Friberg and T.~Sjostrand, in Proc. of the Workshop
{\it Monte Carlo Generators for HERA Physics}, eds. T.A.~Doyle, G.~Grindhammer,
G.~Ingelman and H.~Jung, (DESY, Hamburg, 1999), p.~181. 

\item \label{mecorcol}
J.C.~Collins, hep-ph/0001040 and in these Proceedings.

\item \label{seytop}
G.~Corcella and M.H.~Seymour, \pl{442}{417}{98}.

\item \label{sjow}
G.~Miu and T.~Sjostrand, \pl{449}{313}{99}.

\item \label{mrenna}
S. Mrenna,  preprint UCD-99-4 (hep-ph/9902471).

\item \label{corseyw}
G.~Corcella and M.H.~Seymour, preprint RAL-TR-1999-051 (hep-ph/9908388).

\item \label{resvsps}
C.~Balazs, J.~Huston and I.~Puljak, 
preprint FERMILAB-PUB-00-032-T \\
(hep-ph/0002032), and in these Proceedings. 
 
\item \label{softrev}
G.\ Sterman, in Proc. {\it 10th Topical Workshop on Proton-Antiproton Collider
Physics}, eds. R.\ Raja and J.\ Yoh (AIP Press, New York, 1996), p.~608;
S.\ Catani, Nucl. Phys. Proc. Suppl. 54A (1997) 107, and
in Proc of the {\it 32nd Rencontres de Moriond: QCD and High-Energy
Hadronic Interactions}, ed. J.\ Tran Than Van (Editions Fronti\`eres, Paris,
1997), p.~331. 

\item \label{jetres}
S.\ Catani, G.\ Turnock, B.R.\ Webber and L.\ Trenta\-due, \np{407}{3}{93}.

\item \label{seymour}
M.H.\ Seymour, \np{513}{269}{98}; 
J.R.\ Forshaw and M.H.\ Seymour, \jhep{09}{009}{99}.  

\item \label{DDT}
Yu.L.\ Dokshitzer, D.I.\ Diakonov and S.I.\ Troian, Phys. Rep. 58 (1980) 269,
and references therein.

\item \label{BS}     
N.\ Brown and W.J.\ Stirling, Phys. Lett. 252B (1990) 657.

\item \label{duralg}
S. Catani, Yu.L.\ Dokshitzer, M.\ Olsson, G.\ Turnock and B.R. Webber,
\pl{269}{432}{91}.

\item \label{ktalg}
S.\ Catani, Yu.L.\ Dokshitzer and B.R. Webber, \pl{285}{291}{92};
S.\ Catani, Yu.L.\ Dokshitzer, M.H.\ Seymour and B.R. Webber, \np{406}{187}{93};
S.D.~Ellis and D.E.\ Soper, \pr{48}{3160}{93}. 

\item \label{pp}
G.\ Parisi and R.\ Petronzio, \np{154}{427}{79}.  

\item \label{CSS}
J.C.\ Collins, D.E.\ Soper and G.\ Sterman, \np{250}{199}{85}.

\item \label{S}
G.\ Sterman, \np{281}{310}{87}.

\item \label{CT}
S.\ Catani and L.\ Trentadue, \np{327}{323}{89}, \np{353}{183}{91}.

\item \label{qtback}
J.C.\ Collins and D.E.\ Soper, \np{197}{446}{82}.

\item \label{thback}
S.\ Catani, M.L.\ Mangano, P.\ Nason and L.\ Trentadue,
\np{478}{273}{96}. 

\item \label{qthere}
C.~Bal\'azs, J.C.~Collins and D.E.~Soper, in these Proceedings.

\item \label{dyqt}
J.\ Kodaira and L. Trentadue, \pl{112}{66}{82},
\pl{123}{335}{83};
C.T.H.\ Davies, B.R.\ Webber and W.J. Stirling, \np{256}{413}{85}. 

\item \label{Hqt}
S.\ Catani, E.\ D'Emilio and L.\ Trentadue, \pl{211}{335}{88};
R.P.\ Kauffman, \pr{45}{1512}{92}. 

\item \label{CLS}
H.\ Contopanagos, E.\ Laenen and G. Sterman, \np{484}{303}{97}.

\item \label{Hth}
M. Kramer, E. Laenen and M. Spira, \np{511}{523}{98}.

\item \label{KS}
N. Kidonakis and G. Sterman, \pl{387}{867}{96}, \np{505}{321}{97}.

\item \label{KOS}
N. Kidonakis, G. Oderda and G. Sterman, \np{531}{365}{98}.

\item \label{LM}
E.\ Laenen and S.\ Moch, \pr{59}{034027}{99}.

\item \label{LOS}
E.\ Laenen, G. Oderda and G. Sterman, \pl{438}{173}{98}.

\item \label{CMNg}
S.\ Catani, M.L.\ Mangano and P.\ Nason, \jhep{07}{024}{98}.

\item \label{CMNOV}
S.\ Catani, M.L.\ Mangano, P.\ Nason, C.\ Oleari and W. Vogelsang,
\jhep{03}{025}{99}.

\item \label{kow}
N.~Kidonakis and J.F.~Owens, preprint FSU-HEP-991216 (hep-ph/9912388).

\item \label{cacciari}
M. Cacciari, preprint CERN-TH/99-312 (hep-ph/9910412).

\item \label{kdd}
N.~Kidonakis and V.~Del Duca, preprint FSU-HEP-991123 (hep-ph/9911460).

\item \label{tqll1}
E.~Laenen, J. Smith and W.L. van Neerven, \np{369}{543}{92}, \pl{321}{254}{94}. 

\item \label{tqll2}
E. Berger and H. Contopanagos, \pr{54}{3085}{96},  \pr{57}{253}{98}.

\item \label{tqll3}
S.\ Catani, M.L.\ Mangano, P.\ Nason and L.\ Trentadue, \pl{378}{329}{96}.

\item \label{tqll4}
N.\ Kidonakis, preprint EDINBURGH-99-4 (hep-ph/9904507).

\end{enumerate}


\setcounter{figure}{0}
\setcounter{table}{0}
\setcounter{section}{0}
\setcounter{equation}{0}
\newpage

%
%
\def\kt{$k_T$}
\def\mkt{k_T}
\def\avkt{$\langle k_T \rangle$}
\def\mavkt{\langle k_T \rangle}
\def\avktsq{$\langle k_T^2 \rangle$}
\def\mavktsq{\langle k_T^2 \rangle}
\def\pt{$p_T$}
\def\mpt{p_T}
\def\avpt{$\langle p_T \rangle$}
\def\avptp{$\langle p_T \rangle_{pair}$}
\def\sig{$\sigma$}
\def\sp{$\sigma_{1parton,2D}$}
\def\msp{\sigma_{1parton,2D}}
\def\spp{$\sigma_{2parton,2D}$}
\def\mspp{\sigma_{2parton,2D}}
\def\sone{$\sigma_{1D}$}
\def\msone{\sigma_{1D}}
\def\stwo{$\sigma_{2D}$}
\def\mstwo{\sigma_{2D}}
\def\sg{$\sigma_{\gamma,1D}$}
\def\msg{\sigma_{\gamma,1D}}
\def\sgtwo{$\sigma_{\gamma,2D}$}
\def\msgtwo{\sigma_{\gamma,2D}}
\def\x{$x$}
\def\s{$\sqrt{s}$}
\def\DZERO{D\O}
\def\half{\hbox{${1\over 2}$}}\def\third{\hbox{${1\over 3}$}}
\def\quarter{\hbox{${1\over 4}$}}
\def\smallfrac#1#2{\hbox{${{#1}\over {#2}}$}}
\catcode`@=11 
\def\lsim{\mathrel{\mathpalette\@versim<}}
\def\gsim{\mathrel{\mathpalette\@versim>}}
 \def\@versim#1#2{\lower0.2ex\vbox{\baselineskip\z@skip\lineskip\z@skip
       \lineskiplimit\z@\ialign{$\m@th#1\hfil##$\crcr#2\crcr\sim\crcr}}}
\catcode`@=12 


\begin{center}
\vspace*{1.2cm}
{\Large\sc \bf Partons for the LHC } \\
\vspace*{1.cm} 
{\sc R.D. Ball~\footnote{Royal 
Society University Research Fellow.} and J.~Huston}
\vspace*{1.cm}
\end{center}

\begin{abstract}

We discuss some of the experimental, theoretical and methodological
issues in the determination of parton distributions with meaningful 
error estimates, and their impact on physical cross sections to be 
measured at the Tevatron and LHC. 

\end{abstract}

\section{Introduction}

The calculation of production cross sections at the 
Tevatron and LHC, for both interesting 
physics processes and their backgrounds, relies upon a knowledge of the 
distribution of the momentum fraction $x$ of the partons in a proton at the
relevant scale. These parton distribution functions (pdfs) are 
at present determined by global fits to data from deep inelastic 
scattering (DIS), Drell-Yan (DY), and jet and direct photon 
production at current energy ranges. Two groups, CTEQ and MRS, 
provide semi-regular updates to their best-fit parton distributions 
when new data and/or theoretical developments become available. 
The newest pdfs, in most cases, currently provide the single most 
accurate overall 
description of the world's data, and should be utlilized in 
preference to older pdf sets. The most recent sets from
the two groups are  CTEQ5~\cite{cteq5} and MRST~\cite{mrst98}. 

In this contribution we will discuss the data sets used in the fits, the way in 
which the fits are performed in practice (in particular, issues such as the 
parametrization of initial distributions, the solution of the evolution 
equations, and scheme dependence), and the main uncertainties in the 
fitted pdfs due to uncertain or incomplete experimental data. In particular, we will concentrate on
the difficulties involved in determining the gluon distribution through 
direct photons or jets. We then move on to discuss more general issues
which may affect future pdf determinations: the inclusion of correlated 
systematics and the difficulties involved in combining these for 
different experiments, purely theoretical uncertainties arising from the 
limitations of NLO perturbative QCD, and finally, methodological 
uncertainties such as the dependence on the form of the parametrization 
and the assumption of Gaussian error propagation. We conclude with a 
summary of the progress that might be made before the LHC turns on, and 
the role of LHC data in determining pdfs.

\section{Processes Involved in Global Analysis Fits}

Lepton-lepton, lepton-hadron and hadron-hadron interactions probe
complementary aspects of perturbative QCD (pQCD). Lepton-lepton processes
provide clean measurements of $\alpha_s(Q^2)$ and of the 
fragmentation functions of 
partons into hadrons. Measurements of deep-inelastic scattering (DIS) structure
functions ($F_2,F_3$)
in lepton-hadron scattering and of lepton pair production cross 
sections in hadron-hadron collisions are the main source of information 
on the quark distributions $q^a(x,Q)$ inside hadrons. Scaling violations in 
deep inelastic processes give some information about the gluon distribution 
$g(x,Q)$. Furthermore the gluon distribution
function enters directly (i.e. at leading order) in hadron-hadron 
scattering processes with direct photon and jet final states. 
Modern global parton distribution fits are carried out to next-to-leading
(NLO) order which allows $\alpha_s(Q^2), q^a(x,Q)$ and $g(x,Q)$ to all mix and 
contribute in the  theoretical formulae for all processes. Nevertheless, the 
broad picture described above still holds to some degree in global pdf 
analyses.

        In pQCD, the gluon distribution is always accompanied by a 
factor of $\alpha_s$, in both the hard scattering cross sections and 
in the  evolution equations for parton distributions. Thus, determination of 
$\alpha_s$ and the gluon distribution is, in general, a strongly 
coupled problem. One can determine $\alpha_s$ separately from 
$e^+e^-$ or determine $\alpha_s$ and $g(x,Q)$ jointly in a global 
pdf analysis. In the latter case, though, the coupling of $\alpha_s$
and the gluon distribution may not lead to a unique solution for either (see 
for example the discussion in the CTEQ4 paper where good fits were 
obtained to a global analysis data set, including the inclusive jet data, for 
a wide range of $\alpha_s$ values~\cite{cteq4p}.) 

 Currently, the world average 
value of $\alpha_s(M_Z)$ is $0.119\pm 0.004$~\cite{alpha}. 
This is in agreement with the average value from LEP, while the 
DIS experiments prefer a slightly smaller value (of the order of 
$0.116-0.118$)).  
Since global pdf analyses are dominated by the high 
statistics DIS data, they would tend to favor the  values of $\alpha_s$ closer
to the lower DIS values. The more logical approach is to adopt the world
average value of $\alpha_s(M_Z)$ and concentrate on the determination of the
pdfs. This is what both CTEQ and MRS currently do.~\footnote{One can either 
quote a value of $\alpha_s(M_Z)$ or the value of $\Lambda^{\overline{MS}}$.
In the latter case, however, the number of flavors has to be clearly specified,
since the value of $\alpha_s$ (and not $\Lambda^{\overline{MS}}$) has to be
continuous across flavor thresholds. }

\begin{figure}[th]
\begin{center}
\epsfxsize8cm
\epsfysize=8cm
\mbox{\epsfbox{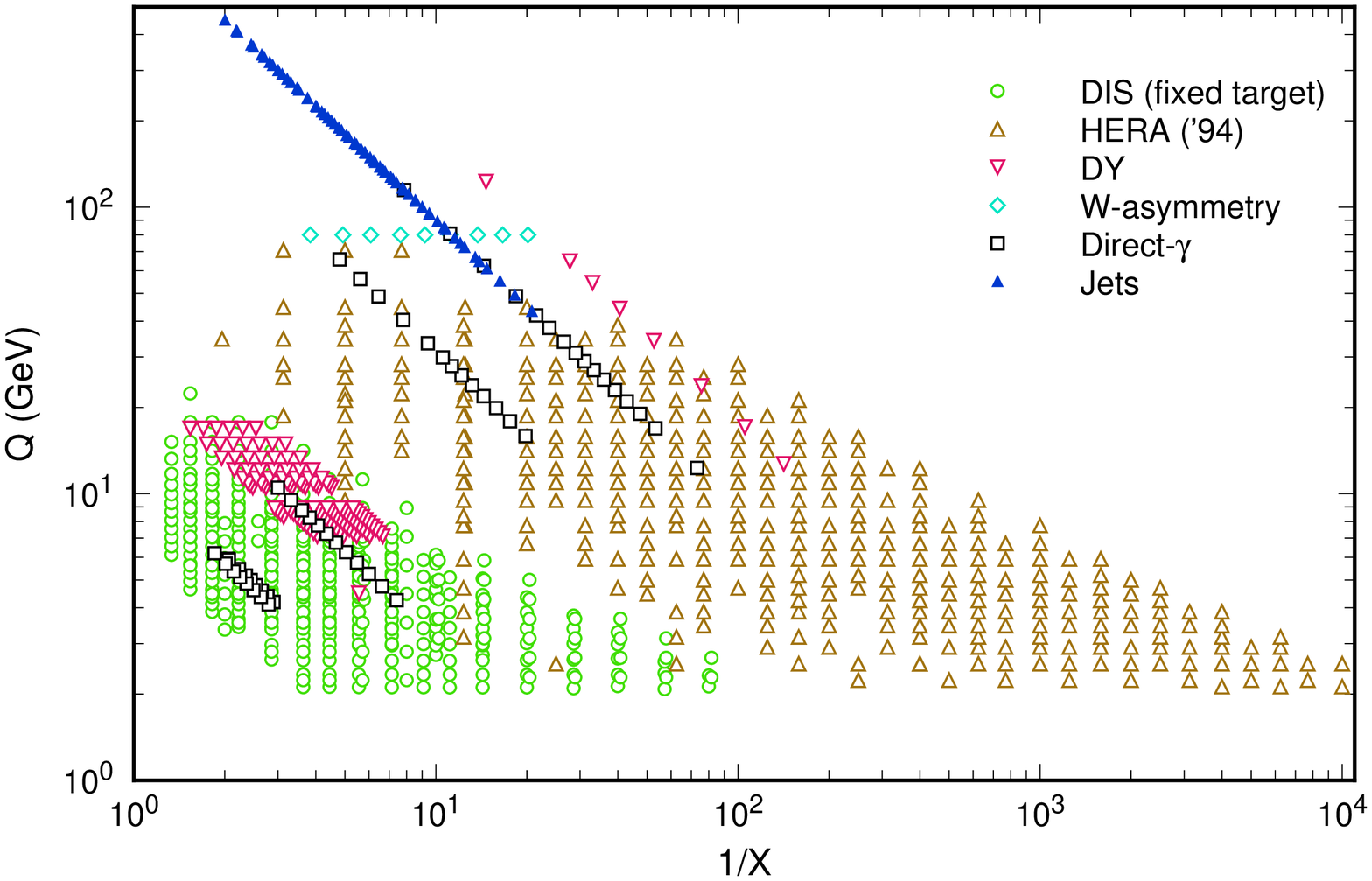}}
\end{center}
\caption{
\sf The kinematic map in the $(x,Q)$ plane of data points used in the CTEQ5
analysis. 
} 
\label{fig:xqall}
\end{figure}

The data from DIS, DY, direct photon and jet processes utilized in pdf
fits cover a wide range in $x$ and $Q\equiv\sqrt{Q^2}$.
The kinematic `map' in the $(x,Q)$ plane of the data points used in a
recent parton distribution function  analyses is
shown in Figure ~\ref{fig:xqall}. 
The HERA data (H1+ZEUS) are predominantly at low $x$, while
the fixed target DIS and DY data are at higher $x$. There is considerable 
overlap, however, with the degree of overlap increasing with time as the
statistics of the HERA experiments increases. DGLAP-based NLO pQCD 
provides an accurate description of the data (and of the evolution of the
parton distributions) over the entire kinematic 
range shown. At very low $x$ and $Q^2$, DGLAP evolution is believed to be no
longer applicable due to unresummed small $x$ logarithms.
Similarly at very large $x$ there are significant contributions from
unresummed soft logarithms (logarithms of $1-x$).
However, no evidence for such corrections is seen in the current range of data; 
thus all global analyses use conventional DGLAP evolution of pdfs.

There is a remarkable consistency between the data in the pdf fits
and the NLO QCD theory fit to them. Over $1300$ data points are shown in 
Figure ~\ref{fig:xqall}  and the $\chi^2$/d.o.f. for the fit of
theory to data is on the order of one.

Parton distributions determined at a given $x$ and $Q^2$ propagate down
to lower $x$ values at higher $Q^2$ values. The accuracy of 
the extrapolation to
higher $Q^2$ depends both on the accuracy of the original measurement and any
uncertainty on $\alpha_s(Q^2)$. For the structure function $F_2$, the typical
measurement uncertainty at medium to large $x$ is on the order of $\pm 3\%$.
At large $x$, the DGLAP  equation for $F_2$ can be approximated as 
$\frac{\partial F_2 }{\partial \log~Q^2} = \alpha_s(Q^2)P^{qq} \otimes F_2$. 
There is an extrapolation uncertainty of around
$\pm 5\%$ in $F_2$ from low to high $Q^2$
($10^5$ $GeV^2$) from the uncertainty in $\alpha_s$. 
Evolved distributions are  also susceptible to 
uncertainties from an anomalously large
contribution to $F_2$ near $x$ values of 1. Such a contribution may not be
evident in fixed target measurements at low $x$ and low $Q^2$, but may
influence higher $Q^2$ measurements~\cite{highx}. 

\begin{figure}[tp]
\begin{center}
\epsfxsize=8cm
\epsfysize=10cm
\mbox{\epsfbox{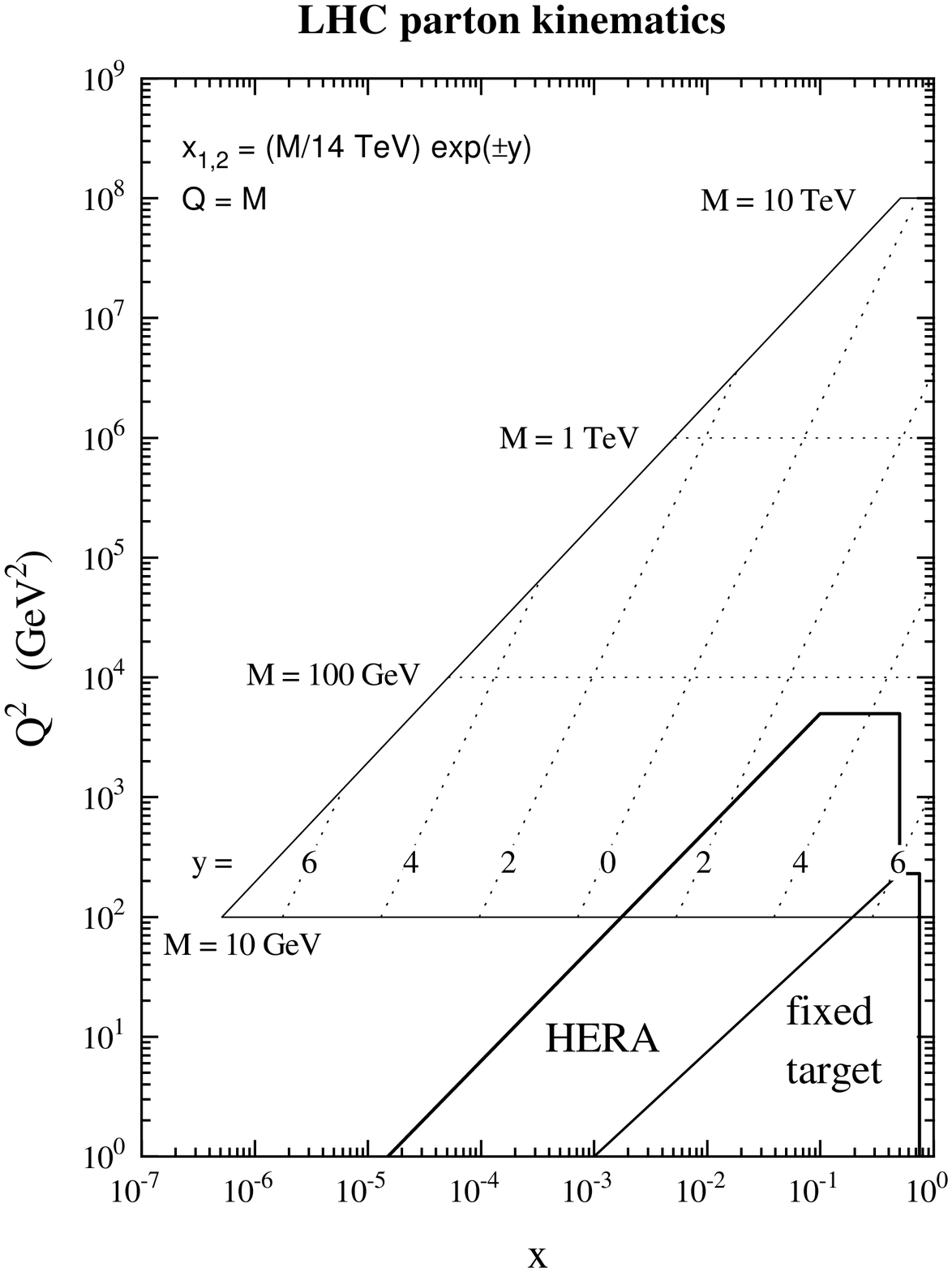}}
\end{center}
\caption{
\sf A plot of LHC parton kinematics in $(x,Q^2)$ space. Also shown are the 
reach of fixed target and HERA experiments.
} 
\label{fig:lhcgridx}
\end{figure}

For comparison, the kinematics appropriate for the production
of a state of mass  $M$ and rapidity $y$ at the LHC is shown in 
Figure~\ref{fig:lhcgridx}~\cite{stirling}.
For example, to produce a state of mass $100$ GeV and rapidity $2$ 
requires partons with $x$ values between $0.05$ and $0.001$ 
at a $Q^2$ value of $10^{4}$ $GeV^2$.
Also shown in the figure is another view of the kinematic coverage 
of the fixed target and HERA experiments used in pdf fits. 
It can be seen that parton distributions determined from these experiments are 
sufficient to predict most LHC cross-sections of interest, provided that 
DGLAP evolution at small and large $x$ is sufficiently reliable.

\section{Evolution, Schemes and Parametrizations}

\subsection{Evolution Codes}

In order to fit the initial pdfs to experimental data they need to be 
evolved up to the correct scale by solving the DGLAP equations either 
to LO or NLO. The evolution can be carried out
in either moment space or configuration space: both MRS and CTEQ use 
configuration space codes. Improvements have been made
in the CTEQ and MRST evolution programs so that both now agree with the 
`DESY standard' evolution prescription \cite{DESYS}. The CTEQ 
and MRST packages 
should be able to carry out the evolution using NLO DGLAP to an
accuracy of a few percent over the LHC kinematic range, except perhaps at
very large and very small $x$. Note that the theoretical predictions 
for the W and Z total cross sections at the LHC may have uncertainties 
of less than 5\%\cite{jamespc}. This
puts a great demand for the pdf evolution to have accuracies 
of better than a few percent, since any error on a pdf gets doubled 
in the cross section calculation. Mellin space codes might be the answer 
here.
 
A global pdf analysis  carried out at next-to-leading order
 needs to be performed in a specific renormalization and factorization  
scheme. The evolution kernels are in a specific scheme and 
to maintain consistency, any hard scattering cross section calculations
used for the input processes or 
utilizing the resulting pdfs need to also have been implemented in that same 
renormalization scheme. Almost universally, the $\overline{MS}$ 
scheme is used: 
pdfs are also available in the DIS scheme, a fixed flavor scheme (as in 
ref.\cite{grv}) and several schemes that differ in their specific treatment
of the charm quark mass ~\cite{ACOT,RT}. 

It is also possible to use only leading-order matrix element
calculations in the global fits which results in leading-order parton 
distribution functions. Such pdfs are preferred when leading order
matrix element calculations (such as Monte Carlo programs like 
{\tt HERWIG}~\cite{herwig1} and {\tt PYTHIA}~\cite{pythia1}) are used. 
The differences between LO and NLO 
pdfs, though, are formally NLO; thus, the additional error 
introduced by using a NLO pdf with  {\tt HERWIG} rather than a LO
pdf, for example, should not be significant, in principle,
 and NLO pdfs can be used
when no LO alternatives are available. 
The accuracy of current DIS/DY data 
is such that the $\chi^2$ values for LO fits are noticeably worse than those
~from the NLO fits: the data are sensitive to the differences 
between LO and NLO partonic cross-sections and evolution kernels. 

\subsection{Parametrization of Initial Distributions}

All current global analyses use a generic form for the 
parametrization of both the quark and gluon distributions at 
some reference value $Q_0$:
\begin{equation}
        f(x,Q_0)=a_0 x^{a_1}(1-x)^{a_2}P(x;a_3,...). 
\label{eq:pdf}
\end{equation}
The reference value $Q_0$ is usually chosen in the range of  $1-2$ GeV. 
The parameter $a_1$ is associated with small-$x$ behaviour 
while $a_2$ is associated with large-$x$ valence counting rules. 
In some pdf fits, $a_1^{\rm gluon}$ has been tied to 
$a_1^{\rm sea quark}$; in more recent fits like CTEQ4,
CTEQ5 and MRST, the two small $x$ exponents are allowed to 
vary independently. The current statistical power
of the low $x$ and $Q^2$ DIS data from HERA warrants this separation.
 
        The first two factors, in general, are not 
sufficient to describe either quark or  gluon distributions. 
The term $P(x; a_3,...)$ is a suitably chosen smooth  function, 
depending on one  or more parameters, that adds
more flexibility to the pdf parametrization. In general, both the 
number of free parameters and the functional form can have an 
influence on the global fit. For example,
the MRS group traditionally uses $P_{MRS}(x; a_3,a_4)=1+a_3 \sqrt{x}+a_4x$. 
The CTEQ3 pdf used $P_{CTEQ3}=1+a_3x$ while CTEQ2, CTEQ4 and CTEQ5 all
use the more general form $P_{CTEQ2,4,5}=1+a_3x^{a_4}$. The flexibility in the 
latter form, for example, makes possible the larger gluon at high $x$ 
observed in the CTEQ4HJ pdf. 

Although the pdfs determined from global analyses should, in principle, be
universal, in practice they could depend on the choice of data sets, and 
in particular on the choice of $Q_{cut}$ values that specify the 
minimum hard physical scale $(Q,p_T,..)$
required for data points to be included in the fit. 

\begin{figure}[th]
\begin{center}
\epsfxsize=8cm
\epsfysize=8cm
\mbox{\epsfbox{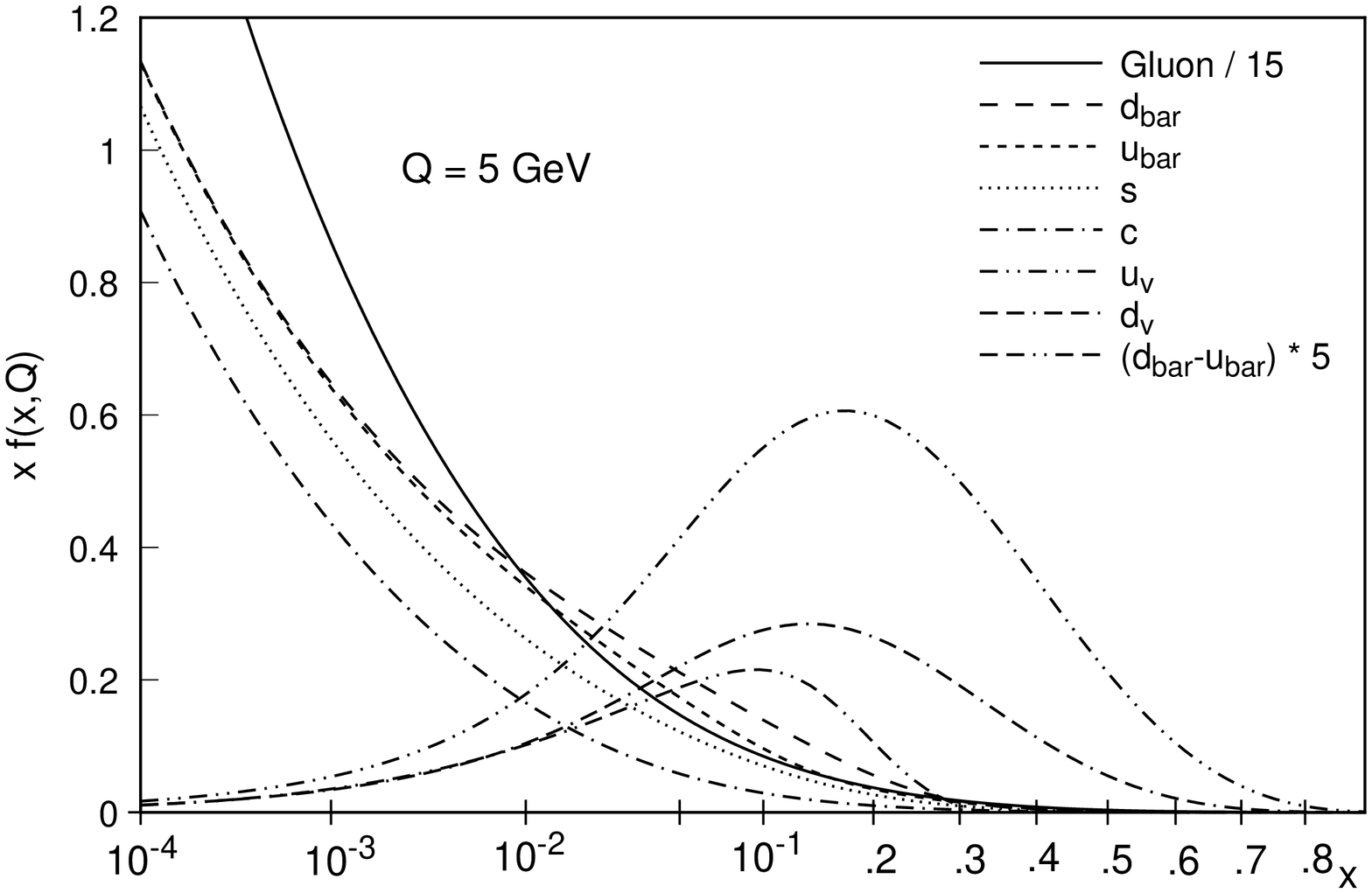}}
\end{center}
\caption{
\sf The parton distributions from the CTEQ5 set plotted at a $Q$ value of
5 GeV.
} 
\label{fig:AllPdf}
\end{figure}

The parton distributions from the recent CTEQ pdf release are 
plotted in Figure ~\ref{fig:AllPdf} at a $Q$ value of 5  $GeV$. The gluon 
distribution is largest at small $x$ values while the valence
quark distributions dominate at higher $x$.

\subsection{Evolution in time and $Q^2$}

        As discussed in the introduction, the MRS and CTEQ groups provide
semi-regular updates to their parton distributions as new data and/or theory
becomes available. The latest parton distributions are the most accurate and 
should be used in preference to previous pdfs.  However, in some cases 
calculations using older pdfs are necessary: for example, until 
recently~\footnote{In the most recent version of {\tt PYTHIA} (6.1), the 
CTEQ5 pdf's are available.} none of the more 
recent pdfs were implemented in {\tt PYTHIA}, and most comparisons in the 
ATLAS TDR have been made with the CTEQ2L pdf (the default pdf 
in {\tt PYTHIA} version 5.7).

\begin{figure}
\begin{center}
\begin{tabular}{cc}
\epsfysize=8cm \epsffile{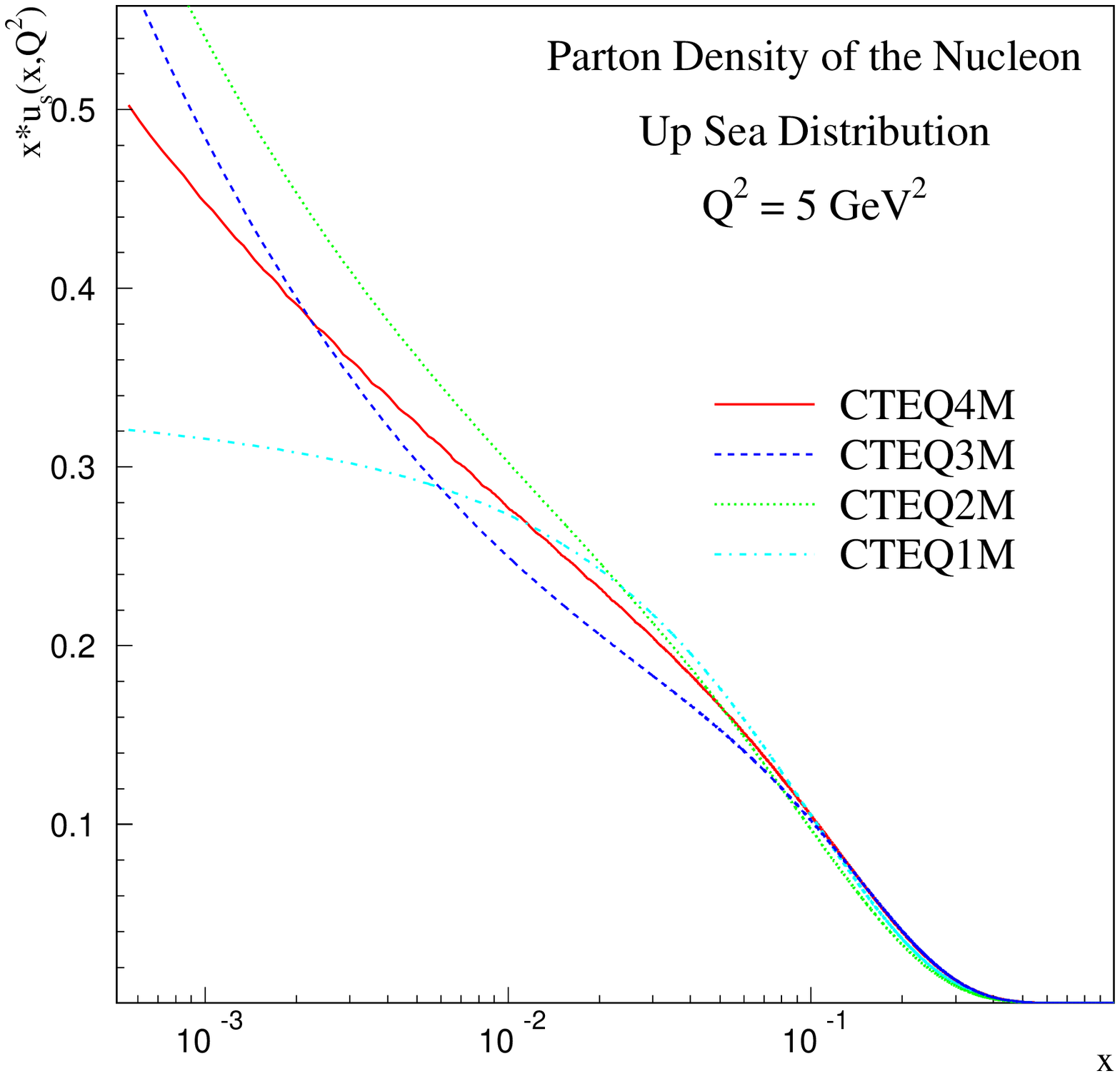} &
\epsfysize=8cm \epsffile{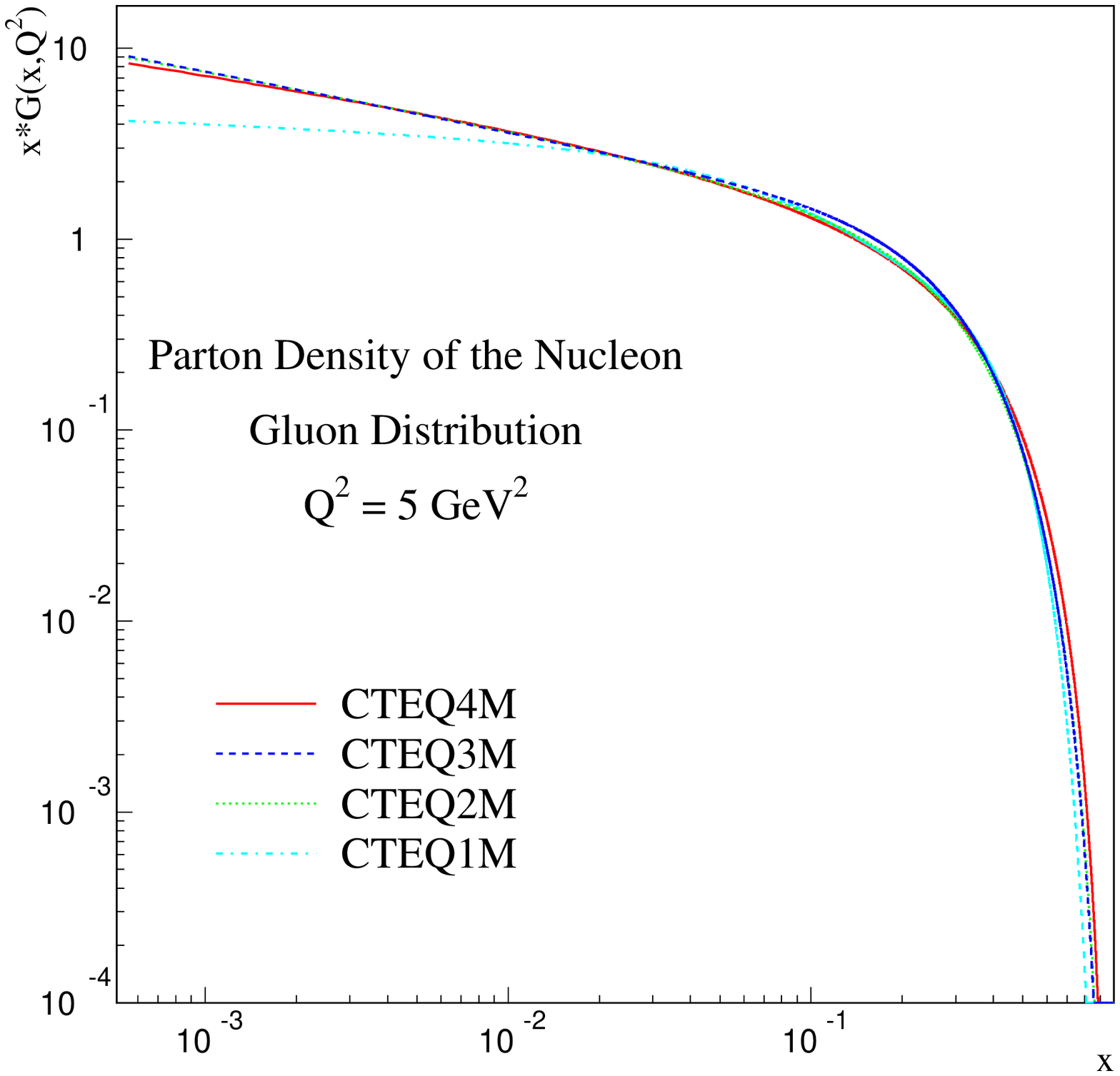}
\end{tabular}
\end{center}
\caption{
\sf The up sea quark and gluon parton distributions from the CTEQ1-4 sets plotted at a $Q^2$ value of
$5$~GeV$^2$.
} 
\label{fig:ctequpsea}
\end{figure}

        A comparison of the  CTEQ1M~\cite{cteq1}, CTEQ2M~\cite{cteq2}, 
CTEQ3M~\cite{cteq3} and CTEQ4M~\cite{cteq4p} parton distributions
(in particular the up sea quark and gluon distributions) are shown in
Figure ~\ref{fig:ctequpsea},
at a $Q^2$ value of $5$ GeV$^2$. The CTEQ2-4 up quark sea distributions 
are substantially steeper than that of CTEQ1,
reflecting the influence of the HERA data. A similar effect is seen with the 
gluon distribution. There is little change in the valence distributions.
\begin{figure}
\begin{center}
\begin{tabular}{cc}
\epsfysize=8cm \epsffile{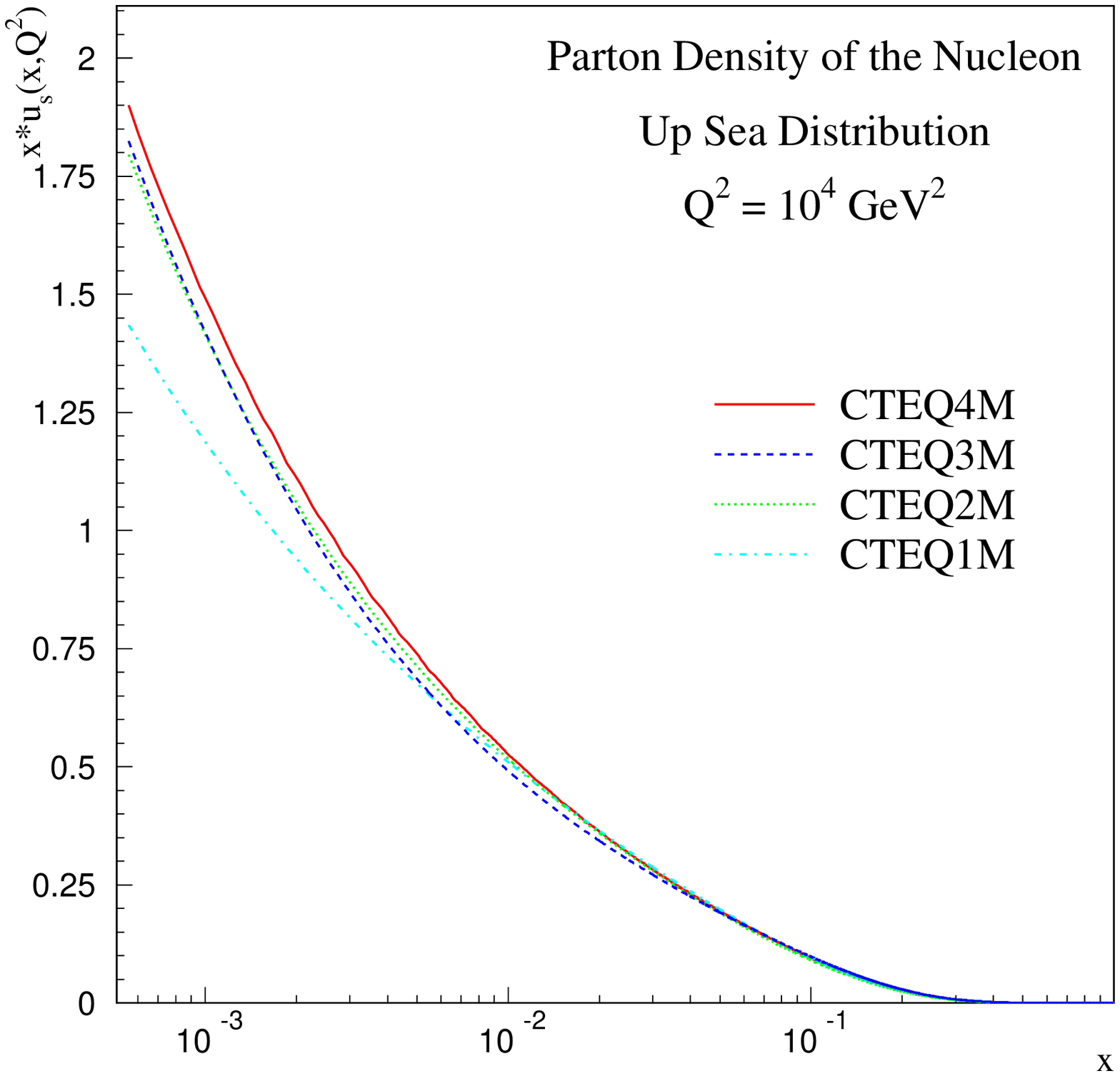} &
\epsfysize=8cm \epsffile{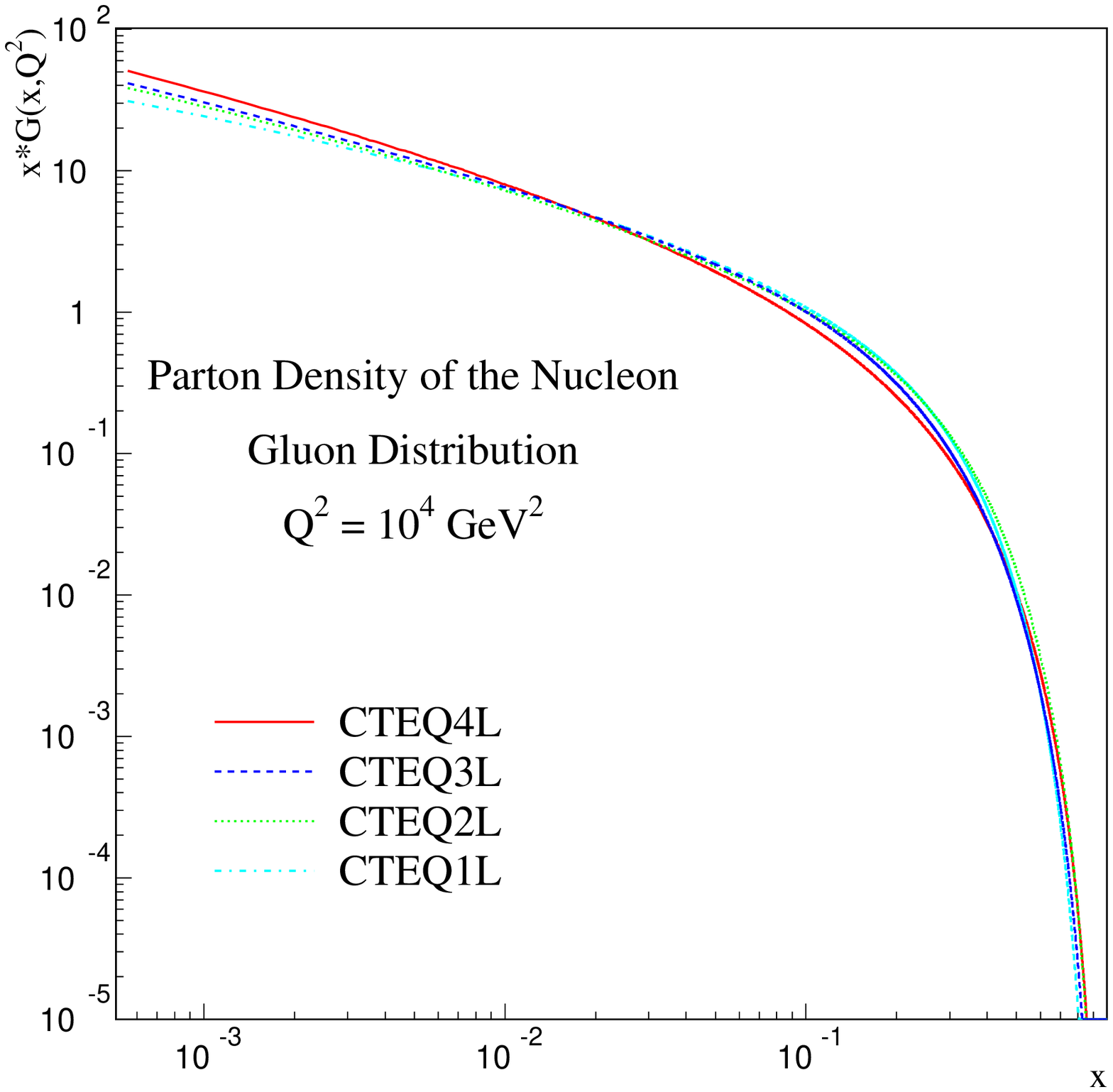}
\end{tabular}
\end{center}
\caption{
\sf The up sea quark and gluon parton distributions from the CTEQ1-4 sets plotted at a $Q^2$ value of
$10^4$~GeV$^2$.
} 
\label{fig:ctequpseaq}
\end{figure}

The up sea quark and gluon distributions are shown in Figure
~\ref{fig:ctequpseaq} at a larger $Q^2$ value of
$10^4$~GeV$^2$. Evolution has evened  out many of the differences observed at
lower $Q^2$ values. A $Q^2$ value of $10^4$~GeV$^2$ corresponds to a mass
scale at the LHC of about $100$~GeV. 

\begin{figure}
\begin{center}
\begin{tabular}{cc}
\epsfysize=6cm \epsffile{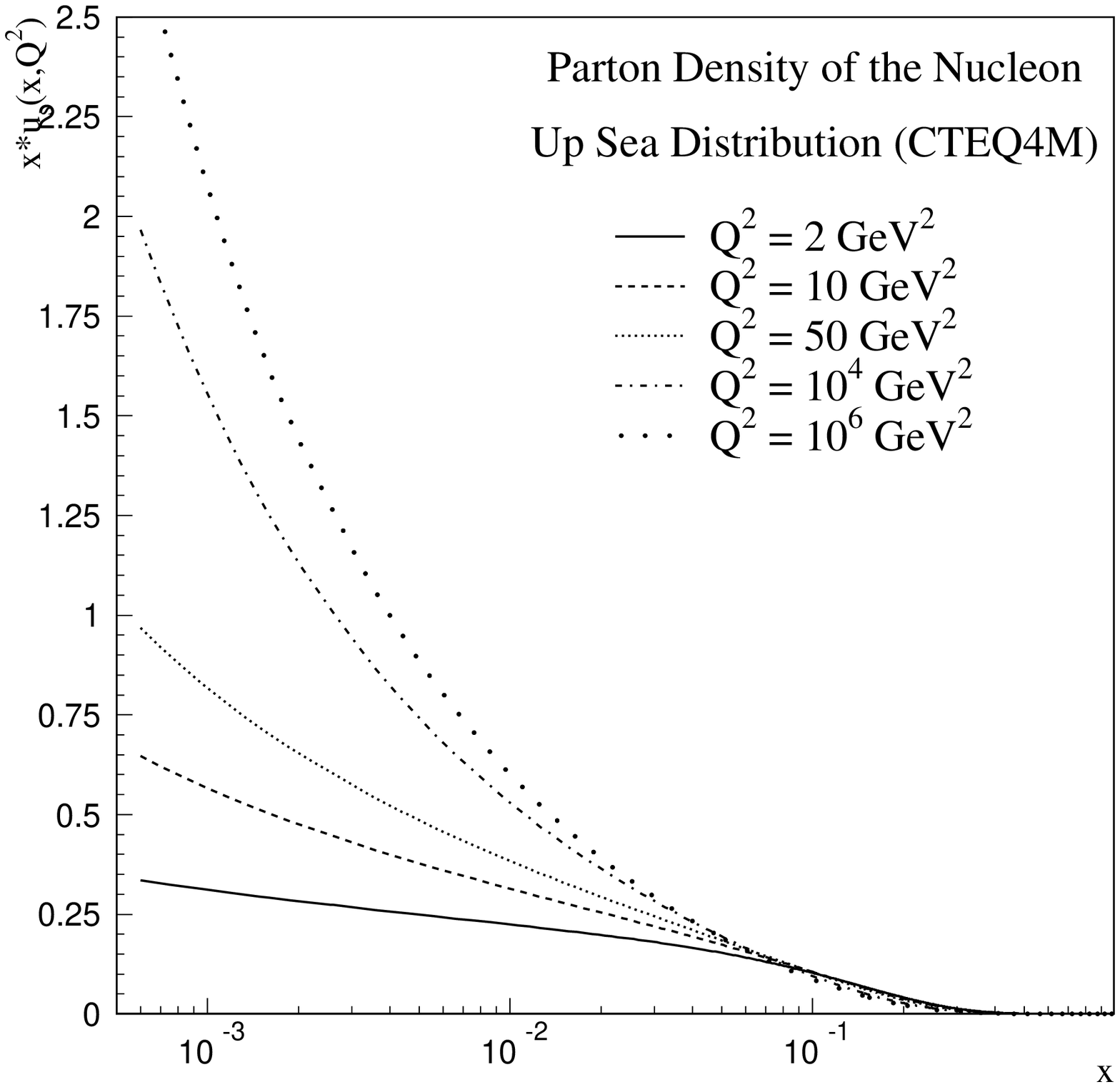} &
\epsfysize=8cm \epsffile{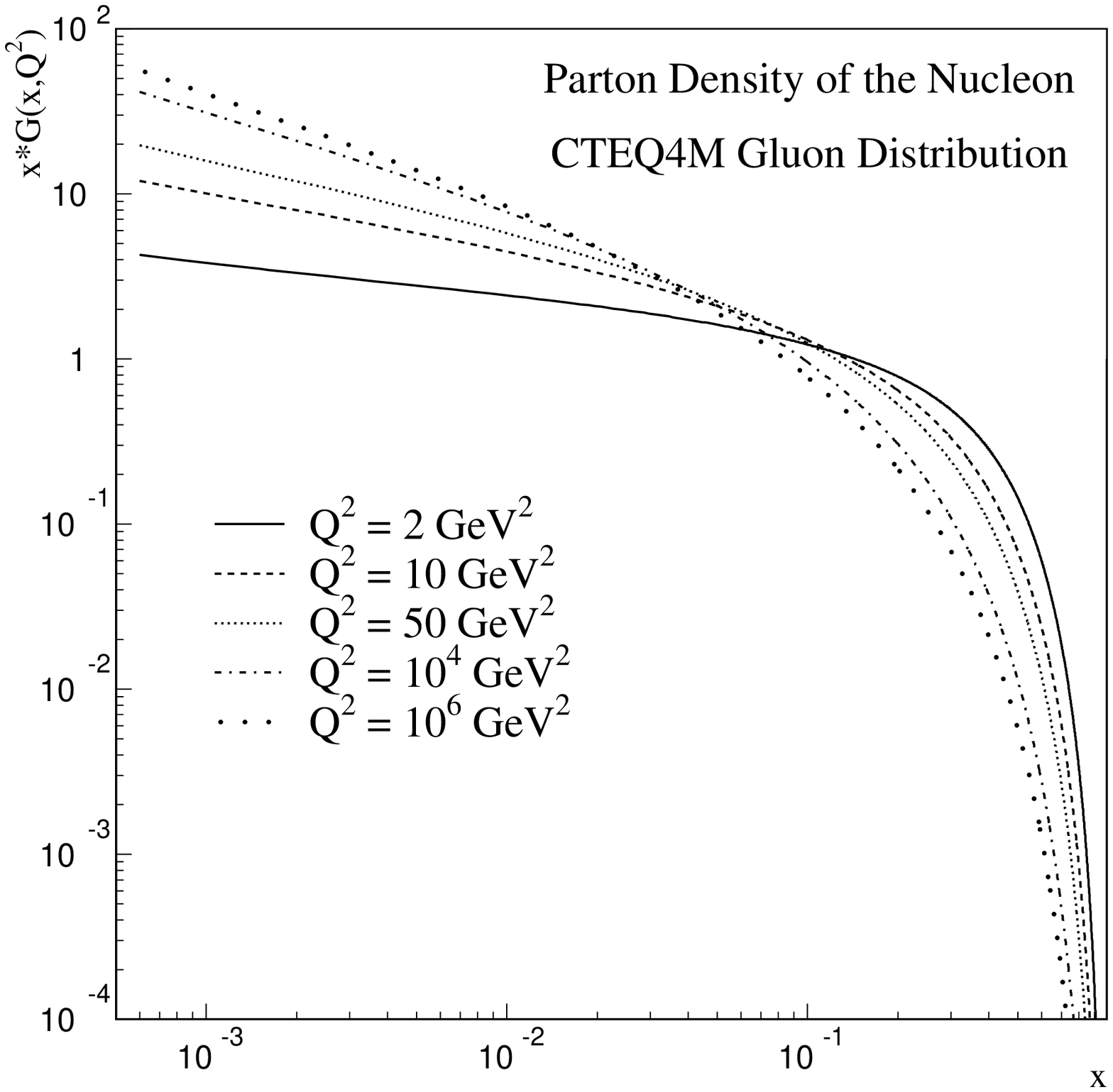}
\end{tabular}
\end{center}
\caption{
\sf The up sea quark and gluon distributions 
~from CTEQ4M shown at five  different $Q^2$ 
} 
\label{fig:cteq_gluon_evolve}
\end{figure}

The effects of evolution are examined in more detail in Figure
~\ref{fig:cteq_gluon_evolve} where the up sea quark and gluon
distributions are plotted at $Q^2$ values of $2$, $10$, $50$, $10^4$ and 
$10^6$ GeV$^2$. There are two interesting features that can be noted. 
Most of the evolution takes place at low $Q^2$ and there is little evolution
for $x$ values in the vicinity of $0.1$. In contrast, at large $x$ 
value the distributions decrease by an order of magnitude from the 
lowest to the highest $Q^2$ value, while at small $x$ they increase by 
an order of magnitude. 

\section{Estimating Uncertainties}

In addition to having the best estimates for the values of the pdfs in a given
kinematic range, it is also important to understand the allowed range 
of variation of the pdfs, i.e. their uncertainties. The crudest 
method of estimating parton
distribution uncertainties is to compare different published parton 
distributions. 
This is unreliable since most published sets of parton distributions (for
example from CTEQ and MRS) adopt 
similar assumptions and the differences between the sets do not fully
explore the full range uncertainties that actually exist.  
Here and in the next section we concentrate on estimating  
the uncertainties due to to the limitations 
of available data sets. 

The sum of the quark distributions $\Sigma (q(x)+\overline{q}(x))$
is, in general, well-determined over a wide range of $x$ and $Q^2$. 
As stated above, the quark distributions are
predominantly determined by the DIS and DY data sets which have large
statistics, and systematic errors in the few percent range ($\pm3\%$ for 
$10^{-4}<x<0.75$). 
Thus the sum of the quark distributions is basically known to a 
similar accuracy. The individual 
quark flavors, though,  may have a greater uncertainty than the sum. This can
be important, for example, in predicting distributions that depend on specific
quark flavors, like the W asymmetry distribution ~\cite{Wasym} and the
W rapidity distribution.

        Information on the $\overline{d}$ and $\overline{u}$ distributions
comes, at small $x$, from HERA and at medium $x$ from fixed target DY 
production on $H_2$ and $D_2$ targets. It is now well-established
~\cite{NA51,e866} that the $\overline{d}$ and $\overline{u}$ 
distributions are not the same. The difference in these distributions 
between the CTEQ4M and CTEQ5M pdfs is due primarily to the influence of the
data from the E866 experiment.  It is worth noting that our 
detailed knowledge of $\overline{d}/\overline{u}$ is limited 
primarily to the $x$ region  (.03-.35) covered by E866.

        The strange quark sea is determined from dimuon production in $\nu$
DIS (CCFR\cite{ccfr}), with the strange quark distribution 
($s +\overline{s}$) being approximately
\half($\overline{u}+\overline{d}$). The charm and bottom quark distributions
are calculated perturbatively from gluon splitting for given masses of $m_c$
and $m_b$. (See also the previous discussion on schemes.)

        Current information on $d/u$ at large $x$ comes from fixed target DY
production on $H_2$ and $D_2$ and the lepton asymmetry in W production at the
Tevatron. In the CTEQ5 and MRST fits, the NMC $D_2/H_2$ data are used to
constrain the large $x$ $d$ quark distribution in this way. Bodek and Yang have
argued that the $D_2$ data need to be corrected for nuclear binding effects,
which would lead to a larger $d/u$ ratio at large $x$ (and thus a larger $d$
quark distribution as the $u$ quark distribution is well-determined from DIS)
~\cite{BY}. The need for the nuclear binding corrections is still an open
question~\cite{stevepaper}.
The larger $d$ quark distribution would lead to an increase in the high 
$E_T$ Tevatron jet cross section of about 10\%. A similar excess would be
expected for high $E_T$ jet production at the LHC. 

The parton distribution with the greatest uncertainty is the 
gluon distribution, simply because it does not couple directly 
to an external probe. The LHC is essentially a 
gluon-gluon collider and  many hadron-collider signatures 
of physics both within and beyond that Standard Model involve gluons in
the initial state. Thus, it is very important to estimate the 
theoretical uncertainty due to the uncertainty in the gluon distribution. 

The gluon distribution can 
be determined indirectly at low $x$ by
measuring the scaling violations in the quark distributions 
(${\partial F_2}/{\partial \log Q^2}$), 
but a direct measurement is necessary at moderate to high 
$x$. Direct photon production 
has long been regarded as potentially the 
most useful source of information on the gluon distribution with  fixed target
direct photon data, especially from the experiment WA70~\cite{wa70}, being
used in a number of global analyses. However, as will be
discussed in the next section, there are a number of theoretical 
complications with the use of direct photon data. 

        The momentum fraction of the proton carried by quarks is determined
very well from DIS data; at a $Q_0$ value of 1.6 GeV, in the CTEQ4
analysis for example, the momentum fraction carried by quarks is 58\% with
an uncertainty of $\pm 2\%$. Thus, the momentum fraction carried by gluons
is 42\% with a similar uncertainty. This constraint is important; if the gluon
distribution increases in one $x$ range, momentum conservation forces it to
decrease in another $x$ range.  Thus, if the gluon flux in the $x$ 
range from $0.01$ to $0.3$ were to decrease by $20\%$, 
the gluon flux would have to increase by a fairly dramatic 
amount in the other $x$ ranges to compensate. For example, if this 
compensation were to come in the high $x$ region, the
gluon distribution would have to double.  

A simple way of estimating the uncertainty in the gluon  distribution 
is to systematically vary the
gluon parameters in a global analysis and then look for incompatibilities 
with the
data sets that make up the global analysis database. This study has 
been 
carried out by CTEQ using only DIS and Drell-Yan data where the 
theoretical and
experimental systematic errors are under good control~\cite{gluonpaper}.
Except at larger values of $x (x > 0.2-0.3)$,
the variation in the gluon distributions is less than 15\% at low values 
of $Q^2$,
decreasing to less than 10\% at high values: as noted earlier, 
evolution is the great equalizer for parton distributions. 
Note that the DIS and DY datasets used
in this analysis do not provide any strong constraints on the 
gluon distribution at
high values of $x$. This study used the CTEQ4 value of $\alpha_s$ 
(i.e. $0.116$). If $\alpha_s$ is varied in the range from 0.113 to 0.122, the 
gluon distribution varies by 3\% for $x < 0.15$.

\begin{figure}[th]
\begin{center}
\begin{tabular}{cc}
\epsfysize=8cm \epsffile{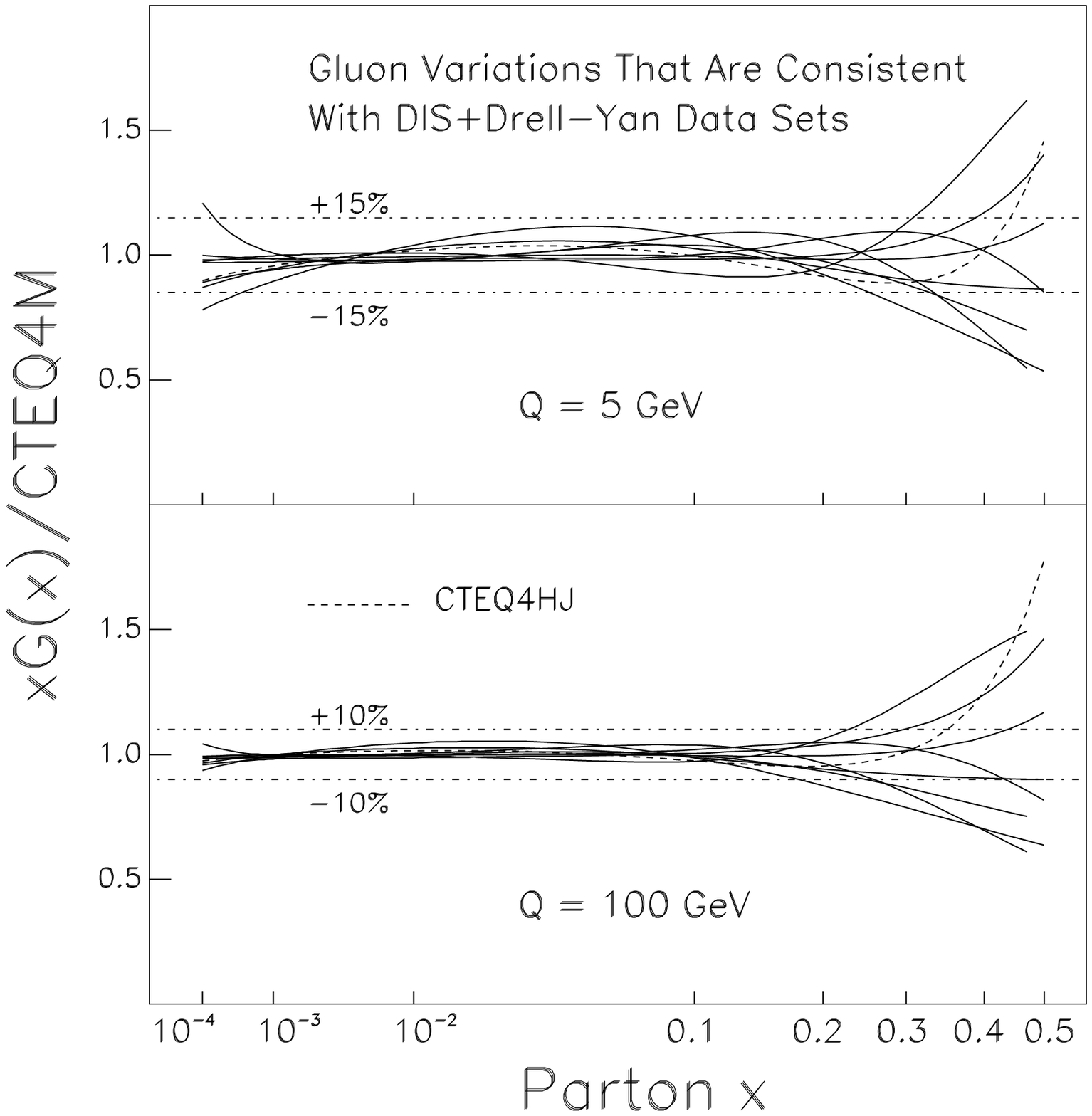} &
\epsfysize=8cm \epsffile{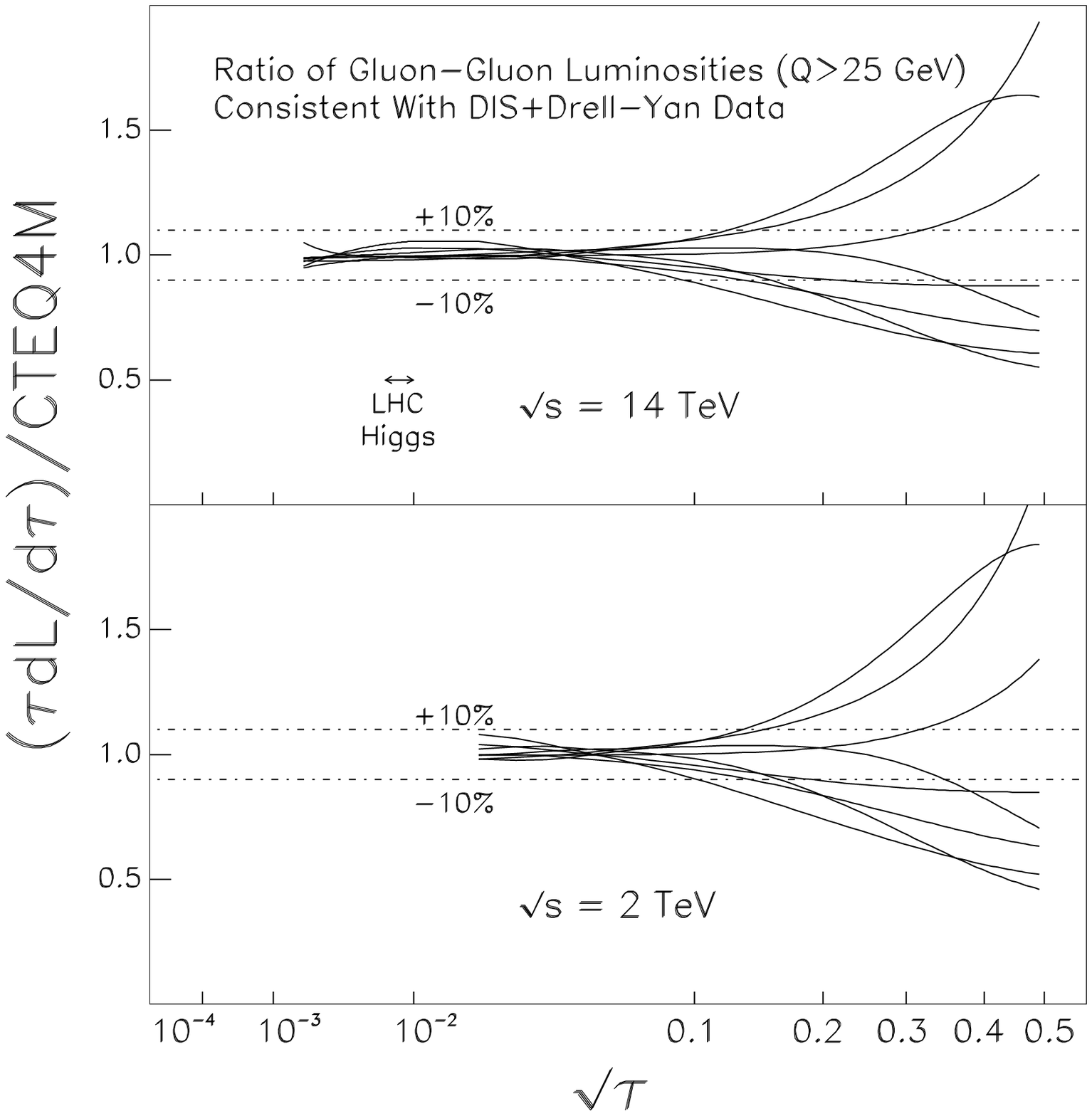}
\end{tabular}
\end{center}
\caption{
\sf The ratio of gluon distributions consistent with the DIS and DY data sets
to the gluon distributions from CTEQ4M. The gluon distribution from CTEQ4HJ is
also shown for comparison. In the second figure are shown the corresponding 
allowable variations in the integrated gluon-gluon luminosity as 
a function of $\protect\sqrt{\tau}$.  
} 
\label{fig:gluonuncert}
\end{figure}

In order to assess the range of predictions for hadronic 
cross sections, it is more 
important to know the uncertainties in the gluon-gluon and gluon-quark 
luminosity functions at the  appropriate kinematic region 
(in $\tau =x_1x_2=\hat{s}/s)$ rather than
the uncertainties in the parton distributions themselves. 
Therefore it is useful to define the relevant integrated
parton-parton luminosity functions: for example the 
gluon-gluon luminosity function can be defined as:
\begin{equation}
\tau {dL\over d\tau} = \int_\tau^1{dx\over x}g(x,Q^2)g(\tau/x,Q^2).
\label{eq:lum}
\end{equation}
This quantity is directly proportional to the cross section for 
s-channel production of
a single particle and it also gives a good estimate for 
more complicated production
mechanisms. In Figure~\ref{fig:gluonuncert} 
is shown the range of allowed gluon-gluon luminosities
(normalized to the CTEQ4M values) for the variations discussed above. 
Here, $Q^2$
is taken to be $\tau s$, which naturally takes the 
$Q^2$ dependence of the gluon 
distribution into account as one changes $\sqrt{\tau}$. 
The top region is for the LHC and
the bottom  is for the Tevatron. 
Above a $\sqrt\tau$ value of $0.1$, the allowed variation grows dramatically;
this indicates the need for more information about the
gluon distribution at large $x$ than provided by the DIS and DY data sets 
used in this analysis. 

In analogy with the discussion of gluon-gluon luminosities, 
one can also study the gluon-quark luminosity (again normalized to 
the CTEQ4M result). The uncertainties on the parton-parton 
luminosities, as a function of $\sqrt{\tau}$, are summarized in Table 1. 
Note that the region of production of a $100-140$ GeV Higgs
at the LHC  lies in the  region where the range of variation 
in the gg luminosity is $\pm 10\%$.

\begin{table}[t!]
\caption{The parton-parton luminosity uncertainty as a function of 
$\protect\sqrt{\tau}$. }
\begin{center}
\footnotesize
\begin{tabular}{lll}

$\sqrt{\tau}$ range &           gluon-gluon &           gluon-quark \\
\hline
$<0.1$ &        $\pm 10\%$ &    $\pm 10\%$ \\
$0.1-0.2$ &     $\pm 20\%$ &    $\pm 10\%$ \\
$0.2-0.3$ &     $\pm 30\%$ &    $\pm 15\%$ \\
$0.3-0.4$ &     $\pm 60\%$ &    $\pm 20\%$ \\

\end{tabular}
\end{center}
\label{table2}
\end{table}

\section{Direct Photons and Jets in Global Fits}

\subsection{Direct Photons}

        As mentioned previously in this section and in Reference~\cite{catani-sec1}, direct photon production has long been 
viewed as an ideal vehicle for measuring the gluon distribution in the 
proton. The quark-gluon Compton scattering subprocess 
$(gq \rightarrow\gamma q)$
dominates photon production in all kinematic regions of $pp$ scattering,
as well as for low to moderate values of parton momentum fraction
$x$ in $\overline{p}p$ scattering. As described previously, 
the gluon
distribution is relatively well constrained at low $x (x < 0.1)$ by DIS and
DY data, but less so at higher $x$. Consequently, fixed target direct photon 
data have been incorporated in several modern global parton 
distribution function analyses with the hope of providing a major
constraint on the gluon distribution at moderate to high $x$. 

A pattern of systematic deviations of direct photon data from NLO predictions 
has been observed~\cite{ktorig, aurenche}, however, these 
being particularly striking for the E706 experiment. The origin of the 
deviations is still quite controversial. One possibility that has been suggested is that the deviations are due to the effects of soft gluon radiation, or $k_T$~\cite{e706,apana}. This view, however, is not universally held; see, for example, the discussion in Reference~\cite{catani-sec1} and in Reference~\cite{aurenche}. 
The $k_T$ values needed to describe the data are too large to be viewed as purely `intrinsic' or non-perturbative in 
origin. But, as discussed in Reference~\cite{catani-sec1}, in
the standard formalism for direct photon production there are no double-logs to be resummed. This is  in contrast to double-arm observables such as Drell-Yan or diphoton production; since
direct photon production is, by definition, a single-arm
observable, there is no restriction of phase space for gluon
emission, and thus no double logarithmic 
enhancement to the $p_T$ distribution. The only enhancement
effects that survive arise from the purely `intrinsic' $k_T$
present in the colliding hadrons. 

	Nonetheless, there is generally a 
substantial amount of $k_T$ that results from the emission of soft gluons in hard scattering processes. 
Direct evidence of this $k_T$ has long been evident in Drell-Yan, 
diphoton and heavy quark measurements. The values of $<k_T>$/parton
for these processes
vary from $1$ GeV at fixed target energies to $3-4$ GeV at the 
Tevatron Collider. The growth is approximately 
logarithmic with center of mass energy.
(The value expected at the LHC for relatively low mass
states ($30-40$ GeV) is in the range of $6.5-7.0$ GeV.) 

Perturbative QCD corrections are insufficient to explain the size of
the observed $k_T$ and fully resummed calculations are required to 
explain Drell-Yan, W/Z and diphoton distributions~\cite{van31}.
These resummed calculations qualitatively describe the growth of the
$<k_T>$ with center-of-mass energy. Currently there is no
rigorous $k_T$-type resummation calculation available for  single photon
production, for the reasons cited above. 
In addition, this calculation is quite challenging in that the final state
parton takes part in soft gluon emission and in color exchange with 
initial state partons, in contrast with the Drell-Yan and diphoton
cases. Also, the calculation is complicated by the fact that several
overlapping power-suppressed corrections can contribute and, at high $x$,
threshold effects are important. 
 
Nevertheless, there
has been recent theoretical progress in single photon resummation~\cite{sterman1,mangano,li,sterman2}. In particular, in Reference~\cite{sterman2}, a technique has been presented for simultaneously treating recoil and threshold corrections
in single photon inclusive cross sections, working within the formalism of collinear factorization. In the preliminary
results, substantial enhancements have been observed, at moderate $p_T$ and $x$, from higher order perturbative and power-law non-perturbative corrections. This approach is still quite new and the efficacy of the formalism still
has to be evaluated. 

	There is an intuitive picture that describes the effects of this soft gluon radiation, both perturbative and
non-perturbative, on the direct  photon cross section. The
presence of soft gluon radiation, or $k_T$, can give a 
`kick' in the photon direction. Due to the steeply falling
cross sections, the $k_T$ kick can lead to the promotion of 
photons from lower $p_T$ to higher values of $p_T$. The
more steeply falling the cross section, the larger the
resulting enhancement. 
Using this intuitive picture,  the effects of soft gluon radiation can be 
approximated by a convolution of the NLO cross section with a 
Gaussian $k_T$ smearing function. The value of $<k_T>$ to be
used for each kinematic regime should be taken directly from
relevant experimental observables, given the lack of  a rigorous
formalism, rather than from a 
theoretical prediction. The behaviour of the $k_T$ smearing
correction is quite different for the Tevatron collider and for
fixed target experiments. 
For the Tevatron, there are
two points to note: (1) the agreement with the data is 
improved if the $k_T$ correction is  taken into account and
(2) the $k_T$ smearing effects fall off roughly as 
$1/p_T^2$~\cite{apana}. The latter behaviour is the expectation for 
such a power-suppressed type of effect and is the behaviour  
expected at the LHC, where the effects of the \kt\ ~smearing 
should not be important beyond $p_T$ values of 
$30$ GeV~\footnote{Similar \kt\  ~smearing 
effects should be present in all hard scattering cross sections, 
for example jet production at the Tevatron. The size of the 
experimental and theoretical
systematic errors in the low  $E_T$ region make such a
confirmation difficult.}. 

\begin{figure}[tp]
\begin{center}
\epsfxsize=8cm
\epsfysize=8cm
\mbox{\epsfbox{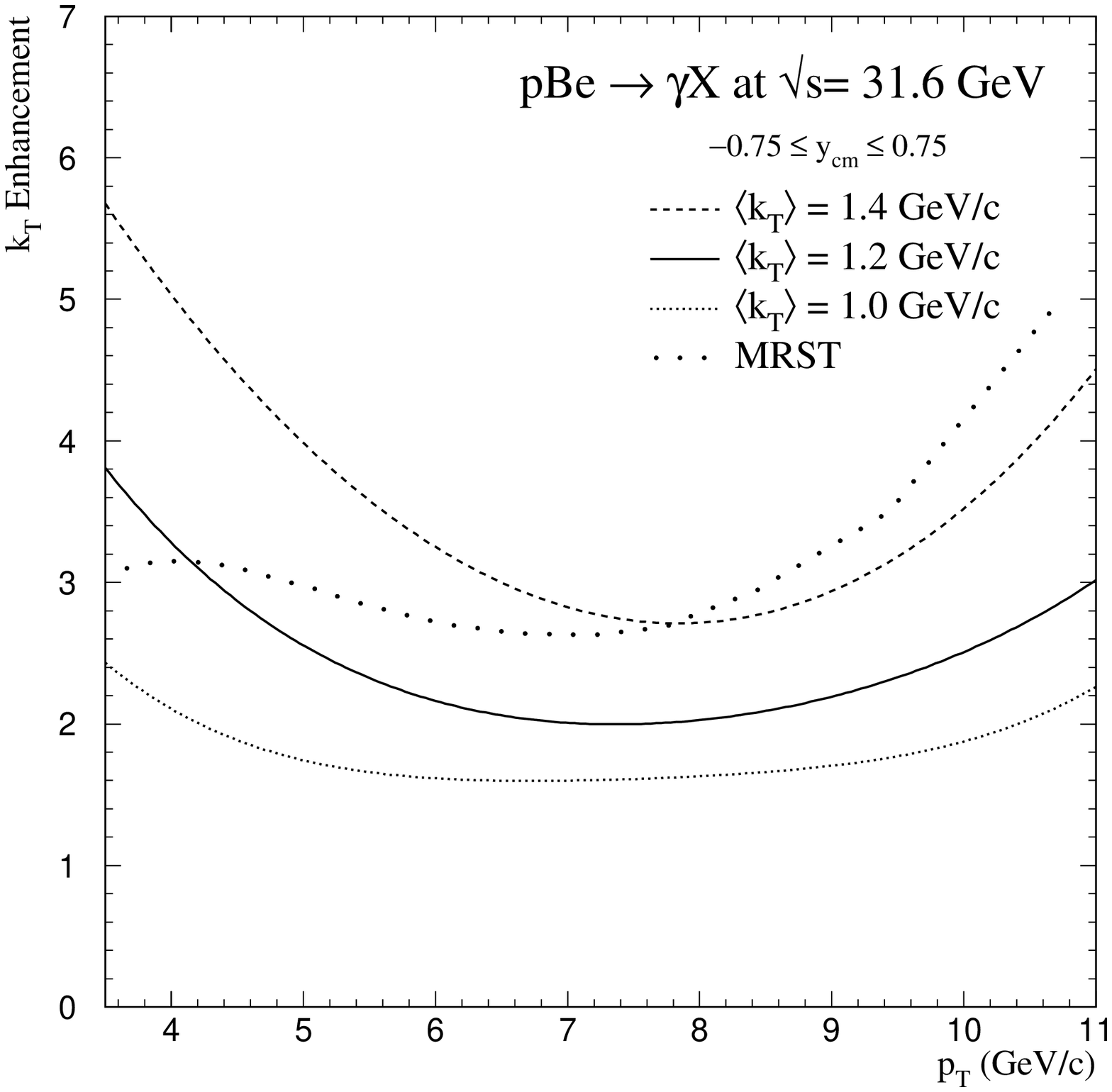}}
\end{center}
\caption{
\sf     The variation of $k_T$ enhancements (ratio of cross
sections with and without the $k_T$ corrections) relevant to
E706 direct photon data at $31.6$ GeV, for different values of
average $k_T$. In addition, the $k_T$ correction for E706 used in the
recent MRST fit is indicated.
} 
\label{fig:e706kt}
\end{figure}

        The $k_T$ ~correction obtained for E706 at a center-of-
mass energy of $31.6$ GeV  is shown in Figure ~\ref{fig:e706kt}.
The value of $<k_T>$ of $1.2$ GeV  was obtained from measurements of 
several kinematic observables in the experiment~\cite{apana}. The
$k_T$ smearing effect is much larger here then observed at the
collider and does not have the $1/p_T^2$ falloff. Also shown are
the $k_T$ corrrections using values of $<k_T>$ of $1.0$ and
$1.4$ GeV (a reasonable estimate of the range of experimental
uncertainty in the $<k_T>$ determination). In addition, the
$k_T$ correction for the E706 data used in the recent MRST pdfs is shown.
The MRST $k_T$ correction, utilizing a different model,  is larger  leading to a smaller
gluon distribution in the relevant $x$ range. (Both the CTEQ4
and MRST pdfs, with their respective $k_T$ corrections, lead to good agreement with the E706
direct photon cross sections.) The differences between the $k_T$ correction
~from Reference~\cite{apana} and that from the MRST pdfs can be taken
as an indication of the uncertainty in the value of this correction. 
Good agreement with the E706 direct photon and cross section 
at $\sqrt{s} = 31.6$ GeV is observed when  the nominal \kt\ ~correction of 
$1.2$ GeV is used; however, the allowed range of variation of  \avkt\ 
($1.0-1.4$ GeV) makes quantitative comparisons, and thus an extraction of the
gluon distribution, difficult~\footnote{NLO QCD predictions for fixed-target
direct photon production (as is also true for other fixed target processes) also contain a non-negligible renormalization
and factorization scale dependence, as discussed in Reference~\cite{catani-sec1}}.
Since the high $p_T$ E706 data
agrees well with CTEQ4M, it would thus disfavor the CTEQ4HJ pdf.
As stated before, however, a definitive conclusion must await a 
more rigorous theoretical treatment.

	Other related fixed target processes, such as $\pi^0$
production, in the same $p_T$ range as the measured direct
photon cross section, may perhaps shed some light on the puzzle. It has been noted~\cite{aurenche2} that essentially
all of the fixed target $\pi^0$ cross sections disagree with 
NLO predictions, by essentially a constant factor. Thus, there may be a common problem causing the deviations, such as uncertainties in the  high $z$ quark and gluon fragmentation functions and possible sizeable higher order corrections. In addition, the importance of the high $z$ fragmentation region implies the need for threshold
resummation techniques to be applied, in processes with non-trivial 
color flow. 

	It is worthwhile pointing out, though, that the
same $k_T$ model used for for single photon production was shown to also provide an adequate description of the experimental $\pi^o$ cross sections~\cite{apana, apana2}. As in the
case of direct photon production, the controversy regarding
the theory/data discrepancies is still open. The $\pi^0$ cross sections may form a crucial role in the ultimate understanding for a number of reasons: if $k_T$ are important for photon production, they should also have a measureable impact on the $\pi^0$ cross sections as well. In addition,  $\pi^0$'s form the 
primary experimental background to direct photon production. 

Finally, it is not clear if any theoretical treatment for photon production is capable of describing all of the current fixed target direct photon data. 
There are discrepancies between the different experiments which
may imply experimental difficulties, which are in addition to any of the
theoretical problems discussed above.

\subsection{Influence of Jets}

An important process that is sensitive to the gluon distribution is jet
production in hadron-hadron collisions. Processes responsible for jet
production include gluon-gluon, gluon-quark and quark-quark(or anti-quark)
scattering. Precise data on 
jet production  at the Fermilab Tevatron are now available over a wide
range of transverse energy, and the theoretical uncertainties in most of 
this range are well-understood. Thus, it is to be expected that  jet production
can provide a good constraint on the gluon distribution.

        The  jet data that  has been utilized in global pdf fits
has been from the CDF and D0 collaborations~\footnote{The experimental
and theoretical errors associated with the UA2 jet cross section make its
use in pdf fits difficult.}.
 The  data cover a wide kinematic range ($E_T$ values
~from $15$ to $450$ GeV corresponding to an $x$ range of $0.02$ to $0.5$).
The  CDF jet data from Run IA were  utilized in the CTEQ4HJ
pdf fit~\cite{cteq4hj}. Here, a large emphasis was given  to
the high $E_T$ data points which show a deviation from NLO
QCD predictions with ``conventional'' pdfs. Given the  lack of
constraints on  the  high $x$ gluon distribution discussed in
Section VI, the extra emphasis on the high $E_T$ region was
enough to cause a significant increase in the gluon 
distribution; for example, the gluon distribution at an $x$
value of $0.5$ ($Q=100$ GeV) increases by a factor of two.
Since the  dominant jet subprocess in this region is $\overline{q}q$
scattering the increase in the gluon distribution of a 
factor  of two causes only a 20\% increase in the jet cross
section. This is sufficient to pass through the bottom of
the CDF high $E_T$ jet error bars. The preliminary
jet cross sections from Run 1B (90 $pb^{-1}$) from both the
CDF and D0 experiments were used in the CTEQ4M fits,
but with statistical errors only and only for $E_T$ in the 
range $50-200$ GeV. The points with $E_T$ lower than $50$ GeV
have substantial systematic errors on both the
theoretical and experimental sides while the 
points with $E_T$ higher than $200$ GeV contain the 
CDF excess. The 
inclusion of the jet data serves to considerably constrain
the gluon distribution over the $x$ range of $0.1$ to $0.2$. 
The resulting  gluon (CTEQ4M) does not
decrease the excess observed by CDF at high $E_T$. 

\begin{figure}[tp]
\begin{center}
\epsfxsize=6cm
\epsfysize=6cm
\mbox{\epsfbox{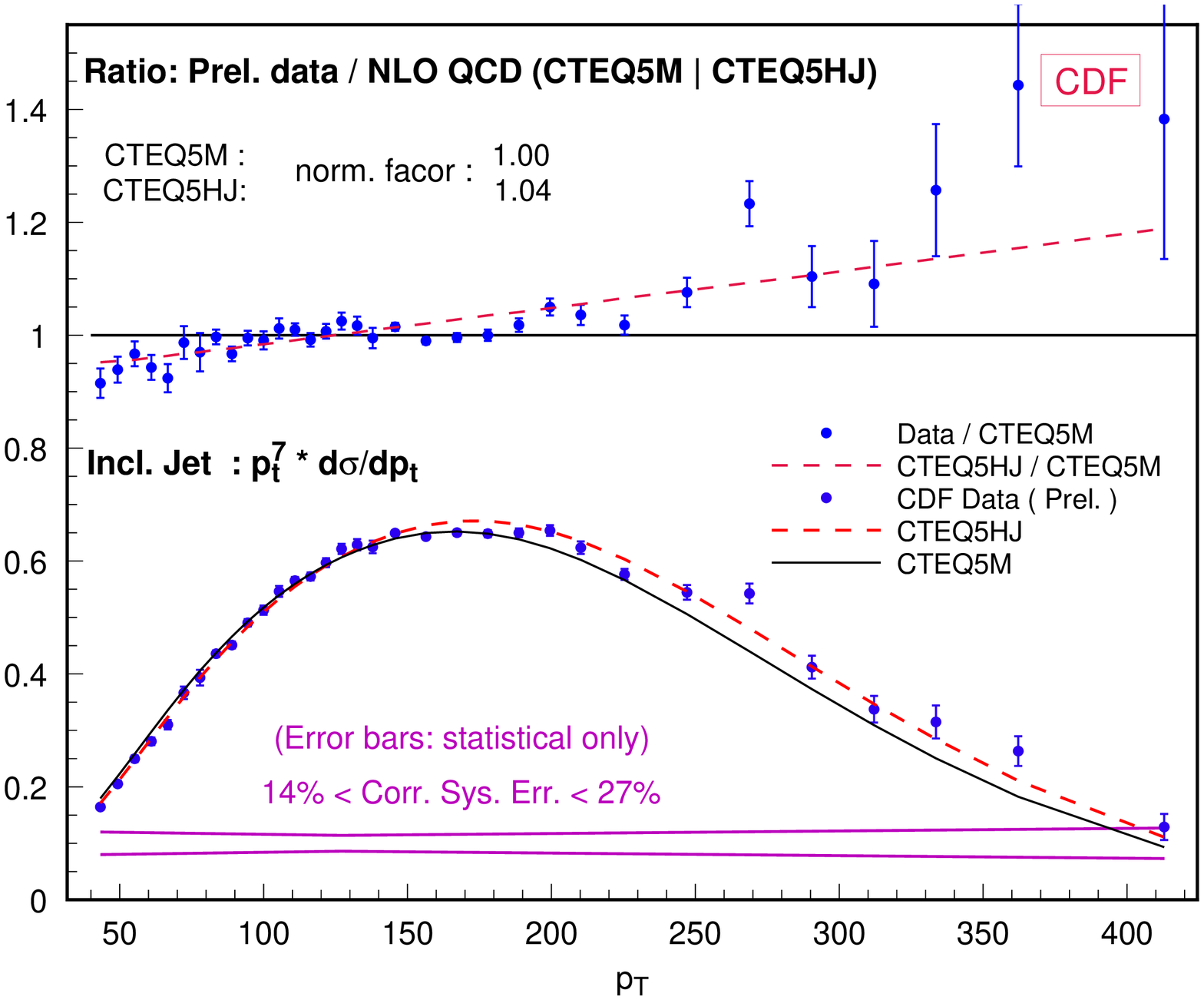}}
\end{center}
\caption{
\sf A comparison of the Run 1B CDF  
inclusive jet cross section to the CTEQ5 fits.
The bottom plot shows the measured cross section multiplied by $p_T^{~7}$ in 
order to allow a linear display. The top plot shows the ratio of the measured
cross section to that calculated with CTEQ5M, as well as the ratios of CTEQ5HJ
to CTEQ5M.
} 
\label{fig:cdfjet}
\end{figure}

        The published D0 jet cross section~\cite{D0jet}  along with the  
(soon-to-be published) CDF jet cross section~\cite{cdfjet}  from Run 1B
were used in the recently released CTEQ5 parton distributions. 
The fits use the full $E_T$ range for  the  cross sections and use
the correlation information on the  systematic errors as  
contained in the covariance matrices for both experiments. 
The two experiments are in agreement
with each other except for a slight normalization shift~\footnote{
A shift on the  order of 3\% is expected since the two experiments
use values for the total inelastic cross section that differ by 
that amount.};
the two highest $E_T$ data points for CDF are above those
for D0, but both experiments have large statistical errors in
this region.  As can be seen in Figure ~\ref{fig:cdfjet}
the NLO QCD prediction with the CTEQ5M
pdf  is in good agreement with the CDF data. The conclusions are
exactly the same for the D0 jet data. 
The CTEQ5M gluon is very similar to CTEQ4M,
except perhaps at very high x. The CTEQ4HJ pdf has been updated to
complement the new CTEQ5M pdf. 
The CTEQ5HJ pdf gives almost as good a global
fit as CTEQ5M to the full set of data on DIS and DY processes, and has the
feature that the gluon distribution is significantly enhanced in the high
$x$ region, resulting in improved agreement with the observed trend of 
jet data at high $E_T$ in both the CDF and D0 experiments.

\section{Systematic Uncertainties}

There is currently an increasing awareness of
the need and possibility of propagating errors in the data into error
estimates on parton distribution functions~\cite{SC,proclhc,run2}.
Ideally, one might hope to perform a full error
analysis and provide correlated errors for all the parton 
distributions determined in a global fit. 
This goal is difficult to carry out for several
reasons. Firstly, there is no established way of quantifying the
theoretical uncertainties for the diverse physical processes that 
are used. More pragmatically,  
only a subset of the experiments usually involved in global analyses provide 
correlation information on their data sets in a way suitable for the analysis.
In these circumstances, comparing data from different experiments 
becomes very difficult.
Furthermore the standard fitting procedure introduces methodological 
uncertainties due in particular to the necessity of choosing specific 
choices of parametrization. All of these uncertainties are of course all 
highly correlated. We discuss each in turn.

\subsection{Theoretical Uncertainties}

The most important theoretical uncertainty in the determination of parton
densities is the truncation of the resummed perturbation series at
NLO. Consistent NNLO determinations will require NNLO splitting
functions: there has recently been some progress in this direction~\cite{NNLO},
and it is hoped that NNLO calculations might be available before the
LHC is turned on. Meanwhile there are some `approximate NNLO'
calculations~\cite{aNNLO}, which attempt to reconstruct the 
NNLO splitting functions from their known integer moments and 
behaviour at large and small $x$:
these analyses suggest that NNLO corrections might reduce theoretical
uncertainties due to truncation of the perturbative expansion by 
at least a factor of two.

One of the most important consequences of the theoretical uncertainty 
~from unknown NNLO corrections is that it currently limits the accuracy of 
most of the experimentally more reliable determinations of $\alpha_s$.
This in turn inevitably limits the accuracy of all extrapolations from
low to high $Q^2$: for example one of the largest uncertainties in the 
prediction of the $W$ and $Z$ cross-sections is that due to the 
uncertainty in $\alpha_s$ \cite{MRSTWZ}.

Uncertainties at low $Q^2$ due to higher twist may be estimated from
phenomenological fits: recent studies~\cite{BY,AK} have shown that there are
important correlations between empirical higher twist and the value of
$\alpha_s$. It has also been shown that the fitted higher twist
contribution drops when estimates of NNLO corrections are
included~\cite{KPS}. The empirical higher twist is qualitatively 
consistent with renormalon estimates. Taken together, these 
observations suggest that 
it is difficult to disentangle genuine higher twist from higher order
perturbative corrections: the true higher twist contribution
may be much smaller than is suggested by the fits.

The correct treatment of heavy quarks close to threshold was developed
some time ago~\cite{ACOT}; more recently it was proven that this procedure 
works to all orders in perturbation theory~\cite{Coll}. This treatment is
now included in some of the CTEQ fits~\cite{cteq5,LT}; a 
closely related but not identical procedure is used by MRS~\cite{mrst98}.
A simpler version of ACOT, which nonetheless accurately reproduces 
its essential features, has also been developed~\cite{KOS}.

An accurate treatment of heavy quark production, and indeed $W$ and
Higgs production, requires the resummation of threshold
logarithms. Recently it has been suggested that
resummation of soft gluons may solve some of the problems with
prompt photons~\cite{sterman1,mangano,li,sterman2}. A fully consistent treatment will require
the inclusion of soft gluon resummations in parton
determinations, but as yet this has not been attempted. Renormalon
studies suggest that such resummations may substantially improve 
the reliability of perturbation theory at large $x$. Again there will 
be strong correlations with higher twist. It would be particularly 
interesting to see the effect of such resummations on the 
predictions for the parton-parton luminosities eq.\ref{eq:lum} in 
the region relevant for Higgs production at the LHC.

The resummation of high energy (small $x$) logarithms is more
problematic. Present data suggest that their effect on inclusive cross
sections must be very small, at least at HERA and the Tevatron 
if not at the LHC. Furthermore, 
conventional theoretical approaches~\cite{ktfac,sums} based on 
summations of LLx and NLLx~\cite{fl} corrections have been shown to break 
down: the NLLx corrections are overwhelmingly large and negative~\cite{brus}. 
Various suggestions for the resummation of these large corrections 
have been put forward~\cite{blm,salami,sch,bfnllx}. Hopefully a 
detailed phenomenological analysis based on one or other of these
procedures will eventually provide a reliable estimate of the 
error due to uncertainties in small $x$ evolution when using parton
distributions measured at HERA to predict those to be used at the
LHC. 

\subsection{Combining Different Experiments} 

On the experimental side, one of the major problems with combining results 
~from different experiments lies in the degree
of `rigour' in the interpretation of the experimental errors. Experimental 
results may be conveniently expressed as probabilities 
$P({\rm data}|{\rm theory})$, i.e. the probabilities of obtaining the 
given set of data given a certain theoretical prediction \cite{agostini}. 
Often these 
probabilities are expressed in terms of predictions and (Gaussian) errors:
for a given experiment, $P(d|t)=\exp(-\half\chi^2(d|t))$, where 
$d$ are the data, $t$ the theoretical predictions
and 
\begin{equation}
\chi^2(d|t) = \sum_{\rm data} (d-t)\Sigma^{-1}(d-t)
\label{eq:chisq}
\end{equation}
where $\Sigma$ is the matrix of correlated errors. Maximizing the 
probability, and thus obtaining the most likely `prediction', then 
corresponds to minimizing the $\chi^2$. It should 
be emphasized that it is not necessary to present experimental results in this
way, and in particular some systematics may be completely non-Gaussian;
however if the experiment is to be useful it must always provide a 
(clear) estimate of $P(d|t)$, otherwise the error analysis is at best 
incomplete and at worst useless.

In the present situation, the predictions will be constrained 
functionals of the input pdfs (the constraints being the result of 
perturbative evolution and cross-sections). If the errors have 
been estimated correctly, and the theory which constrains the 
predictions is sufficiently accurate, then there should be pdfs for 
which the $\chi^2$ per degree of freedom is of order unity. Unfortunately 
for many important datasets this is not the case, and thus if one were to  
insist on the rigour of the statistical method, then many important 
experiments would not be included in the analysis~\cite{proclhc, GK}. Such a 
strict criterion is probably unrealistic: rather the emphasis should 
be placed on using the maximal experimental constraints from experimental 
data \cite{SC}. In this case the standard statistical techniques 
may not apply, but must be supplemented by physical considerations, 
taking into account experimental and theoretical limitations~\cite{run2}.

As an example of how this works in practice, we consider a recent CTEQ 
error analysis of the $W$-production cross-section \cite{proclhc,run2}. 
This uses the standard CTEQ5 analysis \cite{cteq5} as
the starting point: there are fifteen experimental data sets, with a 
total of $\sim 1300$ data points, and experimental errors are generally
treated by ignoring correlations and combining statistical and 
systematic errors in quadrature (so $\Sigma$ in eq.(\ref{eq:chisq}) is
taken to be diagonal, with each diagonal entry set to 
$\sigma_{\rm stat}^2+\sigma_{\rm syst}^2$ of the corresponding data point). 
The initial pdfs are parameterised by $18$ parameters {$a_i, i=1,\dots,18$}:
each theoretical prediction is then a function of these parameters.
The `best-fit' distribution (CTEQ5M1 in this case) is then given by the 
set of parameters $a$ which minimise 
$\sum_{\rm expts}\sum_{\rm data}\chi^2(d|t[f(a)])$,
where $t[f(a)]$ are the theoretical predictions for each data 
point given the pdf $f(a)$ for the fifteen base experimental data sets. 

A natural way to find the limits of a physical observable which depends on 
the pdfs, call it ${\cal O}[f(a)]$, such as the $W$-production 
cross-section $\sigma_{W}$ at $\sqrt{s}=1.8$\,TeV, is then to study the 
dependence of the total $\chi^{2}$ on ${\cal O}$.
An efficient way of doing this is to use Lagrange's method of
undetermined multipliers: one minimizes
\begin{equation}
F(\lambda)
=\sum_{\rm expts}\sum_{\rm data}\chi^{2}(d|t[f(a)])+\lambda {\cal O}[f(a)]
\end{equation}
for fixed $\lambda$, and then varies $\lambda$ in order to map out the 
$\chi^{2}$ as a function of ${\cal O}$.

\begin{figure}[tp]
\centering
\mbox{\includegraphics[width=0.4\textwidth,clip]{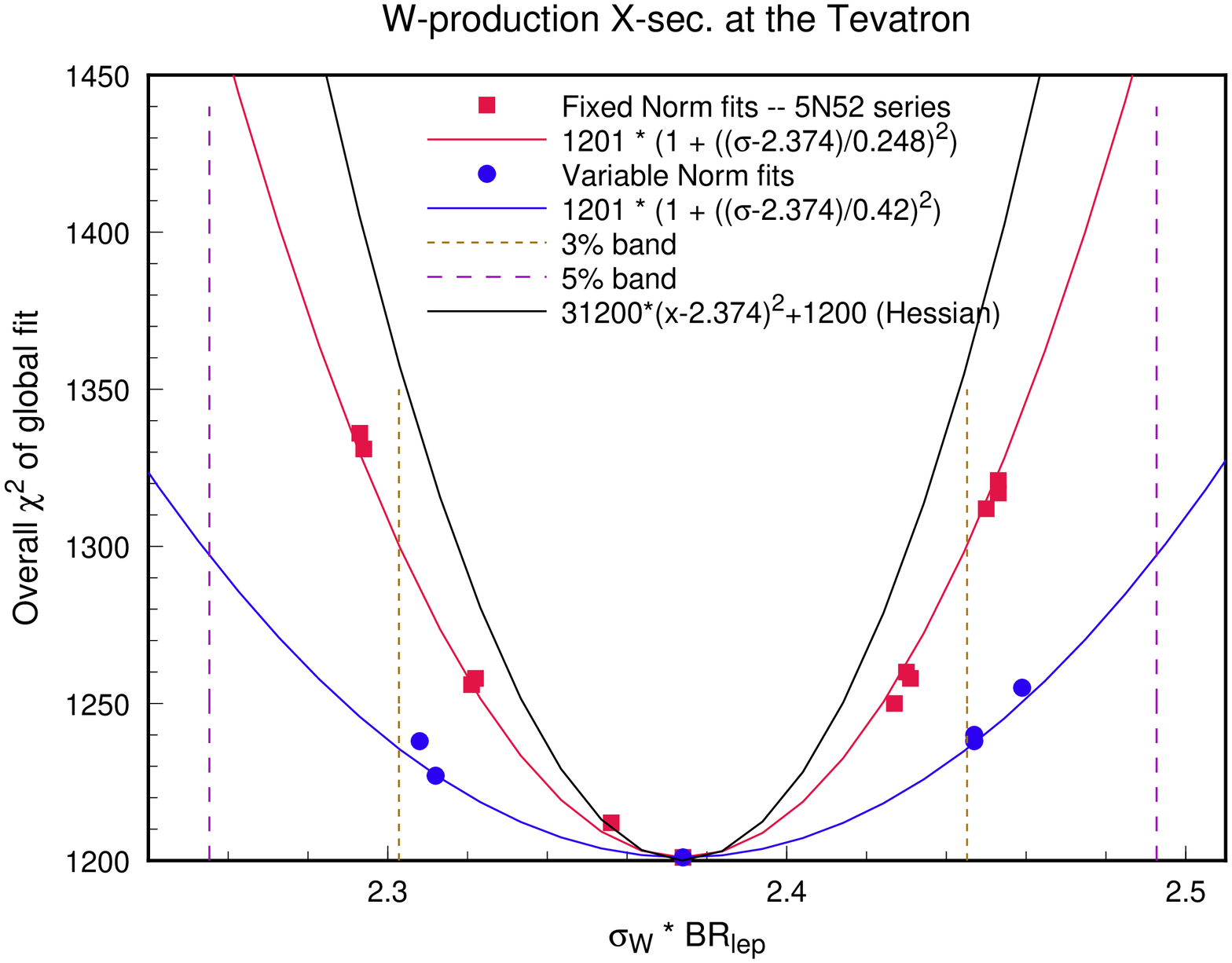}}
\hspace{2cm}
\mbox{\includegraphics[width=0.4\textwidth,clip]{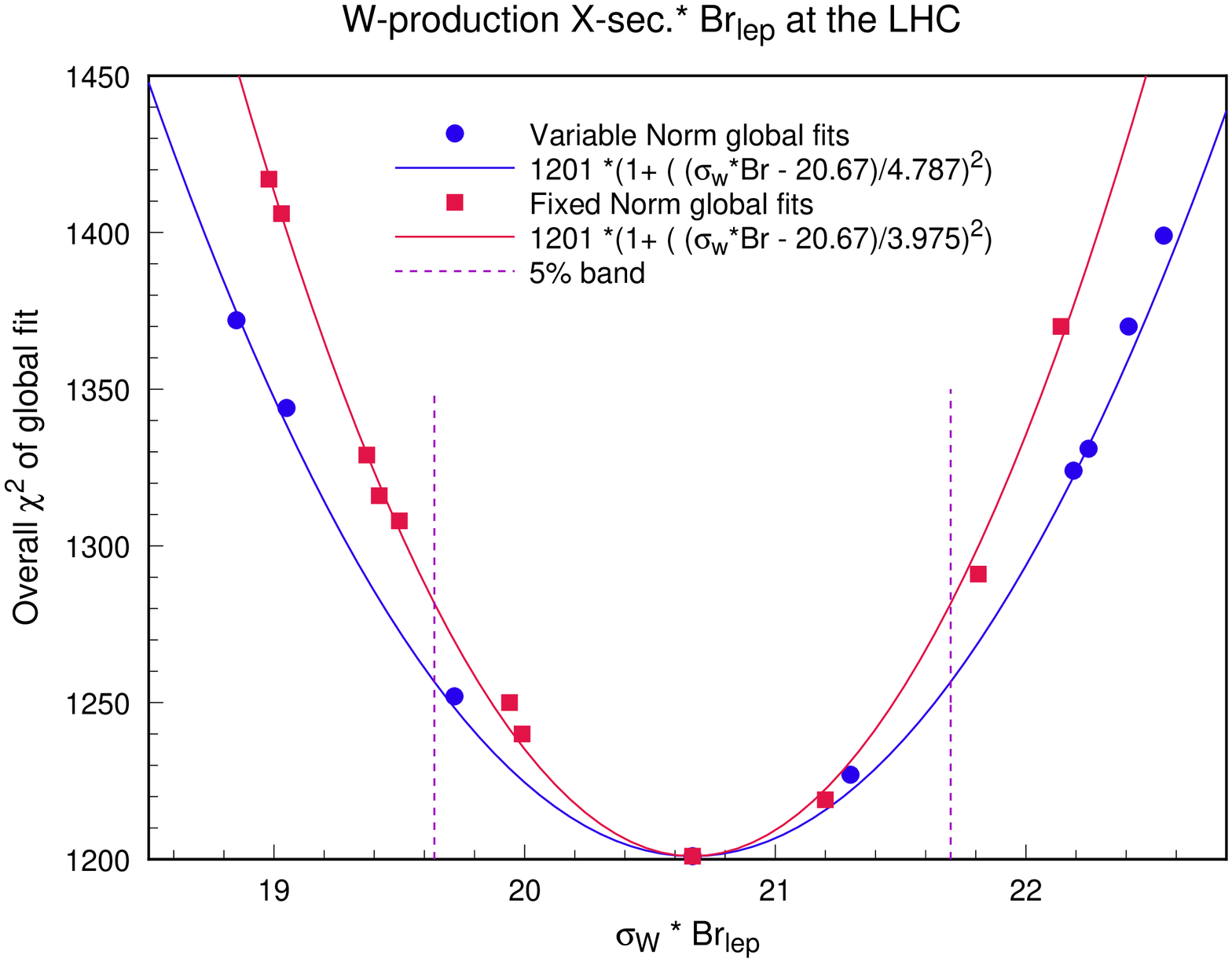}}
 \caption{\sf $\chi^2$ of the base experimental data sets vs. the W
 production cross-section at the Tevatron and LHC.}
 \label{fig:WprodB}
\end{figure}

Figs.~\ref{fig:WprodB}a,b show the $\chi^{2}$ for the fifteen base experimental
data sets as a function of $\sigma_{W}$ at the Tevatron and LHC energies
respectively \cite{proclhc}. Two curves with points corresponding 
to specific global fits are
included in each plot\footnote{The third line in Figs.~\ref{fig:WprodB}a
refers to an alternative technique \cite{proclhc} based on the assumption of 
Gaussian errors in the parameters $a_i$.}: one obtained with all experimental
normalizations fixed; the other with these included as fitting parameters
(with the appropriate experimental errors).  
We see that the
$\chi^2$'s for the best fits corresponding to various values of the W
cross-section are close to being parabolic, as expected. Indicated on the
plots are 3\% and 5\% ranges for $\sigma_W$. The two curves for the Tevatron
case are farther apart than for LHC, reflecting the fact that the W-production
cross-section is more sensitive to the quark/anti-quark distributions and
these are tightly constrained by existing DIS data.

The important question is: how large an increase in $\chi^{2}$ should be taken
to define the likely range of uncertainty in ${\cal O}$? 
The elementary statistical
theorem that $\Delta\chi^{2}=1$ corresponds to one standard deviation of the
measured quantity ${\cal O}$ relies on assuming that the errors are gaussian,
uncorrelated, and with their magnitudes correctly estimated. Because these
conditions do not hold here, this theorem cannot be naively applied
quantitatively: rather one must examine in detail how well the fits along
the parabolas shown in Fig.\ref{fig:WprodB} compare with the individual
precision experiments included in the global analysis, in order to arrive at
reasonable quantitative estimates on the uncertainty range for the W
cross-section. In the meantime, based on past (admittedly subjective)
experience with global fits, it seems that  a $\chi^2$ difference of $40-50$
points represents a `reasonable' estimate of current uncertainty of parton
distributions. This implies that the uncertainty of $\sigma_{W}$ is about 3\%
at the Tevatron, and 5\% at the LHC. 

\subsection{Correlated Experimental Systematics }

There is now an increasing awareness of the necessity and possibility of 
carrying out a careful treatment of correlated systematic errors when 
attempting to determine errors on pdfs. For example 
a systematic study of the uncertainties in the parton distribution
in the small $x$ region has been made recently by experimentalists at H1 and 
ZEUS~\cite{ZEUSglue,Bot}. These studies include a proper treatment
of correlated systematic errors, and some attempt is made to quantify 
parametrization uncertainties. Similar studies of the errors in
polarized parton densities have been made by the SMC~\cite{SMC}. Besides
showing that careful estimates of parton uncertainties are useful 
and necessary, these studies also show that it is possible to include
correlated systematics and combine data sets from different (albeit similar)
experiments in a meaningful way. However they also show that doing
something similar for a global parton determination would be very
difficult and extremely tedious, unless new techniques are developed.

The importance of correlations in experimental systematic errors has
been underlined by a recent reanalysis~\cite{Alek} of the $F_2$ BCDMS data.
A more careful treatment of the correlations between data taken at
different beam energies, and the correlations between the fitted
parton distributions and higher twist, results in a significant increase
in the value of $\alpha_s(M_Z^2)$ extracted from the data: Alekhin
quotes a value of $0.118\pm 0.002$. This is consistent with the current world
average and the value $0.119\pm 0.002$ recently extracted from the
reanalysed CCFR data~\cite{CCFR} (though after a more careful treatment of
correlated higher twist~\cite{AKCCFR} this rises to $0.122\pm 0.005$).  

In this context it should be noted that in
the usual global analyses, in which correlations between systematic
errors are ignored, and higher twist effects are not included, 
neither the BCDMS or the CCFR $F_2$ data show a
minimum in their $\chi^2$ as $\alpha_s$ is varied~\cite{mrst98,cteq4},
despite the fact that when treated separately each is capable of yielding an
excellent determination of $\alpha_s$. Only the minima in the H1 and
ZEUS datasets are strong enough to survive this treatment: this may be
helped by the fact that empirical higher twists are very small at
small $x$~\cite{Rome}. It will be interesting to repeat the preliminary 
determination~\cite{HERAsc} using the 95-97 HERA datasets when these
finally become available.

\subsection{Methodological Issues}

While the issues addressed in the previous three sections are no doubt all 
important, there are also some methodological issues which need to be 
considered if we are to achieve our aim of a reliable determination of 
the errors in a global determination of parton distributions. In 
particular, we need a technique which can give parton distribution 
functions and their errors, such that:

(i) there is no inbuilt methodological bias (for example dependence on
a particular parametrization of the input distributions)

(ii) it is easy to propagate the effects of correlated systematic
errors in the data to correlated uncertainties in the parton
distributions

(iii) it is easy to add new data sets or estimate theoretical errors or 
test models of new physics without redoing the whole of the analysis.

All of these criteria can be met if we `quantise' our parton
distributions: instead of trying to determine a single `best fit' set
of parameterised parton distributions with an associated error matrix,
we construct an ensemble of sets of partons, distributed according to
how well they fit the data~\cite{proclhc,BCS,GK,Kos,rdbmor}. 
The expected result for a parton dependent observable, call 
it ${\cal O}[f]$, would then be 
given by an ensemble average: 
\begin{equation}
\langle {\cal O}[f]\rangle = {\cal Z}^{-1} \int [{\cal D}f]
\,{\cal O}[f]\,J[f]\,s[f]\,
\prod_{\rm expts} P(d|t[f]),
\label{eq:funint}
\end{equation}
where $\int[{\cal D}f]$ means functional integration over all possible
input distributions $f$ (subject to basic constraints such as sum
rules and positivity) and  ${\cal Z}=\langle 1\rangle$ is a normalization
factor. The measure of integration is given essentially by the probability
distributions $P(d|t[f])$ for each of the experiments used as input.
These probabilities are, as explained above, the essential input of 
the experimental data used in the fit: they support distributions which 
fit the data well, and suppress the contribution of distributions which
fit badly. If the errors on the data were assumed Gaussian, these 
probabilities would come in the form of a $\chi^2$, as in 
eq.(\ref{eq:chisq}), though the technique does not depend on such an 
assumption, and non Gaussian errors could also be incorporated.
There is also a Jacobian factor $J[f]$, which turns the integration measure 
from an integration over theoretical predictions $t[f]$ to one over the 
pdfs themselves, and enforces the theoretical constraint that the 
theoretical predictions are related through pQCD. 
It is also necessary to introduce a `smoothness' factor $s[f]$ 
into the measure, to enforce the natural theoretical prejudice 
that the initial pdfs should be smooth functions of $x$, 
without wiggles or jumps: a suitable form for such a factor would be 
$\exp -\half\varepsilon \sum_x (\partial_x f)^2$, where  $\varepsilon$ is a
small parameter which quantifies the extent of this prejudice. 
Final results should be independent of the form of this term, and 
in particular the parameter $\varepsilon$, provided that it is 
varied in a suitable range.

The way in which this procedure works should now be clear, since it is
similar to the quantum mechanics (or more precisely statistical
mechanics) of a particle in a (highly nonlocal) potential~\cite{BCS}: 
the parton distributions may be thought of as quantum fields, with, 
in the case of Gaussian experimental errors, the action 
\begin{equation}
{\cal S}[f] = \half\sum_{\rm expts} \sum_{\rm data} 
(d-t[f])\Sigma^{-1}(d-t[f])
+\half\varepsilon \sum_x (\partial_x f)^2.
\end{equation}
The best fit parton distribution is then the
solution of the classical equations of motion (since it minimises the
action), while the error bands are given by the `quantum' fluctuations
around the classical field. Since the determination of the classical
field is itself nontrivial, the system is best solved numerically: we
discretise the field by introducing a parametrization with a finite
number of parameters $a_i$, $i=1,\ldots,N$, so that $\int[{\cal D}f]\,J[f]\to 
\prod_i d a_i\,J(a_i)$, rather as we would for a lattice field theory.
Here the best discretization would not necessarily be a naive 
discretization in $x_{\rm Bj}$ with spline interpolation: rather it 
might involve expansion of each pdf in sets of orthogonal 
polynomials, or other sets of (orthogonal) functions, for example 
eq.(\ref{eq:pdf}) and its obvious generalizations.
The integration over the parameters $a$ would then be done by Monte Carlo,
using an algorithm such as Metropolis or HMC~\cite{HMC} to generate an
ensemble of configurations distributed according to the measure of
integration, and thus according to its likelihood given the input datasets. 
Finding each such configuration will involve a similar
computational effort to that of finding a best fit. Finally, we would
like to increase the number of parameters $N$ (taking the `continuum
limit') until we are sufficiently close to a truly parametrization
independent ensemble, at which stage we can readily compute
expectation values of observables and their associated errors as averages 
over the ensemble of pdfs.

This procedure has several advantages:

(i) it is intrinsically parametrization independent as the number of
parameters increases, because of the universality of the continuum
limit. Flat directions are no longer the problem that they are in a 
best fit procedure: the total number of parameters is now limited 
only by computational resources. Indeed the flat directions are now
interesting, since they give the most important uncertainties in the
parton distribution functions.

(ii) the propagation of correlated systematics is automatically taken care
of by the procedure. The only limitation is the reliability of the
probabilities $P(d|t)$ produced by experimentalists. This should give added
impetus to the determination of meaningful (and thus comparable) 
estimates of systematic errors by different experimental collaborations,
and their presentation in such a way that they can be readily input into 
such an analysis. Preliminary explorations of the technique \cite{proclhc} 
indicate that the errors in the pdf parameters are not only highly 
correlated, but also in many cases significantly non-Gaussian, even when 
the errors in the data are assumed to be Gaussian.

(iii) Data from new experiments can be added using the old configurations,
since different experiments are (in principle!) statistically
independent, so ${\cal S}_{\rm tot}[f]=\sum_{\rm expts}{\cal S}_{\rm exp}[f]$.
Similarly we could estimate theoretical errors due, for example, to 
NLO truncation, by using the standard configurations reweighted 
by varying renormalization and factorization scales. 
Similarly, we could test the effect of resummations 
by reweighting the configurations generated using 
standard NLO evolution, or indeed test 
models for new physics by reweighting the configurations 
generated using the Standard Model~\cite{GK}.

The main problems to be faced in actually implementing the procedure
are computational: we need a fast evolution code, and high performance
computing. The advantages of parallelization should be obvious. In
fact the computational requirements are very similar to those of 
the lattice gauge theorists: calculating the `action' is more difficult, 
but the `continuum limit' should be reached much more quickly. 

\section{From here to the LHC and Beyond}

\subsection{Progress Before the LHC Turns on}

Perturbative QCD has been extremely
successful in describing data in DIS, DY and jet production, as well as
describing the evolution of parton distributions over a wide range of
$x$ and $Q^2$. From the point of view of pdf determination, the 
primary problem lies in the calculation of the direct photon cross 
sections which could serve as a primary probe of the gluon distribution 
at high $x$. However, a rigorous
theoretical treatment of soft gluon effects (perhaps requiring both  $k_T$ and
Sudakov resummation) will be required before the data can be used  with
confidence in pdf fits. On the experimental side, it will also be 
necessary to resolve the inconsistency between the WA70 and E706 data.

D0 has recently presented a new result for the measurement of the inclusive 
jet cross section as a function of the jet rapidity (up to values of three)
~\cite{levan}. Such a measurement probes a greater kinematic range than
the central inclusive jet cross sections. In addition,
the differential dijet data from the Tevatron explore a wider kinematic
range than the inclusive jet cross section. Both CDF and D0 have 
dijet cross section measurements from Run I which may also serve probe the
high $x$ gluon distribution, in regions where new physics is not
expected but where any parton distribution shifts should be observable. 
The ability to perform such cross-checks is essential.

CDF and D0 will accumulate on the order of 
2-4 $fb^{-1}$ of data in Run II (from 2000-2003), a factor of 20-40 
greater than the current sample. This sample should allow for more 
detailed information on parton distributions to be extracted from direct
photon and DY data, as well as from jet production. Run III (2003-2007)
offers a data sample potentially as large as 30 $fb^{-1}$.

H1 and ZEUS will continue the analysis of the data taken with
positrons in 1991-97. HERA switched to electron running in 1998 and plans
to deliver approximately 60 
In 2000, the  HERA machine will be upgraded for high luminosity
running, with yearly rates of 150 
integrated luminosity of about 1 $fb^{-1}$ by 2005. This will allow for
an error of a few percent on the structure function $F_2$ for $Q^2$ scales
up to $10^4 ~GeV^2$. The gluon density, derived from scaling violations of
$F_2$, should be known to an accuracy of less than 3\% in the kinematic
range $10^{-4}<x<10^{-1}$.

It is also hoped that over the next five years the Monte Carlo 
outlined in the previous section will begin to bear fruit, 
perhaps to the point where they can make a serious contribution to global pdf 
error analysis.

\subsection{Physics cross sections at the LHC and the role of LHC data in pdf determination}

ATLAS measurements of DY (including W and Z), direct photon,
jet  and top production will be extremely useful in determining 
pdfs relevant for
the LHC. The data can be input to the global fitting programs, where it 
will serve to confirm/constrain the pdfs in the LHC range. 
Again, DY production will provide information on the quark (and anti-quark)
distributions while direct photon, jet and top production will provide, in
addition, information on the gluon distribution. 

Other processes might also prove useful. For example diphoton production 
might be useful for determining the gluon distribution, and this in turn 
would lead to an improved knowledge of the relevant parton pdfs and 
parton-parton luminosity functions for the production of the 
Higgs (which is largely due to $gg$ scattering for
low to moderate Higgs' masses). 

Another possibility that has been suggested is to directly determine
parton-parton luminosities (and not the parton distributions per se) by
measuring well-known processes such as W/Z production~\cite{dittmarr}. This
technique would not only determine the product of parton distributions in the 
relevant kinematic range but would also eliminate the difficult
measurement of the proton-proton 
luminosity. It may be more pragmatic, though, to continue to separate out
the measurements of parton pdfs (through global analyses which may contain
LHC data) and of the proton-proton luminosity. The measurement of the latter
quantity can be pegged to well-known cross sections, such as that of the W/Z,
as has been suggested for the Tevatron.  

\section{Conclusions}

The determination of parton distributions and uncertainties is 
an important ingredient of our preparations for physics at the LHC.
The global fitting techniques used for the past fifteen years may soon
be superseded by more sophisticated methods. Developing and exploiting
these techniques will be a great challenge to theorists and
experimentalists alike.

\section{\bf Acknowledgements} 
RDB would like to thank Sergey Alekhin, 
John~Collins, Tony~Doyle, Stefano~Forte, Stefane~Keller, 
Tony~Kennedy, David~Kosower, 
Brian~Pendleton, Dave~Soper, James~Stirling, Wu-Ki~Tung and 
Andreas~Vogt for various stimulating and useful discussions. 
JH would like to thank James Stirling, Steve Mrenna and his CTEQ 
colleagues for useful comments.  
We would also like to thank James Stirling and Lenny Apanasevich for 
providing many of the figures. 
This work was supported in 
part by an EU TMR  contract FMRX-CT98-0194 (DG 12 - MIHT) and by the 
NSF under grant PHY-9901946.



\setcounter{figure}{0}
\setcounter{table}{0}
\setcounter{section}{0}
\setcounter{equation}{0}
\newpage



\def\D0{D{\O}~}
\newcommand{\NPB}[3]{{Nucl.~Phys.}~{\bf B#1}, #2~(19#3)}
\newcommand{\PLB}[3]{{Phys.~Lett.}~{\bf B#1}, #2~(19#3)}
\newcommand{\PRL}[3]{{Phys.~Rev.~Lett.}~{\bf #1}, #2~(19#3)}
\newcommand{\PRD}[3]{{Phys.~Rev.}~{\bf D#1}, #2~(19#3)}
\newcommand{\ZPC}[3]{{Z.~Phys.}~{\bf C#1}, #2~(19#3)}
\newcommand{\CPC}[3]{{Comput.~Phys.~Commun.}~{\bf #1}, #2~(19#3)}
\newcommand{\lesssim}{\stackrel{\scriptscriptstyle<}{\scriptscriptstyle\sim}}

\begin{center}
\vspace*{1.2cm}
{\Large\sc \bf Generalized factorization and resummation} \\
\vspace*{1.cm} 
{\sc C. Bal{\'a}zs, J.C. Collins, D.E. Soper}
\vspace*{1.cm}
\end{center}

\setcounter{footnote}{0}

\begin{abstract}
In this section we summarize the formalism which extends the usual hadronic
factorization theorem to the low transverse momentum region for the
inclusive production of colorless final states, while resumming logarithms
with the ratio of the invariant mass and transverse momentum. Among the
various recent applications the calculation of the $Z^0$ and Higgs boson
transverse momentum distributions are highlighted.
\end{abstract}

\section{The Collins-Soper-Sterman formalism}

The standard factorization formula fails near kinematic boundaries.
We discuss the case of low transverse momentum in $Z_0$ production,
etc; this is an important case because the cross section peaks
there.  The failure of the factorization formula is symptomized by
large corrections involving a factor of $\ln^2 Q/Q_T$ for each power
of $\alpha_s$.  

Although the solutions to the problem are all commonly referred to as
``resummations'', there are in fact two very different approaches.
One is resummation in its strict sense: One performs a selective and
approximation summation of the largest parts of the perturbative
series for the hard scattering in the standard factorization
formalism.

The second approach is that of Collins, Soper and Sterman (CSS) \cite
{CollinsSoper,YTerm,CollinsSoperSterman}. These authors observed that
the conventional factorization formalism is in fact wrong at low
transverse momentum and they derive a correct factorization for this
region.  In an intermediate region of transverse momentum, the
standard factorization with resummation is applicable with somewhat
reduced accuracy, and there is an overlap between the two approaches,
which we will discuss later.

In any case, it is essential to improve on the standard fixed-order
factorization formalism, and the reward is an improved method that
\begin{itemize}
\item  includes large, logarithmic QCD corrections up to all orders in the
strong coupling,

\item  improves the renormalization scale dependence of the prediction,

\item  enables prediction of certain quantities reliably, which cannot be
done in a fixed order calculation,

\item  provides an independent, analytic check for parton shower Monte
Carlo's.
\end{itemize}

\subsection{$k_T$-dependent parton densities}

CSS realized that the failure of the standard factorization when
$Q_T\ll Q$ occurs because it neglects the transverse motion of the
incoming partons in the hard scattering. (Here $Q$ can be the
invariant mass of a colorless particle, or set of particles, created
in a hard partonic collision, and $Q_T$ is the related transverse
momentum.)  The approximation of neglecting parton transverse momentum
is only valid when the cross section is integrated over a large range
of $Q_T$.  But if, for example, $Q_T$ is of order 1 GeV, then we are
outside of the domain in which the factorization is applicable.

A fully satisfactory approach must use a factorization theorem that is
valid for any $Q_T$ that is small compared to $Q$.  CSS's theorem
gives the cross section as a convolution of transverse momentum
distributions
\begin{equation}
\frac{d\sigma }{d^4Q}\propto \int d^2k_TP(x_1,k_T)P(x_2,Q_T-k_T),
\end{equation}
where $P$ is a partonic density distribution that is a function of both
longitudinal ($x$) and transverse ($k_T$) momenta. The partonic recoil
against soft gluons as well as the intrinsic partonic transverse
momentum are included in $P$.

Such a treatment completely formalizes the intuitive notion that partons
must have transverse momentum and that this transverse momentum gives rise
to transverse momentum of the Drell-Yan pair. There is then no need to
convolute a calculated cross section with ``intrinsic transverse momentum''
for the quarks; this manoeuvre is only necessary as an ad hoc correction to
a formalism that is incomplete.

In QCD, complications arise from soft-gluon effects, because these
effects do not cancel, in contrast to the case of the cross section
integrated over $Q_T$.  A consequence, proved by CSS, is a particular
form of the evolution equations for the $k_T$-dependent parton densities.
These equations are \emph{not} the normal DGLAP equations\footnote{
    Although all the physics associated with the DGLAP equations is
    present.
}.
The kernel of the evolution contains a perturbatively calculable part
and non-perturbative part.  The non-perturbative part can be
summarized by saying that there is a fixed amount of gluon radiation
per \textit{unit rapidity}, so that the transverse momentum
distribution of the partons broadens in a characteristic way with
energy.  The non-perturbative part of this energy-dependent radiation
is fitted by the $g_2$ term of Eq.(\ref{Eq:WNP}) below.

This feature may be the dominant reason
why transverse momentum distributions are so broad at high energies, as in $%
Z^0$ production: the transverse momentum of the $Z^0$ has a component due to
the recoil against non-perturbative glue emitted into many units of rapidity.

The CSS formalism clearly entails a phenomenological fitting of the
non-perturbative part of the $k_T$-dependent parton densities and the
evolution kernel. In principle, this can be done at fairly low energy, and
then the evolution equations predict the results for higher energies with no
further adjustable parameters. The more conventional resummation formalism
is compatible with the CSS formalism, but it is not as complete.

\begin{figure}[tp]
\begin{center}
\epsfxsize=12cm
\epsfysize=12cm
\mbox{\epsfbox{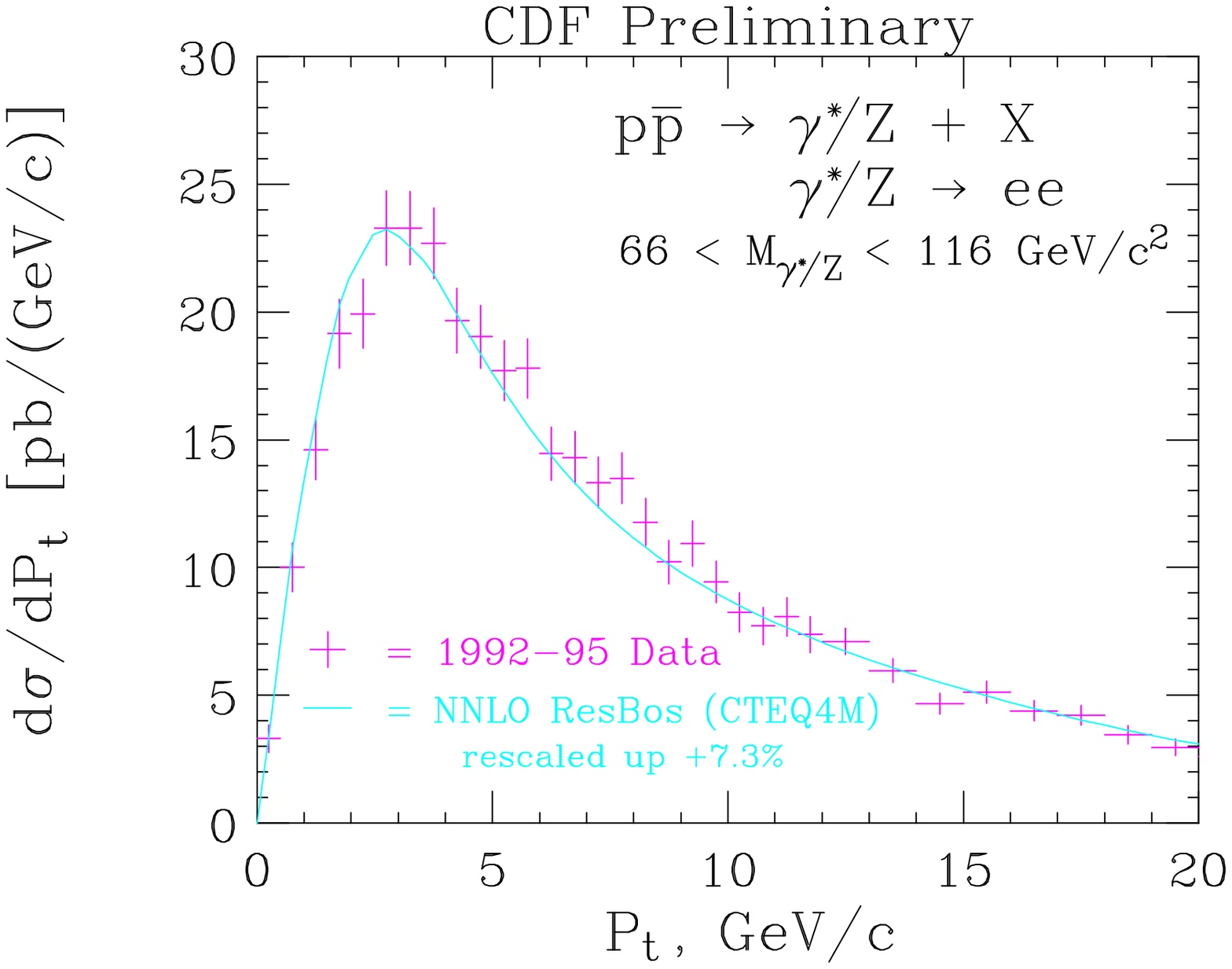}}
\end{center}
\caption{\sf Transverse momentum distribution of electron-positron pairs from
decays of (mostly) $Z^0$ bosons, produced at the Tevatron in $\protect\sqrt{S%
}=1.8$ GeV center of mass proton--anti-proton collisions. The data are CDF
preliminary \protect\cite{CDFEWGWeb}, and the curve is calculated by the
ResBos Monte Carlo event generator \protect\cite{BalazsYuanWZ,Thesis}. }
\label{Fig:ZQT}
\end{figure}

Because the CSS formalism is designed to treat correctly the $Q_T\ll Q$
region, it also provides an appropriate
resummation of the large logarithms, $\ln (Q/Q_T)$ in the standard
factorization formula.  

We can gauge how important these logarithms are in practice by
examining the cross section for $Z$ production at the Tevatron.
The bulk of the cross section is in the low $Q_T$ region, and, as can
be seen from Fig.\ \ref{Fig:ZQT}, there is a peak at around $Q_T=2.7$
GeV, which is much smaller than the invariant mass $Q=m_Z=91.187$ GeV.
This implies that for the bulk of
the events $\ln (Q/Q_T)$ is large enough that 
$\alpha_s(m_Z)\ln^2(Q/Q_T)>1$.
Since we have a double logarithm for each radiated gluon, higher
orders in the perturbative series are not suppressed.

\subsection{From fixed order to resummed}


In this section we show how the results of the standard factorization
theorem are related to a resummation in terms of leading logarithms, etc.

When the $Z^0$ is produced in a hadron-hadron collision its transverse
momentum is balanced by some hadronic activity which stems from
partons emitted by the initial state partons. (In the first order in
the strong coupling a $Z^0$ and a gluon is produced.) The $Q_T$
distribution given by the usual factorization in the low $Q_T$ region
is written as
\begin{equation}
\lim_{{Q_T\to 0}}\frac{d\sigma }{dQ_T^2}=\sum_{n=1}^\infty
\sum_{m=0}^{2n-1}\alpha _s^n~\frac{_nv_m}{Q_T^2}~\ln ^m\left( \frac{Q^2}{%
Q_T^2}\right) +{\mathcal O}\left( \frac 1{Q_T}\right) ,  \label{Eq:QTAsy}
\end{equation}
where the coefficients $_nv_m\,$are perturbatively calculable apart
from some factors of parton densities.  When the two
scales $Q$ and $Q_T$ are very different, the logarithmic terms $\ln
^m(Q^2/Q_T^2)$ are large, and for $Q_T\ll Q$ the perturbative series is
dominated by these terms. For $Q_T\ll Q$ truncation of the perturbative
series, i.e.\ any fixed order calculation, gives an answer which neglects
these important all order logarithmic contributions. At the lowest order, $%
{\mathcal O}(\alpha _s^0)$, the $Z^0$ boson is produced alone, that is with a 
$Q_T$ distribution of $\delta (Q_T)$. The singularity at $Q_T=0$ prevails at
any fixed order in $\alpha _s$, as Eq.~(\ref{Eq:QTAsy}) shows.


One way of reorganizing the perturbation series 
is to make 
the expansion one in terms of $\alpha _s\ln ^2(Q/Q_T)$ instead of 
$\alpha _s$ itself. In this simplified picture, calculating fixed order QCD
corrections means calculating the perturbative series 
\begin{eqnarray}
&& \lim_{{Q_T\to 0}}\frac{d\sigma }{dQ_T^2} = 
\nonumber \\
&&~~~
Q_T^{-2}\left\{\alpha _s({}_1v_1^{\prime }L+{}_1v_0^{\prime%
})+\alpha _s^2({}_2v_3^{\prime }L^3+{}_2v_2^{\prime }L^2{})+\alpha%
_s^3({}_3v_5^{\prime }L^5+{}_3v_4^{\prime }L^4)+~...~~~~~~~~
\right.\nonumber \\
&&~~~\;\;\;\;\;\;\;\;\;\;\;\;\;\;\;\;\;\;\;\;\;\;\;\;\;\;\;\;\;\;\;+\alpha%
_s^2({}_2v_1^{\prime }L_2+{}_2v_0^{\prime }L^0)+\alpha_s^3({}_3v_3^{\prime%
}L^3+{}_3v_2^{\prime }L^2)+~...~~~~~~~~
\nonumber \\
&&~\;\;\;\;\;\;\;\;\;\;\;\;\;\;
\;\;\;\;\;\;\;\;\;\;\;\;\;\;\;\;\;\;\;\;\;\;\;\;\;\;\;\;\;\;\;\;\;\;\;\;\;%
\;\;\;\;\;\;\;\;\;\;\;\;\;\;+\;...~~~~~\;\;\;\;\;\;\;~~~~~~~~~~~~~~...%
\left. {}\right\} ,~~~~~~~~
\nonumber
\end{eqnarray}
column by column. In the leading logarithm  approach, on the other hand, we
calculate the above series line by line \cite{ArnoldKauffman}. While in the
fixed order (column by column) calculation the convergence for low $Q_T$ is
spoiled by the higher order uncalculated logs ($L=\ln (Q/Q_T)$), in the
resummed (line by line) calculation convergence is preserved in each
``order'' (by each line), and higher order corrections are systematically
included.

\subsection{The CSS formula}


The improved factorization theorem of CSS together with their
evolution equation for the $k_T$ dependent parton distributions, leads 
\cite{CollinsSoperSterman} to a useful formula\footnote{
   While solving the RGE, an integro-differential equation, specific
   choices of integration constants were made
   (c.f.\ Ref.\ \cite{CollinsSoperSterman}): 
   $C_1=C_3=2e^{-\gamma _E}\equiv C_0$ and $C_2=C_4=1$, to optimize
   logarithmic contributions. This is similar to the $\mu = Q$ choice
   in case of the ultraviolet renormalization, to make terms like
   $\ln(\mu/Q)$ vanish.} 
for the cross section. For $Z^0$ production it can be written as
\begin{equation}
   \frac{d\sigma (h_1h_2\to Z^0X)}{dQ^2\,dQ_T^2\,dy}
   =
   \sum_j \sigma_{0,j}
      W_{j\bar \jmath}(Q,Q_T,x_1,x_2)
   \, + \, Y(Q,Q_T,x_1,x_2),
\label{Eq:cross.section}
\end{equation}
where the ``resummed'' part, ${W(Q,Q_T,x_1,x_2)}$, is defined as 
\begin{eqnarray}
   W_{j\bar \jmath}(Q,Q_T,x_1,x_2) &=&
\label{Eq:ResummationFormula}
\\
&&\hspace*{-2cm}
    \frac {1}{(2\pi )^2}
    \int d^2b \, e^{i{\vec{Q}_T}\cdot {\vec{b}}}\,
    {\mathcal C}_{j/h_1}(Q,b_,x_1,\mu )
    e^{-{\mathcal S}(Q,b_*)}
    {\mathcal C}_{{\bar{\jmath}}/h_2}(Q,b,x_2,\mu )
\nonumber
\end{eqnarray}
for a given partonic initial state with flavor $j$.\footnote{%
   The lowest order partonic total cross section is $\sigma_{0,j} = \pi^2 g^2
   ((1-4Q_js_w^2)^2 -1)/(48 Q^2 c_w^2)$, where $g$ is the weak coupling
   constant, $s_w^2$ ($c_w^2$) is the sine (cosine) of the weak mixing angle
   squared, and $Q_j$ is the charge of the quark flavor $j$.}
The Fourier integral is introduced
because transverse momentum conservation is explicit in the impact
parameter, $b$, space \cite{ParisiPetronzio}.

All the dangerous logarithms are included in the perturbative
Sudakov exponent 
\begin{equation}
   {\mathcal S}(Q,b_*)
   = \int_{C_0^2/b_*^2}^{Q^2}
     \frac{d\overline{\mu }^2}{\overline{\mu }^2}
    \left[ 
       A\left( \alpha_s(\overline{\mu}) \right)
       \ln \left( \frac{ Q^2 }{ \overline{\mu }^2 }\right)
       +B \left( \alpha_s(\overline{\mu}) \right)
   \right] .
\end{equation}
Here $C_0$ is an arbitrary parameter which cuts off the perturbative
low $Q_T$ region.%
\footnote{%
   In practice $C_0=2e^{-\gamma _E}$ is used, which is related to the values of
   the integration constants of the RGE for the $k_T$ dependent
   PDF's.}
To prevent perturbative calculations from being done in region where
perturbation theory is inapplicable, the ``impact parameter'' $b$ in
the Sudakov exponent was replaced by
\begin{equation}
   b_* = \frac{ b }{ \sqrt{1+(b/b_{\rm max })^2} } .
\label{bstar}
\end{equation}
The errors caused by this replacement are of the same form as the
non-perturbative contributions to be discussed below, and are
therefore correctly treated by being absorbed into the
non-perturbative part of the formula.

The $A$ and $B$ functions are free of large logarithms and can be
reliably calculated perturbatively for a given process as
\begin{equation}
A\left( \alpha _s({\bar{\mu}})\right) =\sum_{n=1}^\infty \left( \frac{\alpha
_s({\bar{\mu}})}\pi \right) ^nA^{(n)},~~~B\left( \alpha _s({\bar{\mu}}%
)\right) =\sum_{n=1}^\infty \left( \frac{\alpha _s({\bar{\mu}})}\pi \right)
^nB^{(n)}.
\end{equation}
The distributions 
\begin{equation}
   {\mathcal C}_{j/h}(Q,b,x,\mu )
=   \sum_a \int_x^1{\frac{d\xi }\xi }
             C_{ja}\left( b_*,\frac{ x }{ \xi }, \mu \right) 
             f_{a/h}(\xi ,\mu )
        \, {\mathcal F}_{a/h}(b,x)
        \, e^{-r(b)\ln{Q}}
\label{C_coeff}
\end{equation}
depend on virtual and real emission contributions for a given process, via
the Wilson coefficients $C_{ja}$. Just as the $A$ and $B$ functions the
Wilson coefficients are expanded in terms of the strong coupling $\alpha_s$ 
\begin{equation}
   C_{ij}\left( b_*,z,\mu \right) 
= \sum_{n=0}^\infty \left( \frac{\alpha_s(\mu) }{ \pi } \right)^n
          C_{ij}^{(n)}\left( z, b_* \right) .
\end{equation}
Since the Sudakov exponent integrates to unity, the $C_{ij}$ function
sets the normalization of the resummed distribution. In particular, if
coefficients up to $C_{ij}^{(n)}$ are included in the calculation then
the resummed rate will equal the rate calculated in fixed order at
${\mathcal O}(\alpha_s^n)$ \cite{BalazsYuanWZ}.  The function
$f_{a/h}(x,\mu)$ is the usual renormalized momentum fraction ($x$)
distribution of parton $a$ in hadron $h$ at the energy scale
$\mu$. Observe that the impact parameter dependence of the perturbative
coefficient functions is cut off by the use of $b_*$ instead of $b$.


Included in Eq.\ (\ref{C_coeff}) are two non-perturbative factors,
${\mathcal F}_{a/h}(b,x)$ and 
$e^{-r(b)\ln{Q}}$.  These implement
the parts of the CSS factorization and evolution equation that cannot
be implemented as a resummation of the standard factorization
theorem.  They also compensate for the errors in the resummation at
large $b$.  The overall effect is that (\ref{C_coeff}) define 
$k_T$-dependent parton densities.  The ${\mathcal F}$ factor can be
interpreted as allowing for intrinsic transverse momentum, and the
$e^{-r(b)\ln{Q}}$ factor allows for the recoil against soft gluon
radiation.  The $\ln{Q}$ in the exponent of the soft-gluon factor
comes from the solution of the CSS evolution equation and can be
interpreted by saying that soft gluons are emitted uniformly in
rapidity. 

The perturbative part of the formula uses $b_*$ instead of $b$, as
defined by Eq.\ (\ref{bstar}). The parameter $b_{\rm max}$ provides an 
infra-red cutoff on the perturbative part of the formula.
In practice the empirically optimal value, $b_{\rm max}=1/2$ GeV$^{-1}$, is
used. 
This arbitrary cutoff of the $b$ integration is compensated by the
parameterization of the non-perturbative part of the formula, which is
\begin{eqnarray}
   W_{ij}^{\rm NP}(Q,b,x_1,x_2) &=& 
   {\mathcal F}_{i/h_1}(Q,b,x_1)
   {\mathcal F}_{j/h_2}(Q,b,x_2)
   e^{-r(b)\ln{Q}}
\label{Eq:WNP}
\\
&=& \exp\left[ -g_1b^2-g_2b^2\ln \left( {\frac Q{2Q_0}}\right)
               -g_1g_3b\ln {(100x_1x_2)}
        \right] ,
\nonumber 
\end{eqnarray}
where $Q_0$ is chosen to be the initial scale of the parton evolution%
\footnote{%
For recent CTEQ PDF's $Q_0 = 1.6$ GeV.} and the $g_i$ parameters have to be
determined using experimental data.\footnote{%
The $\ln \left( Q^2/Q_0^2\right)$ term is introduced to match the
logarithmic term of the Sudakov exponent and its coefficient is expected to
be process independent, depending only on the initial partonic state.}


\subsection{Matching}


The resummed term, defined by Eq. (\ref{Eq:ResummationFormula}), was derived
in the context of a generalized factorization, under the assumption
that 
$Q_T\ll Q$. This assumption will break down within and beyond the
intermediate $Q_T\stackrel{\scriptscriptstyle<}{\scriptscriptstyle\sim }Q$
region. In the high $Q_T$ region (where $Q_T\stackrel{\scriptscriptstyle>}{%
\scriptscriptstyle\sim }Q$) the conventional perturbative factorization
formalism is reliable. To obtain sufficiently accurate results for all
$Q_T$,
it is necessary to combine the formalisms. 

The $Y$ term in Eq.\ (\ref{Eq:cross.section}) was introduced by
CSS \cite{YTerm} to correct the behavior of the resummed piece in the
intermediate and high $Q_T$ regions.\footnote{%
The exact definition of the $Y$ piece for $Z^0$ production can be found in
Refs.\ \cite{CollinsSoperSterman,BalazsYuanWZ}.} It is defined as the
difference of the cross section calculated from the standard
factorization formula 
at a fixed order $n$ of
perturbation theory and the $Q_T\ll Q$ asymptote of this cross section: 
\begin{equation}
Y(Q,Q_T,x_1,x_2)=\left( \frac{d\sigma }{dQ^2\,dQ_T^2dy}\right) _n-\left( 
\frac{d\sigma }{dQ^2\,dQ_T^2dy}\right) _{n,Q_T\ll Q}.  \label{Y.def}
\end{equation}
Thus, the full CSS formula can be written as
\begin{equation}
\frac{d\sigma }{dQ^2\,dQ_T^2dy}=\left( \frac{d\sigma }{dQ^2\,dQ_T^2dy} 
   \right)_{\rm res}
+\left( \frac{d\sigma }{dQ^2\,dQ_T^2dy}\right) _n-\left( \frac{%
d\sigma }{dQ^2\,dQ_T^2dy}\right) _{n,Q_T\ll Q}.  \label{Eq:CSSMatched}
\end{equation}

This method of matching the resummed and fixed order pieces is valid because
the low $Q_T$ asymptote used in Eq.\ (\ref{Y.def}) is the same as the large $%
Q_T$ asymptote of the resummed term $W$. At low $Q_T$ the asymptotic part
dominates the $Q_T$ distribution (the logs are large), and the last two terms
cancel in Eq.(\ref{Eq:CSSMatched}), while the resummed term is significant
near $Q_T=0$. At high $Q_T$ the logs are small, and the expansion of the
resummed term cancels the $Q_T$ singular terms up to higher orders in $%
\alpha _s$.\footnote{%
The cancellation is higher order than the order at which the singular pieces
were calculated.} In this situation the first and third terms cancel and CSS
formula reduces to the fixed order perturbative result. After matching the
resummed and fixed order cross sections in such a ``smooth'' manner, it is
expected that the normalization of the CSS cross section reproduces the
fixed order total rate, since when expanded and integrated over $Q_T$ it
deviates from the fixed order result only in small higher order terms in $%
\alpha _s$ \cite{BalazsYuanWZ}.


Unfortunately the above argument does not completely work in practice.
The problem arises because at large $Q_T$ the $W$ term in Eq.\
(\ref{Eq:cross.section}) is an extrapolation of the cross section from
small $Q_T$.  So it has a $1/Q_T^2$ behavior, modified by
logarithms.  This falls less steeply than the true cross section,
which is subject to kinematic limits.  The errors in the CSS formula
at large $Q_T$ are indeed suppressed by a power of $\alpha_s$.  But the
coefficient of this power is the $1/Q_T^2$ part of the formula, and so
the error can be easily larger than the true cross section.  A symptom
of the problem is that the cross section calculated from
Eq. (\ref{Eq:cross.section}) is typically negative at large enough
$Q_T$.

One possible remedy \cite{Ellis:1997ii} is to abandon the CSS
formalism.  But we 
regard this as undesirable, because it also abandons the important
physical result of CSS that goes beyond mere resummation: their
proper treatment of non-perturbative transverse momentum.  

\begin{figure}[tp]
\begin{center}
\epsfxsize=12cm
\epsfysize=12cm
\mbox{\epsfbox{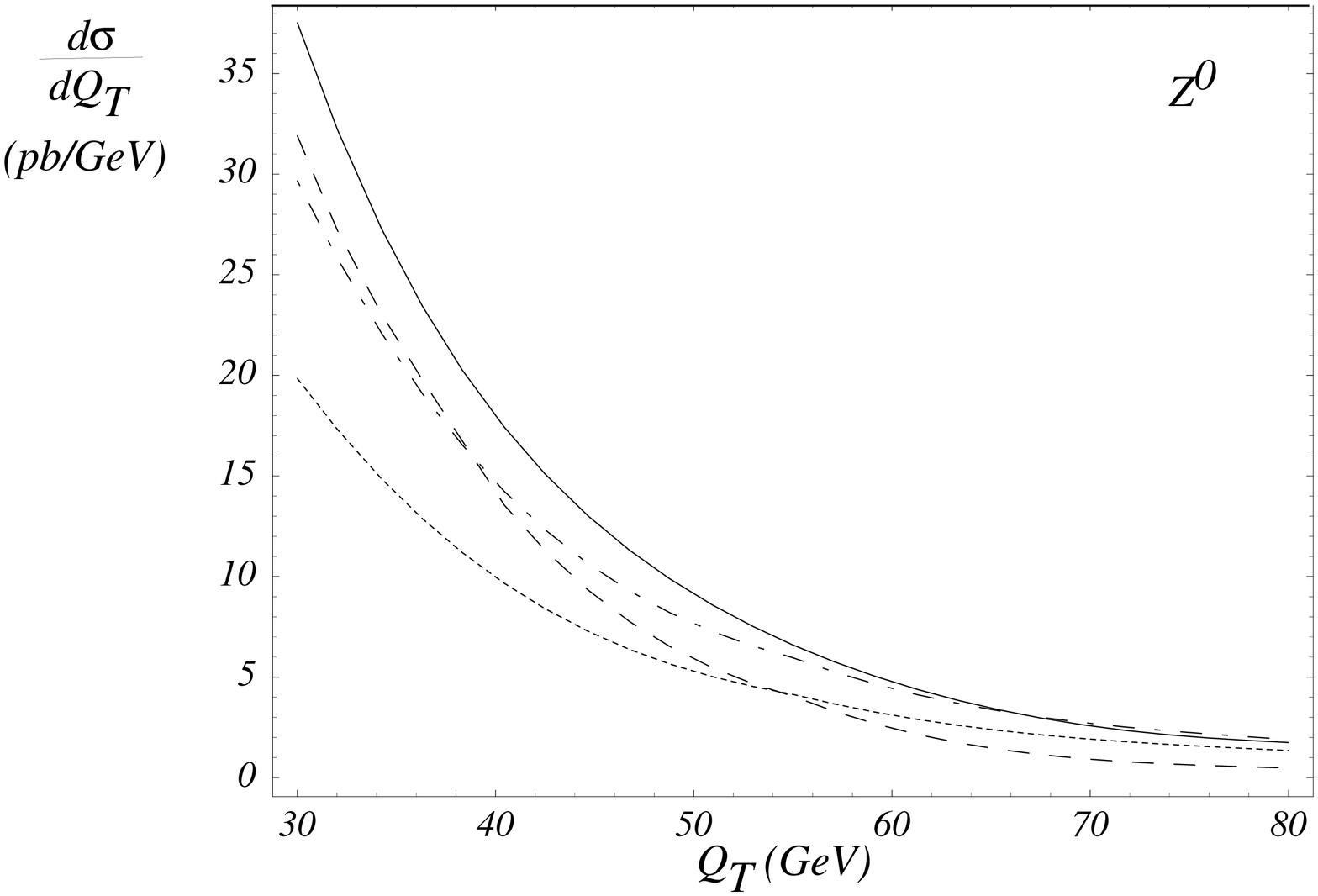}}
\end{center}
\caption{\sf Transverse momentum distributions calculated using the CSS (solid and  
dashed) and the usual factorization (dot-dashed and dotted) formalisms.  The CSS 
calculation is not switched over to the usual formalism to show the typical 
``kink" which occurs at a switching point.  The solid and dot-dashed curves are 
calculated at ${\cal O}(\alpha_s^2)$, while the dashed and dotted curves are at 
${\cal O}(\alpha_s)$, illustrating the improvement of the high $Q_T$ behavior 
of the CSS formula with the perturbative order. }
\label{Fig:Matching}
\end{figure}
\begin{figure}[tp]
\begin{center}
\epsfxsize=12cm
\epsfysize=12cm
\mbox{\epsfbox{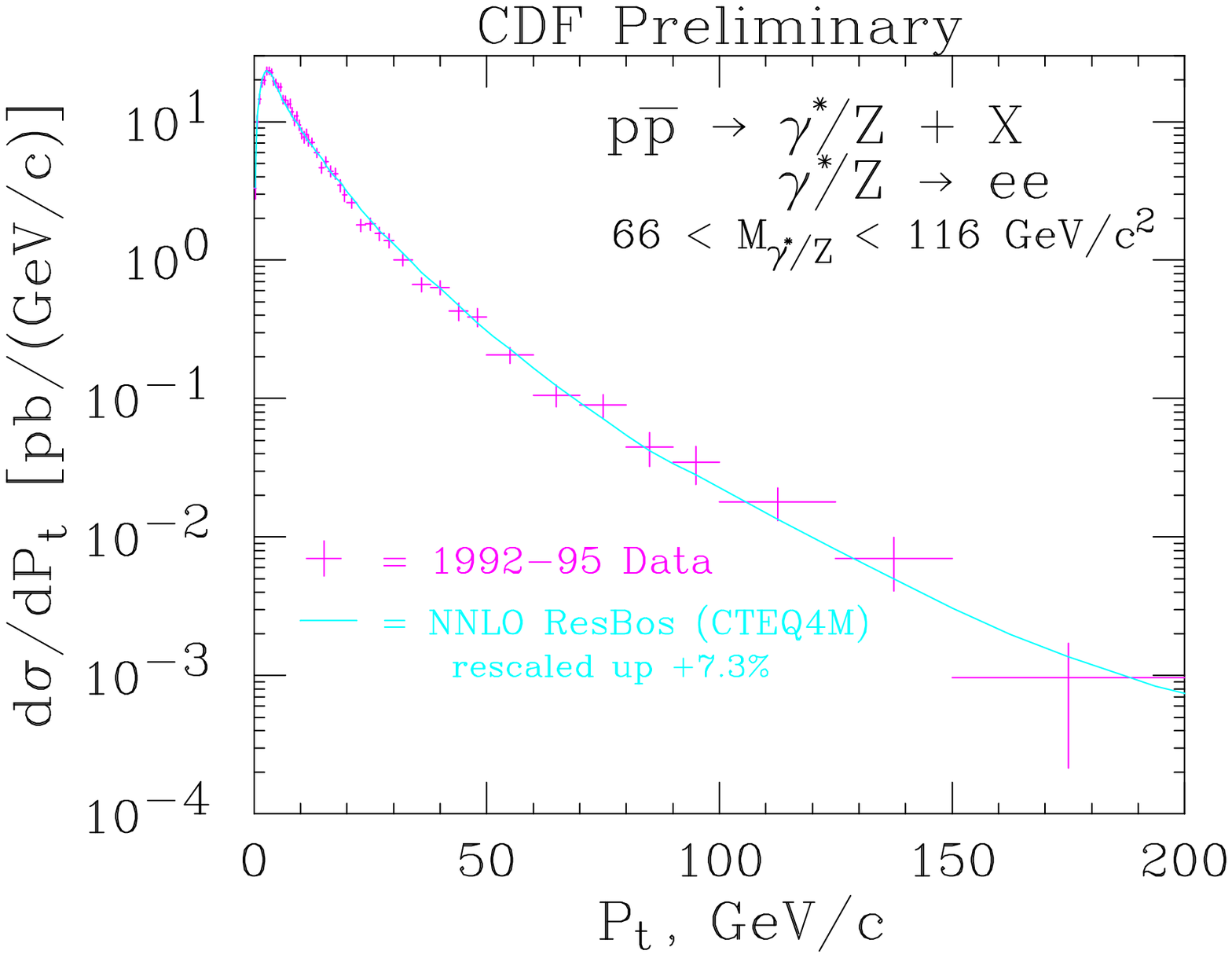}} 
\end{center}
\caption{\sf Same as Fig.\ \ref{Fig:ZQT}, except shown for the full range of 
transverse momentum $Q_T$.  Switching from the CSS to the usual factorization 
formalism in the $60 \lesssim Q_T \lesssim 70$ GeV region 
(c.f. \protect\cite{BalazsYuanWZ}), results in a smooth $Q_T$ distribution
which 
agrees well with the experiment in the full $Q_T$ range.}
\end{figure}
\label{Fig:ZQT_Full}
A second, commonly used remedy, is to utilize the fact that in the high $Q_T$ 
region the fixed order result is a good description of the distribution.  So 
when calculating the $Q_T$ distribution one can simply switch from the CSS to 
the fixed order distribution whenever they cross for high $Q_T$'s.  Since the 
mismatch between the resummed and the asymptotic terms in Eq.(\ref{Eq:CSSMatched}) 
decreases as the perturbative order of the calculation ($n$) increases, it is 
expected that the crossing point shifts toward $Q_T = Q$, and the slope of the 
resummed and fixed order curves approaches each other as $n$ increases 
(cf. Ref.~\cite{BalazsYuanWZ}).  Indeed, calculations at ${\cal O}(\alpha_s^2)$ 
blend closer to $m_Z$, and smoother than at ${\cal O}(\alpha_s)$, as shown in 
Fig.\ \ref{Fig:Matching}.

When this prescription for the switching is followed at the fully differential, 
$d\sigma /dQ_TdQdy$, level the result is a smooth and differentiable $Q_T$ 
distribution, after the invariant mass and rapidity is integrated out.  This 
is illustrated in Fig.\ \ref{Fig:ZQT_Full}.  It was shown in Ref.\ \cite%
{BalazsYuanWZ} that the integral of the $Q_T$ distribution calculated using this
prescription recovers the fixed order total rate within an error which
is the size of the higher orders, as it is expected.

\subsection{Improved matching}

Since the calculations of $W$ and $Y$ are done using truncations of
perturbation theory, the switching between calculational methods introduces
an artificial discontinuity in the slope of the cross section. This
practical problem arises in the matching because of a mismatch of the orders
of perturbation theory at which $W$ and $Y$ are calculated. From the point
of view of a standard factorization calculation, $W$ contains a selective
summation of arbitrarily high orders of perturbation theory. The possibility
of getting such a resummation relies on performing certain approximations
that are only valid at small $Q_T$. The difficulty of performing complete
higher-order calculation means that $Y$ can only be calculated at fixed
order.

At large transverse momentum, $|W|$ is much larger than the actual cross
section, and so the cross section Eq.\ (\ref{Eq:CSSMatched}) is obtained by
the cancellation of two almost equal terms. This is clearly a recipe for bad
numerical work.

Examination of the lowest-order calculation of $Y$, for the $q\bar{q}$
annihilation term \cite{YTerm} shows some of the sources of the problems: 
\begin{eqnarray}
\label{Y.1}
Y &=&\frac C{Q_T^2}\int \frac{d\xi _1}{\xi _1}\frac{d\xi _2}{\xi _2}f_q(\xi
_1)f_{\bar{q}}(\xi _2)  \nonumber \\
&&~~~\left\{ \frac{(Q^2-\hat{t})^2+(Q^2-\hat{u})^2}{\hat{s}}\delta (\hat{s}+%
\hat{t}+\hat{u}-Q^2)\right.  \nonumber \\
&&~~~-2\delta (1-x_1/\xi _1)\delta (1-x_2/\xi _2)\left[ \ln
(Q^2/Q_T^2)-\frac 32\right] \\
&&~~~\left. -\delta (1-x_1/\xi _1)\left[ \frac{1+x_2^2/\xi _2^2}{1-x_2/\xi _2%
}\right] _{+}~-~\left[ \frac{1+x_1^2/\xi _1^2}{1-x_1/\xi _1}\right]
_{+}\delta (1-x_2/\xi _2)\right\} .  \nonumber  
\end{eqnarray}
Here $x_1$ and $x_2$ are the longitudinal momentum fractions of the
Drell-Yan pair. The first term contains the usual perturbative calculation
of the differential cross section, and the other 3 terms give the negative
of its low $Q_T$ asymptote. The intrinsic rate of fall off of the cross
section with $Q_T$ is given by the explicit $1/Q_T^2$ factor which is
present in the parton cross section. But an extra fall off is caused by the
fact that the parton densities are probed at larger fractional momenta when $%
Q_T$ is increased.

Some symptoms of the problems can already be seen. One is that the first
subtraction term, on the second line of Eq.\ (\ref{Y.1}), changes sign at
large $Q_T$: the extrapolation of a positive cross section becomes negative.
The second is the plus distribution in the third line; if the parton
distributions are steeply falling, the plus distributions give a misleading
size for the integrand. This last effect really indicates that there is an
additional scale in the process, so that the relevant scales are:

\begin{itemize}
\item  The transverse momentum $Q_T$ of the Drell-Yan pair.

\item  The invariant mass $Q$ of the pair.

\item  The increase $\Delta Q$ of $Q$ that is necessary to make the typical
parton densities in the factorization formula decrease by a factor 2.
\end{itemize}

We believe the overall approach of a subtraction method is correct: $W$
correctly represents the physics at low $Q_T$, and we do not wish to give up
a method that uses the intuitive notion of $k_T$-dependent parton densities.
We therefore cannot expect to obtain a perfect estimate of the large $Q_T$
cross section from $W$ alone. The idea of adding a correction term $Y$ is a
good way of combining the information in standard fixed order calculations
with the resummed calculations.

But improvements in its implementation are needed. We suggest the following
strategies that could be tried, individually or even in combination:
\begin{itemize}
\item  Multiply $W$ by an ad hoc factor $F(Q_T/M)$. Correspondingly the
formula for the subtraction term in $Y$ will also have the same factor. The
parameter $M$ is in principle arbitrary, and it should be chosen so that the
fall off in the modified $W$ term mimics that of the actual cross section.
The cut-off function obeys $F(0)=1$, so that the small $Q_T$ behavior is
unchanged, and the function should be zero for large $Q_T$.

\item  Change the argument of $W$ from $Q_T$ to some other function of 
$Q_T$. One possible choice would be $Q_T^{\prime }=Q_T/(1-Q_T/M)$, where $M$ is
again a parameter to be chosen. One would replace $W$ by zero if $Q_T>M$.
The effect of the variable change is to leave $W$ unaltered at small $Q_T$
and to give a more rapid fall off at large $Q_T$. Again one would make an
identical redefinition in the subtraction term in $Y$.

\item  Redefine the $+$ distributions such as those in Eq.\ (\ref{Y.1}), by: 
\begin{equation}
\int_0^1dzf(z)\left[ \frac 1z\right] _{+,z_0}=\int_0^1dz\frac 1z\left[
f(z)-f(0)\theta (z_0-z)\right] .
\end{equation}
(The usual definition has $z_0=1$.)
\end{itemize}
In each case we have a generalized renormalization-group invariance of the
exact cross section under changes of the parameter $M$ or $z_0$. But
approximations obtained by truncation of a perturbation series are invariant
only up to a term of order the first uncalculated correction. The aim is to
choose the parameters on physical grounds to be such as to keep these higher
order terms small, to eliminate their reason(s) for being large.

\subsection{Applications}


Beyond $Z^0$ production, in its present form, the CSS formalism can be
applied in hadron-hadron collisions whenever the final state is colorless.
The phenomenological significance of this ''transverse momentum
resummation'' ranges from Drell-Yan pair production, through lepton pair
production via $W^{\pm }$ and $Z^0$ bosons \cite{BalazsYuanWZ}, di-gauge
boson (e.g. photon or $Z^0$ boson pair) production \cite%
{BalazsBergerMrennaYuan,BalazsYuanZZ}, to Higgs production \cite%
{Hinchliffe-Kauffman-Kao,Yuan,BalazsYuanPrep}. In recent years it was
tested in hadronic processes taking place at fixed target (e.g.\ in DY photon
and diphoton production) \cite{Begel} and collider energies (e.g.\ in DY, $%
W^{\pm }$, $Z^0$, and diphoton production). It was applied for different
hadronic initial states in pion--nucleon, proton--nucleon, and proton--anti-%
proton collisions. It was also modified and tested for DIS processes \cite
{NadolskyStumpYuan}. Finally, since was first devised for the calculation of
the energy correlation of jets in $e^{+}e^{-}$ collisions \cite{CollinsSoper}%
, it can be used in jet production at lepton colliders. Such a wide variety
of applicability, and good agreement with existing experimental results for
different processes, colliders, center of mass energies, and initial states
gives us a confidence in the resummed predictions for the LHC.

\section{Higgs production}


At the LHC the SM Higgs boson will be mainly produced through the gluon
fusion subprocess via a top quark loop: $gg$ (top quark loop) $\to HX$ \cite
{Spira:1997zs}. The Higgs boson can be detected in its $H\to \gamma \gamma $
decay mode, if its mass is in the 100-150 GeV range \cite{AtlasCPandCMSTDR}.
If the Higgs mass is higher than about 130 GeV then its $H\to Z^0Z^{0*}$
decay mode is the cleanest and most significant \cite{AtlasCPandCMSTDR}. To
distinguish these signals from the substantial QCD background, besides the
sharp peak in the invariant mass distribution, the most straightforward
measurable to use is the transverse momentum. According to earlier studies,
a statistical significance on the order of 5-10 can be reached for the
inclusive $H\to \gamma \gamma $ signal, actual values depending on
luminosity and background estimates. %
%
Once their transverse momentum distribution is reliably predicted, the
difference in the $Q_T$ of the signal and background can be utilized to
devise kinematic cuts to enhance the statistical significance of the signal.
After the discovery, when determining the properties of the Higgs boson,
besides the total cross section and the invariant mass distribution, the
simplest and most fundamental measurable to use is the transverse momentum.
For a recently proposed new detection mode, $H\to \gamma \gamma \mbox{jet}$, in
Ref.~\cite{Abdullin} it was also found that in order to optimize the
significance it is necessary to impose a 30 GeV cut on the transverse
momentum of the jet, or equivalently (at NLO precision), on the $Q_T$ of the
photon pair. With this cut in place extraction of the signal in the Higgs
plus jet mode requires the precise knowledge of both the signal and
background distributions in the mid- to high-$Q_T$ region.


To reliably predict the $Q_T$ distribution of Higgs bosons at the LHC,
especially in the low to mid $Q_T$ region where the bulk of the rate is, the
effects of the multiple soft--gluon emission have to be included. In
practice, performing soft gluon resummation within the CSS formalism is
equivalent to the determination of the $A^{(n)}$, $B^{(n)}$, and $C^{(n)}$
coefficients and the $Y$ part at some order in $\alpha _s$. One way to
calculate the coefficients is to expand the resummed part in terms of the
strong coupling (expanding the exponent an the Wilson coefficients), and
compare the expansion with a fixed order calculation. Luckily, because of
its significance, there was much work done on fixed order QCD
corrections to Higgs production in the $gg\to HX$ channel. These fixed order
QCD corrections are known to substantially increase the rate: by about 70 to
100 percent, depending on the Higgs mass, at ${\mathcal O}(\alpha _s^3)$\cite
{Graudenz:1993,Djouadi:1991}, and by an additional 50 to 70 percent at $%
{\mathcal O}(\alpha _s^4)$ \cite{FlorianGrazziniKunsztH}. It is expected that
the calculation of even higher order corrections is important to reliably
predict the cross section. In Ref.\ \cite{Kramer:1996iq} it was shown that
multiple soft--gluon emission dominates the higher order corrections.

\subsection{Soft gluon resummation for the $gg \to HX$ channel}


Resummed calculations, taking into account the soft--gluon effect, attempted
to estimate the size of the non-calculated higher order corrections \cite
{Kramer:1996iq}, as well as provide a reliable shape of the Higgs transverse
momentum distribution \cite{Hinchliffe-Kauffman-Kao,Yuan}. 
Our present approach surpasses these by calculating the $Q_T$ distribution
while including ${\mathcal O}(\alpha _s^4)$ terms in the Sudakov exponent,
using the state of the art matching to the latest fixed order distributions,
using a QCD improved gluon-Higgs effective coupling \cite{Kniehl:1995tn},
and using an improved non-perturbative function. We utilize the
approximation that the object which couples the gluons to the Higgs (the top
quark in the SM), is much heavier than the Higgs itself. This approximation
is not essential to our calculation and can be released by the calculation
of the further Wilson coefficients keeping the relevant masses. The heavy
quark approximation in the SM was shown to be reliable within 5 percent for $%
m_H<2m_t$ \cite{FlorianGrazziniKunsztH,Graudenz:1993,Kunszt:1996yp}, and
still reasonable even in the range of $m_H\stackrel{\scriptscriptstyle>}{%
\scriptscriptstyle\sim }2m_t$ \cite{Kramer:1996iq}. It has also been shown
that the approximation remains valid for the $Q_T$ distribution in the large 
$Q_T$ region, provided that $m_H<m_t$ and $Q_T<m_t$ \cite{Baur:1990cm}. In
this work we assume that the approximation is valid in the whole $Q_T$
region. Unlike the authors of Ref.\ \cite{Kramer:1996iq} we do not assume
that the QCD corrections to the $gg\to HX$ cross section can be factorized
into a multiplicative term in the heavy quark limit in all orders of $\alpha
_s$. We can release this approximation because the CSS formalism, by
definition, systematically incorporates higher order fixed order corrections
via the definition of the Sudakov exponent and the Wilson coefficients as
perturbative expansions \cite{BalazsYuanWZ,Thesis}.


Multiple soft--gluon emission affects the $gg\to HX$ cross section when the
transverse momentum of the Higgs is low, while for high transverse momenta
the hard gluon radiation is dominant. Thus, using the CSS formalism we resum
large logs of the type $\ln (Q/Q_T)$ in the low $Q_T$ region, and we match
the resummed result to the fixed order calculation which is valid for high $%
Q_T$ \cite{BalazsYuanWZ}. We also include the $qg$ and $q\bar{q}$
subprocesses which, depending on the Higgs mass, together constitute 0 to 10
percent of the total rate \cite{Graudenz:1993}.

The resummed differential cross section of the Higgs boson production in
hadronic collisions is written as 
\begin{eqnarray}
&&{\frac{d\sigma (h_1h_2\to H^0X)}{dQ^2\,dy\,dQ_T^2}}=\sigma _0\,\frac{Q^2}%
S\,\pi \delta (Q^2-m_H^2)  \nonumber \\
&&\times \left\{ {\frac 1{(2\pi )^2}}\int d^2b\,e^{i{\vec{Q}_T}\cdot {\vec{b}%
}}{\widetilde{W}_{gg}(b_{*},Q,x_1,x_2,C_{1,2,3})}\right.  \nonumber \\
&&~~~~\times \left. \widetilde{W}%
_{gg}^{\rm NP}(b,Q,x_1,x_2)+Y(Q_T,Q,x_1,x_2,C_4)\right\} .  \label{Eq:ResFor}
\end{eqnarray}
The kinematic variables $Q$, $y$, and $Q_T$ are the invariant mass,
rapidity, and transverse momentum of the Higgs boson in the laboratory
frame. The parton momentum fractions are defined as $x_1=e^yQ/\sqrt{S}$, and 
$x_2=e^{-y}Q/\sqrt{S}$, with $\sqrt{S}$ being the center--of--mass (CM)
energy of the hadrons $h_1$ and $h_2$. The lowest order cross section, with
the QCD corrected effective coupling of the Higgs boson to gluons is 
\begin{equation}
\sigma _0=\kappa _\phi (Q){\frac{\sqrt{2}G_F\alpha _s^2(Q^2)}{576\pi }},
\end{equation}
where $G_F$ is the Fermi constant and $\kappa _\phi $ is defined in Ref.\ 
\cite{Kramer:1996iq}. The renormalization group invariant kernel of the
Fourier integral $\widetilde{W}_{gg}$ and the regular terms $%
Y(Q_T,Q,x_1,x_2,C_4)$ (together with the variables $b_{*}$ and $C_1$ to $C_4$%
) are defined in Ref.\ \cite{Yuan}. In addition to Ref.\ \cite{Yuan} we use
the process independent coefficient 
\begin{equation}
A^{(2)}=C_A\left[ \left( {\frac{67}{36}}-{\frac{\pi ^2}{12}}\right) N_C-{%
\frac 5{18}}N_f\right] ,
\end{equation}
in the expansion of the $A$ function ($N_C=3$ the number of colors and $%
N_f=5 $ the number of active quark flavors).

\subsection{Some numerical results}

The resummation formula is coded in the ResBos Monte Carlo event generator 
\cite{BalazsYuanWZ,Thesis}, which uses the following electroweak input
parameters \cite{PDB}: $G_F=1.16639\times 10^{-5}~\mathrm{GeV}%
^{-2},m_Z=91.187~\mathrm{GeV},m_W=80.36~\mathrm{GeV}$. The NLO expressions
for the running electromagnetic and strong couplings $\alpha (\mu )$ and $%
\alpha _S(\mu )$ are used, as well as the NLO parton distribution function
set CTEQ4M (defined in the modified minimal subtraction, $\overline{MS}$,
scheme). The renormalization and factorization scales are set equal to the
Higgs invariant mass. In the choice of the non-perturbative parameters we
follow Ref.~\cite{BrockLadinskyLandryYuan}. Since we are not concerned with
the decays of Higgs bosons in this work, we do not impose any kinematic cuts.


\begin{figure}[tp]
\begin{center}
\epsfxsize=12cm
\epsfysize=12cm
\mbox{\epsfbox{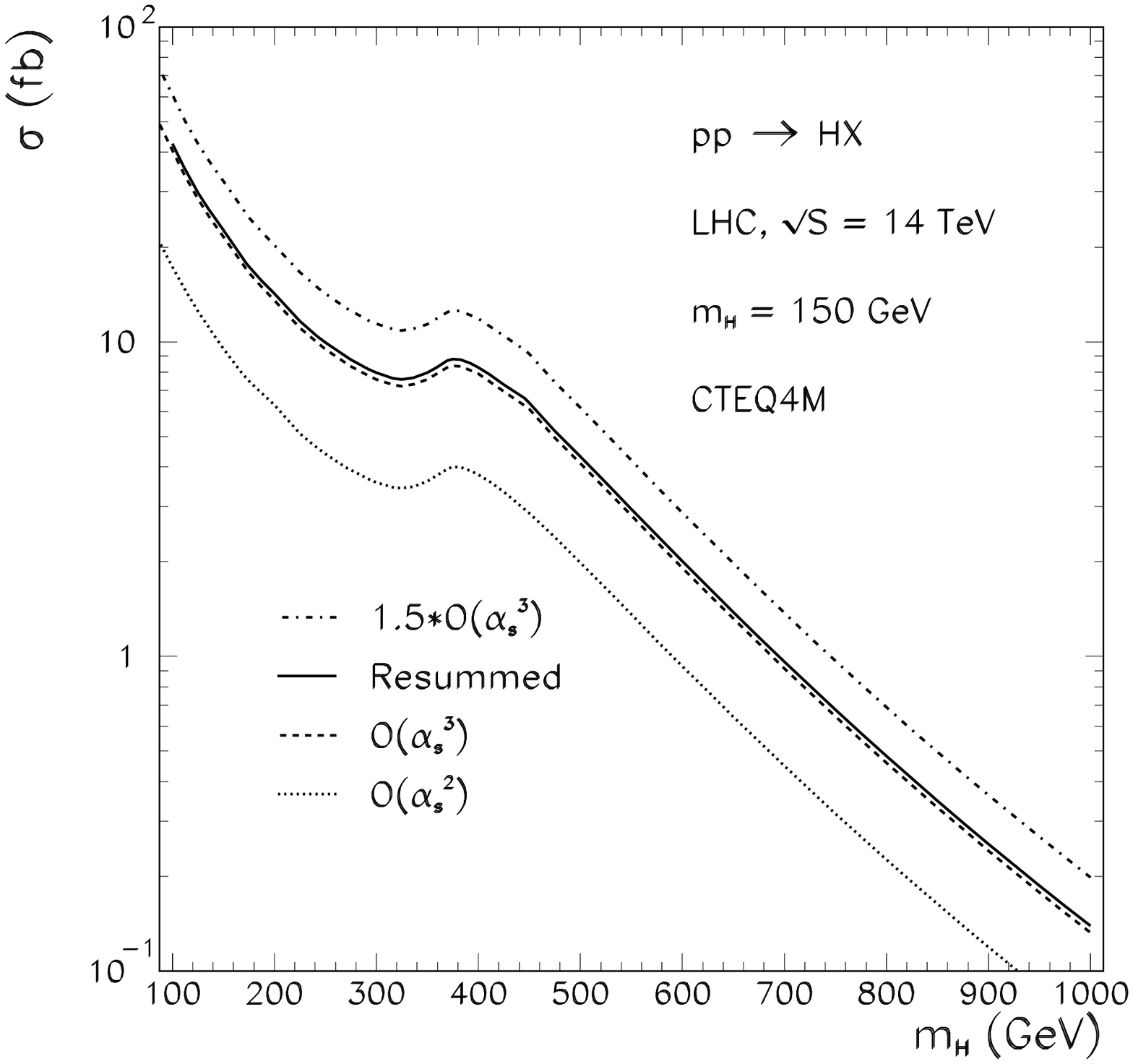}} 
\end{center}
\caption[Fig:Tot]{\sf SM Higgs boson production cross sections at the LHC via
top quark loop as the function of the Higgs mass, with QCD corrections
calculated by soft--gluon resummation (solid), at fixed order ${\mathcal O}%
(\alpha_s^3)$ (dashed), and without QCD corrections at ${\mathcal O}%
(\alpha_s^2)$ (dotted). The ${\mathcal O}(\alpha_s^3)$ curve is scaled by 1.5
(dash-dotted, c.f. Ref.\ \protect\cite{FlorianGrazziniKunsztH}) to estimate
the ${\mathcal O}(\alpha_s^4)$ result. }
\label{Fig:Tot}
\end{figure}

Fig.~\ref{Fig:Tot} displays production cross sections at the LHC, calculated
in the SM as the function of the Higgs mass. Our ${\mathcal O}(\alpha _s^3)$
curve agrees well with the result in Ref.\ \cite{Kramer:1996iq}. The ratio of
the fixed order ${\mathcal O}(\alpha _s^3)$ (dashed) and the lowest order $%
{\mathcal O}(\alpha _s^2)$ (dotted) curves varies between 2.35 and 2.00. We
note that less than 2 percent of the ${\mathcal O}(\alpha _s^3)$ corrections
come from the $qg$ and $q{\bar{q}}$ initial states for Higgs masses below
200 GeV. The resummed curve is slightly (5 to 6 percent) higher than the $%
{\mathcal O}(\alpha _s^3)$ one, as expected based on the findings that the
CSS formalism preserves the fixed order rate within the error of the
matching (which is expected to be higher order) \cite{BalazsYuanWZ}. The
resummed rate is close to the ${\mathcal O}(\alpha _s^3)$, because we used
the ${\mathcal O}(\alpha _s^3)$ fixed order results to derive the Wilson
coefficients which are utilized in our calculation. In Ref.\ \cite
{FlorianGrazziniKunsztH} the ${\mathcal O}(\alpha _s^4)$ corrections were
utilized to show that in the high $Q_T$ region the ${\mathcal O}(\alpha _s^4)$
to ${\mathcal O}(\alpha _s^3)$ $K$-factor is nearly constant and is about 1.5
(for CTEQ4M parton distributions). Based on this finding we also plot the $%
{\mathcal O}(\alpha _s^3)$ curve rescaled by 1.5, to illustrate the size of
the ${\mathcal O}(\alpha _s^4)$ corrections and to establish the
normalization of our resummed calculation among the fixed order results.


\begin{figure}[tp]
\begin{center}
\epsfxsize=12cm
\epsfysize=12cm
\mbox{\epsfbox{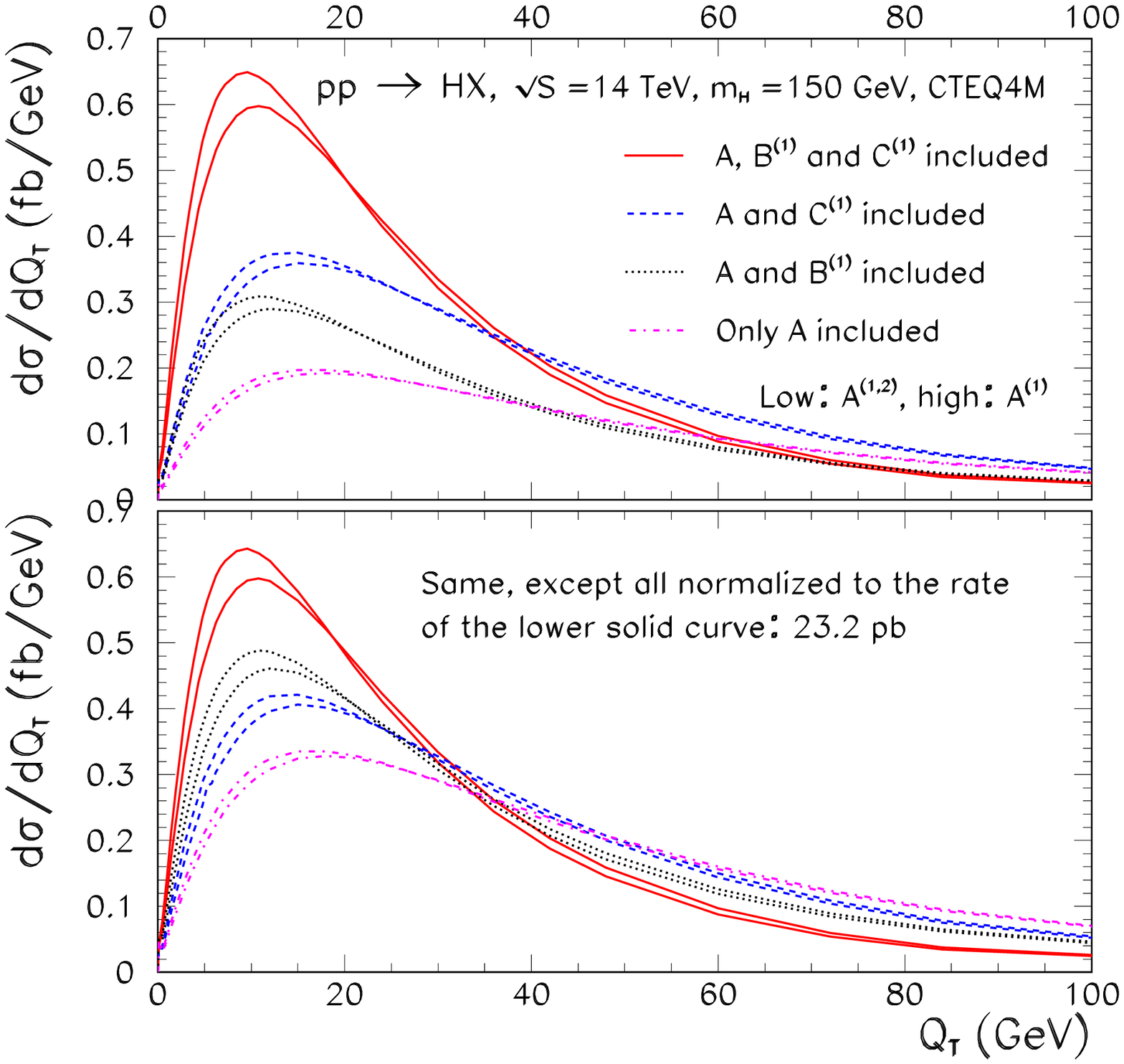}}
\end{center}
\caption[Fig:Logs]{\sf Higgs boson transverse momentum distributions at the
LHC, illustrating the effect of various contributions of the CSS formalism.
Among the pair of curves the ones which peak lower are calculated by using 
$A^{(1,2)}$ and the others by $A^{(1)}$. Additionally, the solid curves 
include $B^{(1)}$ and $C^{(1)}$, the dashed ones $C^{(1)}$, and the dotted 
ones $B^{(1)}$. The dot-dashed curves only include $A$ coefficients.
The lower portion of the figure shows the same curves normalized to the area
under the lower peaking solid curve, to compare the changes in the shape. }
\label{Fig:Logs}
\end{figure}

Fig.~\ref{Fig:Logs} illustrates the effect of the various contributions of the 
CSS formalism on the Higgs boson transverse momentum distribution. The lower 
peaking curves, drawn by the same type line, contain the coefficients 
$A^{(1,2)}$. The others lack the $A^{(2)}$ coefficient. Comparison of pairs of 
curves shows that the log multiplied by the $A^{(2)}$ coefficient increases the 
rate by about 10\% around the peak, and decreases it in the mid-$Q_T$ region. 
The figure also shows that exclusion of the $B^{(1)}$ term leads to about 40\% 
decrease around the peak, and an increase away from it. Finally, the exclusion 
of the $C^{(1)}$ coefficient decreases the overall rate by about a factor of 2, 
coupled with some shape change similar to the $B^{(1)}$ case.


\begin{figure}[tp]
\begin{center}
\epsfxsize=12cm
\epsfysize=12cm
\mbox{\epsfbox{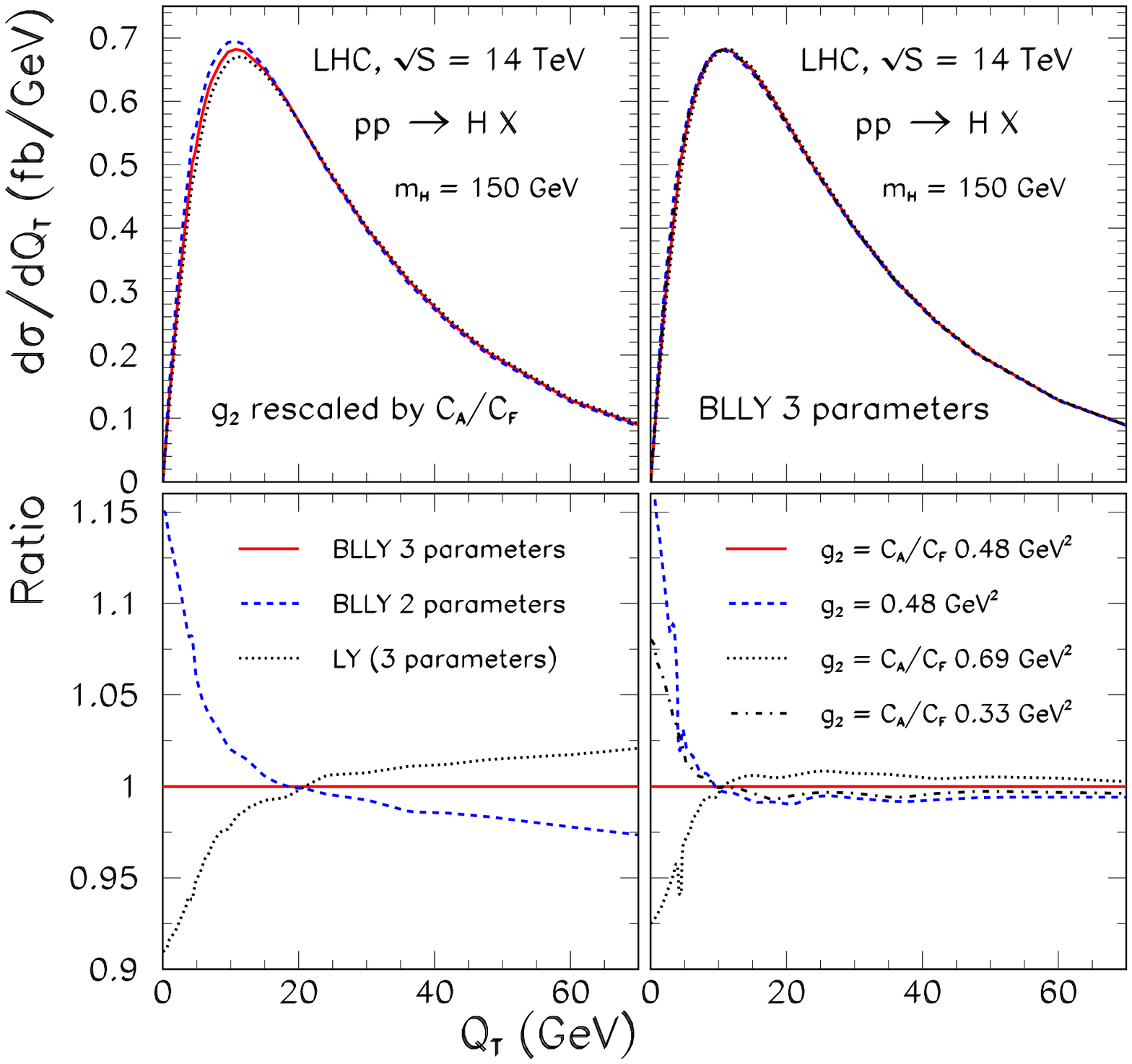}} 
\end{center}
\caption[Fig:NonP]{\sf Higgs boson transverse momentum distributions at the
LHC, displaying uncertainties arising from the non-perturbative sector of
the CSS formalism. a) The solid curve is calculated using the latest
3-parameter fit of the non-perturbative function \protect\cite
{BrockLadinskyLandryYuan}. The dashed curve uses the new 2-parameter fit 
\protect\cite{BrockLadinskyLandryYuan}, and the dotted curve the previous
3-parameter fit \protect\cite{LadinskyYuan}. The lower portion shows the
ratio of the dashed and the dotted curves to the solid one. b) The solid
curve uses the nominal result of the new 3-parameter fit, and the dashed
ones are calculated with $g_2$ parameters deviating by 3 $\sigma$ from the
central value. Also shown in dotted the curve which does not re-scale the $%
g_2$ parameter by $C_A/C_F$. The lower plot shows the ratios with respect to
the solid line. }
\label{Fig:NonP}
\end{figure}

Fig.~\ref{Fig:NonP} displays transverse momentum distributions of Higgs
bosons produced at the LHC. The $Q_T$ distribution is calculated under
several different assumptions for the non-perturbative sector of the CSS
formalism, in order to span the range of scatter of these different
predictions. In Fig.~\ref{Fig:NonP}a the (solid) curve using the result of
the latest 3-parameter fit for the non-perturbative function \cite
{BrockLadinskyLandryYuan} is shown. (The actual values of the parameters
used are: $g_1=0.15$ GeV$^2$, $g_2=(C_F/C_A)*0.48$ GeV$^2$, and $g_3=-0.58$
GeV$^{-1}$.) Also shown the (dashed) curve using the result of the latest
3-parameter fit of Ref.\ \cite{BrockLadinskyLandryYuan}. (The values were
used are: $g_1=0.24$ GeV$^2$, and $g_2=(C_F/C_A)*0.34$ GeV$^2$.) We plotted
the (dotted) curve using the previous 3-parameter fit of Ref.\ \cite
{LadinskyYuan}, as well. In the lower portion of the figure we show the
ratios of the different curves to the solid curve. From this we conclude
that the three different parameterizations differ by about 5 percent, at
most, in the relevant $Q_T$ region. At $Q_T=10$ GeV, in the region of the
peak of the distribution, the difference is about 2 percent.

In Fig.~\ref{Fig:NonP}b the solid curve is the same as in Fig.~\ref{Fig:NonP}%
a. In this figure results using $g_2 = (C_F/C_A)*0.33$ GeV$^2$, and $g_2 =
(C_F/C_A)*0.69$ GeV$^2$ values are plotted (dashed). These $g_2$ values are
3 $\sigma$ deviations from the central value $g_2 = (C_F/C_A)*0.48$ GeV$^2$
of the new 3-parameter fit. Also shown a curve with $g_2 = 0.48$ GeV$^2$,
where the assumption that the non-perturbative parameter $g_2$ scales by $%
C_A/C_F$ for the gluonic initial state was not utilized. The lower portion
of the figure shows that the ratios of the various curves to the solid curve
do not deviate from 1 significantly except in the very low $Q_T$ ($<$ 5 GeV)
region.

\section*{Acknowledgments}

We thank the organizers of the les Houches workshop for their hospitality.
We are indebted for the CTEQ Collaboration for many invaluable discussions
and W. Sakumoto for the CDF results. C.B. thanks M. Spira, and C.-P. Yuan
for discussions. This work was supported in part by the DOE under grant
DE-FG-03-94ER40833.

 
\setcounter{figure}{0}
\setcounter{table}{0}
\setcounter{section}{0}
\setcounter{equation}{0}
\newpage

%
%
%
%

\begin{center}
\vspace*{1.2cm}
{\Large\sc \bf A Comparison of the Predictions from Monte Carlo Programs and 
Transverse Momentum Resummation} \\
\vspace*{1.cm} 
{\sc C. Bal\'azs, J. Huston, I. Puljak}
\vspace*{1.cm}
\end{center}

\setcounter{footnote}{0}

\begin{abstract}
	
	Monte Carlo event generators are being increasingly relied upon for 
predictions of experimental observables at colliders. In this section, the 
parton shower formalism for Monte Carlos is compared to that of analytic
resummation calculations. Predictions for the transverse momentum 
distribution of $Z^0$ bosons, photon pairs, and the Higgs boson are compared
for the Tevatron and the LHC.~\footnote{A more complete treatment of this subject can be found 
in Ref.~\cite{higgs_paper}.}

\end{abstract}

	
\section{Introduction}

	Parton shower Monte Carlo programs such as {\tt PYTHIA}\cite{pythia}, 
{\tt HERWIG}\cite{herwig2} and {\tt ISAJET}\cite{isajet}
are commonly used by experimentalists, both as a way of comparing
experimental data to theoretical predictions, and also as a means of 
simulating experimental signatures in kinematic regimes for which there
is not yet experimental data (such as the LHC). The final output of the 
Monte Carlo programs consists of the 4-vectors of a set of final state
hadrons; this output can either be compared to reconstructed experimental
quantities or, when coupled with a simulation of a detector response, 
can be directly compared to raw data taken by the experiment, and/or
passed through the same reconstruction procedures as the raw data.
In this way, the parton shower programs can be more useful to 
experimentalists than analytic calculations. Indeed, almost all of the 
physics plots in the ATLAS physics TDR~[5]
involve comparisons to {\tt PYTHIA} (version 5.7). 

	For many physical quantities, the predictions from parton shower Monte
Carlo programs should be nearly as precise as those from analytic theoretical 
calculations. This is expected, among others, for calculations which resum logs
with the transverse momentum of partons initiating the hard scattering.
In the recent literature, most calculations of this kind are either based on 
or originate from the formalism developed by J. Collins, D. Soper, and G. 
Sterman (CSS)~\footnote{See, for example, the discussion in the previous section.}, which we choose as the analytic `benchmark' of this section.
In this case, both the Monte Carlo and analytic calculations
should accurately describe the effects of the emission of 
multiple soft gluons from the incoming partons,
an all orders problem in QCD. The initial state soft gluon emission can affect
the kinematics of the final state partons. This may have an impact on the 
signatures of physics processes at both the trigger and analysis levels and thus
it is important to understand the reliability of such predictions. The best 
method for testing the reliability is the direct comparison of the predictions
to experimental data. If no experimental data is available for certain 
predictions, then some understanding of the reliability may be gained from
the comparison of the predictions from the two different methods. 


\section{Parton Showering and Resummation}

For technical reasons, the initial 
state parton shower proceeds by a $\it{backwards}$ evolution, starting at
the large (negative) $Q^2$ scale of the hard scatter and then considering 
emissions 
at lower and lower (negative) virtualities, corresponding to earlier points on
the cascade (and earlier points in time), until a scale corresponding to the 
factorization scale is reached. The transverse momentum of 
the initial state is built up from the  whole series of splittings (and boosts). 
The showering process is independent of the hard scattering process being
considered (as long as one does not introduce any matrix element corrections), and depends only on the initial state partons and the hard scale of
the process.

In the case of
parton showering, the leading order collinear singularities factorize for 
cross sections in the collinear limit
\begin{eqnarray}
   \lim_{p_g \to p_b} |{\cal M}_{n+1}|^2 = 
   g_s^2 (p_b.p_g)^{-1} P_{g \leftarrow a}(z) |{\cal M}_{n}|^2, 
\end{eqnarray}
where ${\cal M}_{n+1}$ is the invariant amplitude for the process producing $n$
partons and a gluon, $g_s$ is the strong coupling constant, $p_b$ and $p_g$ are 
the 4-momenta of the daughters of the 
n'th parton $a$ (i.e. $a$ splits into $b$ and $g$, and when they are collinear 
then $p_b.p_g \to 0$). Finally $P_{g \leftarrow a}(z)$ is the DGLAP splitting 
kernel belonging to the $a \to g$ splitting.
The leading order collinear singularities can be factorized into
a Sudakov form factor: 
$S = 1-P (no ~emission) = exp(-\int dp^2/p^2 \int dz P(z))$. 
The distribution $1-S$ can be used to generate the $Q^2$ for the first emission
and hence
for the whole cascade. The formalism can be extended to soft singularities as 
well
by using angular ordering. 
In this approach, the choice of the hard scattering is based on the use of 
evolved parton distributions, which
means that the inclusive effects of initial-state radiation are already included. 
What remains is therefore to construct the exclusive showers.

Parton showering resums primarily the leading logs, which are
universal, i.e. process independent, and depend only on the given initial state. 
In this lies one of the strengths of Monte Carlos, since parton showering can be
incorporated into a wide variety of physical processes.
An analytic calculation, in comparison, can resum all logs. For example, the CSS
formalism sums all of the logarithms with $Q^2/p_T^2$ in their arguments,
where (for Higgs boson production) $Q$ is the four momentum of the Higgs 
and $p_T$ is its transverse momentum.
As discussed in the previous section on resummation, all of the `dangerous logs' 
are included in the 
Sudakov exponent, which can be written in the impact parameter ($b$) space as:
\begin{eqnarray*}
{\cal S}(p,b)=\int_{1/b^2}^{Q^2}\frac{d\overline{\mu }^2}{%
\overline{\mu }^2}\left[A\left(\alpha_s(\overline{\mu })\right) \ln
\left( \frac{Q^2}{\overline{\mu }^2}\right) {+B}\left(\alpha_s(\overline{%
\mu })\right) \right],
\end{eqnarray*}
with the $A$ and $B$ functions being free of 
large logs and perturbatively calculable:
\begin{eqnarray*}
A\left( \alpha_s({\bar{\mu}})\right) = \sum_{n=1}^\infty \left( 
\frac{\alpha_s({\bar{\mu}})}\pi \right) ^nA^{(n)} , ~~~
B\left( \alpha_s({\bar{\mu}})\right) = \sum_{n=1}^\infty
\left( \frac{\alpha_s({\bar{\mu}})}\pi \right) ^nB^{(n)} .
\end{eqnarray*}

These functions contain an infinite number of coefficients, 
with the $A^{(n)}$ coefficients being universal while the $B^{(n)}$ are
process dependent. 
In practice, the number of towers of logarithms included in the Sudakov exponent
depends on the level to which a fixed order calculation was performed for a 
given process. 
For example, if only a next-to-leading order calculation is available, only the 
coefficients $A^{(1)}$ and $B^{(1)}$ can be included.
If a NNLO calculation is available, then $A^{(2)}$ and $B^{(2)}$ can be 
extracted and incorporated into a resummation calculation, and so on. This is 
the case, for example, for $Z^0$ boson production. So far, only the $A^{(1)}$,
$A^{(2)}$ and $B^{(1)}$ coefficients are known for Higgs production but the 
calculation of $B^{(2)}$ is in progress.~\cite{carlschmidt}
If we try to interpret parton showering in the same language, which is 
admittedly risky, then we can say that the Monte Carlo Sudakov exponent always 
contains a term analogous to $A^{(1)}$. 
It was shown in Reference~\cite{webber} that a suitable modification of the 
Altarelli-Parisi splitting function, or equivalently the strong coupling constant
$\alpha_s$, also effectively approximates the $A^{(2)}$ 
coefficient.~\footnote{This is rigorously true only for the high x or 
$\sqrt{\tau}$ region.}

In contrast with the shower Monte Carlos, analytic resummation 
calculations integrate over the kinematics of the soft gluon emision, with the
result that they are limited in their predictive power for inclusive final
states. While the Monte Carlo maintains an exact treatment of the 
branching kinematics, in the original CSS formalism no kinematic penalty is 
paid for the emission of the soft gluons, although an approximate treatment of 
this can be incorporated into its numerical implementations, like 
ResBos~\cite{csabaref}.
Neither the parton showering process nor the analytic resummation
translate smoothly into kinematic configurations
where one hard parton is emitted (at large $p_T$). In the Monte Carlo matrix 
element corrections, and in the analytic resummation calculation matching
is necessary. This matching is standard procedure for resummation calculations 
and matrix element corrections are becoming increasingly common in 
Monte Carlos~\cite{pythiacor,herwigcor}.
 
With the appropriate input from higher order cross sections,
a resummation calculation has the corresponding higher order normalization and
scale dependence. 
The normalization and
scale dependence for the Monte Carlo, though, remains that of a leading 
order calculation.  The parton showering process redistributes the event 
particles in phase space, but does not change the total cross section (for
example, for the production of a Higgs boson). \footnote{Technically, 
one could add the branching for 
$q\to q$+Higgs in the shower, which would have the 
capability of increasing somewhat the Higgs cross section; however, the main 
contribution to the higher order $K$-factor comes from the virtual 
corrections and the `Higgs Bremsstrahlung' contribution is neglible.}

	In particular, one quantity which should be well-described by both 
calculations is the transverse momentum ($p_T$) 
of the final state electroweak boson in a subprocess such
as $q\overline{q} \to WX$, $ZX$ or $gg \to H X$, where most of the 
$p_T$ is provided by initial state parton showering. The parton showering 
supplies the same sort of transverse kick as the soft gluon radiation
in a resummation calculation. Indeed, very similar 
Sudakov form factors appear in both approaches, with the caveats about
the $A^{(n)}$ and $B^{(n)}$ terms mentioned previously.
This correspondence between the Sudakov form factors in
resummation and Monte Carlo approaches may seem trivial, but there are many 
subtleties between the two approaches relating to both the arguments of the
Sudakov factors as well as the impact of subleading logs~\cite{mrenna}.

At a point in its evolution corresponding to (typically) the virtuality of a few
GeV$^2$, the parton shower is cut off and the effects of gluon emission at
softer scales must be parameterized and inserted by hand.  This is similar to
the (somewhat  arbitrary) division between perturbative and non-perturbative
regions in a resummation calculation. The parameterization is typically
done with a Gaussian formalism similar to that used for the non-perturbative 
$k_T$ in a resummation program. In general, the value for the non-perturbative
$\langle k_T \rangle$ needed in a Monte Carlo program will depend on the particular kinematics
being investigated. 
In the case of the resummation calculation 
the non-perturbative physics is determined from fits to fixed 
target data and then automatically evolved to the kinematic regime of interest.

	A value  for the  average non-perturbative 
$k_T$ of greater than 1 GeV does not imply that there
is an anomalous intrinsic $k_T$ associated with the parton size; rather this
amount of $\langle k_T \rangle$ needs to be supplied to provide what is missing in the 
truncated parton shower. If the shower is cut off at a higher virtuality, more
of the `non-perturbative' $k_T$ will be needed. 

\section{$Z^0$ Boson Production at the Tevatron}

\begin{figure}[thp]
\begin{center}
\epsfxsize=12cm
\epsfysize=12cm
\mbox{\epsfbox{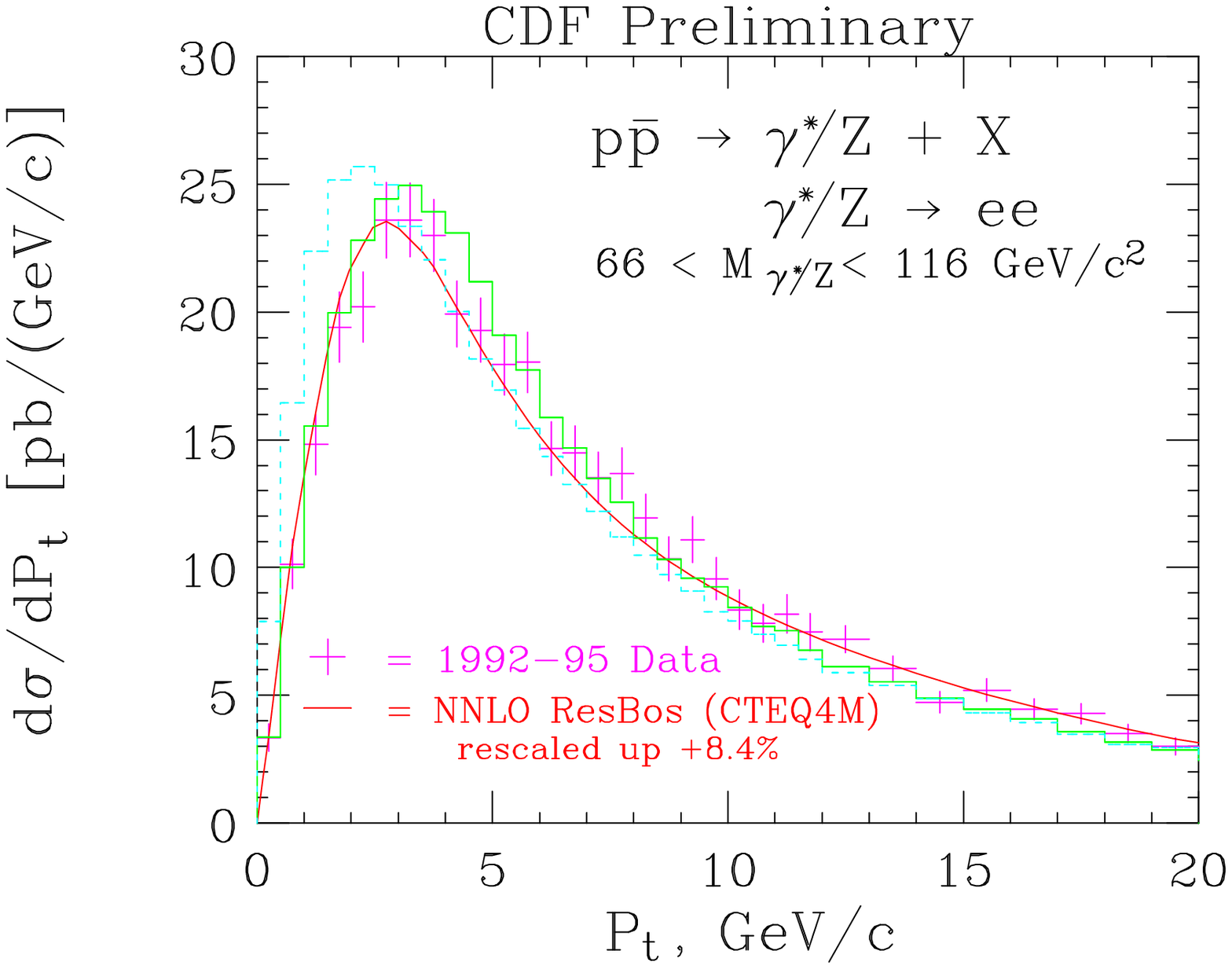}}
\end{center}
\caption{
\sf 
The $Z^0$ $p_T$ distribution (at low $p_T$) from CDF for Run 1 compared to 
predictions from ResBos and from {\tt PYTHIA}. The two {\tt PYTHIA} 
predictions use the default (rms) value for the non-perturbative $k_T$ 
(0.44 GeV) and the value that gives the best agreement with the shape of 
the data (2.15 GeV). 
}
\label{fig:run1_ee_pt}
\end{figure}

The 4-vector of a $Z^0$ boson, and thus its transverse momentum, can be 
measured with great precision in the $e^+e^-$ decay mode. Resolution 
effects are relatively minor and are easily corrected. Thus, the $Z^0$ 
$p_T$ distribution is a great testing ground for both the resummation and 
Monte Carlo formalisms for soft gluon emission. The (resolution corrected) 
$p_T$ distribution for $Z^0$ bosons (in the low $p_T$ region) for the CDF 
experiment\footnote{We thank Willis Sakumoto for providing the figures for 
CDF $Z^0$ production} is shown in Figure~\ref{fig:run1_ee_pt}, compared 
to both the resummed prediction from ResBos, and to two predictions from 
{\tt PYTHIA} (version 6.125). One {\tt PYTHIA} prediction uses the default 
(rms)\footnote{For a Gaussian distribution, $k_T^{rms}=1.13\langle k_T 
\rangle$.} value of  intrinsic $k_T$ of 0.44 GeV and the second a value 
of 2.15 GeV (per incoming parton).~\footnote{A previous 
publication~\cite{pythiacor} indicated the need for a substantially larger 
non-perturbative $\langle k_T \rangle$, of the order of 4 GeV for the 
case of $W$ production at the Tevatron. The data used in the comparison, 
however, were not corrected for resolution smearing, a fairly large effect 
for the case of $W \to e{\nu}$ production and decay.} The latter 
value was found to give the best agreement for {\tt PYTHIA} with the 
data.\footnote{A similar conclusion has been reached for comparisons of 
the CDF $Z^0$ $p_T$ data with {\tt HERWIG}~\cite{corcella}}.
All of the predictions use the CTEQ4M parton distributions~\cite{cteq4}.
The shift between 
the two {\tt PYTHIA} predictions at low $p_T$ is clearly evident. As might have 
been expected, the high $p_T$ region (above 10 GeV) is unaffected by the 
value of the non-perturbative $k_T$. Note that much of the $k_T$  `given' 
to the incoming partons at their lowest virtuality, $Q_0$, is reduced at 
the hard scatter due to the number of gluon branchings preceding the 
collision. The emitted gluons carry off a sizeable fraction of the 
original non-perturbative $k_T$. This point will be investigated in more 
detail later for the case of Higgs production. 

As an exercise, one can transform the resummation formula in order to 
bring it to a form where the non-perturbative function acts as a Gaussian 
type smearing term. Using the Ladinsky-Yuan parameterization~\cite{LY} of 
the non-perturbative function in ResBos leads to an rms value for the 
effective $k_T$ smearing parameter, for $Z^0$ production at the Tevatron, 
of 2.5 GeV. This is similar to that needed for {\tt PYTHIA} and {\tt 
HERWIG} to describe the $Z^0$ production data at the Tevatron. 

In Figure~\ref{fig:run1_ee_pt}, the normalization of the resummed 
prediction has been rescaled upwards by 8.4\%. The {\tt PYTHIA} prediction 
was rescaled by a factor of 1.3-1.4 (remember that this is only a leading 
order comparison) for the shape comparison. 

As stated previously, the resummed prediction correctly describes the 
shape of the $Z^0$ $p_T$ distribution at low $p_T$, 
even with the optimal non-perturbative $k_T$, 
although there is still a noticeable difference in 
shape between the Monte Carlo and the resummed prediction. It is 
interesting to note that if the process dependent coefficients ($B^{(1)}$ and 
$B^{(2)}$) were not incorporated into the resummation prediction, the result 
would be an increase in the height of the peak and a decrease in the rate 
between 10 and 20 GeV, leading to a better agreement with the {\tt PYTHIA} 
prediction~\cite{csaba}.

\begin{figure}[thp]
\begin{center}
\epsfxsize=12cm
\epsfysize=12cm
\mbox{\epsfbox{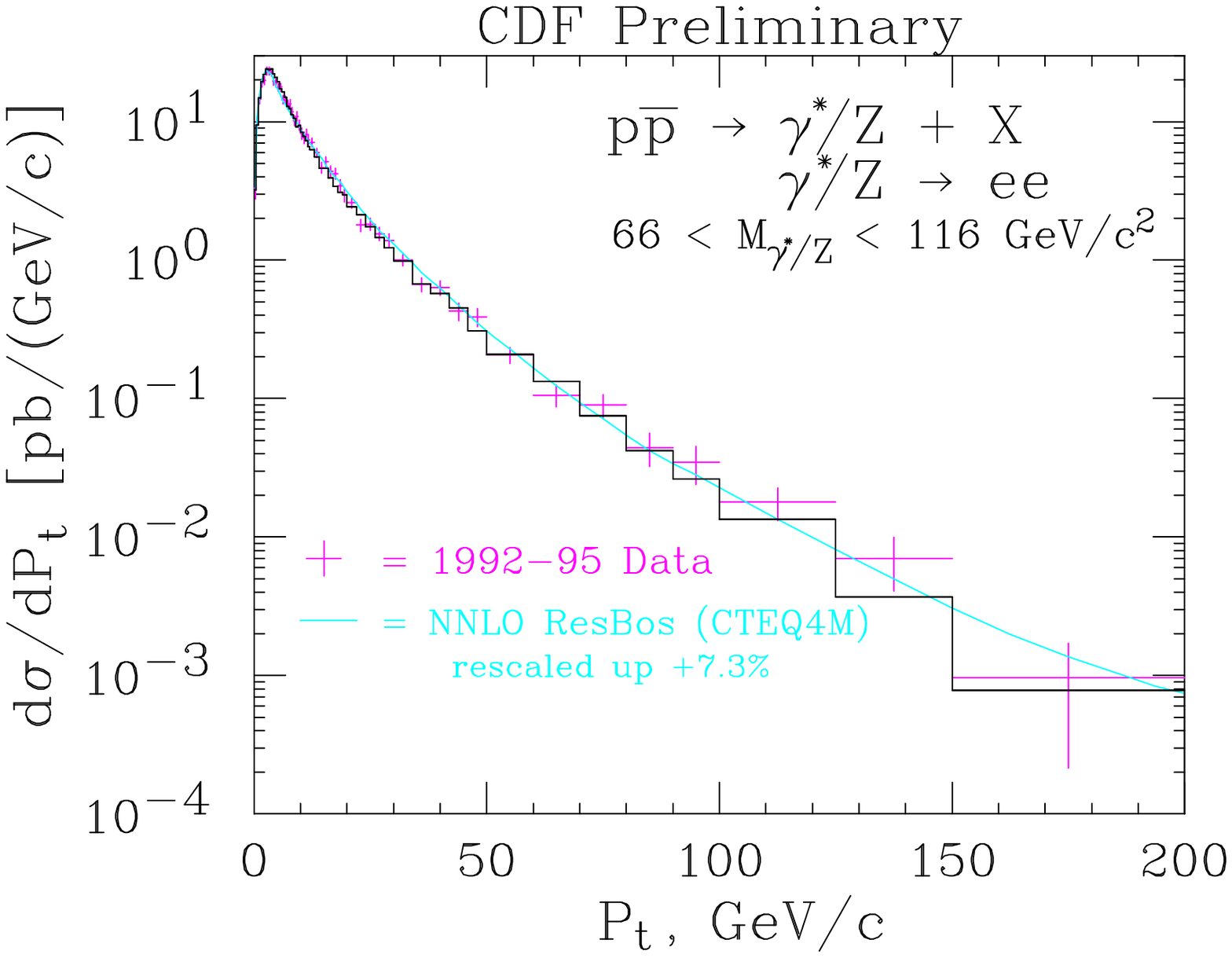}}
\end{center}
\caption{
\sf The $Z^0$ $p_T$ distribution (for the full range of $p_T$) from CDF for 
Run 1 compared to predictions from ResBos (curve) and from {\tt PYTHIA}  
(histogram). 
} 
\label{fig:zpteeall}
\end{figure}
The $Z^0$ $p_T$ distribution is shown over a wide $p_T$ range in  
Figure~\ref{fig:zpteeall}. The {\tt PYTHIA} and ResBos predictions both 
describe the data well. Note especially the agreement of {\tt PYTHIA} with 
the data at high $p_T$, made possible by explict matrix element 
corrections (from the subprocesses $q\overline{q} \to Z^0g$ and 
$gq \to Z^0q$) to the $Z^0$ production process.\footnote{Slightly 
different techniques are used for the matrix element corrections by {\tt 
PYTHIA}~\cite{pythiacor} and by {\tt HERWIG}~\cite{herwigcor}.  In {\tt 
PYTHIA}, the parton shower probability distribution is applied over the 
whole phase space and the exact matrix element corrections are applied 
only to the branching closest to the hard scatter.  In {\tt HERWIG}, the 
corrections are generated separately for the regions of phase space 
unpopulated by {\tt HERWIG} (the `dead zone') and the populated region. In 
the dead zone, the radiation is generated according to a distribution 
using the first order matrix element calculation, while the algorithm for 
the already populated region applies matrix element corrections whenever a 
branching is capable of being `the hardest so far'.}

\section{Diphoton Production}

\begin{figure}[thp]
\begin{center}
\epsfxsize=12cm
\epsfysize=12cm
\mbox{\epsfbox{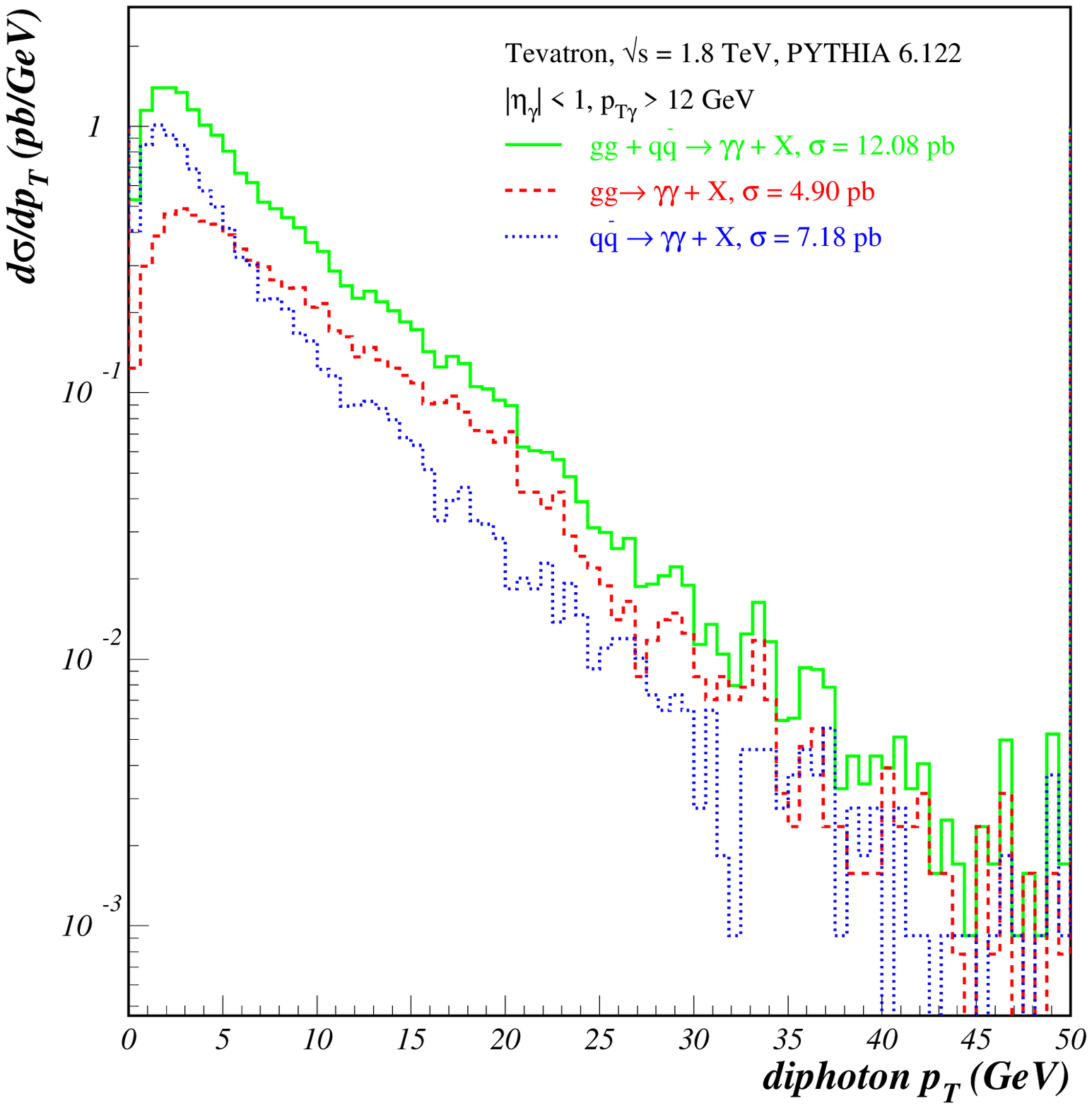}}
\end{center}
\caption{
\sf A comparison of the {\tt PYTHIA} predictions for diphoton production at the 
Tevatron for the two different subprocesses, $q\overline{q}$ and $gg$. 
The same cuts are applied to {\tt PYTHIA} as in the CDF diphoton analysis.
} 
\label{fig:pythiakt}
\end{figure}

Most of the experience that we have for comparisons of data to resummation 
calculations/Monte Carlos  deals with Drell-Yan production, i.e. 
$q\overline{q}$ initial states. It is important then to examine diphoton 
production at the Tevatron, where a large fraction of the contribution at 
low mass is due to $gg$ scattering. The prediction for the diphoton $k_T$ 
distribution at the Tevatron, from {\tt PYTHIA} (version 6.122), is shown 
in Figure~\ref{fig:pythiakt}, using the experimental cuts applied in the 
CDF analysis~\cite{cdfdiphot}. It is interesting to note that about half 
of the diphoton cross section at the Tevatron is due to the $gg$ 
subprocess, and that the diphoton $p_T$ distribution is noticeably broader 
for the $gg$ subprocess than the $q\overline{q}$ subprocess.

\begin{figure}[thp]
\begin{center}
\epsfxsize=12cm
\epsfysize=12cm
\mbox{\epsfbox{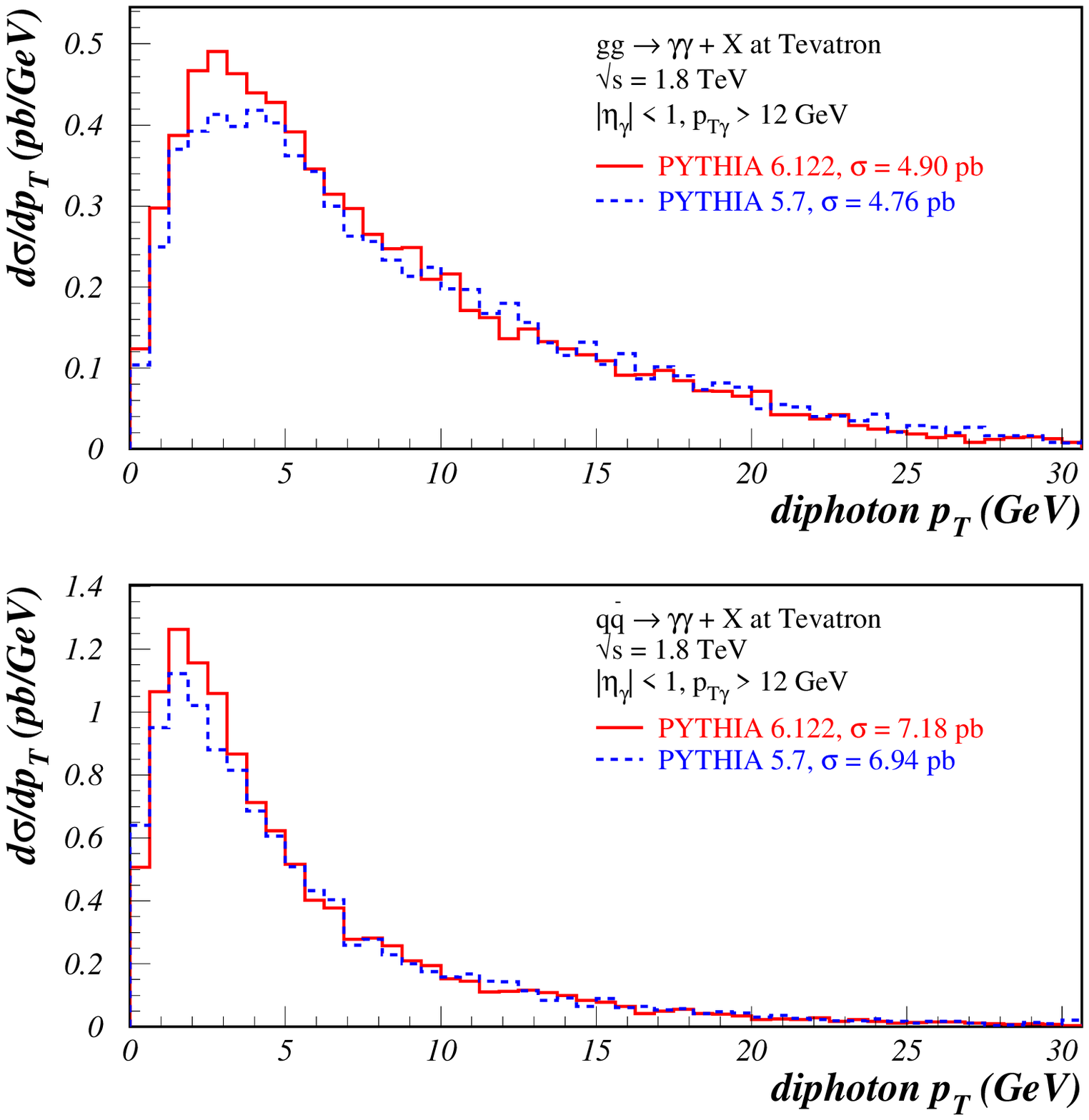}}
\end{center}
\caption{
\sf A comparison of the {\tt PYTHIA} predictions for diphoton production 
at the Tevatron for the two different subprocesses, $q\overline{q}$ and 
$gg$, for two recent versions of {\tt PYTHIA}. The same cuts are applied 
to {\tt PYTHIA} as in the CDF diphoton analysis.
} 
\label{fig:2gamma_tev}
\end{figure}

%
A comparison of the $p_T$ distributions for the two diphoton subprocesses 
$(q\overline{q}, gg)$ in {\tt PYTHIA} versions 5.7 and 6.1 is shown in 
 Figure~\ref{fig:2gamma_tev}. There seems to be
little difference in the 
 $p_T$ distributions between the two versions for both subprocesses.
%

\begin{figure}[thp]
\begin{center}
\epsfxsize=12cm
\epsfysize=12cm
\mbox{\epsfbox{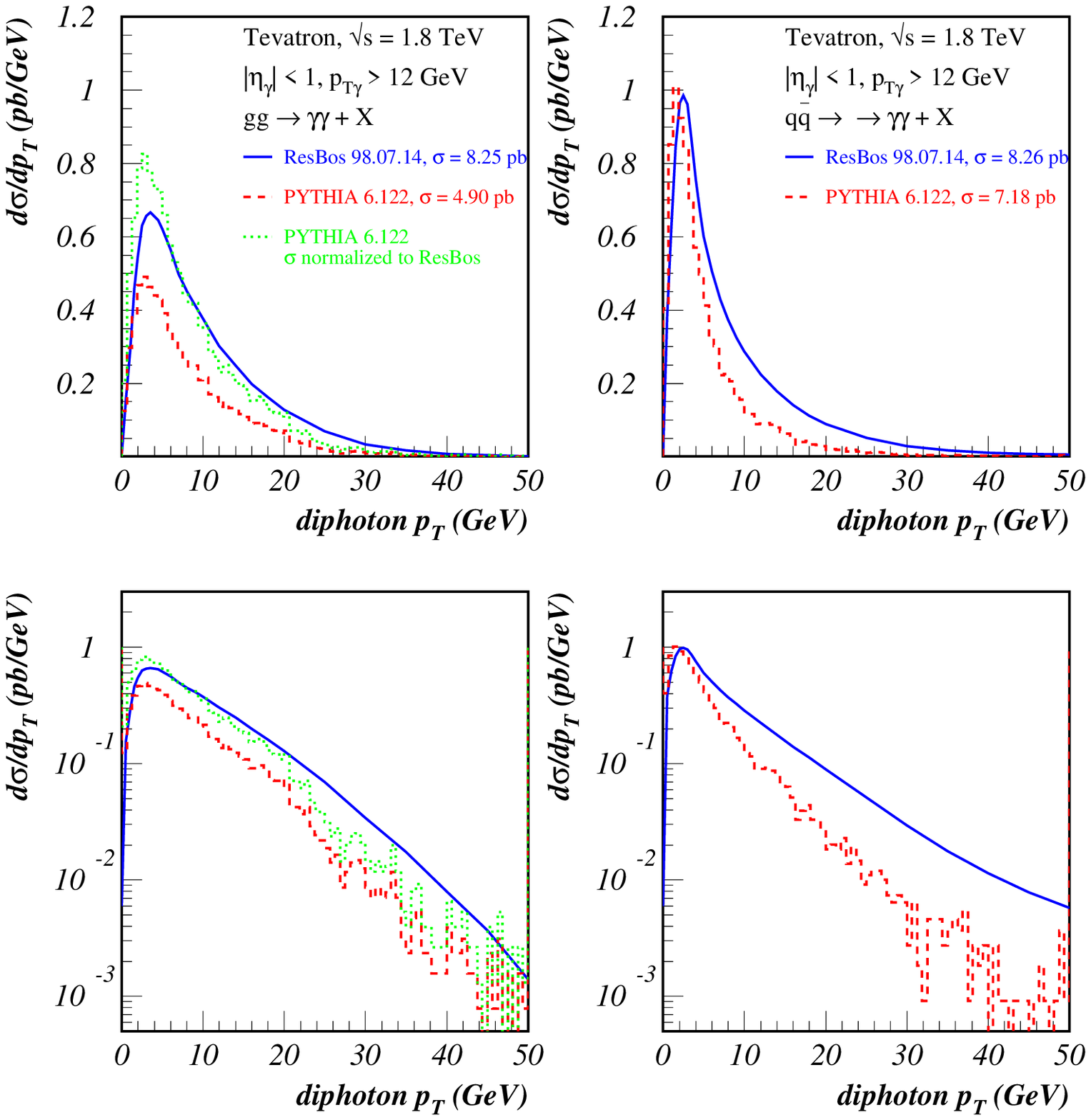}}
\end{center}
\caption{
\sf A comparison of the {\tt PYTHIA}  and ResBos predictions for diphoton 
production at the Tevatron for
the two different subprocesses, $q\overline{q}$ and $gg$. The same cuts are 
applied to {\tt PYTHIA} and ResBos as in the CDF diphoton analysis.
} 
\label{fig:pythrestev}
\end{figure}

In  Figure~\ref{fig:pythrestev} are shown the ResBos predictions for 
diphoton production at the Tevatron from $q\overline{q}$ and $gg$ 
scattering compared to the {\tt PYTHIA} predictions (using the same 
experimental cuts). The $gg$ subprocess predictions in ResBos agree well 
with those from {\tt PYTHIA} 
while the $q\overline{q}$ $p_T$ distribution is noticebly broader in 
ResBos. The latter behavior is due to the presence of the $Y$ piece in 
ResBos at moderate $p_T$, and the matching of the  $q\overline{q}$ cross 
section to the fixed order $q\overline{q} \to {\gamma}{\gamma}g$ at high 
$p_T$. The corresponding matrix element correction is not in {\tt PYTHIA}. 
It is interesting to note that the {\tt PYTHIA} and ResBos predictions for $gg 
\to {\gamma}{\gamma}$ agree in the moderate $p_T$ region, even though the 
ResBos prediction has the $Y$ piece present and is matched to the matrix 
element piece $gg \to {\gamma}{\gamma}g$ at high $p_T$, while there is no 
such matrix element correction for {\tt PYTHIA}. This shows the smallness of 
the $Y$ piece for the $gg$ subprocess, which is the same conclusion that was 
reached in Ref.~\cite{BalazsNadolskySchmidtYuan}.
One way to understand this is recalling that the $gg$ parton-parton 
luminosity falls very steeply with increasing partonic center of mass energy, 
$\sqrt{\hat{s}}$. This falloff 
tends to suppress the size of the $Y$ piece since the production of the 
diphoton pair at higher $p_T$ requires larger $x_1$, $x_2$ values. In the 
default CSS formalism, there is no such kinematic penalty in the resummed 
piece since the soft gluon radiation comes for ``free''. (Larger $x_1$  
and $x_2$ values are not required.)

\begin{figure}[thp]
\begin{center}
\epsfxsize=12cm
\epsfysize=12cm
\mbox{\epsfbox{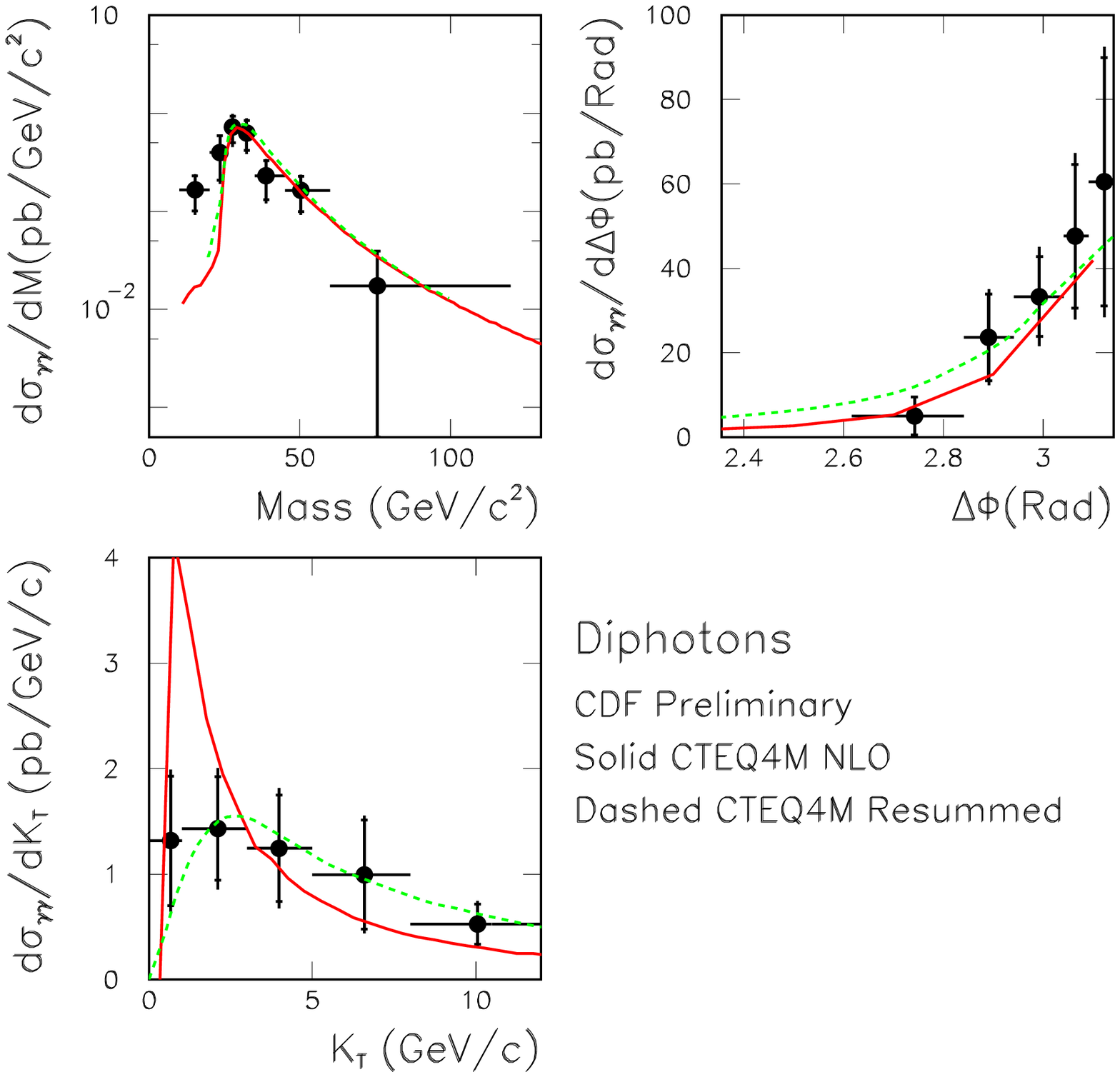}}
\end{center}
\caption{
\sf A comparison of the NLO and ResBos predictions for diphoton production at 
the Tevatron for the diphoton mass, the angle $\Delta\phi$ and the transverse
momentum of the photon pair $K_T$.} 
\label{fig:makemyplot}
\end{figure}

A comparison of the CDF diphoton data to NLO \cite{owens1} and resummed 
(ResBos) QCD predictions is shown in Figure~\ref{fig:makemyplot}. Plotted 
are the diphoton mass, the angle $\Delta{\phi}$ between the two photons 
and the transverse momentum $k_T$ of the diphoton pair. The transverse 
momentum distribution, in particular, is sensitive to the effects of the 
soft gluon radiation and better agreement can be observed with the ResBos 
prediction than with the NLO one. The data shown in this figure is from an 
integrated luminosity of 87 $pb^{-1}$. A much more precise comparison with 
the effects of soft gluon radiation will be possible with the 2 $fb^{-1}$ 
or greater data sample that is expected for both CDF and D0 in Run 2. 

\begin{figure}[thp]
\begin{center}
\epsfxsize=12cm
\epsfysize=12cm
\mbox{\epsfbox{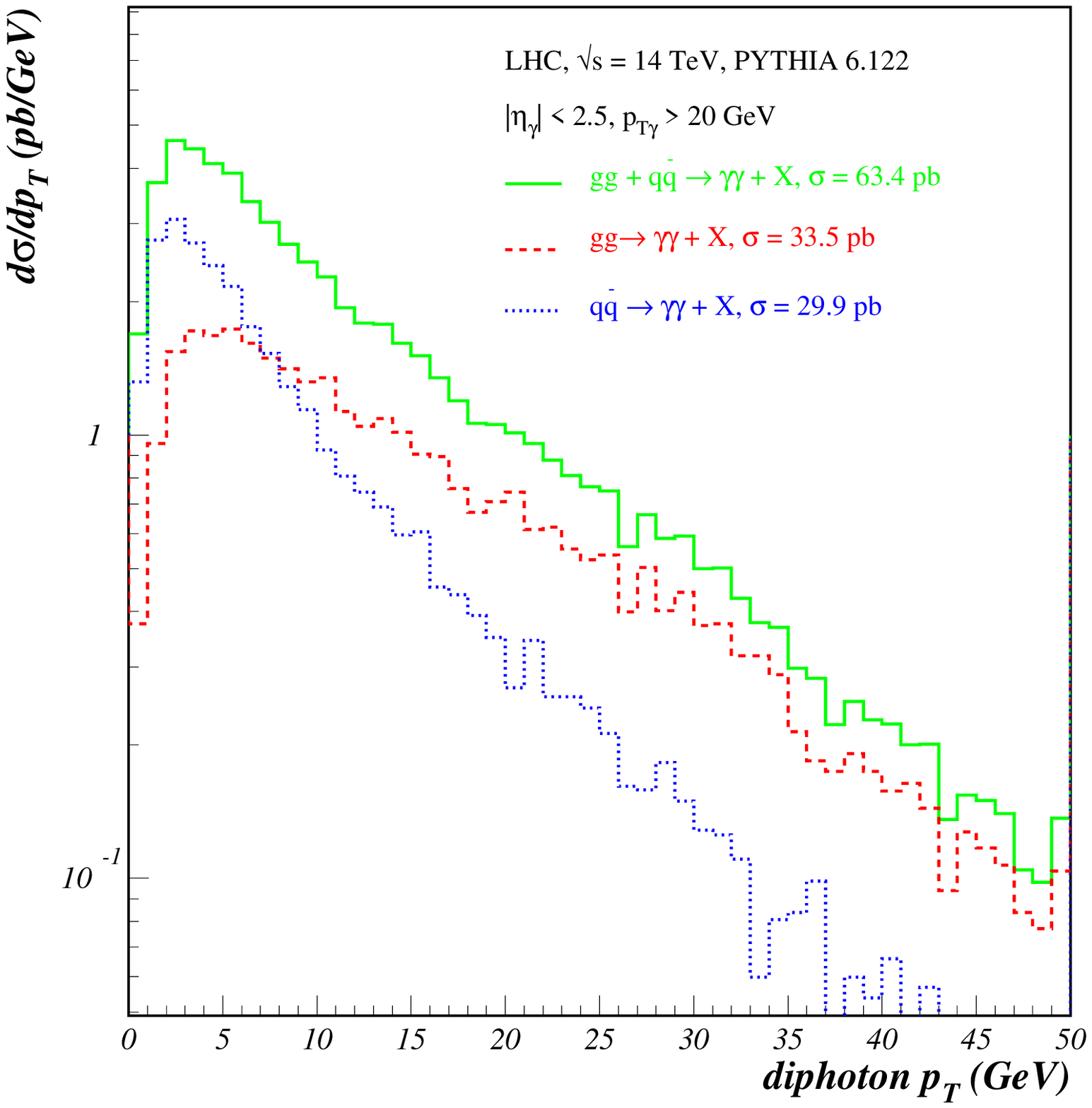}}
\end{center}
\caption{
\sf A comparison of the {\tt PYTHIA} predictions for diphoton production 
at the LHC for the two different subprocesses, $q\overline{q}$ and $gg$. 
Similar cuts are applied to the diphoton kinematics as those used by ATLAS 
and CMS.} 
\label{fig:lhc_diphot}
\end{figure}

The prediction for the diphoton production cross section, as a function
of the diphoton $p_T$ and using cuts appropriate to ATLAS and CMS, is shown
in Figure~\ref{fig:lhc_diphot}. Note  that, as at the Tevatron, about
half of the cross section is due to $gg$ scattering and the diphoton 
$p_T$ distribution from $gg$ scattering is noticeably broader than that from
$q\overline{q}$ production. 
%

\begin{figure}[thp]
\begin{center}
\epsfxsize=12cm
\epsfysize=12cm
\mbox{\epsfbox{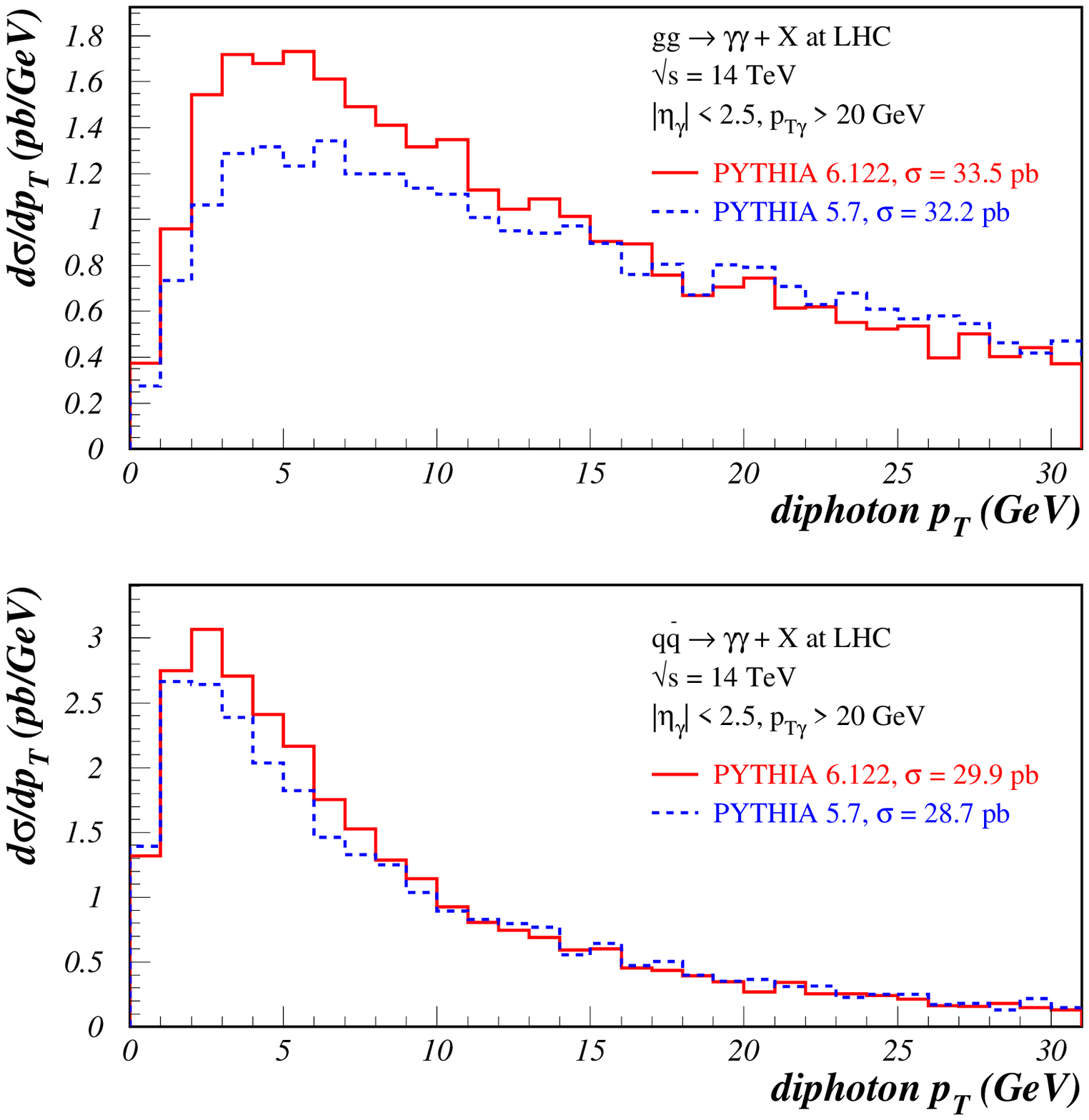}}
\end{center}
\caption{
\sf A comparison of the {\tt PYTHIA} predictions for diphoton production 
at the LHC for the two different subprocesses, $q\overline{q}$ and $gg$, 
for two recent versions of {\tt PYTHIA}. Similar cuts are applied to the 
diphoton kinematics as are used by ATLAS and CMS. 
} 
\label{fig:lhc_diphot_57}
\end{figure}

In Figure~\ref{fig:lhc_diphot_57} is shown a comparison of the diphoton 
$p_T$ distribution for two different versions of {\tt PYTHIA}, for the two 
different subprocesses. Note that the $p_T$ distribution in {\tt PYTHIA} 
version 5.7 is somewhat broader than that in version 6.122 for the case of 
$gg$ scattering. The effective diphoton mass range being considered here 
is lower than the 150 GeV Higgs mass that will be considered in the next 
section. As will be seen, the differences in soft gluon emission between 
the two versions of {\tt PYTHIA} are larger in that case. 
%

\begin{figure}[thp]
\begin{center}
\epsfxsize=12cm
\epsfysize=12cm
\mbox{\epsfbox{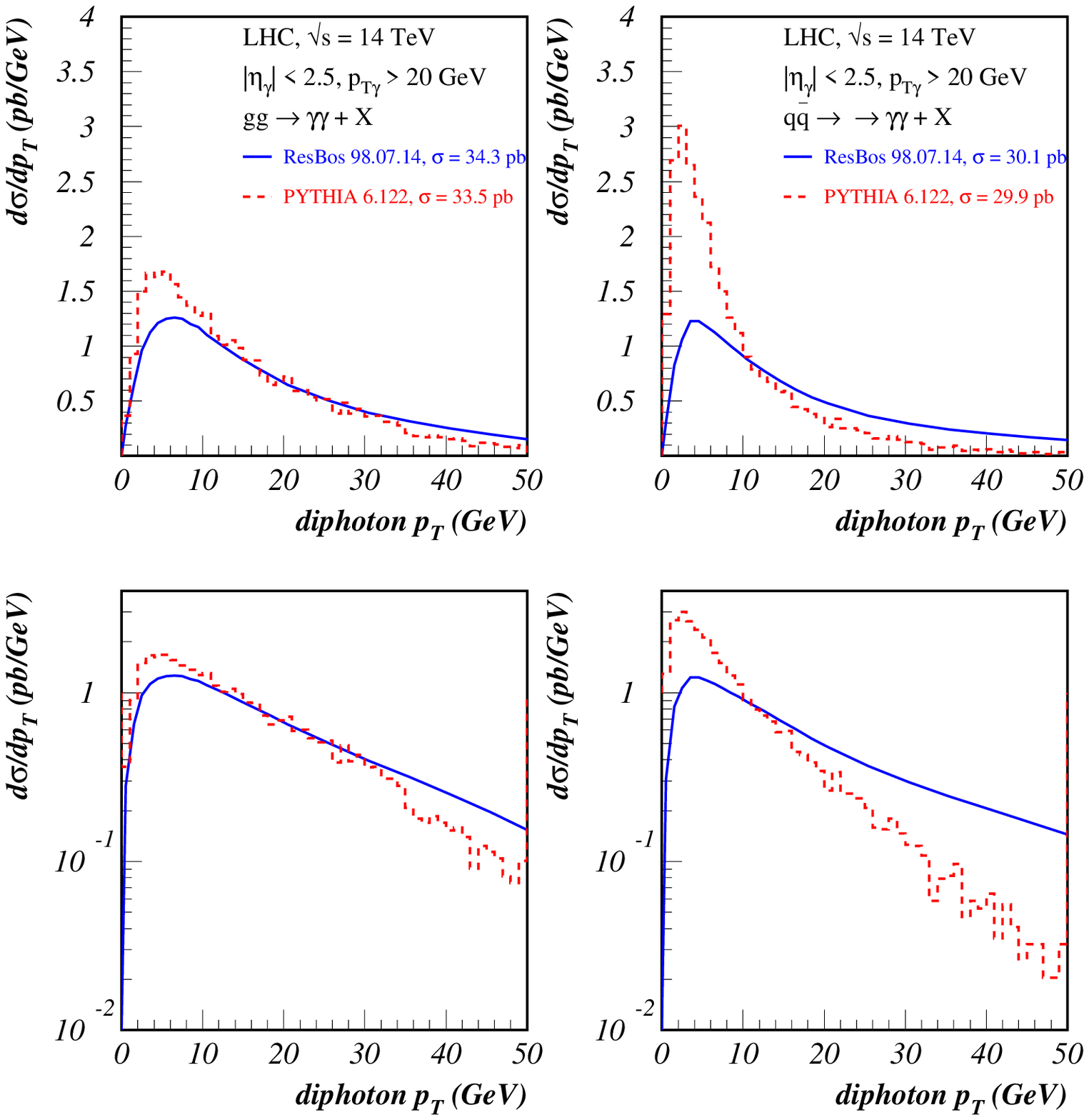}}
\end{center}
\caption{
\sf A comparison of the {\tt PYTHIA}  and ResBos predictions for diphoton 
production at the  LHC for
the two different subprocesses, $q\overline{q}$ and $gg$. Similar cuts are 
applied to {\tt PYTHIA} and ResBos as in the ATLAS and CMS diphoton analyses.
} 
\label{fig:pythreslhc}
\end{figure}

In  Figure~\ref{fig:pythreslhc} are shown the ResBos predictions for 
diphoton production at the LHC from $q\overline{q}$ and $gg$ scattering 
compared to the {\tt PYTHIA} predictions (using the same experimental 
cuts). Again, the $gg$ subprocess predictions in ResBos agree well with 
those from while the $q\overline{q}$ $p_T$ distribution is noticebly 
broader in ResBos, for the reasons cited previously. 
%

\section{Higgs Boson Production}

\begin{figure}[thp]
\begin{center}
\epsfxsize=12cm
\epsfysize=12cm
\mbox{\epsfbox{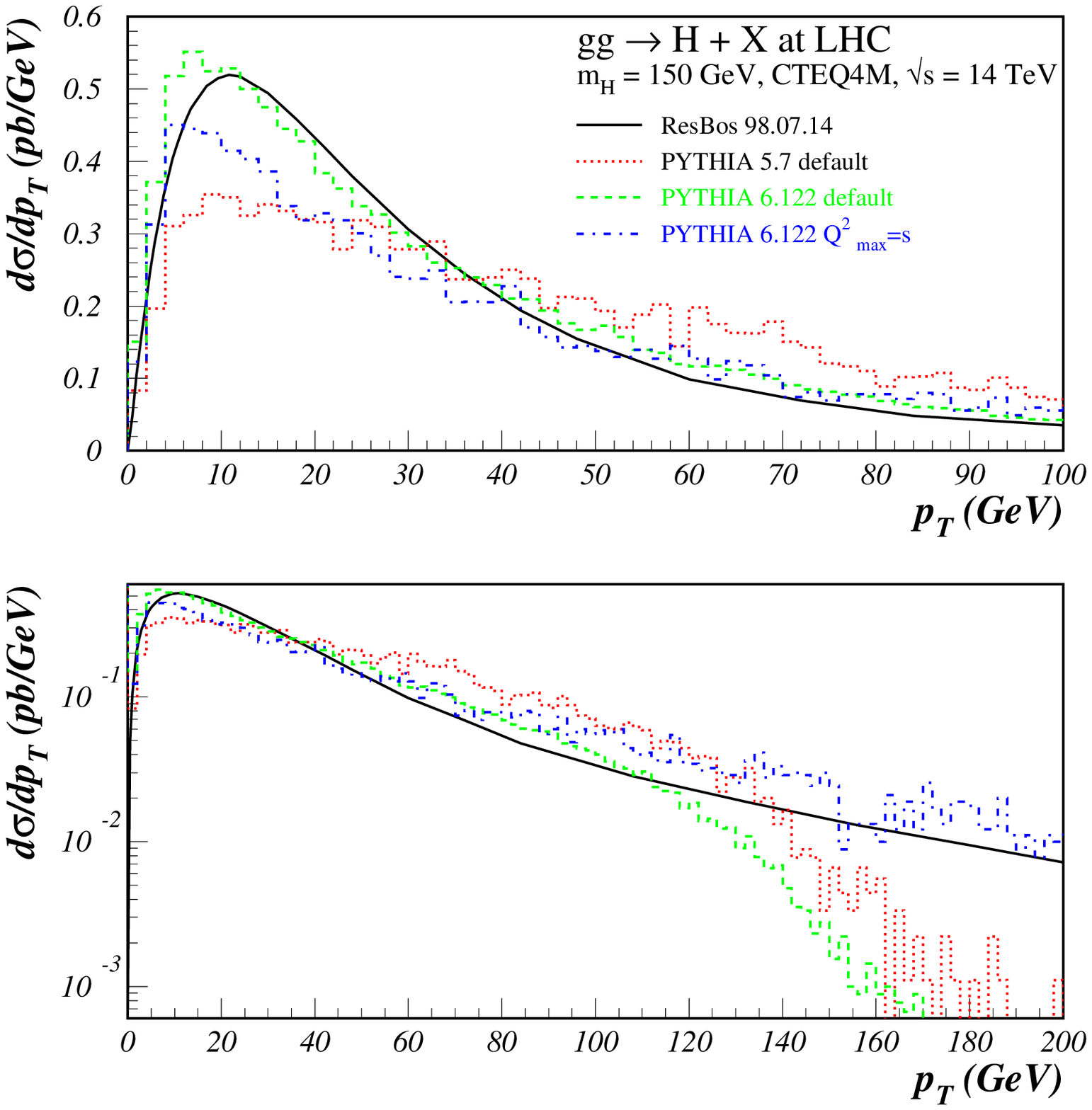}}
\end{center}
\caption{
\sf A comparison of predictions for the Higgs $p_T$ distribution at the LHC from 
ResBos and from two recent versions of {\tt PYTHIA}. The ResBos and {\tt PYTHIA} predictions have
been normalized to the same area. 
} 
\label{fig:resbos_pythia}
\end{figure}

A comparison of the Higgs $p_T$ distribution at the LHC, for a Higgs mass of 
150 GeV, is shown in Figure~\ref{fig:resbos_pythia}, for ResBos and the two 
recent versions of {\tt PYTHIA}. As before, {\tt PYTHIA} has been rescaled 
to agree with the normalization of ResBos to allow for a better shape comparison. 
Note that the peak of the resummed distribution has moved to $p_T \approx$ 
11 GeV (compared to about 3 GeV for $Z^0$ production at the Tevatron). 
This is  partially due to  the larger mass (150 GeV compared to 90 GeV), 
but is primarily because of the larger color factors associated with 
initial state gluons ($C_A = 3$) rather than quarks ($C_F = 4/3$), and also 
because of the larger phase space for initial state gluon emission at the LHC. 
%

The newer version of {\tt PYTHIA} agrees well with ResBos at low to 
moderate $p_T$, but falls below the resummed prediction at high $p_T$. 
This is easily understood: ResBos switches to the NLO Higgs + jet matrix 
element at high $p_T$ while the default {\tt PYTHIA} can generate the 
Higgs $p_T$ distribution only by initial state gluon radiation, using as 
maximum virtuality the  Higgs mass squared. High $p_T$ Higgs production is 
another example where a $2 \to 1$ Monte Carlo calculation with parton 
showering can not completely reproduce the exact matrix element 
calculation, without the use of matrix element corrections. The high $p_T$ 
region is better reproduced if the maximum virtuality $Q_{max}^2$ is set 
equal to the squared partonic center of mass energy, $s$, rather than 
$m_H^2$. This is equivalent to applying the parton shower to all of phase 
space. However, this has the consequence of depleting the low $p_T$ region 
as `too much' showering causes  events to migrate out of the peak.  The 
appropriate scale to use in {\tt PYTHIA} (or any Monte Carlo) depends on 
the $p_T$ range to be probed.  If matrix element information is used to 
constrain the behavior, the correct high $p_T$ cross section can be 
obtained while still using the lower scale for showering. The 
incorporation of matrix element corrections to Higgs production (involving 
the processes $gq \to qH$,$q{\overline{q}} \to gH$, $gg \to gH$) is the 
next logical project for the Monte Carlo experts, in order to accurately 
describe the high $p_T$ region.

%
\begin{figure}[thp]
\begin{center}
\epsfxsize=12cm
\epsfysize=12cm
\mbox{\epsfbox{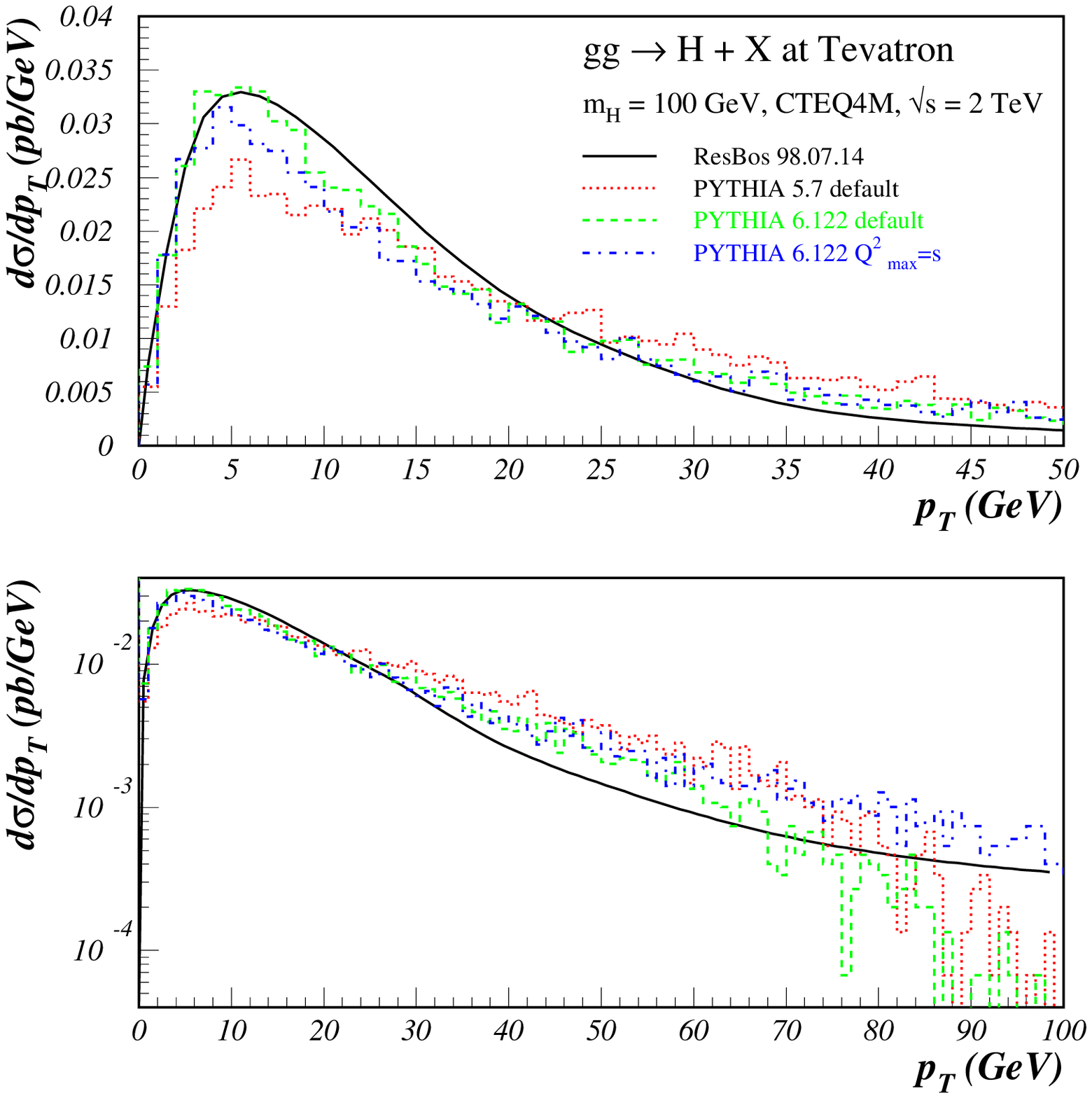}}
\end{center}
\caption{
\sf A comparison of predictions for the Higgs $p_T$ distribution at the 
Tevatron from ResBos and from two recent versions of {\tt PYTHIA}. The 
ResBos and {\tt PYTHIA} predictions have been normalized to the same area. 
} 
\label{fig:resbos_pythia_higgs_tev}
\end{figure}
A comparison of the two versions of {\tt PYTHIA} and of ResBos is also 
shown in Figure~\ref{fig:resbos_pythia_higgs_tev} for the case of Higgs 
production (at a Higgs mass of 100 GeV) at the Tevatron with center-of-mass 
energy of 2.0 TeV. The same qualititative  features are observed as at 
the LHC: the newer version of {\tt PYTHIA} agrees better with ResBos in 
describing the low $p_T$ shape, and there is a falloff at high $p_T$ 
unless the larger virtuality is used for the for the parton showers. The 
default (rms) value of the non-perturbative $k_T$ (0.44 GeV)  was used for 
the {\tt PYTHIA} predictions for Higgs production. 

The older version of {\tt PYTHIA} produces too many Higgs events at 
moderate $p_T$ (in comparison to ResBos) at both the Tevatron and the LHC. 
Two changes have been implemented in the newer version. The first change 
is that a cut is placed on the combination of $z$ and $Q^2$ values in a 
branching: $\hat{u} = Q^2-\hat{s}(1-z) < 0$, where $\hat{s}$ refers to the 
subsystem of the hard scattering plus  the shower partons considered to 
that point.  The association with $\hat{u}$ is relevant if the branching 
is interpreted in terms of a $2 \to 2$ hard scattering. The corner of 
emissions that do not respect this requirement occurs when the $Q^2$ value 
of the spacelike emitting parton is little changed and the $z$ value of 
the branching is close to unity. This effect is mainly for the hardest 
emission (largest $Q^2$). The net result of this requirement is a 
substantial reduction in the total amount of gluon radiation 
\cite{pythiaman}.~\footnote{Such branchings are kinematically allowed, but 
since matrix element corrections would assume initial state partons to 
have $Q^2=0$, a non-physical $\hat{u}$ results (and thus  no possibility 
to impose matrix element corrections). The correct behavior is beyond the 
predictive power of leading log Monte Carlos.} In the second change, the 
parameter for the minimum gluon energy emitted in spacelike showers is 
modified by an extra factor roughly corresponding to the $1/\gamma$ factor 
for the boost to the hard subprocess frame~\cite{pythiaman}. The effect of 
this change is to increase the amount of gluon radiation. Thus, the two 
effects are in opposite directions but with the first effect being 
dominant. 

This difference in the $p_T$ distribution  between the two versions of 
{\tt PYTHIA} could have an impact on the analysis strategies for Higgs 
searches at the LHC. For example, for the CMS detector, the higher $p_T$ 
activity associated with Higgs production in version 5.7 would have 
allowed for a more precise determination of the event vertex from which 
the Higgs (decaying into two photons) originated. Vertex pointing with the 
photons is not possible in the CMS barrel region, and the large number of interactions 
occuring with high intensity running will mean a substantial probability 
that  at least one of the interactions will produce jets at low to 
moderate $E_T$. This could lead to the wrong vertex being chosen for the Higgs, leading to a significant degradation in the $\gamma\gamma$ effective mass resolution.~\cite{denegri}
In principle, this problem could affect the $p_T$ distribution for all 
{\tt PYTHIA} processes. In practice, it affects only $gg$ initial states, 
due to the enhanced probability for branching with such an initial state. 

As an exercise, an 80 GeV $W$ and an 80 GeV Higgs were generated at the 
Tevatron using {\tt PYTHIA}5.7~\cite{mrennarun2}. A comparison of the 
distribution of values of $\hat{u}$ and the virtuality $Q$ for the two
processes indicates a greater tendency for the Higgs virtuality to be near
the maximum  value and for there to be a larger number of Higgs events
with positive $\hat{u}$ (than W events).
%
%

\section{Comparison with {\tt HERWIG}}

The variation between versions 5.7 and 6.1 of {\tt PYTHIA} gives an 
indication of the uncertainties due to the types of choices that can be 
made in Monte Carlos. The requirement that $\hat{u}$ be negative for all 
branchings is a choice rather than an absolute requirement.  Perhaps the 
better agreement of version 6.1 with ResBos is an indication that the 
adoption of the $\hat{u}$ restrictions was correct. Of course, there may 
be other changes to {\tt PYTHIA} which would also lead to better agreement 
with ResBos for this variable. 

%
\begin{figure}[thp]
\begin{center}
\epsfxsize=12cm
\epsfysize=12cm
\mbox{\epsfbox{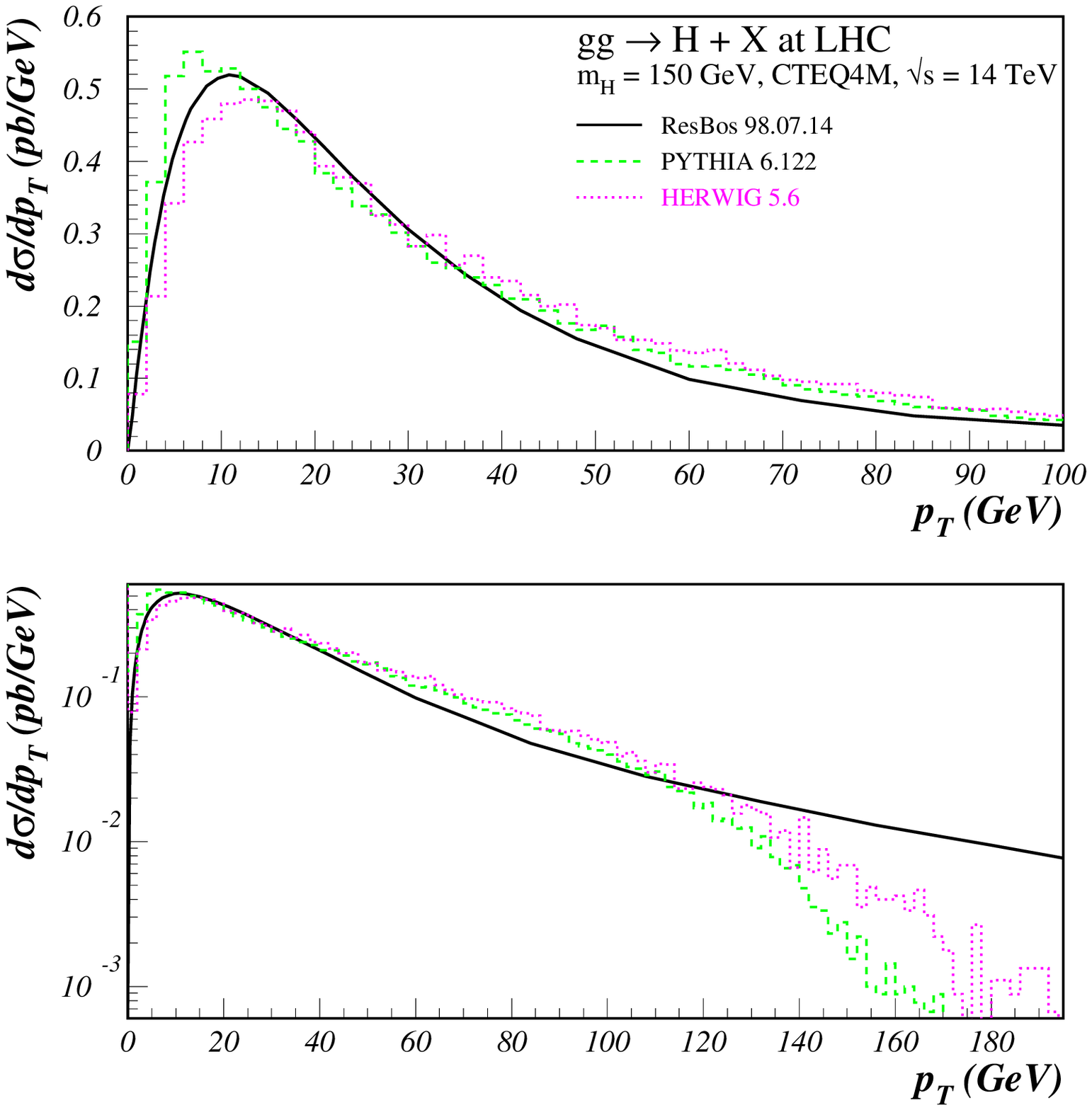}}
\end{center}
\caption{
\sf  A comparison of predictions for the Higgs $p_T$ distribution at the LHC from 
ResBos, two recent versions of {\tt PYTHIA} and {\tt HERWIG}. The ResBos, {\tt PYTHIA} and {\tt HERWIG}
 predictions have been normalized to the same area.
} 
\label{fig:comparison_lhc}
\end{figure}
\begin{figure}[thp]
\begin{center}
\epsfxsize=12cm
\epsfysize=12cm
\mbox{\epsfbox{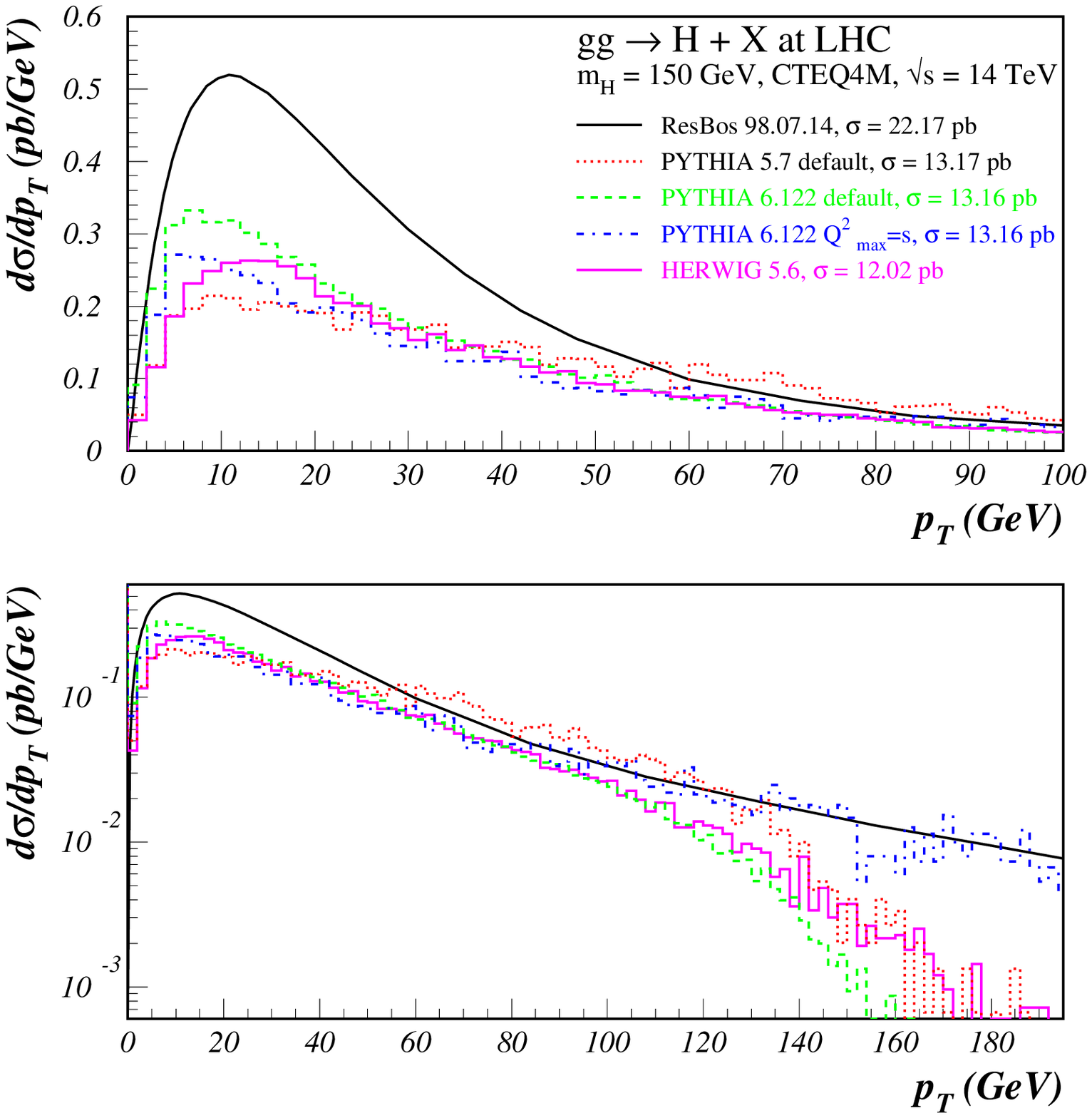}}
\end{center}
\caption{
\sf A comparison of predictions for the Higgs $p_T$ distribution at the 
LHC from ResBos and from two recent versions of {\tt PYTHIA}. The ResBos 
and {\tt PYTHIA} predictions have their absolute normalizations. 
} 
\label{fig:comp_lhc_no_norm}
\end{figure}

Since there are a variey of choices that can be made in Monte Carlo 
implementations, it is instructive to compare the predictions for the 
$p_T$ distribution for Higgs production from ResBos and {\tt PYTHIA} with 
that from {\tt HERWIG} (version 5.6, also using the CTEQ4M parton 
distribution functions). The {\tt HERWIG} prediction is shown in 
Figure~\ref{fig:comparison_lhc} along with the {\tt PYTHIA} and ResBos 
predictions, all normalized to the ResBos prediction.~%
\footnote{The normalization factors (ResBos/Monte Carlo) are {\tt 
   PYTHIA} (both versions)(1.68) and {\tt HERWIG} (1.84).} 
(In all cases, the CTEQ4M parton distribution was used.) The predictions 
from {\tt HERWIG} and {\tt PYTHIA} 6.1 are very similar, with the {\tt 
HERWIG} prediction matching the ResBos shape somewhat better at low $p_T$. 
For reference, the absolutely normalized predictions from ResBos,  {\tt 
PYTHIA}  and {\tt HERWIG} for the $p_T$ distribution of a 150 GeV Higgs at 
the LHC are shown in Figure~\ref{fig:comp_lhc_no_norm}.

%

\section{Non-perturbative $k_T$}

\begin{figure}[thp]
\begin{center}
\epsfxsize=12cm
\epsfysize=12cm
\mbox{\epsfbox{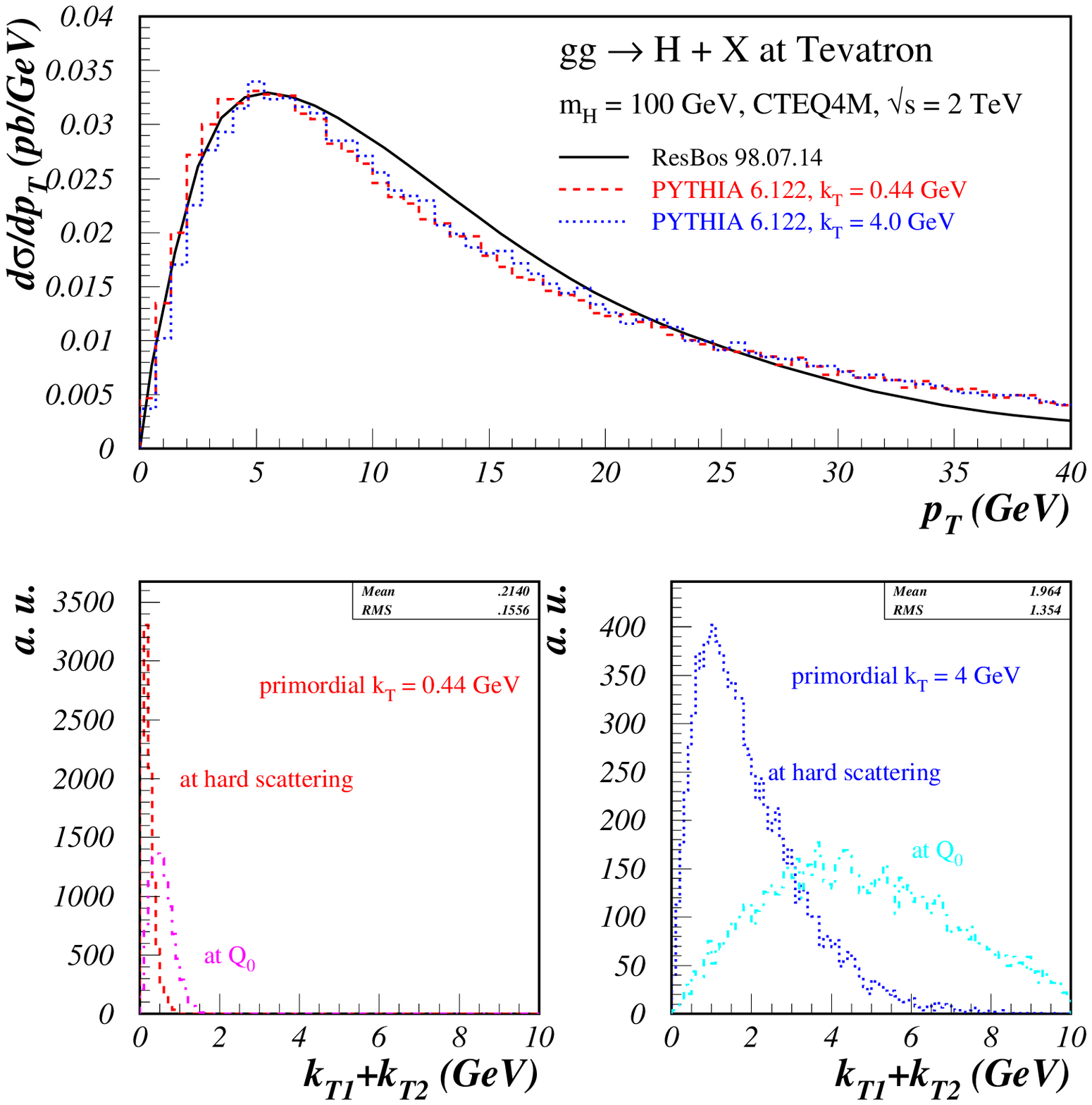}}
\end{center}
\caption{
\sf (top) A comparison of the {\tt PYTHIA} predictions for the $p_T$ 
distribution of a 100 GeV Higgs at the Tevatron using the default (rms) 
non-perturbative $k_T$ (0.44 GeV) and a larger value (4 GeV), at the 
initial scale $Q_0$ and at the hard scatter scale. 
Also shown is the ResBos prediction 
(bottom) The vector sum of the intrinsic $k_T$ ($k_{T1}$+$k_{T2}$) for the 
two initial state partons at the initial scale $Q_0$ and at the hard 
scattering scale for the two values of intrinsic $k_T$.
}
\label{fig:kt_higgs_tev}
\end{figure}

\begin{figure}[thp]
\begin{center}
\epsfxsize=12cm
\epsfysize=12cm
\mbox{\epsfbox{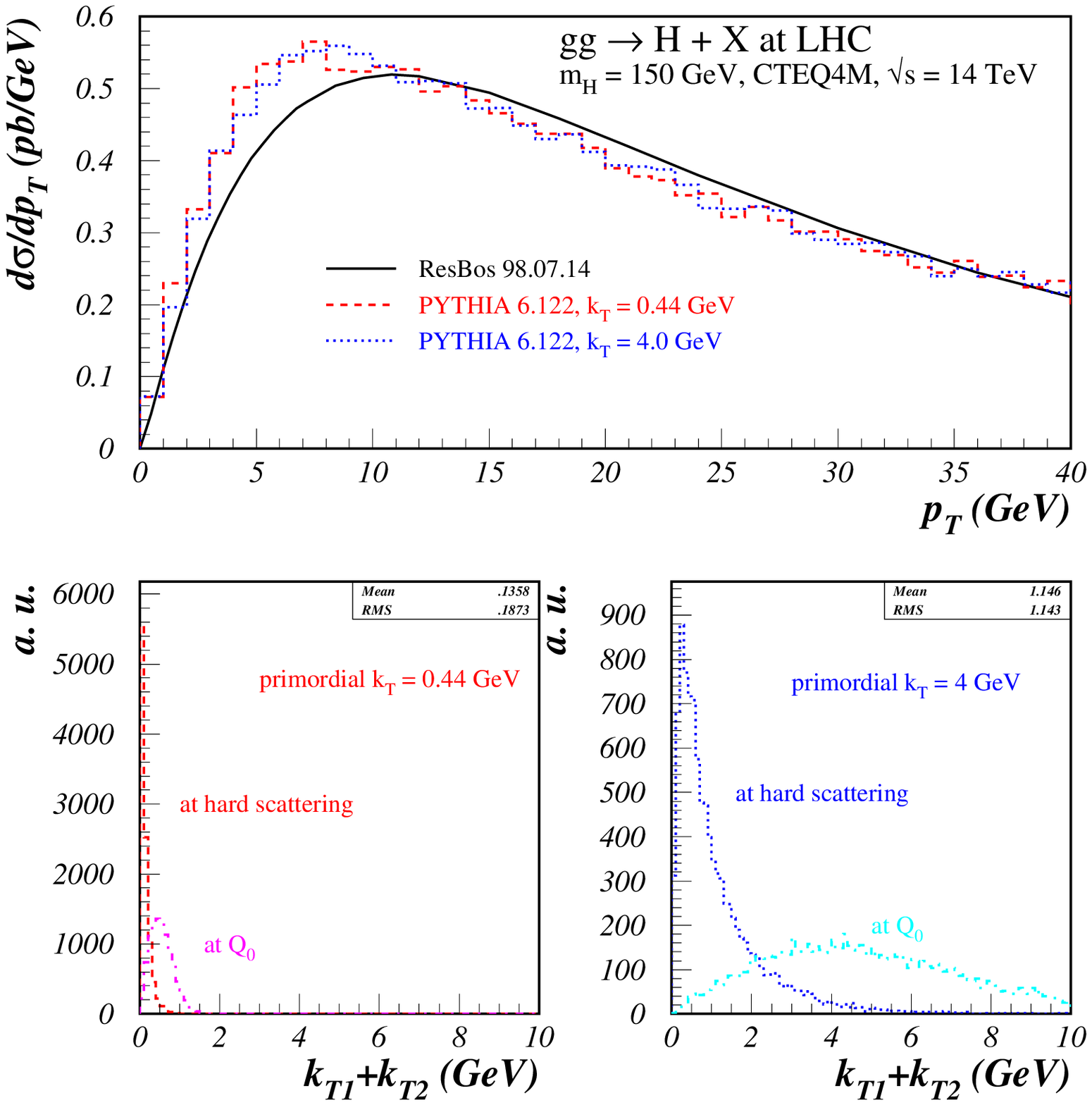}}
\end{center}
\caption{
\sf (top) A comparison of the {\tt PYTHIA} predictions for the $p_T$ 
distribution of a 150 GeV Higgs at the LHC using the default (rms) non-
perturbative $k_T$ (0.44 GeV) and a larger value (4 GeV), at the initial 
scale $Q_0$ and at the hard scatter scale. Also shown is the ResBos 
prediction. (bottom) The vector sum of the intrinsic $k_T$ 
($k_{T1}$+$k_{T2}$) for the two initial state partons at the initial scale 
$Q_0$ and at the hard scattering scale for the two values of intrinsic 
$k_T$.
}
\label{fig:kt_higgs_lhc}
\end{figure}

A question still remains as to the appropriate value of non-perturbative 
$k_T$ to input in the Monte Carlos to achieve a better agreement in shape, 
both at the Tevatron and at the LHC. In Figures~\ref{fig:kt_higgs_tev} 
and~\ref{fig:kt_higgs_lhc}  are shown  comparisons of ResBos and {\tt 
PYTHIA} predictions for the Higgs $p_T$ distribution at the Tevatron and 
LHC. The {\tt PYTHIA} prediction (now version 6.1 alone) is shown with 
several values of non-perturbative $k_T$. Suprisingly, no difference is 
observed between the predictions with the  different values of $k_T$, with 
the peak in {\tt PYTHIA} always being somewhat below that of ResBos. This 
insensitivity can be understood from the plots at the bottom of the two 
figures which show the sum of the non-perturbative initial state $k_T$ 
($k_{T1}$+$k_{T2}$) at $Q_0$ and at the hard scatter scale $Q$. Most of 
the $k_T$ is radiated away.  with this effect being larger (as expected) 
at the LHC. The large gluon radiation probability from a gluon-gluon 
initial state (and the greater phase space available at the LHC) lead to a 
stronger degradation of the non-perturbative $k_T$ than was observed with 
$Z^0$ production at the Tevatron.
%
%
%
%
%
%
%
%

%
%
%
%

\begin{figure}[thp]
\begin{center}
\epsfxsize=12cm
\epsfysize=12cm
\mbox{\epsfbox{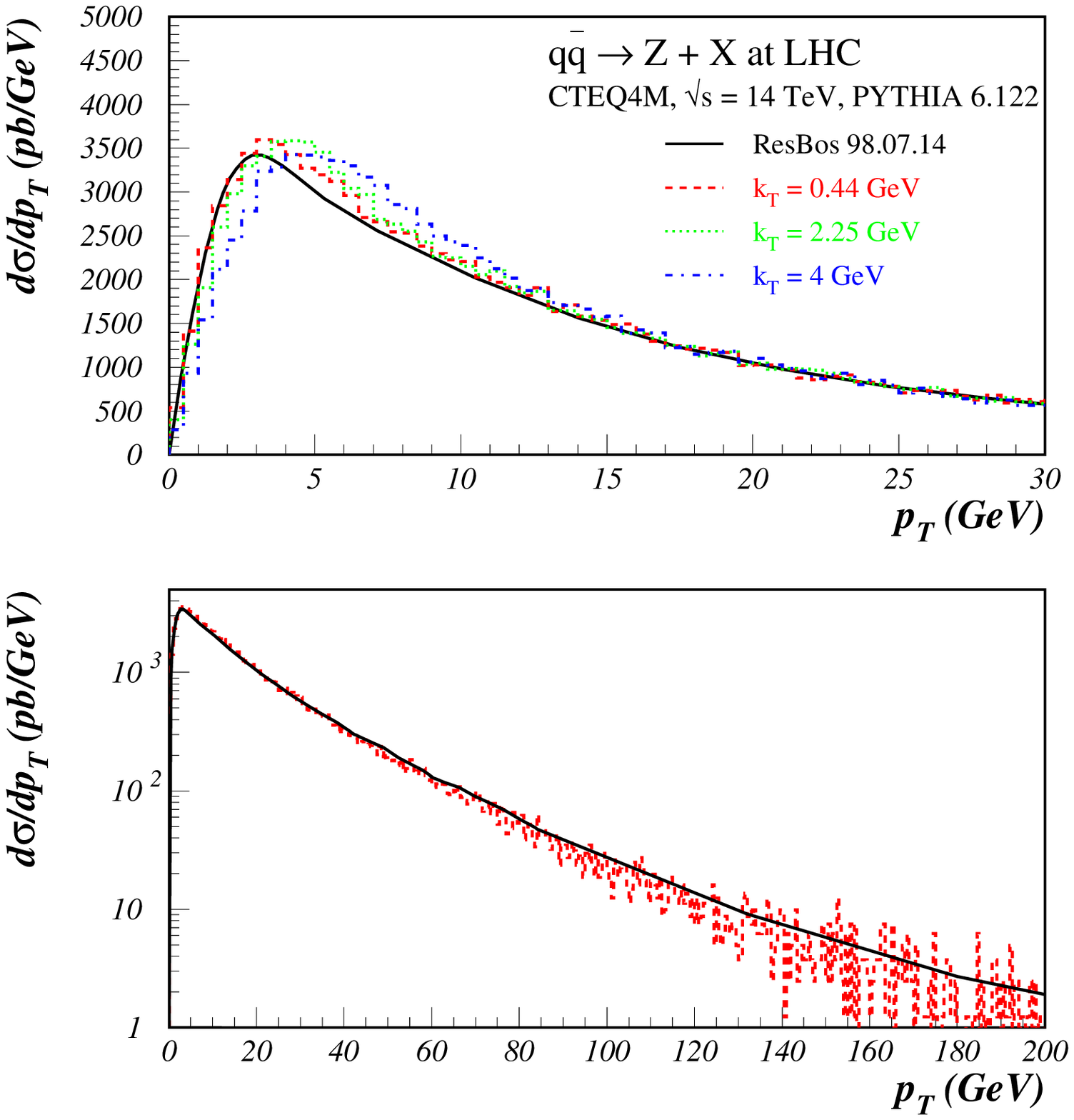}}
\end{center}
\caption{
\sf A comparison of the predictions for the $p_T$ distribution for $Z^0$ 
production at the LHC from {\tt PYTHIA} and ResBos, where several values 
of $k_T$ have been used to make the {\tt PYTHIA} predictions. 
} 
\label{fig:z_lhc}
\end{figure}

For completeness, a comparison of {\tt PYTHIA} and ResBos is shown in 
Figure~\ref{fig:z_lhc} for $Z^0$ boson production at the LHC.  There are two 
points that are somewhat surprising. There is still a very strong 
sensitivity to the value of the non-perturbative $k_T$ used in the 
smearing, and the best agreement with ResBos is obtained with the default 
value (0.44 GeV), in contrast to the 2 GeV needed at the Tevatron. Note 
again the agreement of {\tt PYTHIA} with ResBos at the highest values of 
$Z^0$ $p_T$ due to the explicit matrix element corrections applied. 
%

\begin{figure}[thp]
\begin{center}
\epsfxsize=12cm
\epsfysize=12cm
\mbox{\epsfbox{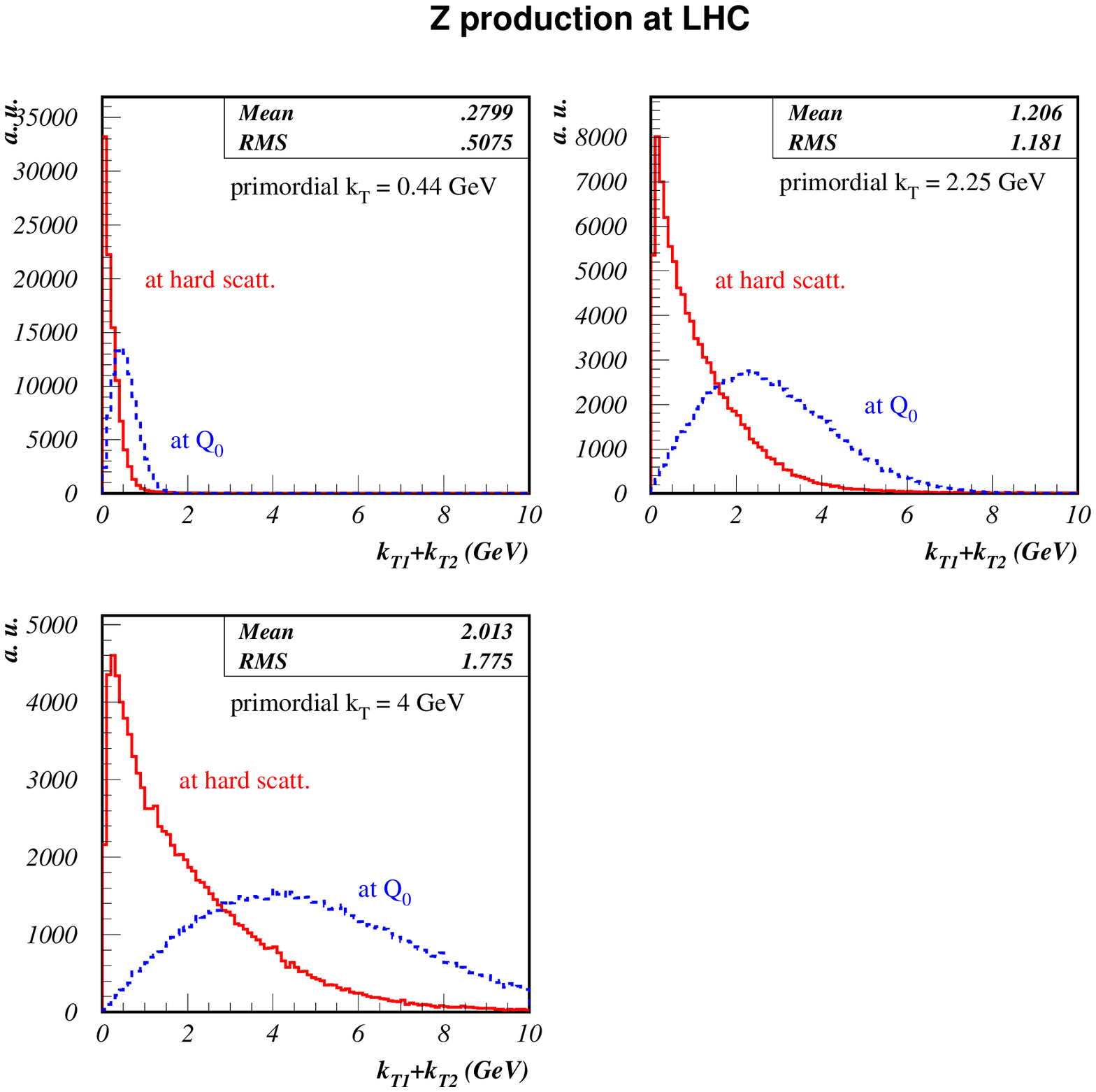}}
\end{center}
\caption{
\sf A comparison of the total initial state $k_T$ ($k_{T1}+k_{T2}$) distributions 
for $Z^0$ production at the
LHC from {\tt PYTHIA}, both at the initial scale $Q_0$ and at the hard scattering
scale, for several (rms) values of the initial state $k_T$. The mean and rms 
numbers refer to the values at the hard scattering scale.
} 
\label{fig:z_lhc_kt}
\end{figure}

The sum of the incoming parton $k_T$ distributions, both at the scale 
$Q_0$ and at the hard scattering scale, are shown in 
Figure~\ref{fig:z_lhc_kt} for several different starting (rms) values of 
primordial $k_T$ (per parton). There is substantially less radiation for a 
$q\overline{q}$ initial state than for a gg initial state (as in the case 
of the Higgs), leading to a noticeable dependence of the $Z^0$ $p_T$ 
distribution on the primordial $k_T$ distribution. 
%

\section{Conclusions}

An understanding of the signature for Higgs boson production at either the 
Tevatron or LHC depends upon the understanding of the details of soft 
gluon emission from the initial state partons.  This soft gluon emission 
can be modelled either in a Monte Carlo or in a $k_T$ resummation program, 
with various choices possible in both implementations.  A comparison of the 
two approaches is useful to understand the strengths and weaknesses of 
each. The data from the Tevatron that either exists now, or will exist in 
Run 2, will be extremely useful to test both approaches. 

\section{Acknowledgements}

We would like to thank Claude Charlot, 
Gennaro Corcella, Steve Mrenna, Willis Sakumoto, Torbjorn Sjostrand and  
Valeria Tano for useful conversations and plots. This work was supported in 
part by the NSF under grant PHY-9901946 and by the DOE under grant DE-FG-03-94ER40833.





\setcounter{figure}{0}
\setcounter{table}{0}
\setcounter{section}{0}
\setcounter{equation}{0}
\newpage









\begin{center}
\vspace*{1.2cm}
{\Large\sc \bf Automatic Computation of LHC Processes} \\
\vspace*{1.cm} 
{\sc E.~Boos, V.~Ilyin, K.~Kato, A.~Pukhov, A.~Semenov, A. Skatchkova}
\vspace*{1.cm}
\end{center}

\setcounter{footnote}{0}

Automatic computation is a new approach to HEP computing. The first such
systems, GRACE \cite{GRACE}, FeynArt/FeynCalc \cite{Feyn} and CompHEP
\cite{CompHEP}, were reported at the 1st International Workshop AIHENP
held on
March, 1990 in Lyon-Villeurbanne (France). Under this terminology, {\it
automatic computation system (ACS)}, we assume, as a distinguishing feature,
the generation of the computing code for a specific collision process with the
aid of another code. 

ACS's are now used  widely by phenomenologists for the calculation of many collision
processes. For example, the GRACE and CompHEP systems were used in the
LEP2 Workshop
\cite{yellow}, and for evaluation of processes at TeV linear colliders
\cite{TESLA}. With ACS one can calculate {\it all} collision processes within
a given physical model, where by physical model we mean the set of Feynman
rules. Recent developments  with the LanHEP package \cite{LanHEP} have opened
a possibility to derive Feynman rules in the form of the ACS intrinsic physical
model in a fully automatic way, starting from the Lagrangian. Now, not
only the
Standard Model but a number of its extensions, like SUSY models, are implemented
in ACS. 
%
A general review of this new approach is given in these Proceedings by
K.Kato together with discussion of main directions of the ACS
development. 
Here, we discuss in more detail specifics of the ACS applications in LHC
phenomenology, and in particular to the evaluation of QCD processes.

To close this preview we list below the main ACS options in order to provide an idea for users of
what is available: 

\begin{enumerate}
\item[i)]   selection of physical model (Lagrangian) and hard subprocess,
\item[ii)]  Feynman diagram generation, 
\item[iii)] generation of the code for matrix element, 
\item[iv)]  convolution with parton distributions, 
\item[v)]   generation of kinematics (phase space parameterization) with 
            regularization of kinematical peaks, 
\item[vi)]  integration over the phase space (evaluation of cross
section), 
\item[vii)] generation of events at partonic level, including the
interface to hadronization tools. 
\end{enumerate}

\section{The problem of multiparticle final states:
 why automatic computations?}

We start from the problem of the accurate evaluation of hard
subprocesses in the case of multiparticle final states.

When physicists simulate HEP processes with such generators as PYTHIA
\cite{PYTHIA}, ISAJET \cite{ISAJET} or HERWIG \cite{HERWIG} they use a
data base
of hard subprocesses implemented in these packages. It means that a) matrix
elements are stored as formulas, and  b) a knowledge about the behaviour of
matrix elements as phase space integrands are coded in the form of modelling
functions in order to get a fast generation of the partonic events. One can note that
these data bases include a rather simple variety of subprocesses, mainly
of the $2\to
2$ type.

If one tries to include a hard subprocesses with 3, 4 and more particles
in the final
state, large problems appear. Indeed, the size of matrix elements
increases very
fast. For example, in the $2\to 4$ case, the size of the code for
evaluation of
helicity amplitudes for one subprocess is at the 100's  of Kbyte level.
However, the
main problem lies elsewhere;  it is impossible to construct an analytical
formula for
matching peaks and other structures of the rather singular behaviour of matrix
elements. In the $2\to 3$ case, phase space has 4 dimensions plus two for
convolution with the PDF's; in the $2\to 4$ case 7+2 dimensions are
present, and so on. As a
result, the set of kinematical singularities has, as a rule, a very complicated
positioning in the multidimensional phase space. This particular problem was
not solved accurately, e.g., when the $Zbb$ final state was implemented
in PYTHIA
5.7.

Let us discuss further this somewhat delicate point. It is necessary to integrate
the squared matrix element over the phase space in order to
obtain the cross section. Precise information about
the behaviour of the integrand then is necessary  for further event generation.
This information  can be obtained at the step of the phase space integration.
The problem is that the
integrand, as a rule, has a singular behaviour with sharp kinematical peaks
connected with different denominators (propagators) of Feynman diagrams. This
problem is caused, in particular, by the circumstance that one has to
take into
account nonzero masses of particles in many important cases, especially if
accurate calculations are needed. The masses of elementary particles can have
extremely small values, e.g. the  masses of the 1st generation quarks
(few MeV), and
can also be zero (for the photon and gluon). At the same time, other
parameters are of the order
of a hundred GeV, e.g. masses of $W$ and $Z$ bosons and $t$-quark. Moreover
the collision energy can also have a very large value, e.g. 14000 GeV
for LHC processes,
and some other important variables, like the transverse energy of jets,
are at the
hundred GeV scale or even greater. This huge scale interval for different
parameters causes
serious computational problems which result in the appearance of sharp
peaks for the
integrand. So, at the step of phase space parameterization, one has to
include a
regularization of the integration measure in order to smooth the
singularities of the
integrand.

LHC phenomenology requires the computation of a wide spectrum of hard subprocesses
with 3,4 and even more particles (partons) in the final state. This is a common need for all of the physics working groups: QCD, EW, Higgs, SUSY
etc. 
These requirements are especially common for new physics searches.
Furthermore, a major challenge results from background analyses,
where QCD subprocesses play a major role with, in many cases,  multiparton
final states. As a rule, for each LHC discovery reaction, one should calculate
several QCD processes giving both irreducible and reducible backgrounds. 
The parton-shower generation of multiparticle final
states is usually utilized in this situation.
 However,  this can be too crude an
approximation for many
important studies leading to sometimes grossly unreliable predictions.

We emphasize that ACS can give the possibility to compute accurately a
variety of LHC
processes (and in particular QCD processes) with 3, 4 and more bodies in
the final
partonic state. Indeed, the first problem (the size of the matrix
element computing
code and the difficulty to obtain the  exact matrix elements) is solved
in ACS by the
automatic generation of the corresponding code. This step is fast and
pain-free from
the viewpoint of the user.
 The second problem (the accurate integration over the
multidimensional phase space) is solved in ACS by the generation of
kinematics where the
necessary regularizations are included.
%
%
For example, in CompHEP  the user has to list
a set of singular propagators using the menu system. After that,  the
code for 
kinematics (with regularizations) is generated automatically. In GRACE,
a library
of kinematics (with regularizations) is used and the user has to make
the necessary
choices. Thus, the high art (mathematics and programming), needed to
elaborate the
sharp peaks, is enclosed in a form hidden from the user, giving him a possibility
to compute complicated processes.

At the step of integration over the phase space, ACS uses adaptive Monte Carlo
integrators (VEGAS \cite{VEGAS} in CompHEP, and BASES
\cite{BASES-SPRING} in
GRACE). To match the complete set of singularities, the multichannel MC approach
\cite{multichannel} is utilized. As a result, the phase space grid is
created with an
accurate mapping of the singular behaviour of the matrix element. This
complex body of
information (let us call it {\it MEgrid}) has a rather large size that
rapidly increases with the number of phase space dimensions. One can consider
{\it MEgrid} as a multidimensional analog of the modelling function used in
PYTHIA and other similar packages for the effective generation of
partonic events.
Of course, this information can not be expressed in analytical form. It is
necessary to point out also that the convolution with parton distributions
should be made at the same stage as the integration over the phase space.
Indeed, the contributions of different subspaces (in particular different
kinematical peaks) can depend largely on the partonic collison energy, $\hat
s$, resulting from the information stored in {\it MEgrid}. 

ACS can be considered as a tool for the automatic generation of the data
base of
hard subprocesses for physical generators like PYTHIA, ISAJET and HERWIG.
However, it is difficult to imagine that the data base  created can be
implemented in
the code of these generators. This is due,  first of all, to the size of
the generated codes.
Thus, we propose a {\it two stage} approach. At the first stage, ACS is used
resulting in a  cross section and {\it MEgrid} for the subprocess under
evaluation. This can be stored in a special LHC data base. This data
base can be
used for the effective generation of partonic events. In GRACE, it is available
with the SPRING \cite{BASES-SPRING} generator, and in CompHEP
 by a relatively straightforward procedure 
%
and an effective generator is under construction). The output is a
partonic event flow that can be used as an input for physical generators like
PYTHIA, ISAJET and HERWIG; this is second stage of the full simulations. At
this stage partons (quarks and gluons) should be hadronized and unstable
particles decayed. We note that in PYTHIA there exists a rather flexible
interface for  such a  {\it two stage} approach, the option for
inclusion of
external processes through the routine {\sf PYUPEV}.

This is a general view on the way in which ACS (GRACE and CompHEP in
particular) can
be used for the simulation of LHC processes.  Below we discuss some specific
features of this technology with special attention to QCD aspects.

\section{General Considerations about GRACE and CompHEP}

With CompHEP and GRACE the user can evaluate hard subprocesses at the tree
level, i.e.
Feynman diagrams are generated without loops.  This corresponds to
the basic request for LHC phenomenology. However, it is well known that QCD
next-to-leading corrections are large, as a rule, for LHC processes. In many
cases these corections can be accounted for in the form of so-called
K-factors and
one can include them easily in tree level calculations. Nevertheless, in many
important cases an explicit evaluation of higher order corrections is necessary.
At this moment it is not clear how to automate calculations of LHC
processes at
NLO level. The problem is connected, in particular, with the
circumstance that
different resummations of large logarithms should be included in order
to get reliable
NLO predictions. The interface between resummation techniques and event
generators is under intensive discussion now, and at the present
Workshop also. We
note in this respect, that the GRACE package includes the code for the
generation and
evaluation of one-loop diagrams.

The user interface should provide the  possibility to calculate
complicated processes for users not experienced in programming. CompHEP
has a
(graphical) menu driven system where the user proceeds through all steps
of the
calculation without any programming. In GRACE, the user  needs to write
a few simple
interface routines.

The information on the GRACE system and its products can be found at

\centerline{\sf http://www-sc.kek.jp/minami/ }

\noindent

The code of CompHEP is free for users and one can take it from the following
Web page

\centerline{\sf http://theory.npi.msu.su/comphep}

\noindent 
where the user's manual is available in PS format (see also
hep-ph/9908288). The CompHEP package, adapted for LHC processes (see
next section)
is installed on the SUN platform  

\centerline{\sf /afs/cern.ch/cms/physics/COMPHEP/v33-SUN}

\noindent
and on the PC/Linux platform

\centerline{\sf /afs/cern.ch/cms/physics/COMPHEP/v33-Linux}

The interface between CompHEP and PYTHIA has beencreated with the corresponding
code  available from the address:

\centerline{\sf /afs/cern.ch/cms/physics/comp-pyth}

\noindent
where one can find a short description in the file README.
With this interface, the partonic event flow for any processes
calculated with CompHEP
can be sent to PYTHIA to generate physical events.

\section{QCD aspects in automatic computations}

In this section we discuss the treatment of QCD effects in the case of
automatic computations, and consider CompHEP options as an example. As has
been discussed above, CompHEP calculates only at tree level, and so at
leading order (LO).
Thus, the main problems concerning an accurate accounting of QCD effects
are outside the
discussion. Nevertheless, some important QCD dependencies can not be avoided
even at tree level and the corresponding options are available for
users. These
aspects are: a) parton distributions, b) QCD scale, and  c) running strong
coupling constant.

\vspace{0.3cm}

\noindent
{\it Parton distributions.}

In CompHEP the specification of initial states in the collision process under
evaluation can include the convolution with structure function. So, in
the case
of hadron collisions, the cross section is evaluated as an integral

$$ \sigma(s) \;=\; \int^1_0 dx_1 dx_2 f_i(x_1,Q) f_j(x_2,Q) 
                 \hat \sigma_{ij}(x_1 x_2 s) $$
                 
\noindent
where $f_i$ are the corresponding parton distributions, $\hat\sigma$ is the
partonic cross section and $Q$ is the QCD scale.

In CompHEP v.33, installed at CERN (see address above), parton
distributions from two pdf families are implemented, MRS and CTEQ, and in particular the
following versions:  1)
MRS(A') and MRS(G) \cite{mrs}, 2) CTEQ4l and CTEQ4m \cite{cteq}. Note that
CTEQ4l is a LO parametrization, while in all others the evolution of parton
distributions is realized in the next-to-leading (NLO) approximation.

In addition, a special interface is available to include a user's
defined parton
distribution. By this way one can implement the most recent
parametrizations (at
this moment CTEQ5 and MSRT). See the CompHEP user's manual for the corresponding
procedure (section 3.6.2).

\vspace{0.3cm}
\noindent
{\it Choice of QCD scale.}

The factorization theorem states that parton distributions depend not
only on the
Bjorken variable $x$ but also on some parameter $Q$ which characterizes
the energy
(or momentum) scale at which the QCD effects give the main contribution
to the hard
subprocess. This parameter is set by the user for each specific QCD
process. It is possible to set a fixed scale or a running scale. In the 
later case,  $Q^2$
can be a squared linear combination  of any set of initial and outgoing
particles momenta, for example, $(p_1-p_3)^2$, $(p_1-p_3-p_4)^2$, $(p_3+p_4)^2$
and so on (initial and outgoing momenta enter with opposite signs). The
corresponding settings are made through the option {\sf QCD SCALE} of the
numerical menu.

\vspace{0.3cm}
\noindent
{\it Running $\alpha_s$.}

It is the nature of strong interactions that there is no absolute normalization
of the corresponding coupling constant. This is in contrast to the value
$1/137$ for the electromagnetic constant known with high accuracy  from
classical electrodynamical
experiments. Instead, we have a function for $\alpha_s$
rather than a constant.  Even in the leading order approximation,
$\alpha_s^{LO}=6\pi/[(33-2n_f) \log{Q/\Lambda^{(n_f)}}]$, where $Q$ is
the QCD scale
of the hard subprocess under the evaluation with $\Lambda$ the so-called QCD
fundamental parameter. Then, $n_f$ is the number of parton flavours with masses
lower than $Q$. The $n_f$ dependence in QCD parameter $\Lambda$ matches the
quark mass threshold effects.

In the version of  CompHEP installed at CERN, the running $\alpha_s$ is
realized in LO, NLO and
NNLO. All of the corresponding formulas are based on the choice of $\Lambda^{(6)}$
(see {\it Review of Particle Physics} \cite{PDG} p.81. The user can find
the corresponding
switch in the option {\sf QCD SCALE} in the numerical menu.

\vspace{0.5cm}

Therefore, to evaluate QCD processes with CompHEP,  one has, first of
all, to 
fix the normalization of $\alpha_s$. The popular normalization point is
the mass of
$Z$ boson, $Q=M_Z$. By changing the parameter $\Lambda^{(6)}$, the user
should set
the strong coupling at the appropriate value, say $\alpha_s(M_Z)=0.118$.
Then, the user has
to choose the order for the running $\alpha_s$ (LO, NLO or NLO).
Finally, the  user has
to define the QCD scale $Q$, which will be used both for the evaluation
of $\alpha_s$ at this scale and in the parton distributions.

Thus, the complete LO calculations of LHC processes are available, with
the 
matrix element, parton distributions and running strong coupling
constant 
calculated in the lowest order of the perturbation theory.
 This is a self-consistent
starting point in the phenomenological analysis; when/where higher order corrections are available, all elements of the calculation can be calculated
at the higher order and then compared to the leading order result. 
%

However, it is also common for phenomenologists use a mixed
approach, with the matrix element evaluated at LO but the parton
distributions and
running $\alpha_s$  taken in NLO approximation. Surely, only a part of
the NLO
corrections is accounted for in this case. We note that this option  is also
available for users in CompHEP calculations.

\section{Partonic Subprocesses}

	When hadronic collision processes are evaluated,
especially in the case of a large number of final state particles, one serious problem
 is the  large number of
contributing partonic subprocesses. This occurs because of the quark and
gluon content of the initial hadrons and CKM quark mixing. For example, at
LHC energies,  180 subprocesses contribute to the $W+2jets$ and 292
subprocesses to
the $W+3jets$ production (taking into account only quarks of  the first two generations ). 
During this workshop a new method has been proposed
to avoid a multiplication of channels due to the mixture of quark
states~\cite{qQ}. The method leads to a simple modification of the rules for the
evaluation
of the cross sections and distributions. It is based on the unitary
rotation of
down quarks, thus providing the transportation of mixing matrix elements
from
vertices of Feynman diagrams to the parton distribution functions. As a
result,
one can calculate cross sections with a significantly smaller number of
subprocesses contributing. For the examples mentioned above, one needs to
evaluate (with the new rules) only 21 and 33 subprocesses, respectively, in order
to compute the cross sections for the  $W+2jets$ and $W+3jets$ processes. The matrix
elements
of the subprocesses are calculated without quark mixing, but with a
modified PDF
convolution which now depends  on the quark mixing angle and the  topologies of
the gauge
invariant classes of diagrams contributing to the subprocesses. The method
proposed has been incorporated into the CompHEP program and checked with
many examples.

\section{PEVLIB - library of LHC processes}

Now the library of CompHEP based event generators for LHC processes has been
started at the address:

\centerline{\sf /afs/cern.ch/cms/physics/PEVLIB}

The following QCD processes are stored already in this library: $Zb \bar b$,
$Wb \bar b$, $t\bar t b\bar b$ and some others. In the corresponding
directories (with the names literally corresponding to the final states)
unweighted events are stored (see the files {\sf README} in these directories
for details about evaluation of the corresponding samples of events).

Together with the CompHEP-PYTHIA interface code (see discussion above) these
event files can be used for full LHC simulations with the help of PYTHIA package
and detector simulation software in the standard way.

Let us discuss the process $Zb \bar b$ in order to supply more details.
In the directory

\centerline{\sf /afs/cern.ch/cms/physics/PEVLIB/Z\_b\_b}

\noindent  the file {\sf \_\_pevZbb} includes about 200000 unweighted events
with the final state $Zb \bar b$. Each event includes the Lorentz
momenta of all
particles in the initial and final states. In the present version of
this library, there
is no information about the color flow in the event. Thus, only the {\sf Independent
Fragmentation Model} can be used for the hadronization. Of course, the
user can use
the Lund model;  for this one has to define the corresponding color
flows by hand
in the routine {\sf PYUPEV}. The same remark is valid also for FSR
(final state
radiation), what is switched off by default in CompHEP-PYTHIA interface.
In the
same time ISR (initial state radiation) is switched on automatically.

Note that the user can generate more events than stored in the library.
In the
corresponding subdirectories (indicated in the file {\sf README}) the
generators are stored in the form of the executable code (at this moment for
SUN platform only). These generators are the corresponding CompHEP codes
for the
process with the proper set of kinematical regularizations.

The library PEVLIB is under construction now. New processes will be
added. The
structure and user's interface will be developed.

\section{Acknowledgements}

The work of V.I., A.P., A.S, E.B. and A.S. was partially supported by
the  CERN-INTAS grant 377 and  RFBR-DFG grant 99-02-04011.


 
\setcounter{figure}{0}
\setcounter{table}{0}
\setcounter{section}{0}
\setcounter{equation}{0}
\newpage



\begin{center}
\vspace*{1.2cm}
{\Large\sc \bf Monte Carlo Event Generators at NLO} \\
\vspace*{1.cm} 
{\sc J. Collins}
\vspace*{1.cm}
\end{center}

\setcounter{footnote}{0}



\begin{abstract}
I am concerned here with QCD calculations for processes with a hard
scattering --- production of heavy particles, jets, etc.  The most
accurate calculations are by the ``analytic'' methods.  But the most
useful calculations for direct comparison with data are done by
Monte-Carlo event generators, and these are limited in accuracy.  In
particular, there is as yet no known method of systematically
improving the Monte-Carlo calculations by incorporating the
non-logarithmic parts of higher order perturbative corrections.  This
creates limitations on the analysis of future data.  Therefore I
summarize some ideas for remedying the situation.
\end{abstract}


\paragraph{Factorization for inclusive processes}
\begin{itemize}
\item Normal proofs of factorization are for inclusive cross sections.
\item To get the simplifications in the factorization formula, as
      compared with the exact cross section, one makes suitable
      approximations.
\item The approximation is to the hard scattering part $H$ of a cross
      section, as in: 

	\begin{figure}[h]
     	\centering \includegraphics[scale=0.35]{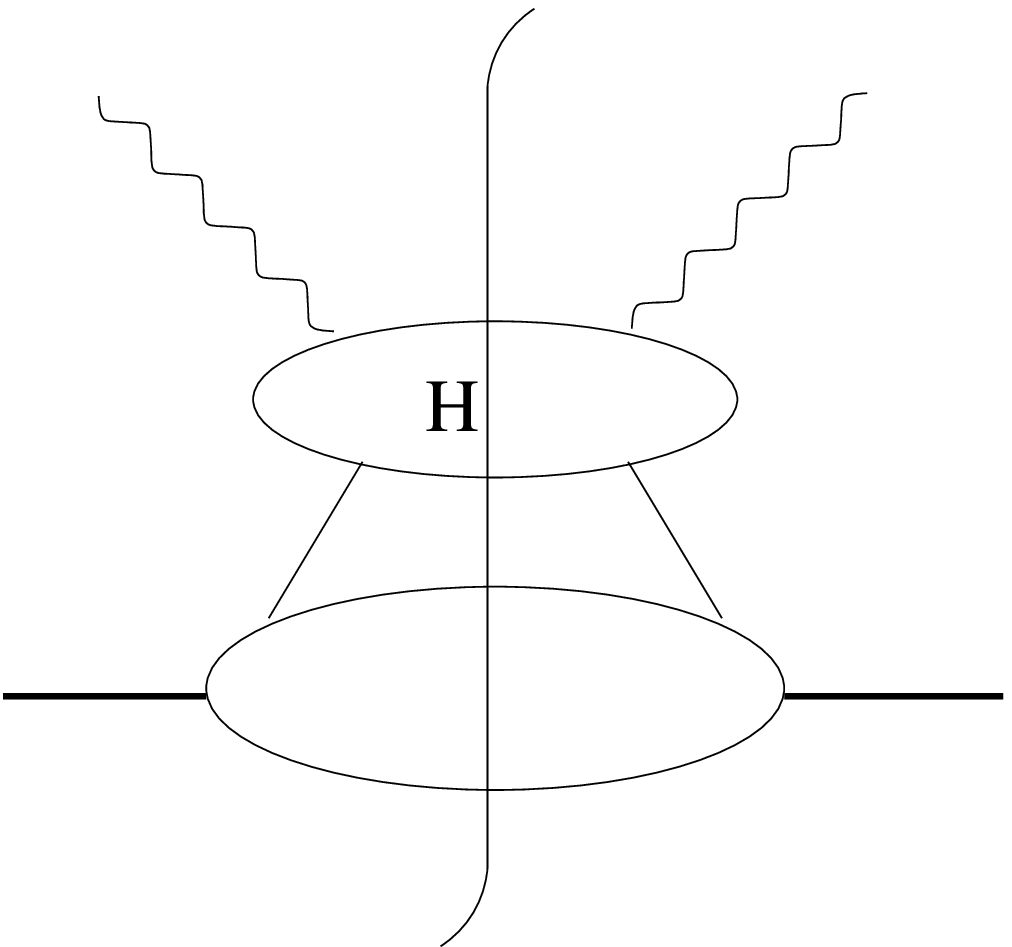} 
	\caption{}\label{fact}
	\end{figure}
      
which is a graph for the DIS cross section.  The kinematics of
      the external lines of $H$ are changed to massless on-shell
      partons with zero transverse momentum.  Also the internal (light
      parton) lines of $H$ are made massless.
\item The approximation is correct to the leading power of $m^2/Q^2$,
      where $m$ is a typical hadronic scale and $Q$ is the scale of
      the hard scattering (e.g., $Q^2$ is the virtuality of the
      virtual photon in DIS).
\item Subtractions are applied to the hard scattering to cancel double
      counting.  A high-order graph for the hard scattering has
      subtractions that correspond to smaller hard subgraphs (and
      hence smaller regions of momentum space).
\item The exact form of the subtractions corresponds to the
      approximations made for the smaller regions.
\end{itemize}


\paragraph{Monte-Carlo event generators at NLO (and beyond)}
Current event generators essentially use the leading order for the
hard-scattering coupled to an algorithm that approximates the
exclusive structure of the low virtuality parts of graphs for the
cross section.  The algorithm is in an improved leading logarithm
approximation. 

With the exception of the recent paper by Friberg and Sj{\"o}strand
\cite{FS}, previous attempts, e.g., \cite{MC.NLO}, at incorporating
NLO corrections have tended to implement them by a reweighting of the
events generated by showering from the LO matrix elements.  Normal NLO
``analytic'' calculations for the hard-scattering coefficients for
inclusive scattering are not usable as they stand in a Monte-Carlo
event generator, because they involve singular distributions.

Here I summarize some ideas \cite{JCC.MC} to remedy this situation.  I
have applied them to one specific case in DIS in Ref.\
\cite{JCC.MC.lepto}, but I think they can be generalized:
\begin{itemize}
\item Separate groups of events are generated with LO and NLO hard 
      scattering coefficients.
\item To obtain the NLO coefficients, the same general ideas are used
      as in inclusive hard scattering. 
\item The methods are applied both to the hard scattering itself and
      to the showering kernels.
\item However, because we are now working with {\em exclusive}
      processes, the form of the approximations is different.  The
      approximated graphs must satisfy the following requirements:
      \begin{itemize}
      \item Exact 4-momentum conservation must be obeyed for each
            subprocess (hard scattering or one stage of the
            showering).  I.e., $p_i^\mu = p_f^\mu$, where $p_i^\mu$
            and $p_f^\mu$ are the total initial and final momenta the
            the subprocess.
      \item The approximation may change the momenta of the internal
            lines but it must preserve the momenta of the external
            lines.  This avoids the problem with having singular
            distributions.
      \end{itemize}
\item A cut-off is applied, for otherwise the approximated graphs give
      ultra-violet divergences when integrated to large transverse
      momentum. The kinematics associated with exact momentum
      conservation do provide a cutoff, but such a cutoff tends to
      violate factorization of the momentum-space integrals. So a
      separate artificial cutoff is better, and probably makes for a
      better implementation of the algorithm.
\item Conventionally, in a Monte-Carlo a sharp cutoff is used.  But a
      smooth cut-off will probably be better for numerical work.  It
      will also make it easier to get positive cross sections.
\item The dependence on the cutoff is a generalized
      renormalization-group transformation, and the exact cross
      section is independent of the form of the cutoff.
\item Separate explicit soft factors are needed, as in the
      factorization theorem for the $q_T$ distribution for the
      Drell-Yan process; unlike the case of an inclusive cross
      section the cancellation of the soft region is not complete.
      This issue is not treated in Refs.\ \cite{JCC.MC.lepto,JCC.MC},
      but will need further work, which is in progress.
\end{itemize}


\paragraph{Example of application to DIS \cite{JCC.MC.lepto}}

\begin{figure}
   \centering \includegraphics[scale=0.4]{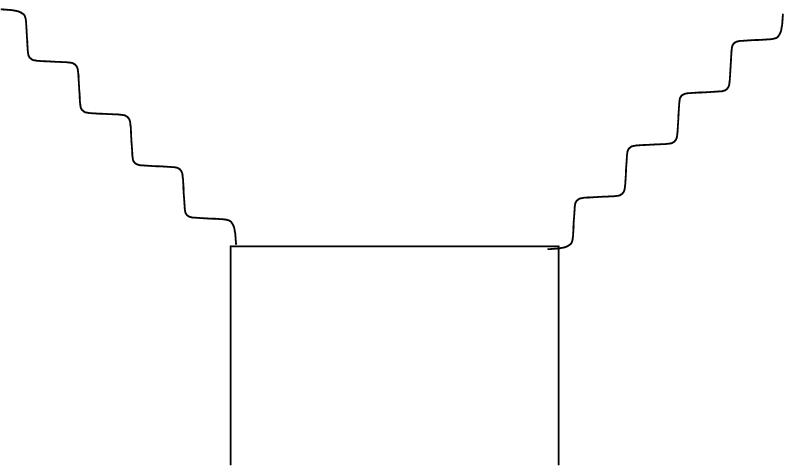}
   \caption{Born graph for DIS}\label{Born}
\end{figure}

The parton model graph of Fig.\ \ref{Born} is combined with showering
to give a LO cross section that can be summarized as
\begin{equation}
   \sigma_{\rm LO} = \mbox{Born~graph} \times \mbox{initial-state showering} 
       \times \mbox{final-state showering}.
\end{equation}

The NLO cross section is obtained from subtracted one-loop graphs, and 
the hard-scattering coefficient is of the form
\begin{eqnarray}
   \raisebox{-0.5\height}{\includegraphics[scale=0.35]{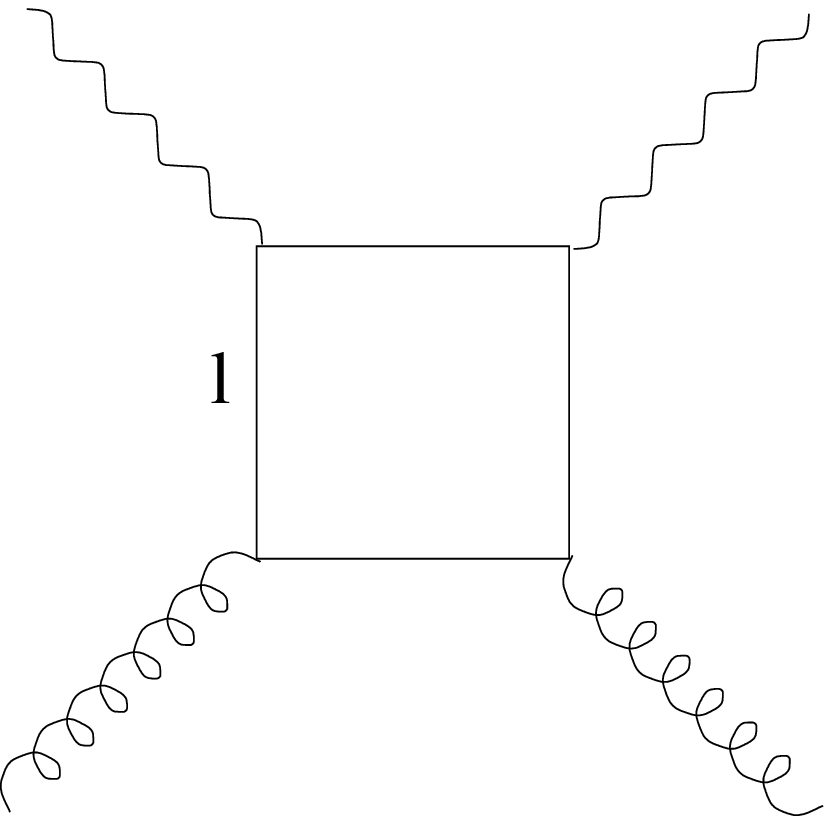}}
   ~~ - ~~
    \raisebox{-0.5\height}{\includegraphics[scale=0.35]{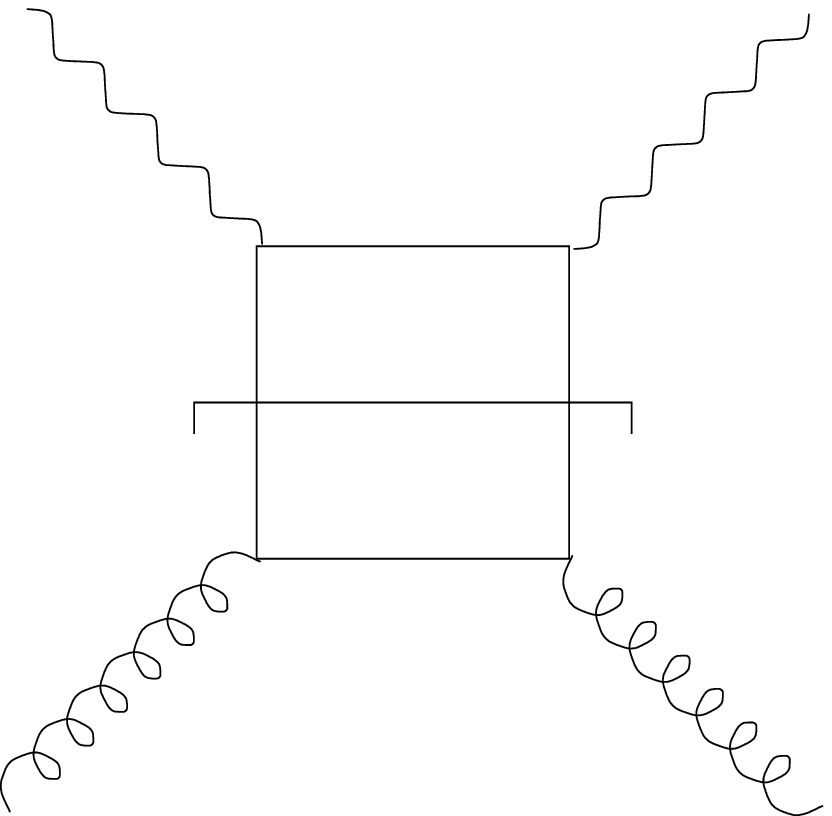}}
    ~ \times ~ \mbox{Jacobian}
    ~ \times ~  \mbox{cutoff function} .
\end{eqnarray}
The first term is an unsubtracted NLO graph.  The subtraction
corresponds to the approximation made in LO.  Above the horizontal
line, the following replacement for the internal momentum $l^\mu$ is
made:
\begin{eqnarray}
  l &\to& \mbox{massless on-shell, zero transverse momentum}
\\
    &\to& \mbox{Lorentz-transformed momentum with correct final state}.
\end{eqnarray}
The first step is the standard approximation.  The second step is
needed to obtain the correct kinematics with conservation of
4-momentum.  It is somewhat non-trivial to implement consistently
since the subtraction term is needed when it is far from the collinear
region.  The bulk of my work in Refs.\
\cite{JCC.MC.lepto,JCC.MC} is about constructing a definite correct
and consistent implementation.  Correctness here means that the
subtraction term is the order $\alpha_s$ approximation to the showering in 
the LO Monte Carlo algorithm.


\paragraph{Differences between conventional ``analytic'' method and
Monte-Carlo method} 
These differences can be illustrated by the following mathematical
example. {\em Warning: In a number of respects this example is
over-simplified.}  For example, it does not take account of the parton 
densities.  As explained in Ref.\ \cite{JCC.MC.lepto}, the parton
densities in the Monte Carlo are not in the usual $\overline{\rm MS}$
scheme.  However, they are related to them by definite formulae.

Let the unapproximated unsubtracted integrand at NLO be
\begin{equation}
   \frac{d\hat\sigma }{d^2k_T}
   =
   \frac{Q^2}   {(Q^2+k_T^2+m^2) (k_T^2+m^2)} .
\end{equation}
The conventional approach obtains the hard-scattering coefficient by
setting the mass $m$ to zero.  The integral is then infinite, and a
subtraction is inserted which consists of a delta function at $k_T=0$
with an infinite coefficient.  The result is a $+$ distribution, which 
has a finite integral:
\begin{equation}
    \frac{d\hat\sigma }{d^2k_T}
    ~=~ \frac{Q^2}{Q^2+k_T^2}
      \left( \frac{1}{k_T^2} \right)_{+}
    ~=~ \frac{Q^2}{(Q^2+k_T^2) \,k_T^2} - C \delta^{(2)}(k_T) ,
\label{singular}
\end{equation}
where $C$ is an infinite constant, defined with the aid, for example,
of dimensional regularization.

In my new approach the subtracted integrand is
\begin{equation}
    \frac{d\hat\sigma }{d^2k_T}
    ~=~ \frac{Q^2}{(Q^2+k_T^2) \,k_T^2} - \frac{f(k_T/\mu) J(k_T) }{k_T^2} ,
\label{non-singular}
\end{equation}
where $f(k_T/\mu)$ is the previously mentioned cutoff function, which
is unity for small $k_T$ and zero for large $k_T$.  As usual $\mu$ is
the factorization scale.  The factor $J(k_T)$ symbolizes the Jacobian
that is necessary in the transformation from the variables appropriate 
for generation of events and the variables for the measured particles.




\setcounter{figure}{0}
\setcounter{table}{0}
\setcounter{section}{0}
\setcounter{equation}{0}
\newpage



\def\eqn#1{Eq.~(\ref{#1})}
\def\eqns#1#2{Eqs.~(\ref{#1}) and~(\ref{#2})}
\def\eqnss#1#2{Eqs.~(\ref{#1})-(\ref{#2})}
\def\fig#1{Fig.~{\ref{#1}}}
\def\sec#1{Section~{\ref{#1}}}
\def\app#1{Appendix~\ref{#1}}
\def\tab#1{Table~\ref{#1}}


 

\newskip\humongous \humongous=0pt plus 1000pt minus 100pt
\def\caja{\mathsurround=0pt}
\def\eqalign#1{\,\vcenter{\openup1\jot \caja
       \ialign{\strut \hfil$\displaystyle{##}$&$
        \displaystyle{{}##}$\hfil\crcr#1\crcr}}\,}
\newif\ifdtup
\def\panorama{\global\dtuptrue \openup1\jot \caja
        \everycr{\noalign{\ifdtup \global\dtupfalse
        \vskip-\lineskiplimit \vskip\normallineskiplimit
        \else \penalty\interdisplaylinepenalty \fi}}}
\def\eqalignno#1{\panorama \tabskip=\humongous
        \halign to\displaywidth{\hfil$\displaystyle{##}$
        \tabskip=0pt&$\displaystyle{{}##}$\hfil
        \tabskip=\humongous&\llap{$##$}\tabskip=0pt
        \crcr#1\crcr}}

\makeatletter
\def\@eqnnum{\hbox{\reset@font\rm(\theequation)}}
\let\make@eqnnum=\@eqnnum %
\def\eqnum#1{\dec@eqnnum \global\def\make@eqnnum{\reset@font\rm(#1)}%
\def\@currentlabel{#1}%
}
\def\inc@eqnnum{\addtocounter{equation}{1}}
\def\dec@eqnnum{\addtocounter{equation}{-1}}
\@definecounter{equation}%
\@addtoreset{equation}{section} %
\def\theequation@prefix{{\thesection}.} %
\def\theequation{\theequation@prefix\arabic{equation}}%
\makeatother



\begin{center}
\vspace*{1.2cm}
{\Large\sc \bf NLO and NNLO Calculations} \\
\vspace*{1.cm} 
{\sc V.~Del Duca and G.~Heinrich }
\vspace*{1.cm}
\end{center}


\setcounter{footnote}{0}

\section{The NLO and NNLO program}
\label{sec:program}

QCD calculations of multijet rates beyond the leading order (LO) in
the strong coupling constant $\alpha_s$ are usually quite involved.
Nowadays we know (see \sec{sec:nlo}) how to
perform in general calculations of the next-to-leading order
(NLO) corrections to multijet rates, and almost every process of
interest has been computed to that accuracy. Instead, the calculation
of the next-to-next-to-leading order (NNLO) corrections is still
at an organizational stage and represents a main challenge.
Why should we perform calculations which are technically so complicated ?

The general motivation is that the calculation of the NLO 
corrections allows us to estimate reliably a given production rate,
while the NNLO corrections allow us to estimate the theoretical uncertainty 
on the production rate. This is achieved by reducing
the dependence of the cross section
on the renormalization scale, $\mu_R$, and for processes with
strongly-interacting incoming particles the dependence on the
factorization scale, $\mu_F$, as well.

An example is the determination of $\alpha_s$ from
event shape variables in $e^+ e^- \rightarrow 3$ jets~\cite{schm}.
Although the NLO contributions to $e^+ e^- \rightarrow 3\, jets$ have been 
computed for some time now~\cite{ert,nk}, the NNLO
contributions have yet to be obtained.  A calculation of these
NNLO contributions would be needed to further reduce the theoretical
uncertainty in the determination of $\alpha_s$.

We present in \sec{sec:higgs} an additional motivation for performing
QCD calculations at NNLO, which is specific to the LHC program,
and we outline in \sec{sec:nlo}
how QCD calculations at NLO are implemented and in \sec{sec:nnlo} how
QCD calculations at NNLO could be performed.

\subsection{Higgs production}
\label{sec:higgs}

The main goal of the LHC physics program is the investigation of the 
mechanism of the electroweak symmetry breaking, and namely
the search and detection of the Higgs boson.
If the Higgs boson is light (100 GeV $\le m_H \le$ 140 GeV), 
the rare decay channel in two photons, H$\to \gamma\gamma$,
provides the best signature \cite{atlastp,atlastdr,cmstp,cmstdr}.
Since the signal-to-background ratio is quite low ($\sim 7\%$),
the analysis of this channel promises to be demanding. Our theoretical
understanding of signal and background is still preliminary:
the NLO QCD corrections to the signal are known to be quite large 
$({\cal O} (100\%))$ \cite{zerwas}.
Also the QCD background $pp\to\gamma\gamma$, given at LO by
the parton subprocess $q\bar q \to \gamma\gamma$, is known to
NLO \cite{owens}, with the full NLO fragmentation 
contributions having just 
been evaluated~\cite{pilon}. However, $pp\to\gamma\gamma$ receives a 
sizeble 
contribution from NNLO corrections because of the large gluon luminosity of the
subprocess $gg\to \gamma\gamma$ appearing first at NNLO~
\cite{kunszt}.
Thus in order to have a reliable theoretical estimate both the signal and the
background need to be determined at least to NNLO accuracy.

In order to improve the signal-to-background ratio, Higgs production
in association with a high transverse energy ($E_T$) jet, 
$p\,p\to H\, jet \to \gamma\gamma\, jet$, has been considered
~\cite{abdullin}. 
This production rate offers the advantage of being
more flexible in choosing suitable acceptance cuts to curb the
background. $p\,p\to H\, jet$ is known to LO exactly~\cite{ellis}, while
the NLO corrections~\cite{grazz} have been computed in the infinite top-mass
limit. The NLO corrections to the signal are large. However, it is 
believed that the background,
$p\,p\to\gamma\gamma\, jet$, can be more reliably calculated because 
LO production is dominated by the parton subprocess $q\, g \to q\,
\gamma\gamma$, which benefits from the large gluon luminosity, while the
subprocess $gg\to g \gamma\gamma$, 
which is believed to dominate the NNLO contribution, yields a comparatively
smaller contribution~\cite{kunszt,yuan}. 
Thus, even though the signal,
$p\,p\to H\, jet$, likely needs be computed at NNLO accuracy, it should
suffice to evaluate the background, $p\,p\to\gamma\gamma\, jet$, at NLO.
The NLO corrections to the background, though, have yet to be computed,
with the appropriate QCD amplitudes having just 
been evaluated~\cite{dkm}.

\subsection{NLO algorithms and one-loop amplitudes}
\label{sec:nlo}

In recent years it has become clear how to construct 
general-purpose algorithms for the calculation of multijet rates
at NLO accuracy.
The crucial point is to organise the cancellation of the infrared
(i.e. collinear and soft) singularities of the QCD amplitudes
in a universal, i.e.
process-independent, way. The universal terms in a NLO calculation are
given by the tree-level splitting~\cite{ap} and eikonal
~\cite{bcm,bg} 
functions, and by the universal structure of the poles of the one-loop
amplitudes~\cite{gg,ks,kst}. 
The universal NLO terms and the 
process-dependent amplitudes are combined into 
effective matrix elements, which are devoid of singularities. The various NLO 
algorithms (phase-space 
slicing~\cite{gg,slicing}, 
subtraction method
~\cite{ks,subtr},
dipole formalism~\cite{dipole} and 
subtraction-improved slicing~\cite{gk}) provide different methods to
construct the effective matrix elements. These
can be integrated, analytically or otherwise numerically, in four dimensions. 
The integration can be performed with arbitrary experimental acceptance cuts.

\begin{figure}[tbp]
\epsfxsize=10cm \epsfbox{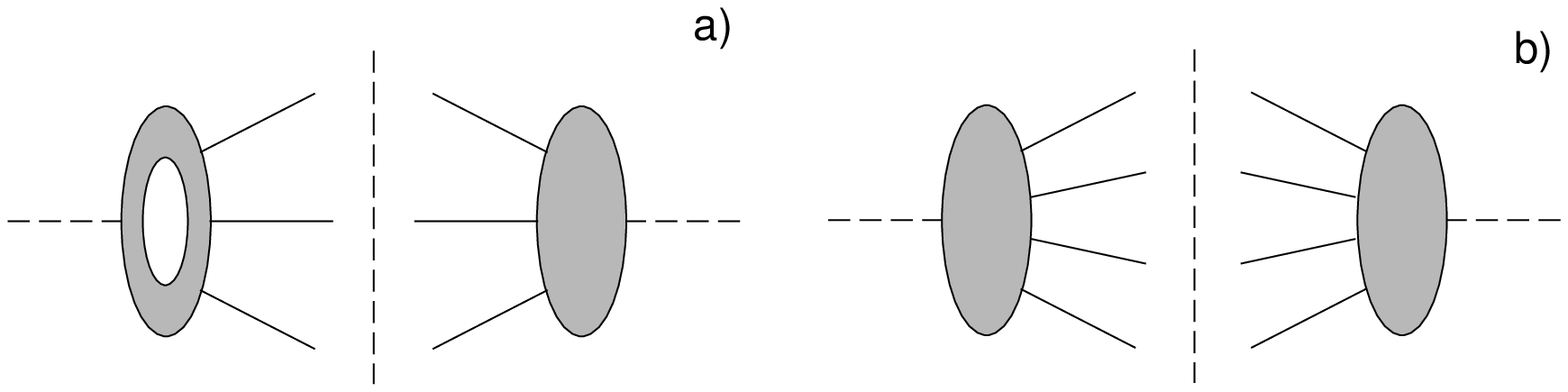}
\caption{\sf Squared matrix elements which contribute the NLO corrections to
$e^+ e^- \rightarrow 3\, jets$. The dashed line represents a massive
vector boson, $\gamma^*, W, Z$.  $a)$ interference term between
one-loop and tree amplitudes. The final-state partons are
a $q{\bar q}$ pair and a gluon. $b)$ square of a tree amplitude. 
The final-state partons are a $q{\bar q}$ pair and two gluons, or 
two $q{\bar q}$ pairs. In figure $b)$ one of the partons is unresolved.}  
\label{fig:one}
\end{figure}

Then the remaining work to be performed to
calculate a production rate at NLO is to compute the
appropriate tree and one-loop amplitudes. 
To compute $n$-jet production at NLO, two sets of amplitudes
are required: {\it a}) $n$-particle production amplitudes at tree
level and one loop; {\it b}) $(n+1)$-particle production
amplitudes at tree level. If the one-loop amplitudes are regularised 
through dimensional regularization (DR) by evaluating them in $d=4-2\epsilon$
dimensions, it suffices at NLO to compute them to ${\cal O}(\epsilon^0)$.
As an example, in Fig.~\ref{fig:one} we show
the squared matrix elements which are required to calculate the NLO 
corrections to $e^+ e^- \rightarrow 3\, jets$.

Efficient methods based on the 
color decomposition
~\cite{Color,bk,dfm,ddm} of an amplitude in color-ordered
subamplitudes, which are then projected onto the helicity states of the
external partons, have largely enhanced the ability of computing 
tree~\cite{mpReview} and one-loop~\cite{bdkReview} amplitudes.
Accordingly, tree amplitudes with up to seven massless partons~\cite{mpReview,
bgk} and with a vector boson and up to five massless partons~\cite{bgkboson}
have been computed analytically. In addition, efficient
techniques to evaluate numerically tree multi-parton amplitudes have been 
introduced~\cite{dkp,cmmp}, and have been used to compute tree amplitudes
with up to eleven massless partons~\cite{cmmp}.
The calculation of one-loop amplitudes can be reduced to the
calculation of one-loop $n$-point scalar integrals~\cite{pv, neervan}.
The reduction method~\cite{pv} allowed the computation
of one-loop amplitudes with four massless 
partons~\cite{kes} and with a vector boson and three massless 
partons~\cite{kert}.
However, one-loop scalar integrals present infrared divergences, 
induced by the massless external legs.
For one-loop multi-parton amplitudes, the infrared divergences hinder 
the reduction methods of 
ref.~\cite{pv,neervan}.
This problem has been overcome in ref.~\cite{bdkint}.
Accordingly, one-loop amplitudes with five massless 
partons~\cite{bdk,kts} and with a vector boson and four massless 
partons~\cite{bdkcg} have been computed analytically. 
The reduction procedure of ref.~\cite{bdkint} 
has been generalised
in ref.~\cite{bgh}, where it has been shown that any 
one-loop $n$-point scalar integral, with $n> 4$, can be reduced
to box scalar integrals, and that in the reduction of $n$-point tensor
integrals, all higher dimensional ($d>4-2\epsilon$) $n$-point integrals 
with $n> 4$ drop out. 
The calculation of one-loop multi-parton amplitudes thus can be
pushed a step further in the near future.

\subsection{NNLO calculations}
\label{sec:nnlo}

Eventually, a procedure similar to the one followed at NLO will permit the 
construction of general-purpose algorithms at NNLO accuracy.
It is mandatory then to fully investigate the infrared structure of the
phase space at NNLO. The universal pieces needed to organise the cancellation
of the infrared singularities are given by the tree-level 
double-splitting~\cite{glover,cat,dfm}, 
double-eikonal~\cite{bg,catgra} and 
splitting-eikonal~\cite{glover,catgra} functions, 
by the one-loop 
splitting~\cite{bds,dk} and 
eikonal~\cite{bds} functions,
and by the universal structure of the poles of the two-loop
amplitudes~\cite{catani}. These universal pieces have yet to be 
assembled together, to show the cancellation of the infrared divergences
at NNLO.

\begin{figure}[th]
\epsfxsize=10cm \epsfbox{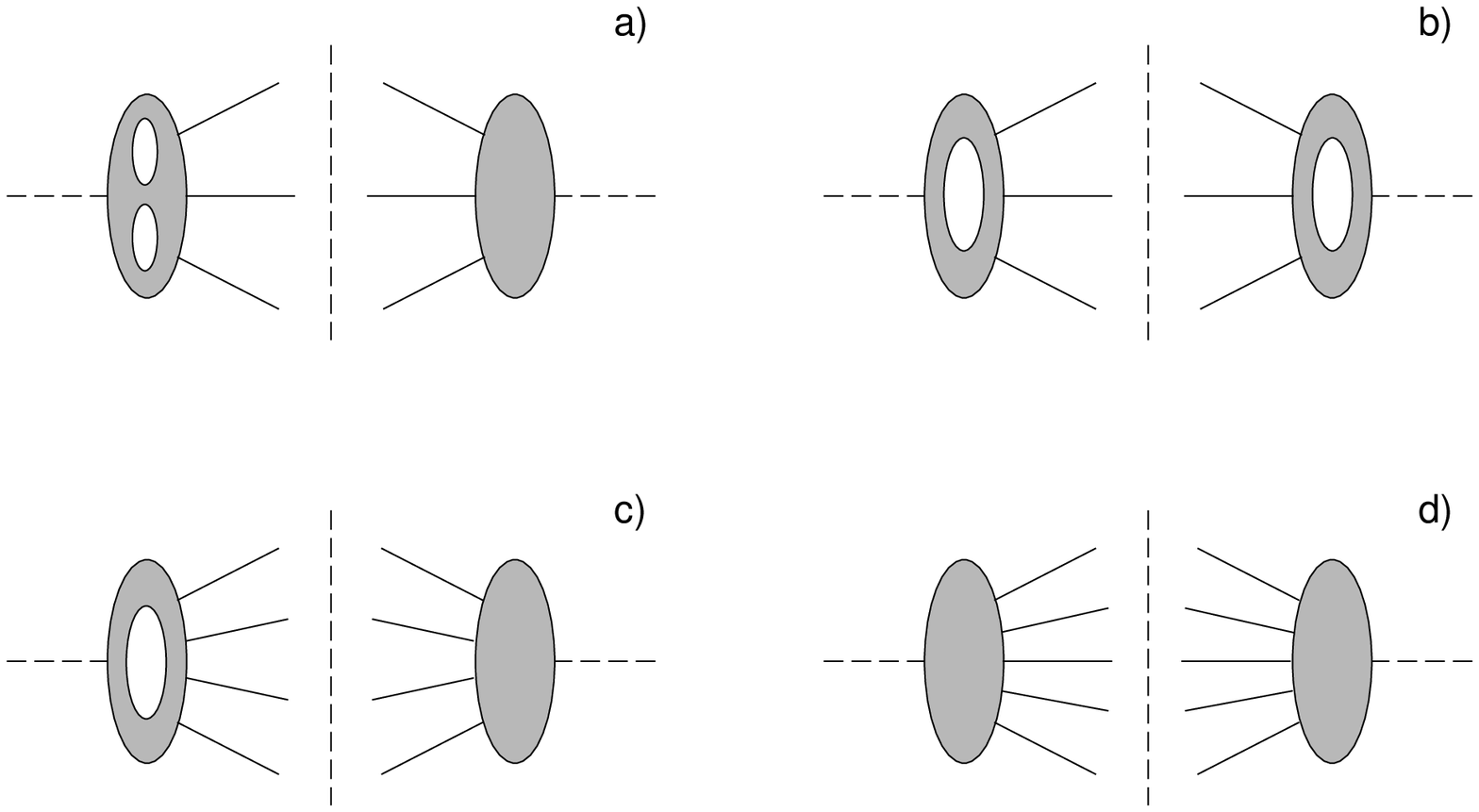}
\caption{\sf Squared matrix elements which contribute the NNLO corrections to
$e^+ e^- \rightarrow 3\, jets$. The dashed line represents a massive
vector boson, $\gamma^*, W, Z$.  $a)$ interference term between
two-loop and tree amplitudes, and $b)$ square of a one-loop amplitude.
In figures $a)$ and $b)$ the final-state partons are
a $q{\bar q}$ pair and a gluon. $c)$ interference term between
one-loop and tree amplitudes. The final-state partons are
a $q{\bar q}$ pair and two gluons or two $q{\bar q}$ pairs. One of the
partons is unresolved. $d)$ square of a tree amplitude. The final-state 
partons are a $q{\bar q}$ pair and three gluons, or two $q{\bar q}$ pairs
and a gluon. Two of the partons are unresolved.}  
\label{fig:two}
\end{figure}

Then to compute $n$-jet production at NNLO, three sets of amplitudes
are required: {\it a}) $n$-particle production amplitudes at tree
level, one loop and two loops; {\it b}) $(n+1)$-particle production
amplitudes at tree level and one loop; {\it c}) $(n+2)$-particle
production amplitudes at tree level. 
In Fig.~\ref{fig:two} we show
the squared matrix elements which are required to calculate the NNLO 
corrections to $e^+ e^- \rightarrow 3\, jets$.
In DR at NNLO, the two-loop amplitudes 
need be computed to ${\cal O}(\epsilon^0)$, while the one-loop amplitudes 
must be evaluated to ${\cal O}(\epsilon^2)$~\cite{bds,bc}. 
The main challenge is the calculation of the two-loop amplitudes.
At present, the only amplitude known at two loops is the one for $V 
\leftrightarrow q\bar q$~\cite{mvann}, 
with $V$ a massive vector boson,
which depends only on one kinematic variable. It has been used to evaluate
the NNLO corrections to Drell-Yan production~\cite{neerdy} and to 
deeply inelastic scattering (DIS)~\cite{neerdis}.
No two-loop computations exist
for configurations involving more than one kinematic variable, except
in the case of maximal supersymmetry~\cite{bry}. 
One of the main obstacles 
for configurations involving two kinematic variables
is the analytic computation of the two-loop four-point functions
with massless external legs, where significant progress has just been 
achieved. 
These consist of planar double-box 
integrals~\cite{Smirnov}, 
non-planar double-box integrals~\cite{Tausk:1999vh},
single-box integrals with a bubble insertion on one of the 
propagators~\cite{Anastasiou:1999cx} and 
single-box integrals with a vertex correction~\cite{Anastasiou:1999bn}. 
The two-loop four-point functions with massless external legs are needed
for the computation of two-loop amplitudes in parton-parton scattering.
Finally, the topical processes 
considered above, i.e. $e^+ e^- \rightarrow 3\, jets$ and $p\,p\to H\, jet$
sport configurations involving three kinematic variables and require
the analytic computation of two-loop four-point functions  with a massive 
external leg. The two-loop four-point functions of this kind
with up to five different denominators have been derived 
recently~\cite{Gehrmann:1999as}, while those with six and seven different 
propagators are still missing. 
Another obstacle is the color decomposition of
two-loop amplitudes, which is not known yet. 
Substantial progress is expected in the next future on all the 
issues outlined above, which should make the present note soon outdated.

Finally, we mention that in the factorization of collinear 
singularities~\cite{css} for strongly-interacting incoming 
particles,
the evolution of the parton distribution functions ($pdf$'s) in the jet
cross section should be 
determined to an accuracy matching the one of the parton cross section.
For hadroproduction of jets computed at NLO, one needs the NLO, or two-loop,
evolution of the $pdf$'s~\cite{frs,cfp,hmsvn}. Accordingly for
hadroproduction  at NNLO the evolution of the $pdf$'s
should be computed to NNLO, or three-loop, accuracy. Except for the 
lowest five (four) even-integer moments of the three-loop non-singlet (singlet)
splitting functions~\cite{lnrv},
no calculation of the NNLO evolution of the $pdf$'s exists yet. 
However, NNLO analysis based on the finite set of known moments 
have been performed for $xF_3$~\cite{kkps,kps} and $F_2$
(non-singlet~\cite{pkk} and singlet~\cite{sy}). 
Furthermore, in ref.~\cite{vanNeerven:1999ca} a quantitative assessment of the 
importance of the yet unknown higher-order terms has been performed, 
with the conclusion that they should be numerically significant only for 
Bjorken-scaling $x < 10^{-2}$. 

The computation of the evolution of the 
$pdf$'s at NNLO accuracy is a main challenge in QCD.
The NLO computation was performed with two different methods, one using
the  operator product expansion (OPE) in a covariant
gauge~\cite{frs}, the other using the 
light-cone axial (LCA) gauge with principal value prescription~
\cite{cfp}.
However, the prescription used in ref.~\cite{cfp} 
has certain shortcomings.
Accordingly, the calculation has been repeated in the LCA
gauge using a generally correct prescription~
\cite{bhkv}, which makes it
amenable to extensions beyond NLO. 
On the other hand, using the OPE method, there had been a problem with 
operator mixing in the singlet sector, which  has been 
fixed\cite{hmsvn} only recently, 
and the result finally coincides with  
the one obtained in the LCA gauge in ref.~\cite{cfp}.
Thus the calculation of the $pdf$ evolution at NLO accuracy is fully
under control. Recent proposals for a calculation beyond  NLO
include extensions of the OPE technique, which have been used 
to recompute the 
NNLO corrections to DIS~\cite{Moch:1999eb}, and a
computation of the $pdf$ evolution by combining the universal 
gauge-invariant collinear pieces~\cite{uwer}.
For the two-loop $pdf$ evolution, e.g., they are the collinear 
pieces mentioned at the beginning of this section.




\setcounter{figure}{0}
\setcounter{table}{0}
\setcounter{section}{0}
\setcounter{equation}{0}
\newpage

\begin{center}
\vspace*{1.2cm}
{\Large\sc \bf Jet Algorithms } \\
\vspace*{1.cm} 
{\sc S. Catani and D. Zeppenfeld}
\vspace*{1.cm}
\end{center}

\setcounter{footnote}{0}

\section{Jet algorithms}

Jet algorithms have the task to assign streams of hadrons in hard scattering 
processes to a jet, who's energy, mass and momentum can then be related 
to a collection of partons in a perturbative QCD calculation. 
Although, 
at the experimental level, jets can be defined by using rather general
and intuitive procedures, if we would like to compute jet cross sections and
properties by using QCD perturbation theory, the definition of jets should
fulfil stronger constraints to guarantee its perturbative safety. Perturbative
safety means that the definition has to be infrared safe (jet properties cannot
depend on the presence of arbitrarily soft partons), collinear safe (jet 
properties cannot change by replacing a parton with a set of collinear partons
carrying the same total momentum) and collinearly factorizable (jet 
properties should be insensitive to partons radiated collinearly to the
beam direction). If the jet definition is not perturbative safe, we
cannot perform calculations order-by-order in perturbation theory because
they are affected by uncancelled infrared divergences. Of course, in the full
QCD theory (i.e. beyond perturbation theory) the perturbative divergences
are regularized by small physical cutoffs related to
hadron masses and the finite experimental
resolution (size of calorimeter cells, energy thresholds, etc.).
The physical cutoffs are always present, independently of the jet definition.
However, in the case of a perturbative safe definition, their effects are
suppressed by some inverse power of the jet transverse energy $E_T$, and thus 
they can be made small by sufficiently increasing $E_T$. This power suppression
is not at work in perturbative unsafe jet definitions, where the 
effects of the small physical cutoffs can amount to large corrections 
(of order unity) to the perturbative results. Thus, perturbative safe 
definitions are preferred.
 
Jet algorithms start from a list of ``particles'' which we would like 
to freely associate with calorimeter cells or hadrons at the experimental 
level,
and with partons in a QCD calculation. Each particle $i$ carries a 
4-momentum $p^\mu_i$, which we take to be massless. The task is to select
a set of particles which are emitted close to each other in angle and 
combine their momenta to form the momentum of a jet. The selection process 
is called the ``jet algorithm'', the momentum addition rule is called the 
``recombination scheme''.

Let us start with a discussion of recombination schemes.
In a hadron collider environment the arbitrary boost of the hard scattering
system along the beam axis needs to be taken into account in the definition 
of angles to ensure collinear factorizability. This is achieved by using 
transverse momentum, 
$p_T=\sqrt{p_x^2+p_y^2}$, rapidity $y=1/2 \log (E+p_z)/(E-p_z)$ and
azimuthal angle $\phi$ of the massless particles as the kinematic variables.
When adding the massless 4-vectors of particles we obtain 
massive objects which only approximately correspond to the massless partons
which we would like to associate with jets at tree level. 

One popular choice, the Snowmass convention~\cite{snowmass}, 
leaves the question of jet mass open, by only 
defining total transverse energy, rapidity and azimuthal angle of a set of
parton momenta, as the $E_T$ weighted sums of the individual particle
variables. For the original massless particles $E_{Ti}=p_{Ti}$ and 
$\eta_i=y_i$. The corresponding recombined variables for a cluster of
particles are then given by the total transverse energy
\bq
\label{etsnowmass} 
E_T = \sum_i E_{Ti}\; ,
\eq
the cluster pseudorapidity
\bq
\eta = \sum_i {E_{Ti}\over E_T}\; \eta_i\;,
\eq
and the azimuthal angle of the cluster
\bq
\phi = \sum_i {E_{Ti}\over E_T}\; \phi_i\;.
\eq
Note that the designation of $\eta$ as pseudorapidity is purely conventional.
It corresponds to neither the pseudorapidity nor the rapidity of the massive
cluster and is approximately equal to either only in the limit of small 
cluster mass ($<< E_T$). The concomitant loss of Lorenz invariance 
is a serious disadvantage of the Snowmass convention. Another serious problem
appears in resummation calculations (see Sect.~\ref{sec:res}): 
the kinematic boundary of jet $E_T$
shifts (from $\sqrt{\hat s}/2$ in e.g. dijet kinematics) when including 
additional final state partons.

Because of these shortcomings we formulate all jet algorithms in the 
4-momentum recombination scheme (also called E-scheme) in the following,
i.e. the kinematic variables of a cluster of particles is given by
direct addition of the 4-momenta of the individual massless particles:
\bq
\label{4mom}
p^\mu = (E,p_x,p_y,p_z) = \sum_i p_i^\mu\; .
\eq
Since the resulting clusters have a clearly defined mass, we must distinguish
transverse energy $E_T$ from transverse momentum $p_T$, and pseudorapidity
$\eta$ from rapidity $y$. We define 
\bq
\label{truept}
p_T = \sqrt{p_x^2+p_y^2}\;, \qquad \phi = \tan^{-1} {p_x\over p_y}\; ,
\qquad y = {1\over 2}\log {E+p_z\over E-p_z}\; , 
\eq
The rapidity $y$ and azimuthal $\phi$ should be used as the legoplot
position of the jet when calculating its separation from other particles 
or jets. Auxiliary quantities are
\bq
\label{auxquant}
\theta = \cos^{-1}{p_z\over \sqrt{p_x^2+p_y^2+p_z^2}}\;, \qquad
E_T = E\;\sin\theta\;, \qquad \eta=-\log\tan{\theta\over 2}\;.
\eq

\subsection{The $k_T$ algorithm}\label{sec:alg.kt}

The $k_T$ algorithm~\cite{es_kt} 
is a successive recombination algorithm. The idea is to
recombine particles with nearly parallel momenta, beginning with the softest
particles in the sample. This recombination stops once all clusters of 
particles are separated by a distance larger than $D$ in the legoplot.
The $k_T$ algorithm starts from a list of protojets and their momenta, 
$p_i^\mu$, which in the beginning consists of the list of all particles:
 
\begin{itemize}

\item[(1)] For each protojet $i$ define 
\bq
d_i = E_{Ti}^2
\eq
and for each pair of protojets $i,j$ define a distance 
\bq
d_{ij}= \min (E_{Ti}^2,E_{Tj}^2) { (y_i-y_j)^2+(\phi_i-\phi_j)^2\over D^2}\;.
\eq

\item[(2)] Find the smallest of all $d_i$ and $d_{ij}$ and call it $d_{min}$.

\item[(3)] If $d_{min}$ is $d_{ij}$ then merge protojets $i$ and $j$ to 
form a new protojet $k$ of momentum
\bq
p_k^\mu = p_i^\mu + p_j^\mu
\eq

\item[(4)]
If $d_{min}$ is $d_i$ then remove protojet $i$ from the protojet list
and move it to the list of completed jets.

\item[(5)]
Continue with step 1 until the list of protojets is empty.

\end{itemize}

The algorithm is infrared safe because it renders all soft partons harmless:
it either takes them off the protojet list or it combines them with nearby 
harder partons. The only effect of the soft parton then is a shift in
the momentum of the recombined cluster. However, this shift is small, 
disappearing in the infrared limit, which guarantees infrared safety.
Also for collinear emission the algorithm is safe, because in the limit of
zero angle between two partons, these two will have the smallest
$d_{ij}$ and will thus be combined early on in the recombination process,
thus restoring the momentum of the almost on-shell parton from which they
originated by splitting. Finally, every original particle is assigned to 
exactly one jet, i.e. there are no splitting/merging issues to be resolved 
for the $k_T$ algorithm.

\subsection{ILCA: an infrared safe cone algorithm}\label{sec:alg.ilca}

Cone algorithms are intended to cluster all energy within a given radius,
$R$, around a point in the legoplot, to form jets. Naively, this procedure 
is both infrared and collinear safe: the effect of infrared radiation on the 
cluster momentum vanishes in the infrared limit, and the energy measured 
for the jet is the same whether a single particle is at the core of the 
cone or whether there has been collinear splitting. This naive 
expectation can easily be violated, however, by the prescription for
selecting cones. Two examples illustrate this point~\cite{seymour}.

\begin{figure}[th]
\begin{center}
\epsfxsize=14cm
\epsfysize=5cm
\mbox{\epsfbox{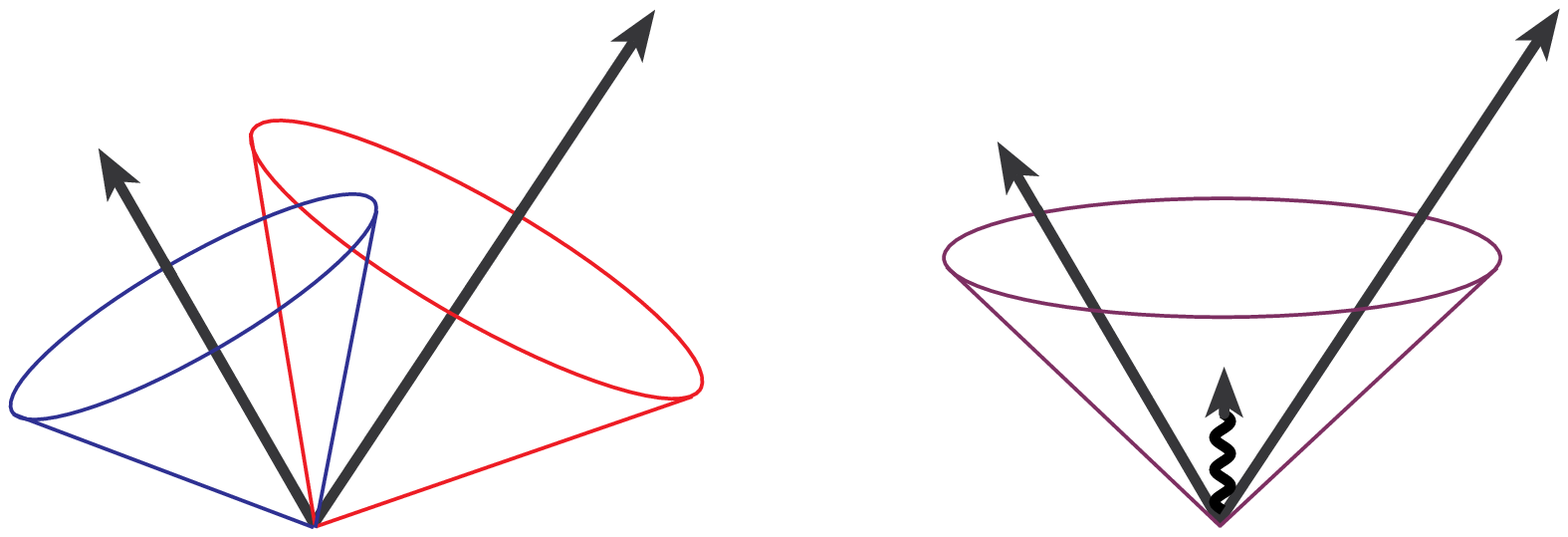}}
\end{center}
\caption{
\sf 
Example for a situation which can lead to infrared problems in an unsafe 
cone algorithm.
}
\label{fig:coneSSir}
\end{figure}

Assume that cones are constructed around actual energy depositions only.
In Fig.~\ref{fig:coneSSir}(a) two particles are emitted at a distance 
greater than the cone radius $R$ but smaller than $2R$
and therefore are assigned to separate cones, 
which are then identified as two distinct jets. The only difference 
in  Fig.~\ref{fig:coneSSir}(b) is the emission of a third soft particle 
(a ``soft gluon'') between the original two particles. Now the additional
cone around the soft energy deposition encompasses all three particles and
they will be classified as a single jet. The presence of a soft particle
changes the classification of a hard event: this is an example for an 
infrared unsafe
algorithm. In perturbation theory, at sufficiently high order, an arbitrary
number of soft gluons will be radiated, hence, cones should be allowed
anywhere in phase space to anticipate this feature of higher order 
corrections. An arbitrary restriction on allowed cone positions may lead to
an infrared unsafe algorithm.

\begin{figure}[th]
\begin{center}
\epsfxsize=14cm
\epsfysize=8cm
\mbox{\epsfbox{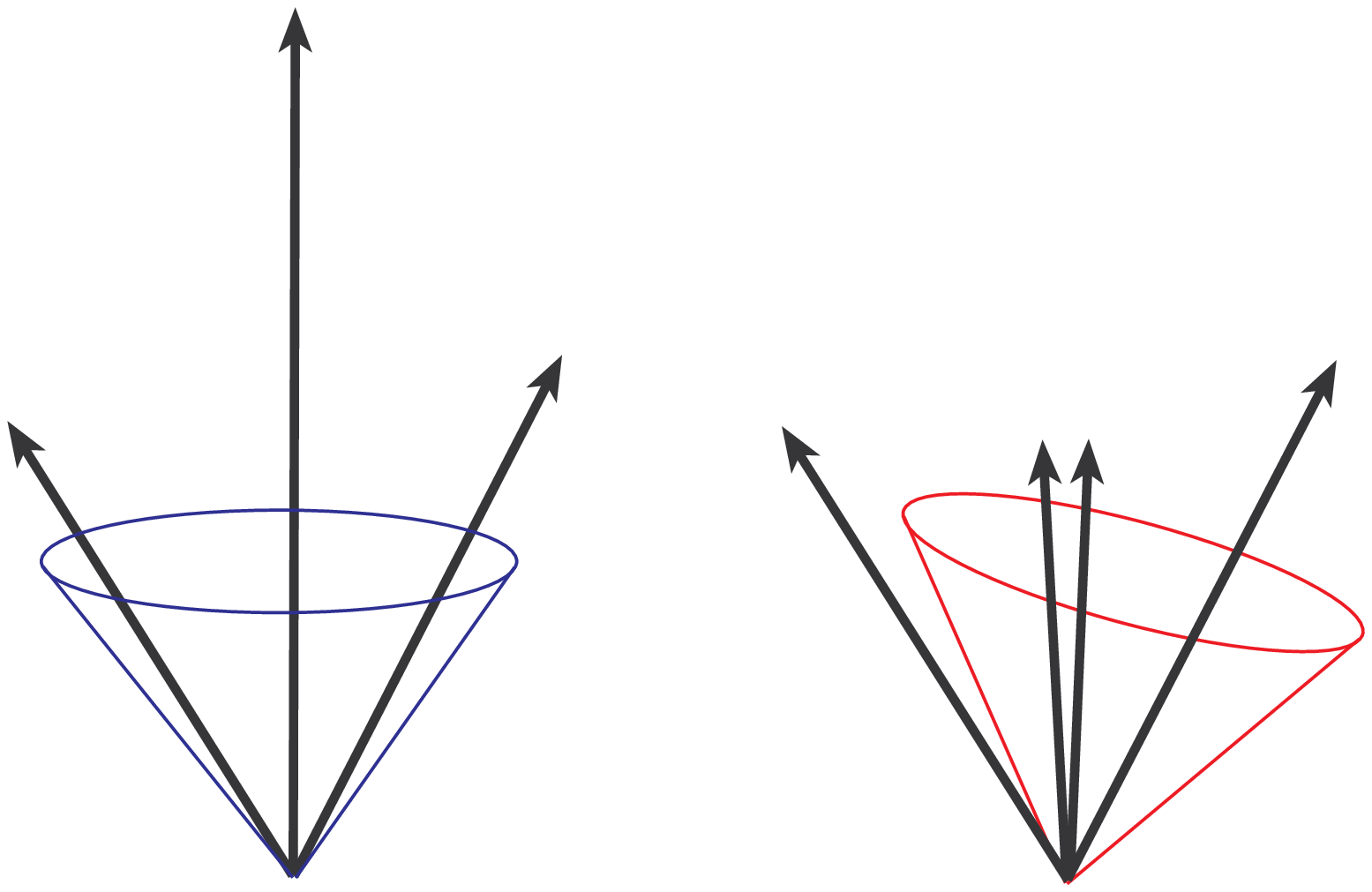}}
\end{center}
\caption{
\sf 
Example for a situation which can lead to collinear problems in an unsafe 
cone algorithm.
}
\label{fig:coneSScol} 
\end{figure}

Similarly, an infinite number of collinear splittings occurs at higher order 
in perturbation theory. A possible collinear problem, resulting from $E_T$
ordering of particles, is illustrated in
Fig.~\ref{fig:coneSScol}. The difference between the two situations is that
the central (hardest) parton may split into two almost collinear partons.
On the left-hand-side the distance between the lateral partons is larger than
$R$ but
the three hard partons all fall within a cone of radius 
$R$ around the central parton, which happens to have the largest $E_T$. 
As a result, all three partons are recombined to a single jet. Collinear 
splitting renders the right hand parton to be the one with the largest $E_T$.
Drawing the first cone around the highest $E_T$ parton will recombine it 
with the two central partons and a separate jet is likely to be assigned 
to the remaining fourth parton. 
A differing jet number which depends on the presence or absence of collinear 
splitting must be avoided because the incomplete cancellation of the 
logarithmic divergences of real emission and virtual contributions will lead
to a collinear unsafe jet algorithm. One can eliminate such ambiguities
by making the selection or ordering of 
jet definition cones independent of the $E_T$ of individual particles.  
Also, allowing trial cones anywhere in phase space would have made the two
situations in Fig.~\ref{fig:coneSScol} more similar: allowing cones 
centered between the central and the outside partons from the start
would lead to a more similar jet identification in the two cases.

The above considerations lead us to consider a cone algorithm which allows
trial cones to be positioned anywhere in phase space, irrespective of the 
transverse momentum carried by individual particles or calorimeter cells.
We start by formulating this seed-less algorithm at the calorimeter level,
where the basic entities are calorimeter towers.

\begin{itemize}

\item[(0)]
Make a list of all calorimeter towers.

\item[(1)]
Select the next tower on the list as the center of a trial cone of
radius $R$.\\
Goto (4) if the list of towers is exhausted.

\item[(2)] 
Add the momenta of all towers inside the trial cone and determine the 
legoplot position $(y,\phi)$ corresponding to this momentum.

\item[(3)] 
If this position is outside the selected tower, discard the trial cone and
go to (1).\\
If $(y,\phi)$ is inside the selected tower, add the set of towers inside the
trial cone as a new entry to the list of protojets.

\end{itemize}

\noindent
At this stage we have a list of protojets, and we need to split/merge them
to make jets.

\begin{itemize}

\item[(4)]
Select the highest $E_T$ protojet remaining on the list. (If the list 
is exhausted jet identification for the event is complete.)

\item[(5)] 
Does the selected protojet share any towers with other protojets?
   \begin{itemize}
      \item[(5.a)]
      No: Move protojet to list of jets and continue with (4).
      \item[(5.b)]
      Yes: Find the highest $E_T$ protojet that shares towers with the
      selected protojet. (Call this the neighbor protojet.) Decide 
      whether the $E_T$ in the shared cells is greater than a fraction $f$ 
      of the $E_T$ in the neighbor protojet.

      \begin{itemize}
         \item[(5.b.1)]
         No: Split the shared towers.

         \begin{itemize}
            \item[(5.b.1.a)]
            Allocate shared towers to either the selected or the neighbor 
            protojet depending on which jet center is closer.
            \item[(5.b.1.b)]
            Calculate the new momenta for the modified protojets, i.e. 
            their $E_T$, and legoplot positions $(y,\phi)$. 
            Continue with step (4).
         \end{itemize}

         \item[(5.b.2)]
         Yes: Merge the selected and neighbor protojets to form a new 
         protojet. \\
         Add the momenta of both protojets and determine the total $E_T$ and 
         and the legoplot position $(y,\phi)$. Continue with step (4).
      \end{itemize}
  \end{itemize}
\end{itemize}

\noindent
The procedure that defines the list of 
protojets is infrared and collinear safe.
The additional steps completely define how to solve the problem of 
overlapping cones. The critical overlap fraction $f$ is a free parameter 
of the algorithm and may be chosen as 50\%, similar to the D0 choice in
run I of the Tevatron (CDF uses 75\%). The $E_T$-ordering of protojets in this 
split/merge
step does not introduce collinear problems provided the cone size $R$
is chosen sufficiently large. 

The definition of calorimeter towers, i.e. a discretization of $(y,\phi)$
space, would be cumbersome in a theoretical calculation, and is indeed 
not necessary. In a perturbative calculation at fixed order, the maximal 
number, $n$, of partons is fixed. The only possible positions of stable 
cones are then given by the partitions of the $n$ parton momenta, i.e. 
there are at most $2^n-1$ possible locations of protojets. They are given 
by the legoplot
positions of individual partons, all pairs of partons, all combinations of
three partons etc.  In a perturbative calculation, e.g. via a NLO Monte Carlo
program, the protojet selection of the seedless algorithm 
(steps (0) to (3) above) can then be replaced as follows:

\begin{itemize}

\item[(0)] 
Make a list of all possible cone centers. These are the legoplot
coordinates of all parton momenta $p_i$, of all pairs of parton momenta
$p_i+p_j$, of all triplets of parton momenta $p_i+p_j+p_k$, etc.
For each cone center record which set of partons defines it.

\item[(1)]
Select the next cone center on the list as the center of a trial cone of
radius $R$.\\
Goto (4) if the list of cone centers is exhausted.

\item[(2)] 
Add the momenta of all partons inside the trial cone and determine the 
legoplot position $(y,\phi)$ corresponding to this momentum.

\item[(3)] 
If this position is different from the trial cone center, i.e. if the
cone center record and the list of partons inside the trial cone disagree,
discard the trial cone and go to (1).\\
If $(y,\phi)$ is the trial cone center, add the set of partons inside the
trial cone as a new entry to the list of protojets.

\end{itemize}

\noindent
As before, different protojets may share partons, i.e. they may overlap.
The required split/merge step is then identical to the calorimeter level
steps (4) and (5), with towers replaced by partons as elements of protojets.

In an actual experiment the number of calorimeter towers may be very large 
(order 6000 for tower sizes of $\Delta\eta\times\Delta\phi=0.1\times 0.1$
and an $\eta$ coverage of $\pm 5$ units of pseudorapidity). The calorimeter
level algorithm may then be rather slow computationally. The question 
arises whether an acceptable approximation of the seedless algorithm can be 
constructed, analogous to the parton level short-cut, by considering only
those towers which have energy depositions above a minimal seed threshold.
One would like to replace the list of parton momenta above by the list
of tower momenta with
\bq
p_{Ti}> E_{T,seed}\; .
\eq
Since the algorithm is infrared and collinear safe when $E_{T,seed}=0$, it is
always possible to chose
the seed threshold $E_{T,seed}$ low enough so that
variations of $E_{T,seed}$ lead to negligeable variations in any observable
under consideration. 

One would like to include in the determination of jet momenta all 
towers, of course, which lie inside the cone of radius $R$ around the protojet
axis. This requires an additional iteration of the cone axis in the parton
level algorithm when a seed threshold is imposed. The steps leading to the 
definition of protojets can then be modified as follows:

\begin{itemize}

\item[(0)] 
Make a list of all possible cone centers. These are the legoplot
coordinates of all parton/tower momenta $p_i$ with $p_{Ti}>E_{T,seed}$, of 
all pairs of such parton/tower momenta
$p_i+p_j$, of all triplets $p_i+p_j+p_k$, etc.

\item[(1)]
Select the next cone center on the list as the center of a trial cone of
radius $R$.\\
Goto (3) if the list of cone centers is exhausted.

\item[(2)] 
Add the momenta of all partons/towers inside the trial cone (also those with 
$p_{Ti}<E_{T,seed}$) and determine the legoplot position $(y,\phi)$ 
corresponding 
to this cone momentum. Use $(y,\phi)$ as the new center of the trial cone and
iterate this step until the position is stable. The set of all
towers/partons inside the final trial cone constitutes a new protojet.
Continue with step (1).

\item[(3)] 
Eliminate all duplicate protojets, i.e. protojets with an identical set of
towers/partons.

\end{itemize}

With these changes, the resulting algorithm 
(named Improved Legacy Cone Algorithm or ILCA) 
is quite close to those used 
in run I of the Tevatron. The main change is the inclusion of midpoints
of seeds (the $p_i+p_j$ pairs) and of centers of larger numbers of seeds
as additional seed locations for trial cones. Including these additional
midpoints is absolutely crucial in perturbative calculations in order
to achieve infrared safety (see discussion on Fig.~\ref{fig:coneSSir}).
When dealing with data, these effects are somewhat diminished, because
with sufficiently low seed thresholds $E_{T,seed}$, a large number of 
trial cones will be generated from actual soft energy depositions in 
the calorimeter. However, because these soft energy depositions will 
decide how many jets are reconstructed, one potentially 
introduces a high sensitivity 
of jet observables to soft hadrons, Monte Carlo modelling of soft 
particles etc. The inclusion of the extra midpoints eliminates 
these soft effects because observables no longer depend on whether 
soft emission actually took place.

\section{Resummed calculations}
\label{sec:res}

The ILCA and $k_T$-algorithm eventually lead to jets whose topology is
not extremely different from that expected on the basis of a naive definition 
in terms of cones in azimuth-rapidity space. This is obviously true for the
ILCA, where jets can contain particles whose distance is smaller than $2R$
and have a shape that differs from a cone-shape only because of the 
merging/splitting procedure. In the $k_T$-algorithm, jets have no sharp
boundaries, but opening angles of particles within each jet are, typically, 
smaller than $D$ and all opening angles between jets are larger than $D$. 
The detailed jet structure is, however, different in the two
algorithms. Although both algorithms are perturbative safe, the differences
show up in higher-order perturbative calculations.

Higher-order perturbative computations and, in particular, resummed 
calculations can be necessary in special kinematics configurations
that lead to large logarithmically-enhanced contributions at any fixed order
in perturbation theory. Typical examples are
calculations of jet cross sections near the phase-space boundary and
of the fine internal structure of jets (shape variables, subjets, etc.).
These quantities can be strongly dependent on the jet definition.
The corresponding perturbative calculations do strongly depend on the jet
definition, because they are the result of the integration of the
QCD matrix elements (which do not depend on jets) over phase-space regions
whose boundaries depend on the fine details of jet kinematics.

As an example of this strong sensitivity, we can consider the one-jet
inclusive cross section as a function of the transverse momentum $p_T$
of the jet. {\em If} the jet variable 
$p_T$ is defined by using the 4-momentum recombination scheme (see 
Eqs.~(\ref{4mom})--(\ref{auxquant})), the kinematical boundary is 
$x_T \leq 1$, where $x_T=2p_T/{\sqrt S}$.
Close to the boundary $x_T \sim 1$, the perturbative contributions are
enhanced by large logarithmic corrections $(\alpha_S \log^2(1-x_T))^n$
that need to be resummed to all orders in $\alpha_S$. Techniques to
perform this resummation can be developed (see below). However, {\em if}
the jet variable $p_T$ is defined by using a recombination scheme that
does not conserve the 4-momentum (e.g. the true $p_T$ in Eq.~(\ref{truept})
is replaced by the variable $E_T$ in Eq.~(\ref{etsnowmass})), the kinematical
boundary for $x_T$, although close to $x_T=1$, is not fixed: the corresponding
large logarithms cannot be resummed because the $x_T$-boundary shifts in a
complicated manner depending on the number of final-state partons in the
calculation~\cite{Catani:1997xc}. 

The feasibility of resummed calculations depends not only on the recombination
scheme but also on the jet algorithm, as is well known for jets in $e^+e^-$
annihilation~\cite{eejets}. The jet definition of the $k_T$-algorithm
is inspired by the parton shower picture of jet fragmentation~\cite{es_kt}.
Thus resummed calculations can be carried out by using the analytic version
of the recurrence techniques used to generate multiparton final states in
Monte Carlo parton showers. This is demonstrated by explicit calculations
of subjet multiplicity and rates in hadron collisions~\cite{Seymour:1996tf}.
The ILCA has still to be investigated in this respect. The two algorithms
differ only slightly at highly inclusive level~\cite{es_kt}. Thus, in 
these cases (such as the large-$x_T$ behaviour of the one-jet inclusive cross
section), resummed calculations in the ILCA should be feasible as in the
$k_T$-algorithm. Studies of the internal structure of ILCA jets may instead be
more difficult since it depends on the procedure to merge/split overlapping
jets.

\section{Conclusions}

During the Les Houches workshop discussions were centered on the general 
properties of jet algorithms, in particular their infrared and collinear 
safety at the perturbative level. The $k_T$-algorithm and the seedless 
cone algorithm described in Section~\ref{sec:alg.ilca} fulfil these 
requirements. Beyond these theoretical concerns there are many experimental
issues which need to be addressed to obtain a practical algorithm. Among these
are ease of energy calibration, effects of underlying event and overlapping 
events in a high luminosity hadron collider environment, high jet 
reconstruction efficiency, and efficient use of computer resources in 
reconstructing jets. These issues have been addressed in a parallel study,
during the Run II QCD Workshop at Fermilab~\cite{runIIjets}. In particular 
it has been shown that the ILCA produces small corrections only, when
compared with the jet algorithms used in run I of the Tevatron.
We refer the reader to the Proceedings of the Run II Workshop
for a detailed study of these effects.


\bibliographystyle{plain}


\setcounter{figure}{0}
\setcounter{table}{0}
\setcounter{section}{0}
\setcounter{equation}{0}
\newpage

%
%
%
%
%
%

\begin{center}
\vspace*{1.2cm}
{\Large\sc \bf A Study of the Underlying Event in Jet and Minimum Bias Events} \\
\vspace*{1.cm} 
{\sc J. Huston and V. Tano}
\vspace*{1.cm}
\end{center}

\setcounter{footnote}{0}

\begin{abstract}
In order to determine more accurately the energy contribution in a jet cone
due to the underlying event, and in order to understand better the ambient event
environment at both the Tevatron and the LHC, we have studied the energy distribution in a cone of radius 0.7
in both jet  and in minimum bias events. 
We have compared the results from CDF data 
from Run 1b with results from {\tt HERWIG} passed through the detector
simulation program QFL~\cite{cdfqcd}. 
\end{abstract}

\section{Introduction}

Due to the importance of the inclusive jet cross section as a test of perturbative QCD over a wide range of $Q^{2}$ values, it is necessary to carefully consider all systematic effects that influence its measurement. 
In addition to the hard interaction that produces the jets in the final state 
there is also an underlying event, 
 originating mostly from 
$soft$ spectator parton interactions. Because of the softness of the scale, their contribution cannot be perturbatively calculated.  
There may also be a contribution due to $semi-hard$ interactions between spectator partons, which create $mini-jets$ at transverse momenta almost large enough for perturbative calculations, but much smaller than that of the  primary interaction responsible for the highest $E_{T}$ jets in the event. This process is known as double parton scattering. Both of the above processes,
as well as higher order radiation from the $2{\rightarrow}2$ hard subprocess~\cite{ue_hard},
 contribute to the underlying event.

	The experimental cross sections are most commonly compared
to theoretical calculations at next-to-leading order (NLO) in the coupling constant
$\alpha_s$, such as JETRAD \cite{JETRAD} or EKS \cite{EKS}. At NLO, there can be at most 3 partons in
the final state, leading to the presence of either 2 or 3 jets, depending on
whether the third parton is present in the final state and whether it ends up
being clustered with one of the other two partons.
As the jet clustering is based on a fixed cone algorithm, the contribution due to the underlying event must be subtracted from the jet cone, in order to compare the results with NLO QCD calculations. As will be seen below, one of the largest sources of systematic error for the inclusive jet cross section at low $E_T$ is due to the uncertainty on the subtraction of this underlying event. 

	The current `paradigm' \footnote{Dave Soper claims that this term is vastly overused but we choose to employ it anyway.} is that the underlying event in jet events is similar to the average energy level found in `active' minimum bias events. Thus,
this energy level needs to be determined and subtracted from the energy in a jet cone
before the jet data is compared to NLO theory predictions. 
CDF assumes an uncertainty of 30$\%$ on this underlying event subtraction which makes it the dominant error at low $E_{T}$.

The flip side of the above `paradigm' is that the underlying event energy in a  jet event (once the two leading jets have been subtracted) should be the sum of  the minimum bias level contribution and the third parton in the NLO calculation. The preliminary results of a study in CDF designed to test the accuracy  of these assumptions are described in this section. 

	In this section, the experimental results will be  compared to the {\tt HERWIG} \cite{HERWIG3} Monte Carlo which has the $2\rightarrow2$ matrix elements for jet
production, parton showering in the initial and final state, and a  model for 
the underlying event. 
The ultimate result from {\tt HERWIG}
consists of the 4-vectors of the final state hadrons.  
In {\tt HERWIG} the soft underlying event in the  hadron-hadron collision is assumed to be a soft collision between the two `beam clusters', which contain the spectators from the incoming hadrons. 
The model for the simulation of underlying event uses the $p\bar{p}$ event generator from  the UA5 Collaboration, which is modified to make use of the {\tt HERWIG} fragmentation algorithm. 

	The results from {\tt HERWIG} can be quoted at the parton level (excluding the
soft underlying event portion), 
the hadron level and/or comparisons can be made at the detector level after the
Monte Carlo hadrons are passed through the CDF detector simulation program QFL ~\cite{QFL}. For most of the results that we will be reporting, the QFL 
comparisons will be crucial. 

To summarize,	
the purpose of this analysis is to examine in detail the underlying event in 
jet events and to understand whether the amount subtracted from the jet cones is
correct and whether the uncertainty assumed can be reduced. In addition, a
test will be made as to how well {\tt HERWIG} models the underlying event energy in jet events as well as in minimum bias events. 

\section{Underlying Energy at $90^{0}$ in jet events}

Events generated with {\tt HERWIG} were passed through QFL.
The {\tt HERWIG} code was  adapted to produce  the same information as found in the data samples. 
This information includes the energy, the position and the number of calorimeter towers of the jets in the event, together with the energy and the number of towers in two cones situated at $\pm 90^{0}$ in $\phi$ and at the same $\eta$ as the leading jet. $\phi$ and $\eta$ are respectively the azimuthal and polar angle. 

\begin{figure}[tp]
\centerline{\includegraphics[height=8cm]{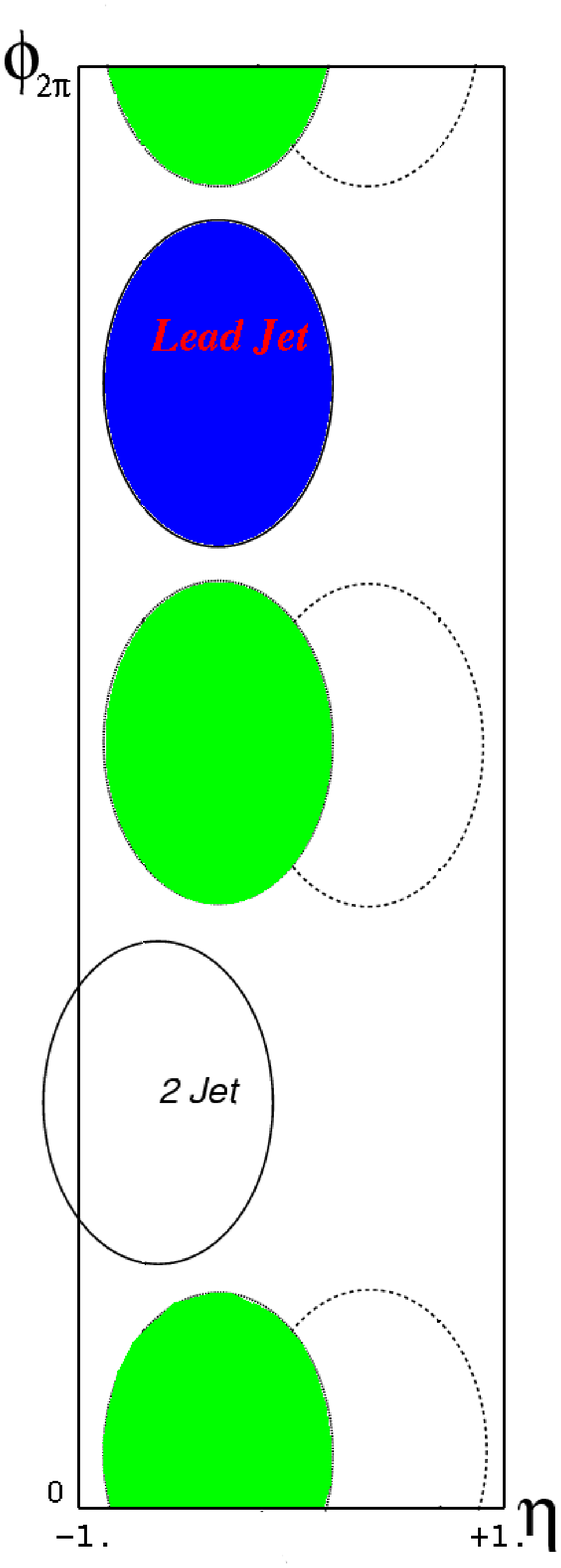}}
\caption{\sf An example of the  cones under investigation in the central calorimeter region. The dotted cones are at $\eta = \eta_{LeadJet}$ and $\phi = \phi_{LeadJet} \pm {\pi\over{2}}$, while dashed cones are at $\eta = - \eta_{LeadJet}$ and $\phi = \phi_{LeadJet} \pm {\pi\over{2}}$. $|\eta| < 1.$, $0^{0} < \phi  < 360^{0}$.}
\label{Fig-phase-space}
\end{figure}

For each jet event,  two cones of radius $0.7$ at $\eta = \eta_{LeadJet}$  and $\phi = \phi_{LeadJet} \pm {\pi\over{2}}$ were examined.
The energy in each cone was determined for two different calorimeter tower thresholds: 50 and 100 MeV. A cut of 100 MeV on tower energies is typically
used for jet analyses. For most of the comparisons to follow, a 50 
MeV cut was used, though, since we are interested in possible contributions to the
tower energies from a number of different sources. 
The two cones were used to study the underlying event energy because they are supposed to be in a semi-quiet region, far away from the two leading jets, but
 still in the central rapidity region. Given the non-uniform response of the
CDF detector as a function of rapidity, the latter criterion is essential.
 The leading jet was  required to be in the central region, $|\eta|< 0.7$, the same as in the
inclusive jet analysis. No requirement was made on the location of the second jet. In Fig~\ref{Fig-phase-space} the calorimeter central region is shown as `unrolled'; $\eta$ ranges are between 
-1 and +1, while $\phi$ goes from $0^{0}$ to $360^{0}$. The leading jet cone and the two cones under study are shown. 

The $E_T$ distributions inside the two cones provide an idea of the contribution of the  underlying event in the jet cone.
For each event the cone which has the maximum energy and the cone 
with the minimum energy were labelled. This is useful because NLO perturbative corrections to the $2\rightarrow2$ hard scattering can contribute only to one of these two regions \cite{Marchesini}. The difference between the maximum and the minimum cone provides information on this contribution, while the minimum cone  gives an  indication of the amount of underlying event. The (roughly constant) underlying event contribution  should be suppressed in the difference.       

\begin{figure}[tbp]
\centerline{\includegraphics[height=8cm]{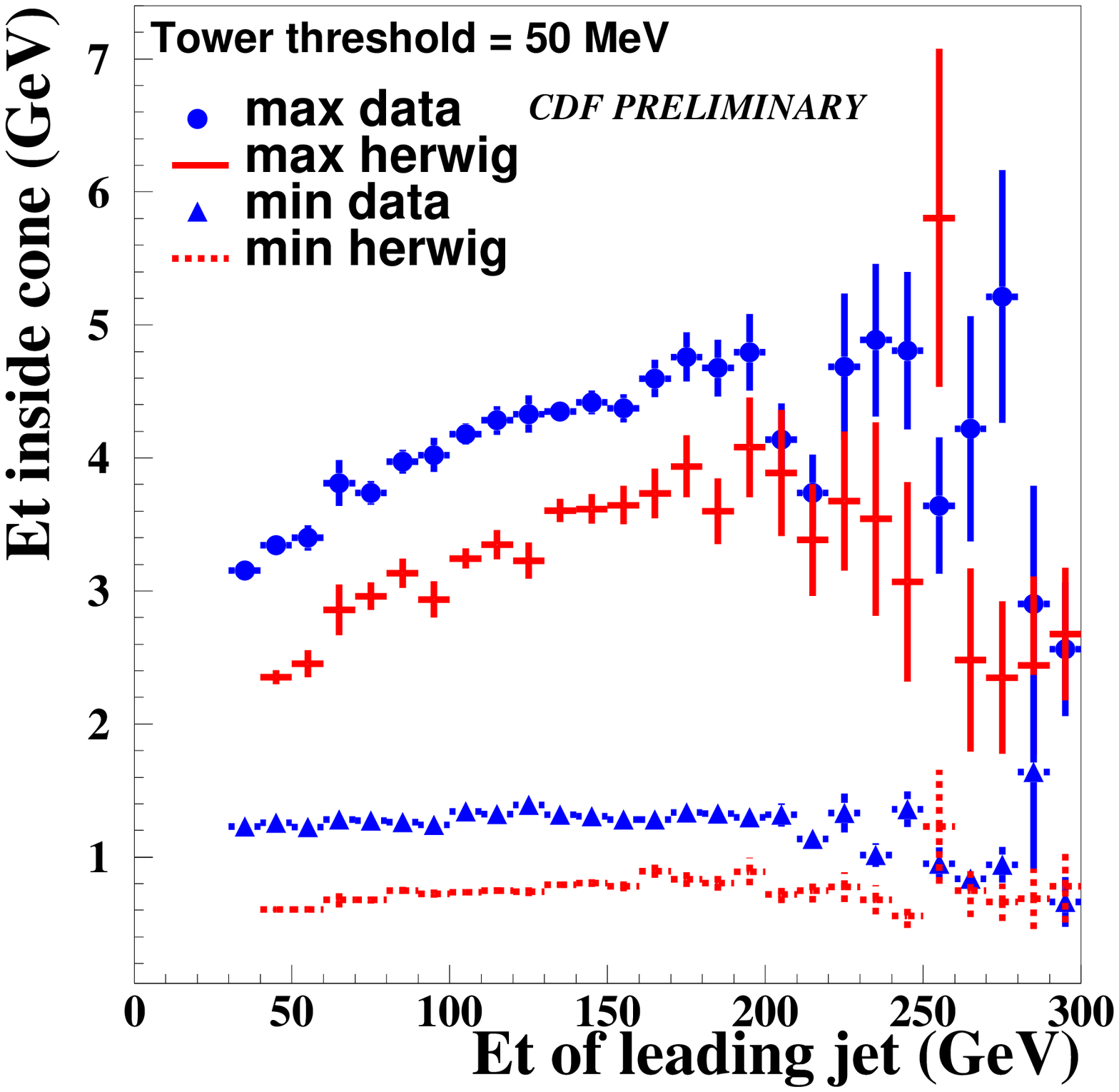}} 
\caption{\sf $E_{T}$ inside the max and min cones as a function of the $E_{T}$ of leading jet. Both the data and {\tt HERWIG} distributions are plotted.}
\label{Fig-max-min-Et}
\end{figure}

The data were required to have one and only one vertex 
in order to insure that there is only one interaction per event. A similar cut was made in the simulation.
In Figure~\ref{Fig-max-min-Et} the transverse energy inside  the two cones (max and min) is plotted as a function of the $E_{T}$ of the leading jet.  
It can be clearly observed that {\tt HERWIG} and the data have a similar behaviour for the max and min cone; the min cone stays flat while the max cone increases with the  $E_{T}$ of the leading jet.
The increase of the max cone energy with increasing jet $E_T$ is easily 
understandable.   What may be surprising is the flatness of the min cone energy 
as the lead jet transverse energy increases. Contrary to the pronouncements of
some of the politicians of our day, a rising tide does not raise all boats (or cones), but instead favors the cone in the highest tax bracket. 
Of course, the division into a max and min cone partially encourages this 
effect through selection. However, the level of flatness is still somewhat 
surprising.

It is evident that there is an offset between data and the {\tt HERWIG}+QFL simulation of about $800$ MeV for the max cone and $500$ MeV for the min cone.
If the tower threshold is increased from $50$ to 100 MeV, the transverse energy decreases by about $180$ MeV in the data (both cones), while in {\tt HERWIG} the transverse energy decreases  by about  $70$ MeV in the max cone and $40$ MeV in the min cone.

\begin{figure}[tbp]
\centerline{\includegraphics[height=8cm]{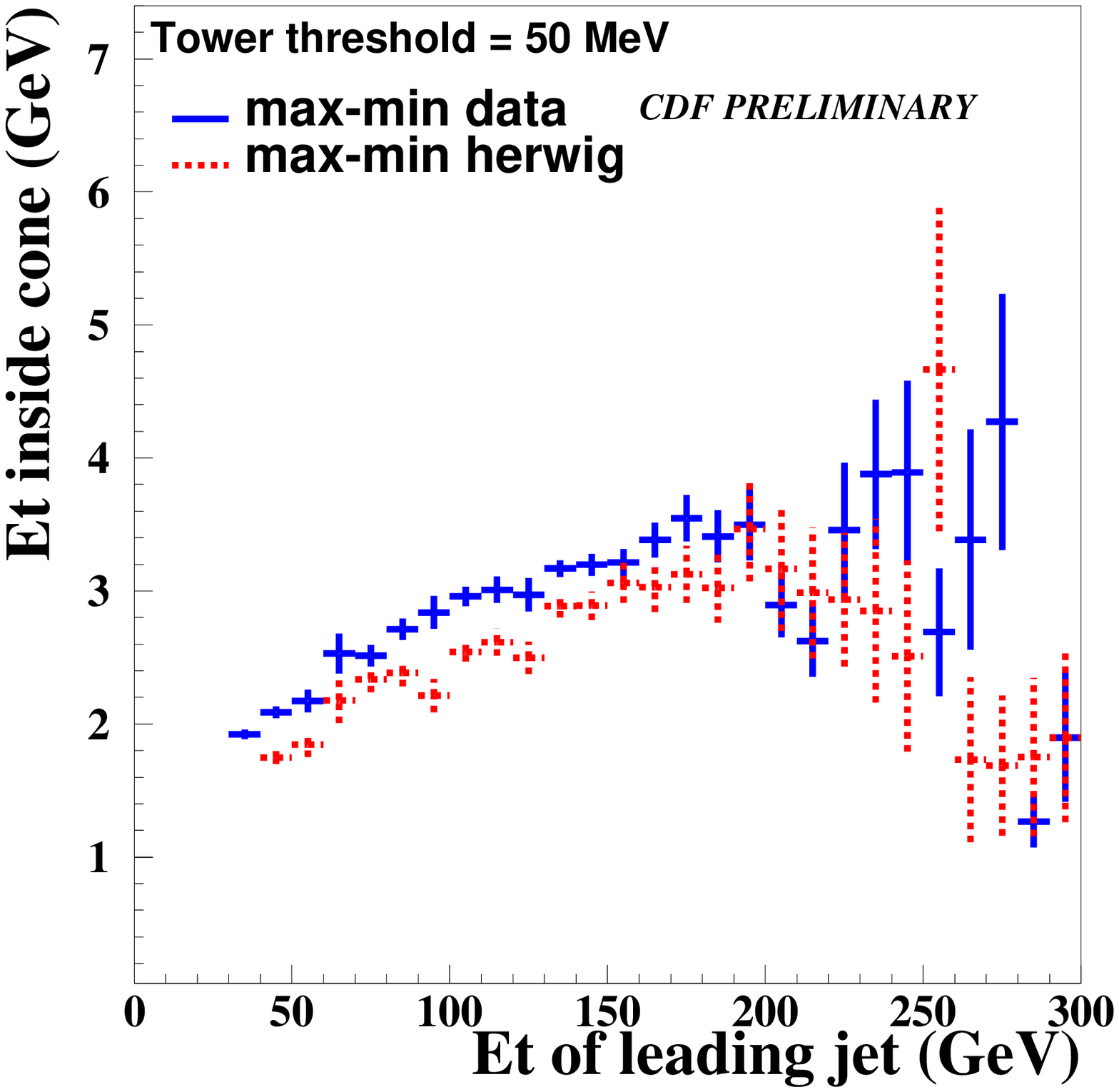}}
\caption{\sf The difference between $E_{T}$ inside the max and min cones as a function of the $E_{T}$ of the leading jet. Both the data and {\tt HERWIG} distributions are plotted. }
\label{Fig-diff-Et}
\end{figure}

The difference between the transverse energy in the max and in the min  cones has a similar trend in both data and simulation(Figure~\ref{Fig-diff-Et}).  There 
is still an offset but the offset decreases to about $300$ MeV. It appears that the max-min distribution starts going down again at very high $E_{T}$ (perhaps due to kinematic suppression), although the statistics become poor. 

\begin{figure}[tbp]
\centerline{\includegraphics[height=8cm]{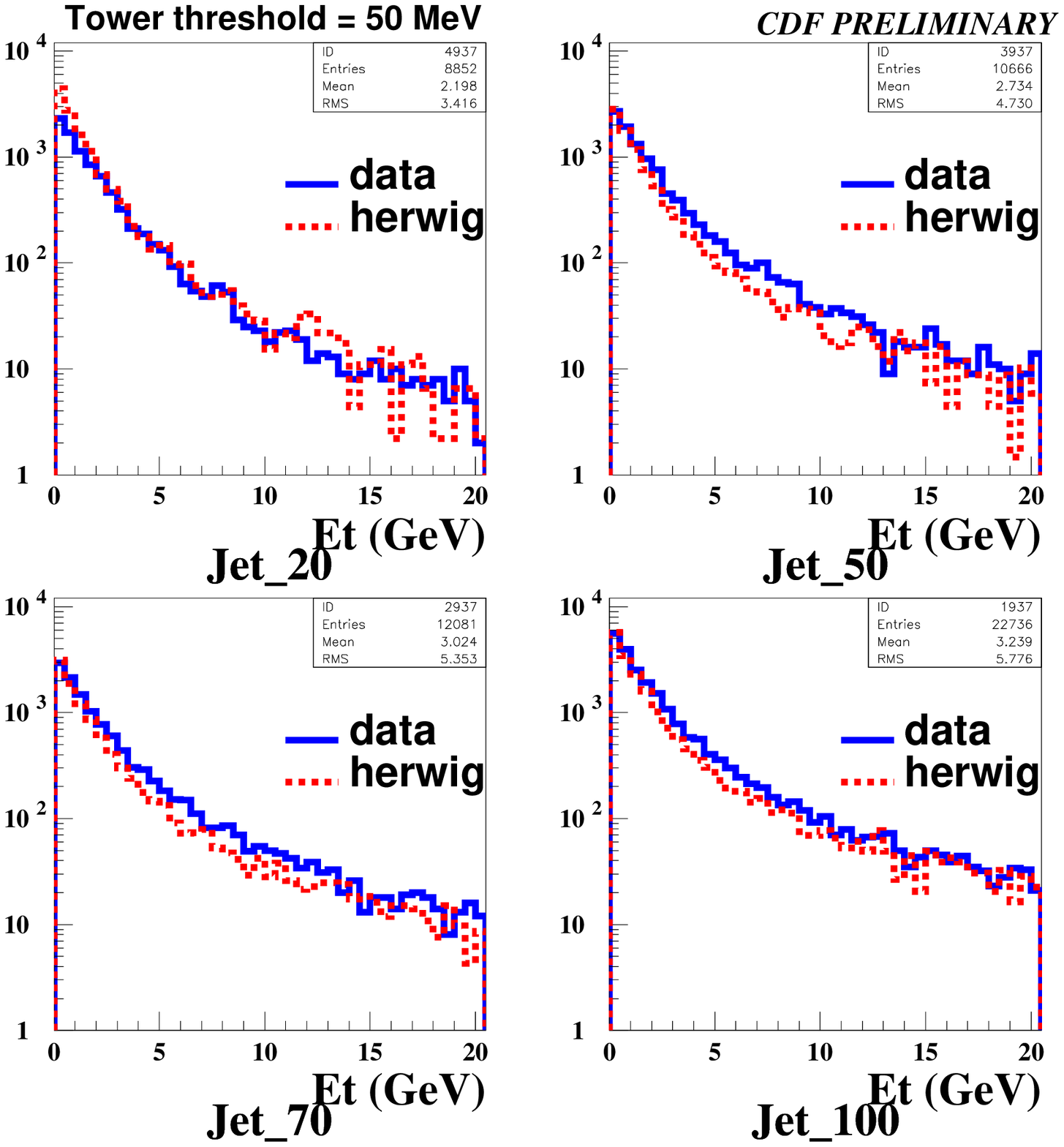}}
\caption{\sf The frequency distribution for the difference between $E_{T}$ in the max cone and $E_{T}$ in the min cone. Solid line: data, dashed line: {\tt HERWIG}. The calorimeter tower energy threshold is 50 MeV.}
\label{Fig-freq-diff}
\end{figure}

In Figure~\ref{Fig-freq-diff} the $E_{T}$ frequency distributions  for data and {\tt HERWIG}+QFL are compared for four different jet sub-samples. In this plot, the $E_{T}$ values for max-min  are plotted, for both data and {\tt HERWIG}+QFL. The number of entries is scaled to easily allow a direct comparison. The $E_{T}$ distribution of the max-min cone  for {\tt HERWIG}+QFL looks very similar to that of the data. Here the contribution of the underlying event as minimum bias data  is presumably removed.

\subsection{Parton-Hadron-Detector level}


	With {\tt HERWIG} (unlike the data), we have the advantage of being able to 
examine the energy distributions not only at the detector level, but also at
the hadron and parton levels. 
The {\tt HERWIG} model for the soft underlying event, though, does not show any effect at the parton level because the energy contribution is calculated directly at the hadron level. In the following discussion, in order to examine the differences 
between hadron, detector and parton level, 
the underlying event in {\tt HERWIG} has been switched off.

\begin{figure}[tbp]
\centerline{\includegraphics[height=8cm]{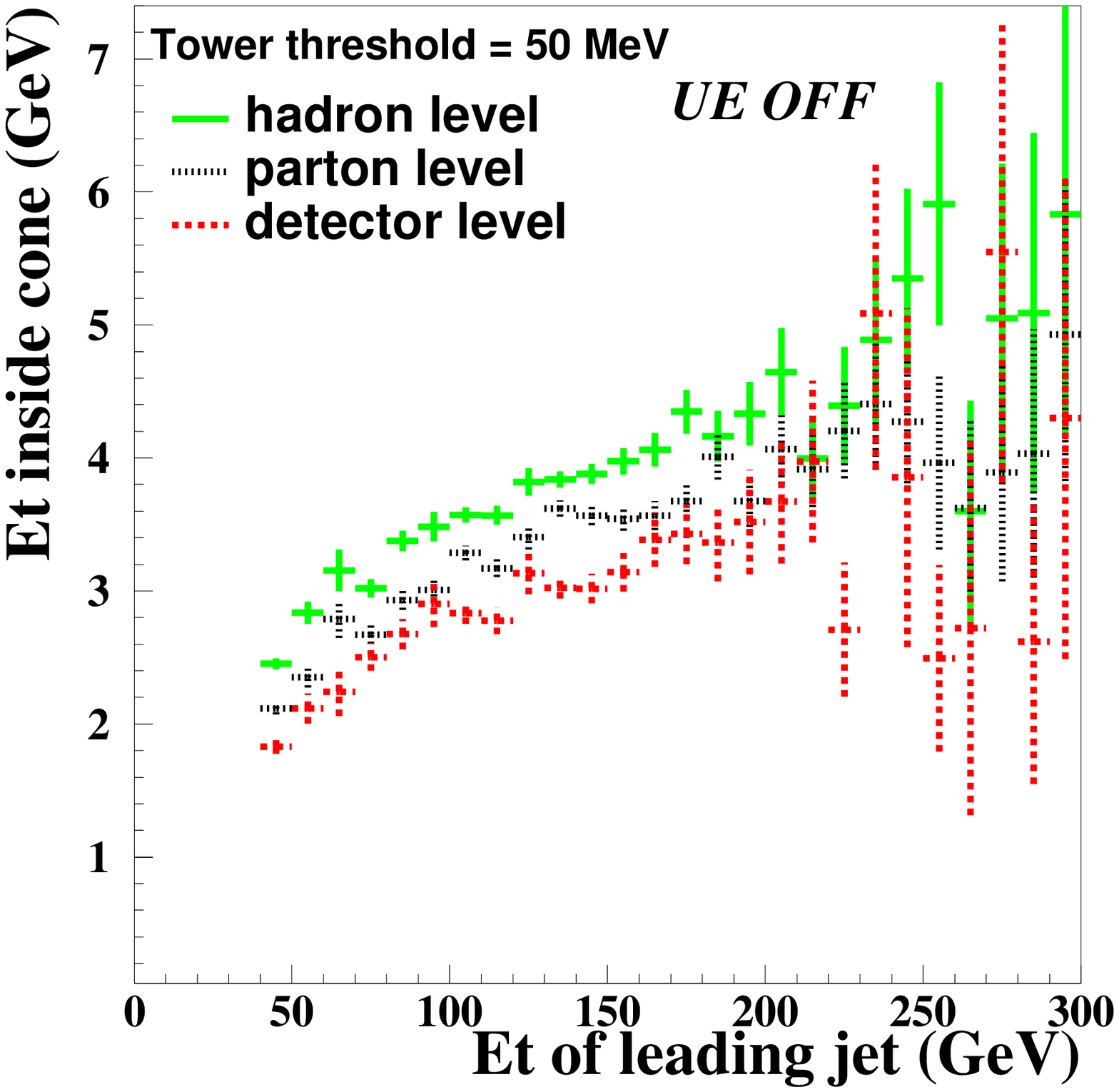}}
\caption{\sf The parton-hadron-detector level for $E_{T}$ in the max cone is plotted as a function of the $E_{T}$ of the leading jet. The underlying event in {\tt HERWIG} is switched off.}
\label{Fig-noue-max}
\end{figure}

\begin{figure}[tbp]
\centerline{\includegraphics[height=8cm]{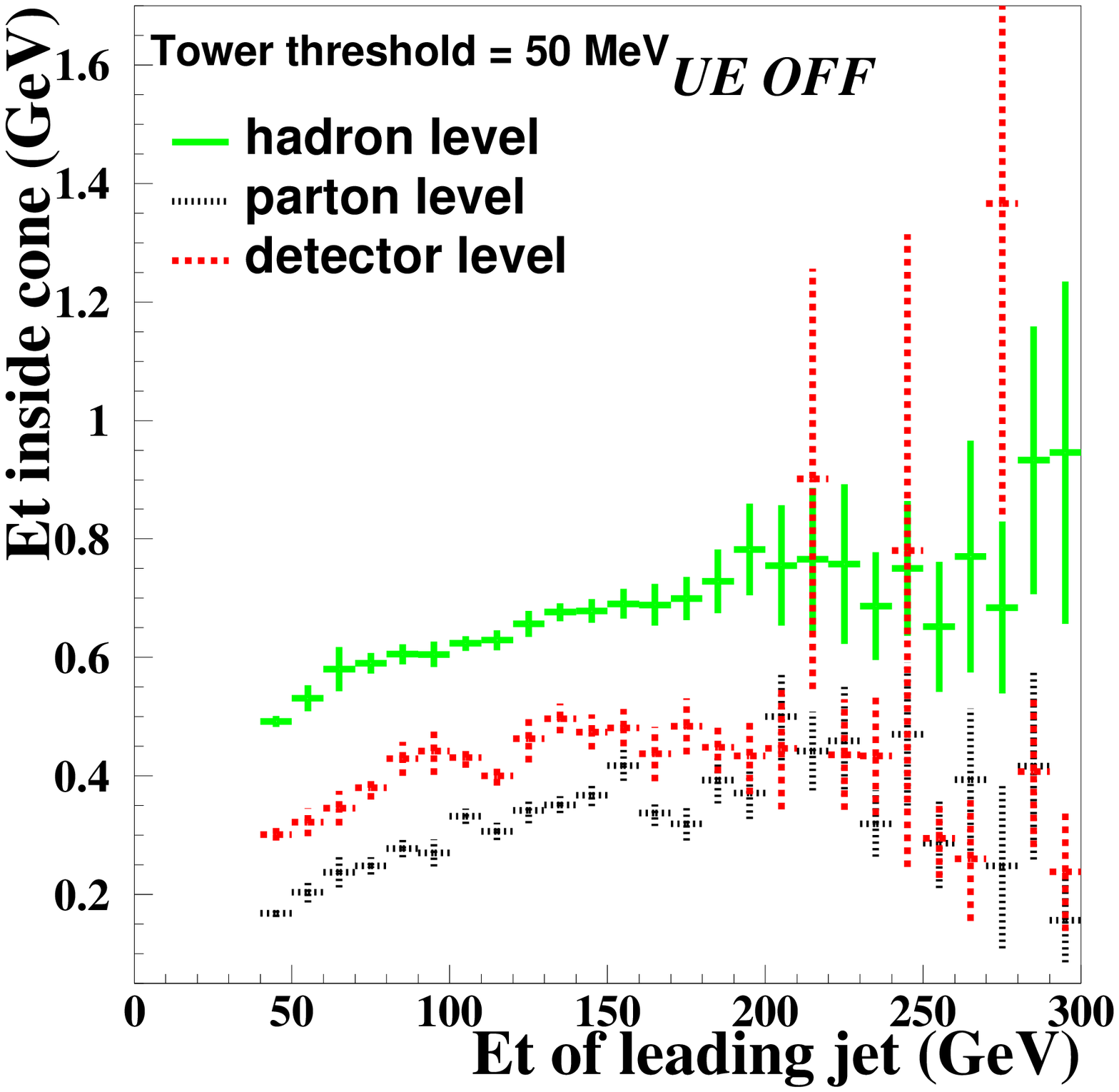}}
\caption{\sf The hadron-parton-detector level for $E_{T}$ in the min cone is plotted as a function of the $E_{T}$ of the leading jet. The underlying event in {\tt HERWIG} is switched off. Note that the $E_{T}$ at the hadron level is always greater than the $E_{T}$ at the parton level.}
\label{Fig-noue-min}
\end{figure}

Figures~\ref{Fig-noue-max} and ~\ref{Fig-noue-min} show the transverse energy inside the max and min cones at $\eta = \eta_{LeadJet}$ and $\phi = \phi_{LeadJet} \pm {\pi\over{2}}$ as a function of the leading jet transverse
energy at the parton, hadron and detector level. The lead jet is always in the central region. Because of the degradation due to the detector response, the amount of energy is higher at the hadron level  than at the detector level.

It is also interesting to note that 
the hadron level energy is larger than the parton level energy, by the order of
several hundred MeV. This is due to hadronization effects of the partons 
produced in or near the lead and second jet cones. Most of the hadronization
effects come from resonance production ($\rho, A_1, A_2,...$) and their
subsequent decays.  The hadronization effects from the partons inside the jet
cone have previously been termed ``splashout''.
It is important to note that this splashout is not currently taken into account 
in either the CDF or D0 jet analyses. Both experiments implicitly assume 
that the hadron and parton levels produce the same energy in the jet cone. This
is especially relevant for low $E_T$ jet production.

In order to evaluate to what level resonance decays influence the energy inside the two cones at $90^{0}$ from the leading jet,  all  resonance decays are 
switched off and the  energy in the cones were examined at both the parton and  the hadron level.
The difference of $E_{T}$ inside  the min cone (between hadron  and parton level) decreases from an average of 300 MeV to 100 MeV, while the difference in the max cone  goes from 500 MeV to 100 MeV.

\section{Underlying Energy in minimum bias events}


The model used in {\tt HERWIG} to simulate minimum bias events  is the same as used  for the soft underlying energy in hard scattering
events.  Minimum bias events were studied in order to see if the reason for the offset observed between the data and simulation results from the {\tt HERWIG} description of the soft underlying event.
The minimum bias events generated with {\tt HERWIG} were passed through the detector simulation program QFL and  the information on the energy released in the calorimeter towers stored. 

The amount of transverse energy in the calorimeter, in a random cone of radius 0.7 that is required to be in the central region ($|\eta| < 0.7$), was determined.
The transverse energy distributions for the two different tower thresholds are 
summarized in Table~\ref{Tab-min-bias} where a comparison with data also can be found. 

\begin{table}[tbp]
\begin{center}
\caption{\sf A comparison of data and {\tt HERWIG} for minimum bias events. The average  amount of transverse energy in a cone of radius 0.7 is shown. Thresholds are in Mev, results  are in GeV.}

\label{Tab-min-bias}
\begin{tabular}{|c|c|c|c|}
\hline
\hline
\multicolumn{1}{|c|}{Thresholds} &
\multicolumn{1}{|c|}{DATA} &
\multicolumn{1}{|c|}{{\tt HERWIG}} &
\multicolumn{1}{|c|}{DATA-{\tt HERWIG}} \\
\hline
  50  & 1.05 & .37 & .68 \\
  100 & .92  & .35 & .57 \\
\hline
\end{tabular}
\end{center}
\end{table}

The offset of about 650 MeV between data and {\tt HERWIG} is slightly higher than the one found comparing the min cones in the jet events. 

\section{$E_{T}$ summed in the central region (Swiss Cheese)}

For these comparisons, 
the transverse energy in every calorimeter tower in the central region ($|\eta| < 1$) is summed, excluding the towers in a radius 0.7 from the center  of the two (or three) most energetic jets in the event: $$ Sum~of~E_{T} = \sum_{towers}E_{T}^{towers} - \sum_{2/3 jets}\left[\sum_{towers}E_{T}^{towers_{jet}}\right] $$where $E_{T}^{towers_{jet}}$ are all the towers in a radius 0.7 from the center of the jet. We require $E_{TJet} >  5$ GeV. This configuration has been labelled 
`Swiss cheese'~\footnote{Or specifically Emmental}. 

\begin{figure}[tbp]
\centerline{\includegraphics[height=8cm]{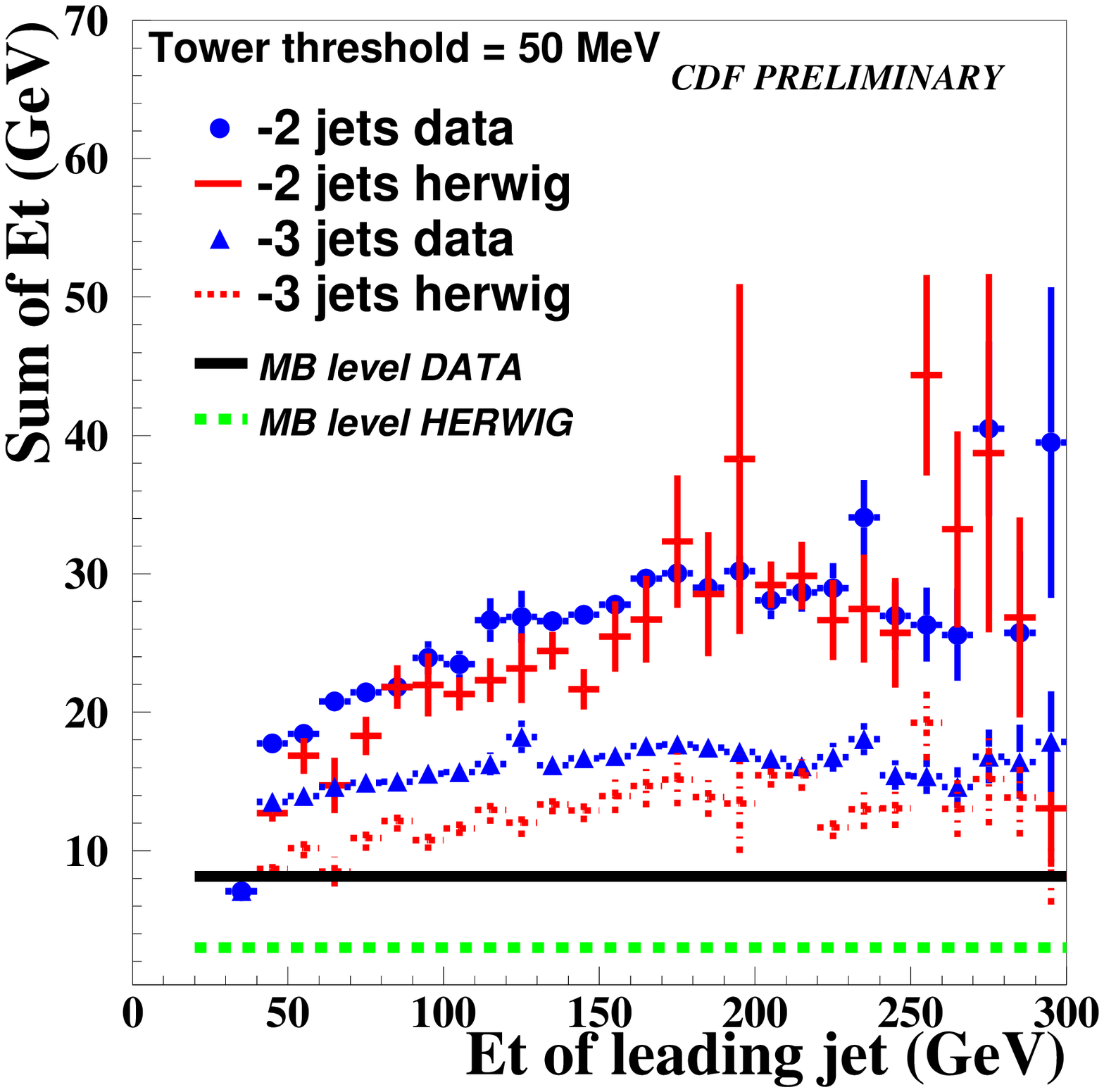}}
\caption{\sf Sum of $E_{T}$. The two and three most energetic jets in the events
are subtracted from the total transverse energy in the central calorimeter region. Both  data and {\tt HERWIG} results are shown.}
\label{Fig-ue-sum}
\end{figure}

There are an average of between 2 and 2.5 jets in the central rapidity region, with this
average having a slight slope as a function of the lead jet transverse energy.
The Swiss cheese energy in the central region is plotted  in Fig~\ref{Fig-ue-sum} 
at the hadron, parton and detector level. The approximate minimum bias level for {\tt HERWIG} and data is shown with a flat line on the picture.
In the simple picture presented earlier, and on which the CDF and D0 jet analyses 
are based, the difference between the Swiss cheese energy with two jets 
subtracted and the minimum bias level should be proportional to the NLO (third parton) contribution. The Swiss cheese level with three jets subtracted should have little or no NLO contribution and can be directly compared to the minimum
bias data level. The 3-jet subtracted Swiss cheese energy is larger than the
minimum bias level and there is a small slope as a function of the lead jet $E_T$ (the offset varies from 6-8 GeV over the $E_T$ range). This indicates perhaps that there is more complexity here than in the simple picture. Other 
possible contributions to the Swiss cheese energy include hadronization from the
jets (``splashout''), double parton scattering and higher order radiation effects.  

As was done for the min and max cone studies,  the underlying event in {\tt HERWIG} can be switched off and  the hadron/detector level in the Swiss Cheese plots compared when the resonance decay is not allowed. 
At the hadron level, when the resonance decay is not allowed, we find about 1.5 GeV energy less then when allowing the resonance decay. This implies a 600-700 MeV contribution of splashout per jet (again at the detector level) to the Swiss cheese energy.

	The comparison of the Swiss cheese results for the data and {\tt HERWIG}+QFL is complex and its interpretation is continuing. 

\section{Conclusions and where do we go from here}

The energy from the underlying event is not perturbatively calculable and must 
be subtracted from a jet cone in order for comparisons to be made to NLO 
calculations. Because of the ambiguous definition of what constitutes this
underlying event, a relatively  large uncertainty has been assigned to the
value of this subtraction. 
In order to study the underlying event, we have considered two cones in the calorimeter  far away from the leading jet and we examined the energy in the cones,  both in the CDF data and with the {\tt HERWIG} simulation. We discovered that both the data and {\tt HERWIG} exhibited a similar behaviour for the max and the min cone; the min cone stays flat, while the max cone increases as a function of the leading jet $E_{T}$. There is  an offset, however,  of about 500 MeV for the min and of 800 MeV for the max cone between data and {\tt HERWIG}. If we examine  the difference between the max and min cones, where the underlying event energy contribution should be minimized, we find  very similar distributions for data and {\tt HERWIG}. In minimum bias,  the {\tt HERWIG} model predicts a level of energy substantially below the one found in minimum bias data (400 MeV compared to 1 GeV). Part of this difference is due to the lack of any kind of hard interaction in the  
minimum bias model.

With {\tt HERWIG} we investigated max/min cone distributions at the parton, hadron and detector level  and we found out that the energy inside the cones is higher at the hadron then at the parton level. This is mainly due to resonance decay.

	An improved understanding of the underlying event is desired for a 
number of reasons: 

\begin{itemize}
\item	The underlying event subtraction is the largest uncertainty for  the jet cross section at low transverse energy (below 60 GeV). In order to have  a good comparison of the data with theory, a better understanding of the proper level of this subtraction must be obtained. This uncertainty is
especially important for the measurement of the jet cross section at 630 GeV,
since most of the data points are below 60 GeV, and similar considerations to
those at 1800 GeV also apply.

\item This analysis probes the interface between perturbative and non-perturbative QCD, an arena where a great deal of work still needs to be done.

\item	The authors of the Monte Carlo programs are trying to predict the environments for physics measurements at the LHC. This can be 
difficult/uncertain without the proper understanding of what is happening at the
Tevatron.
\end{itemize}

	This analysis will be extended to the jet and minimum bias data taken
at 630 GeV by CDF. It will be especially interesting to observe the level of
agreement of the {\tt HERWIG} minimum bias predictions with the CDF data, given that
the {\tt HERWIG} model parameters were determined from the UA5 taken at a similar
energy. It may be that there is an increase in the semi-hard component of 
minimum bias energy when going from 630 to 1800 GeV. After  the comparisons 
at 630 and 1800 GeV are complete, extrapolations will made made to LHC energies 
for the underlying event in both jet and minimum bias events. 



\section{Acknowledgements}

	This work was performed in conjunction with our colleagues on CDF, Anwar Bhatti and Eve Kovacs. 



\setcounter{figure}{0}
\setcounter{table}{0}
\setcounter{section}{0}
\setcounter{equation}{0}
\newpage


 


\def\tr{\mathop{\rm tr}}
\def\Tr{\mathop{\rm Tr}}
\def\Im{\mathop{\rm Im}}
\def\Re{\mathop{\rm Re}}
\def\bR{\mathop{\bf R}{}}
\def\bC{\mathop{\bf C}{}}
\def\C{\mathop{\rm C}}
\def\bra#1{\left\langle #1\right|}
\def\ket#1{\left| #1\right\rangle}
\def\VEV#1{\left\langle #1\right\rangle}
\def\gdot#1{\rlap{$#1$}/}
\def\abs#1{\left| #1\right|}
  \newcommand{\ccaption}[2]{
    \begin{center}
    \parbox{0.85\textwidth}{
      \caption[#1]{\small{\it{#2}}}
      }
    \end{center}
    }
\def\beq{\begin{equation}}
\def\eeq{\end{equation}}
\def\eq{\beq\eeq}
\def\beqn{\begin{eqnarray}}
\def\eeqn{\end{eqnarray}}
\relax
\let\h=\hat
\newcommand\sss{\scriptscriptstyle}
\newcommand\gs{g_{\sss S}}
\newcommand\ep{\epsilon}
\newcommand\Th{\theta}
\newcommand\epb{\overline{\epsilon}}
\newcommand\aem{\alpha_{\rm em}}
\newcommand\refq[1]{$^{[#1]}$}
\newcommand\avr[1]{\left\langle #1 \right\rangle}
\newcommand\lambdamsb{\Lambda_5^{\rm \sss \overline{MS}}}
\newcommand\qqb{{q\bar{q}}}
\newcommand\qb{\bar{q}}
\newcommand\xto{\tilde{x}_1}
\newcommand\xtt{\tilde{x}_2}
\newcommand\aoat{a_1 a_2}
\newcommand\Oop{{\cal O}}
\newcommand\Sfun{{\cal S}}
\newcommand\Pfun{{\cal P}}
\newcommand\mug{\mu_\gamma}
\newcommand\mue{\mu_e}
\newcommand\muf{\mu_{\sss F}}
\newcommand\mufp{\mu_{\sss F}^\prime}
\newcommand\mufs{\mu_{\sss F}^{\prime\prime}}
\newcommand\mur{\mu_{\sss R}}
\newcommand\murp{\mu_{\sss R}^\prime}
\newcommand\murs{\mu_{\sss R}^{\prime\prime}}
\newcommand\muh{\mu_{\sss H}}
\newcommand\muhp{\mu_{\sss H}^\prime}
\newcommand\muhs{\mu_{\sss H}^{\prime\prime}}
\newcommand\muo{\mu_0}
\newcommand\mua{\mu_{\sss A}}
\newcommand\MSB{{\rm \overline{MS}}}
\newcommand\DIG{{\rm DIS}_\gamma}
\newcommand\CA{C_{\sss A}}
\newcommand\DA{D_{\sss A}}
\newcommand\CF{C_{\sss F}}
\newcommand\TF{T_{\sss F}}
\newcommand\ptg{p_{{\sss T}\gamma}}
\newcommand\xtg{x_{\sss T}^\gamma}
\newcommand\etag{\eta_\gamma}
\newcommand\phig{\phi_\gamma}
\newcommand\ptj{p_{{\sss T}j}}
\newcommand\etaj{\eta_j}
\newcommand\epg{\epsilon_\gamma}
\newcommand\epc{\epsilon_c}
\newcommand\epem{e^+e^-}
\def\rightrightarrows{\rlap{\lower 2.5 pt \hbox{$\mathchar\rightarrow$}} 
                      \raise 1pt \hbox {$\mathchar\rightarrow$}}
\def\rightleftarrows{\rlap{\lower 2.5 pt \hbox{$\mathchar\leftarrow$}} 
                     \raise 1pt \hbox {$\mathchar\rightarrow$}}
\begin{center}
\vspace*{1.2cm}
{\Large\sc \bf Isolated Photon Production} \\
\vspace*{1.cm} 
{\sc S. Frixione and W. Vogelsang}
\vspace*{1.cm}
\end{center}

\section{Isolated-photon production\label{sec:phiso}}
\vskip 0.3cm
\vskip 0.3cm

\setcounter{footnote}{0}

\subsection{General features of photon production at colliders}

When mentioning the photon in the framework of high-energy collider
physics, one is immediately led to think -- with good reasons --
to Higgs searches through the gold-plated channel $H\to\gamma\gamma$. 
However, the production of photons also deserves attention on its own.
Firstly, a detailed understanding of the continuum two-photon production
is crucial in order to clearly disentangle any Higgs signals from the
background. Secondly, in hadronic collisions, where a very large number
of strong-interacting particles is produced, photon signals
are relatively clean, since the photon directly couples only to quarks.
Therefore, prompt-photon data can be used to study the underlying parton 
dynamics, in a complementary way with respect to analogous studies performed 
with hadrons or jets. For the same reason, these data represent a
very important tool in the determination of the gluon density in the 
proton, $g(x)$. Indeed, in recent years almost all the {\em direct} 
information (that is, not obtained through scaling violations as predicted 
by Altarelli-Parisi equations) on the intermediate- and high-$x$ behaviour
of $g(x)$ came from prompt-photon production, $pp\rightarrow \gamma X$ 
and $pN \rightarrow \gamma X$, in fixed-target experiments.
The main reason for this is that, at leading order, a photon in the
final state is produced in the reactions $qg\to\gamma q$ and
$q\bar{q}\to\gamma g$, with the contribution of the former subprocess
being obviously sensitive to the gluon and usually dominant over that
of the latter. It is the `point-like' coupling of the photon to the
quark in these subprocesses that is responsible for a much cleaner
signal than, say, for the inclusive production of a $\pi^0$, which
proceeds necessarily through a fragmentation process.

There is, however, a big flaw in the arguments given above. In fact,
photons can also be produced through a fragmentation process, in which
a parton, scattered or produced in a QCD reaction, fragments into a
photon plus a number of hadrons. The problem with the fragmentation
component in the prompt-photon reaction is twofold: first, it
introduces in the cross section a dependence upon non-perturbative
fragmentation functions, similar to those relevant in the case of
single-hadron production, which are not calculable in perturbative QCD
and are, at present, very poorly determined by the sparse LEP data
available. Secondly, {\em all} QCD partonic reactions contribute to
the fragmentation component; thus, when addressing the problem of the
determination of the gluon density, the advantage of having a priori
only one partonic reaction ($q\bar{q}\to\gamma g$) competing with the
signal ($qg\to\gamma q$) is lost, even though some of the subprocesses
relevant to the fragmentation part at the same time result from a
gluon in the initial state.

The relative contribution of the fragmentation component with respect
to the direct component (where the photon participates in the
short-distance, hard-scattering process) is larger the larger the
center-of-mass energy and the smaller the final-state transverse
momentum ~\footnote{Actually, in the fixed-target $pp\to\gamma X$
reaction, one can see the fragmentation component increasing
relatively to the direct one also at very {\em large} $\ptg$,
because of the direct cross section dying out very quickly at such
momenta. This effect is of no phenomenological relevance at the
LHC.}: at the LHC, for transverse momenta of the order of few tens of
GeV, it can become dominant.  However, here the situation is saved by
the so-called `isolation' cut, which is imposed on the photon signal
in experiments. Isolation is an experimental necessity: in a hadronic
environment the study of photons in the final state is complicated by
the abundance of $\pi^0$'s, eventually decaying into pairs of
$\gamma$'s. The isolation cut simply serves to improve the
signal-to-noise ratio: if a given neighbourhood of the photon is free
of energetic hadron tracks, the event is kept; it is rejected
otherwise. Fortunately, by requiring the photon to be isolated, one
also severely reduces the contribution of the fragmentation part to
the cross section. This is because fragmentation is an essentially
collinear process: therefore, photons resulting from parton
fragmentation are usually accompanied by hadrons, and are therefore
bound to be rejected after the imposition of an isolation cut.

Thus, the fragmentation contribution, that threatened to spoil the
cleanliness of the photon signals at colliders, is relatively well under 
control in the case of isolated-photon cross sections. There is of course
a price to pay for this gain: the isolation condition poses additional
problems in the theoretical computations, which are not present in the
case of fully-inclusive photon cross sections. This topic will be
the argument of the next subsection.

\subsection{Isolation prescriptions}

Consistently with what written above, we write the cross section for
the production of an isolated-photon in hadronic collisions as follows:
\beqn
&&d\sigma_{AB}(K_A,K_B;K_\gamma)=
\nonumber \\*&&\phantom{aa}
\int dx_1 dx_2 f^{(A)}_a(x_1,\muf) f^{(B)}_b(x_2,\muf) 
d\hat{\sigma}_{ab,\gamma}^{isol}(x_1 K_A,x_2 K_B;K_\gamma;\mur,\muf,\mug)
\nonumber \\*&&
+\int dx_1 dx_2 dz f^{(A)}_a(x_1,\mufp) f^{(B)}_b(x_2,\mufp) 
d\hat{\sigma}_{ab,c}^{isol}(x_1 K_A,x_2 K_B;K_\gamma/z;\murp,\mufp,\mug) 
D^{(c)}_\gamma (z,\mug),\phantom{aaa}
\label{factth}
\eeqn
where $A$ and $B$ are the incoming hadrons, with momenta $K_A$ and $K_B$
respectively, and a sum over the parton indices $a$, $b$ and $c$ is 
understood. In the first term on the RHS of eq.~(\ref{factth}) 
(the direct component) the subtracted partonic cross sections 
\mbox{$d\hat{\sigma}_{ab,\gamma}^{isol}$} get contributions from all 
the diagrams with a photon leg. On the other hand, the subtracted
partonic cross sections \mbox{$d\hat{\sigma}_{ab,c}^{isol}$}
appearing in the second term on the RHS of eq.~(\ref{factth}) 
(the fragmentation component), get contribution from the pure 
QCD diagrams, with one of the partons eventually fragmenting 
in a photon, in a way described by the parton-to-photon fragmentation 
function $D^{(c)}_\gamma$. As the notation in eq.~(\ref{factth}) indicates, 
the isolation condition is embedded into the partonic cross sections. 

It is a well-known fact that, in perturbative QCD beyond leading order,
and for all the isolation prescriptions known at present, with the
exception of that of ref.~\cite{stefano}, neither the direct nor
the fragmentation components are {\em separately} well defined
at any fixed order in perturbation theory: only their sum is
physically meaningful. In fact, the direct component is affected
by quark-to-photon collinear divergences, which are
subtracted by the bare fragmentation function that appears in
the unsubtracted fragmentation component. Of course, this subtraction 
is arbitrary as far as finite terms are concerned. This is formally 
expressed in eq.~(\ref{factth}) by the presence of the same scale 
$\mug$ in both the direct and fragmentation components: a finite piece
may be either included in the former or in the latter, without affecting
the physical predictions. The need for introducing a fragmentation 
contribution is physically better motivated from the fact that a QCD hard 
scattering process may produce, again through a fragmentation process, 
a $\rho$ meson that has the same quantum numbers as the photon and can 
thus convert into a photon, leading to the same signal. 

As far as the isolation prescriptions are concerned, here we will
restrict to those belonging to the class that can be denoted as `cone
isolations'~\cite{cone}. In the framework of hadronic collisions,
where the need for invariance under longitudinal boosts suggests not to
define physical quantities in terms of angles, the cone is drawn in
the pseudorapidity--azimuthal angle plane, and corresponds to the set
of points
\beq
{\cal C}_{R}=\left\{(\eta,\phi)\mid
\sqrt{(\eta-\etag)^2+(\phi-\phig)^2}\le R\right\},
\label{coneRz}
\eeq
where $\etag$ and $\phig$ are the pseudorapidity and azimuthal angle
of the photon, respectively, and $R$ is the aperture (or half-angle)
of the cone. After having drawn the cone, one has to actually impose
the isolation condition. We consider here two sub-classes of cone
isolation, whose difference lies mainly in the behaviour of the
fragmentation component. Prior to that, we need to define the total
amount of hadronic transverse energy deposited in a cone of half-angle
$R$ as
\beq
E_{T,had}(R)=\sum_{i=1}^n E_{Ti}\theta(R-R_{\gamma i}),
\eeq
where
\beq
R_{\gamma i}=\sqrt{(\eta_i-\etag)^2+(\phi_i-\phig)^2},
\eeq
and the sum runs over all the hadrons in the event (or, alternatively,
$i$ can be interpreted as an index running over the towers of a
hadronic calorimeter). For both the isolation prescriptions we are 
going to define below, the first step is to draw a cone of fixed 
half-angle $R_0$ around the photon axis, as given in eq.~(\ref{coneRz}). 
We will denote this cone as the isolation cone.

\begin{description}

\item[Definition A.] The photon is isolated if the total amount of 
hadronic transverse energy in the isolation cone fulfils the 
following condition:
\beq
E_{T,had}(R_0)\le \epc \ptg,
\label{iscondA}
\eeq
where $\epc$ is a small number, and $\ptg$ is the transverse momentum
of the photon. 

\item[Definition B.] The photon is isolated if the following inequality 
is satisfied:
\beq
E_{T,had}(R)\le \epg\ptg {\cal Y}(R),
\label{iscondB}
\eeq
for {\it all} the cones lying inside the isolation cone, that is for
$R\le R_0$. The function ${\cal Y}$ is arbitrary to a large extent, 
but must at least have the following property:
\beq
\lim_{R\to 0} {\cal Y}(R)=0,
\label{limY}
\eeq
and being different from zero everywhere except for $R=0$.

\end{description}

\noindent
Definition A was proven to lead to an infrared-safe cross section
at all orders of perturbation theory in ref.~\cite{CFP}. 
The smaller $\epc$, the tighter the isolation. Loosely
speaking, for vanishing $\epc$ the direct component behaves like
\mbox{$\log\epc$}, while the fragmentation component behaves like
\mbox{$\epc\log\epc$}. Thus, for $\epc\to 0$ eq.~(\ref{factth})
diverges. This is obvious since the limit $\epc\to 0$ corresponds
to a fully-isolated-photon cross section, which cannot be a meaningful
quantity, whether experimentally (because of limited energy resolution)
or theoretically (because there is no possibility for soft particles 
to be emitted into the cone).

Definition B was proposed and proven to lead to an infrared-safe cross 
section at all orders of perturbation theory in ref.~\cite{stefano}. 
Eq.~(\ref{limY}) implies that the energy of a parton falling into the isolation
cone ${\cal C}_{R_0}$ is correlated to its distance (in the $\eta$--$\phi$
plane) from the photon. In particular, a parton becoming collinear to
the photon is also becoming soft. When a quark is collinear to the photon,
there is a collinear divergence; however, if the quark is also soft,
this divergence is damped by the quark vanishing energy. When a gluon is
collinear to the photon, then either it is emitted from a quark, which is
itself collinear to the photon -- in which case, what was said previously
applies -- or the matrix element is finite. Finally, it is clear that
the isolation condition given above does not destroy the cancellation
of soft singularities, since a gluon with small enough energy can be 
emitted anywhere inside the isolation cone. The fact that this prescription
is free of final-state QED collinear singularities implies that the 
direct part of the cross section is finite. As far as the fragmentation
contribution is concerned, in QCD the fragmentation mechanism is purely
collinear. Therefore, by imposing eq.~(\ref{iscondB}), one forces the
hadronic remnants collinear to the photon to have zero energy. This
is equivalent to saying that the fragmentation variable $z$ is restricted
to the range $z=1$. Since the parton-to-photon fragmentation functions
do not contain any $\delta(1-z)$, this means that the fragmentation
contribution to the cross section is zero, because an integration over
a zero-measure set is carried out. Therefore, only the first term on the 
RHS of eq.~(\ref{factth}) is different from zero, and it does not contain 
any $\mug$ dependence.

We stress again that the function ${\cal Y}$  can be rather freely 
defined. Any sufficiently well-behaved function, fulfilling
eq.~(\ref{limY}), could do the job, the key point being the correlation
between the distance of a parton from the photon and the parton energy,
which must be strong enough to cancel the quark-to-photon
collinear singularity. Throughout this paper, we will use
\beq
{\cal Y}(R)=\left(\frac{1-\cos R}{1-\cos R_0}\right)^n,\;\;\;\;\;\;
n=1.
\label{Yfun}
\eeq
We also remark that the traditional cone-isolation prescription, 
eq.~(\ref{iscondA}), can be recovered from eq.~(\ref{iscondB}) by 
setting ${\cal Y}=1$ and $\epg=\epc$.

\subsection{Isolated photons at the LHC}

In this section, we will present results for isolated-photon cross
sections in $pp$ collisions at 14 TeV. These results have been obtained
with the fully-exclusive NLO code of ref.~\cite{Vancouver}, and are
relevant to the isolation obtained with definition B; the actual
parameters used in the computation are given in eq.~(\ref{Yfun}),
together with $\epg=1$. We let $R_0=0.4$. We will comment in the 
following on the outcome of definition A.

Any sensible perturbative computation should address the issue of
the perturbative stability of its results. A rigorous estimate of
the error affecting a cross section at a given order can be given
if the next order result is also available. If this is not the case,
it is customary to study the dependence of the physical observables
upon the renormalization ($\mur$) and factorization ($\muf$) scales.
It is important to stress that the resulting spread should not be
taken as the `theoretical error' affecting the cross section;
to understand this, it is enough to say that the range in which
$\mur$ and $\muf$ are varied is arbitrary. Rather, one should
compare the spread obtained at the various perturbative orders;
only if the scale dependence decreases when including higher
orders the cross section can be regarded as perturbatively stable
and sensibly compared to data.

Usually, $\mur$ and $\muf$ are imposed to have the same value, $\mu$,
which is eventually varied. However, this procedure might hide some
problems, because of a possible cancellation between the effects 
induced by the two scales. It is therefore desirable to vary $\mur$
and $\muf$ independently. Here, an additional problem arises at
the NLO. The expression of any cross section in terms of $\mu$
(that is, when $\mur=\muf$) is not ambiguous, while {\em it is}
ambiguous if $\mur\ne\muf$. In fact, when $\mur\ne\muf$, the cross
section can be written as the sum of a term corresponding to the
contribution relevant to the case $\mur=\muf$, plus a term of the kind:
\beq
\as(\mua)\,{\cal B}(\as(\mur))\,\log\frac{\mur}{\muf},
\label{addterm}
\eeq
where ${\cal B}$ has the same power of $\as$ as the LO contribution,
say $\as^k$. The argument of the $\as$ in front of eq.~(\ref{addterm}),
$\mua$, can be chosen either equal to $\mur$ or equal to $\muf$, since
the difference between these two choices is of NNLO. Thus, it follows
that the dependence upon $\mur$ or $\muf$ of a NLO cross section
reflects the arbitrariness of the choice made in eq.~(\ref{addterm}),
which is negligible only if the NNLO ($\as^{k+2}$) corrections are
much smaller than the NLO ones ($\as^{k+1}$). This leads to the
conclusion that a study of the dependence upon $\mur$ or $\muf$ 
{\em only} can be misleading. In other words: ${\cal B}$ in
eq.~(\ref{addterm}) is determined through RG equations in order to
cancel the scale dependence of the cross section up to terms of order 
$\as^{k+2}$. This happens regardless of the choice made for $\mua$ in 
eq.~(\ref{addterm}). However, here we are not discussing the cancellation 
to a given perturbative order of the effects due to scale variations; we 
are concerned about the coefficient in front of the ${\cal O}(\as^{k+2})$
term induced by such variations, whose size is dependent upon the
choice made for $\mua$ and therefore, to some extent, arbitrary. We
have to live with this arbitrariness, if we decide to vary $\mur$ or
$\muf$ only. However, we can still vary $\mur$ and $\muf$
independently, but eventually putting together the results in some
sensible way, that reduces the impact of the choice made for $\mua$.
In this section, we will consider the quantities defined as follows:
\beqn
\left(\frac{\delta\sigma}{\sigma}\right)_\pm&=&
\pm\left\{\,\,
\left[\frac{\sigma(\mur=\muo,\muf=\muo)-\sigma(\mur=a_\pm\muo,\muf=\muo)}
           {\sigma(\mur=\muo,\muf=\muo)+\sigma(\mur=a_\pm\muo,\muf=\muo)}
\right]^2\right.
\nonumber \\&&\phantom{\pm}+\left.
\left[\frac{\sigma(\mur=\muo,\muf=\muo)-\sigma(\mur=\muo,\muf=a_\pm\muo)}
           {\sigma(\mur=\muo,\muf=\muo)+\sigma(\mur=\muo,\muf=a_\pm\muo)}
\right]^2\right\}^{\frac{1}{2}},
\label{delsigdef}
\eeqn
where $a_+$ and $a_-=1/a_+$ are two numbers of order one, which we
will take equal to 1/2 and 2 respectively; the $\pm$ sign in front of
the RHS of eq.~(\ref{delsigdef}) is purely conventional. We can
evaluate \mbox{$(\delta\sigma/\sigma)_\pm$} by using $\mua=\mur$ or
$\mua=\muf$ in eq.~(\ref{addterm}). The reader can convince himself,
with the help of the definition of the QCD $\beta$ function, that the
difference between these two choices is of order $\as^4$ in the
expansion of {\em the contribution to}
\mbox{$(\delta\sigma/\sigma)_\pm^2$} {\em due to eq.~(\ref{addterm})};
on the other hand, this difference is only of order $\as^3$ in each of
the two terms under the square root in the RHS of
eq.~(\ref{delsigdef}). This is exactly what we wanted to achieve: a
suitable combination of the cross sections resulting from independent
$\mur$ and $\muf$ variations is less sensitive to the choice for
$\mua$ made in eq.~(\ref{addterm}) with respect to the results
obtained by varying $\mur$ or $\muf$ {\em only}.

\begin{table}
\begin{center}
\begin{tabular}{|l||c|c|c|c|c||c|c||c|} \hline
& \multicolumn{5}{c||}{MRST99} 
& \multicolumn{2}{c||}{CTEQ5} 
& 
\\ \hline
& 1 & 2 & 3 & 4 & 5
& M & HJ
& $(\delta\sigma/\sigma)_\pm$
\\ \hline\hline
NLO, $\abs{\etag}<2.5$ 
  & 23.78 & 23.20 & 24.19 & 22.07 & 25.49 
  & 25.10 & 24.61 & $^{+0.068}_{-0.057}$
\\ \hline
 LO, $\abs{\etag}<2.5$ 
  & 10.34 & 10.07 & 10.52 & 9.875 & 10.78 
  & 10.91 & 10.66 & $^{+0.090}_{-0.072}$
\\ \hline
NLO, $\abs{\etag}<1.5$ 
  & 14.59 & 14.23 & 14.88 & 13.66 & 15.53 
  & 15.35 & 15.01 & $^{+0.068}_{-0.056}$
\\ \hline
 LO, $\abs{\etag}<1.5$ 
  & 6.457 & 6.270 & 6.583 & 6.212 & 6.657 
  & 6.771 & 6.596 & $^{+0.091}_{-0.073}$
\\ \hline
\end{tabular} 
\end{center}                                                            
\ccaption{}{\label{tab:xsec}
Isolated-photon cross sections (nb), with $40<\ptg<400$~GeV, in two
different rapidity ranges, for various parton densities. The scale
dependence, evaluated according to eq.~(\ref{delsigdef}), is also shown.
}
\end{table}                                                               
In table~\ref{tab:xsec} we present the results for the total isolated-photon
rates, both at NLO and at LO. The latter cross sections have been obtained
by retaining only the lowest order terms (${\cal O}(\aem\as)$)
in the short-distance cross section, and convoluting them with
NLO-evolved parton densities. Also, a two-loop expression for 
$\as$ has been used. There is of course a lot of freedom in
the definition of a Born-level result. However, we believe that
with this definition one has a better understanding of some 
issues related to the stability of the perturbative series.
In order to obtain the rates entering table~\ref{tab:xsec}, we
required the photon transverse momentum to be in the range
$40<\ptg<400$~GeV, and we considered the rapidity cuts $\abs{\etag}<1.5$ and
$\abs{\etag}<2.5$, in order to simulate a realistic geometrical acceptance 
of the LHC detectors. We first consider the scale dependence of our
results (last column), evaluated according to eq.~(\ref{delsigdef}).
We see that the NLO results are clearly more stable than the LO
ones; this is reassuring, and implies the possibility of a sensible
comparison between NLO predictions and the data. Notice that the 
size of the radiative corrections ($K$ factor, defined as the
ratio of the NLO result over the LO result) is quite large. 
From the table, we see that the cross sections obtained with
different parton densities differ by 6\% at the most (relative
to the result obtained with MRST99-1~\cite{MRST99}, which we take as 
the default set). MRST99 sets 2 and 3 are meant to give an estimate 
of the effects due to the current uncertainties affecting the gluon density, 
whereas sets 4 and 5 allow to study the sensitivity of our predictions to
the value of $\as(M_{\sss Z})$ (sets 1, 4 and 5 have 
$\lambdamsb=$220, 164 and 288~MeV respectively). On the other
hand, the difference between MRST99-1 and CTEQ5M~\cite{CTEQ5} results 
is due to the inherent difference between these two density sets
(CTEQ5M has $\lambdamsb=$226~MeV, and therefore the difference in
the values of $\as(M_{\sss Z})$ plays only a very minor role).

From inspection of table~\ref{tab:xsec}, we can conclude that
isolated-photon cross section at the LHC is under control, both 
in the sense of perturbation theory and of the dependence upon
non-calculable inputs, like $\as(M_{\sss Z})$ and parton densities.
The relatively weak dependence upon the parton densities, however,
is not a good piece of news if one aims at using photon data to
directly access the gluon density. On the other hand, the expected
statistics is large enough to justify attempts of a direct measurement 
of such a quantity. In the remainder of this section, we will
concentrate on this issue. We will consider 
\beq
{\cal R}_x=\frac{d\sigma_0/dx-d\sigma/dx}{d\sigma_0/dx+d\sigma/dx},
\eeq
where $x$ is any observable constructed with the kinematical variables
of the photon and, possibly, of the accompanying jets. $\sigma$ and
$\sigma_0$ are the cross sections obtained with two different sets
of parton densities, the latter of which is always the default
one (MRST99-1). We can imagine a gedanken experiment, where it is
possible to change at will the parton densities; in this way, we can assume
the relative statistical errors affecting $\sigma$ and $\sigma_0$ to 
decrease as $1/\sqrt{N}$ and $1/\sqrt{N_0}$, $N$ and $N_0$ being
the corresponding number of events. It is then straightforward to
calculate the statistical error affecting ${\cal R}_x$; by imposing
${\cal R}_x$ to be larger than its statistical error, one gets
\beq
{\cal R}_x\,>\,\left({\cal R}_x\right)_{min}\,\equiv\,
\frac{1}{\sqrt{2{\cal L}\ep\sigma(x,\Delta x)}},
\label{Rmin}
\eeq
where ${\cal L}$ is the integrated luminosity, $\ep\le 1$ collects
all the experimental efficiencies, and 
\beq
\sigma(x,\Delta x)=\int_{x-\Delta x/2}^{x+\Delta x/2} dx\,\frac{d\sigma}{dx}
\eeq
is the total cross section in a range of width $\Delta x$ around $x$.

\begin{figure}
\centerline{
   \epsfig{figure=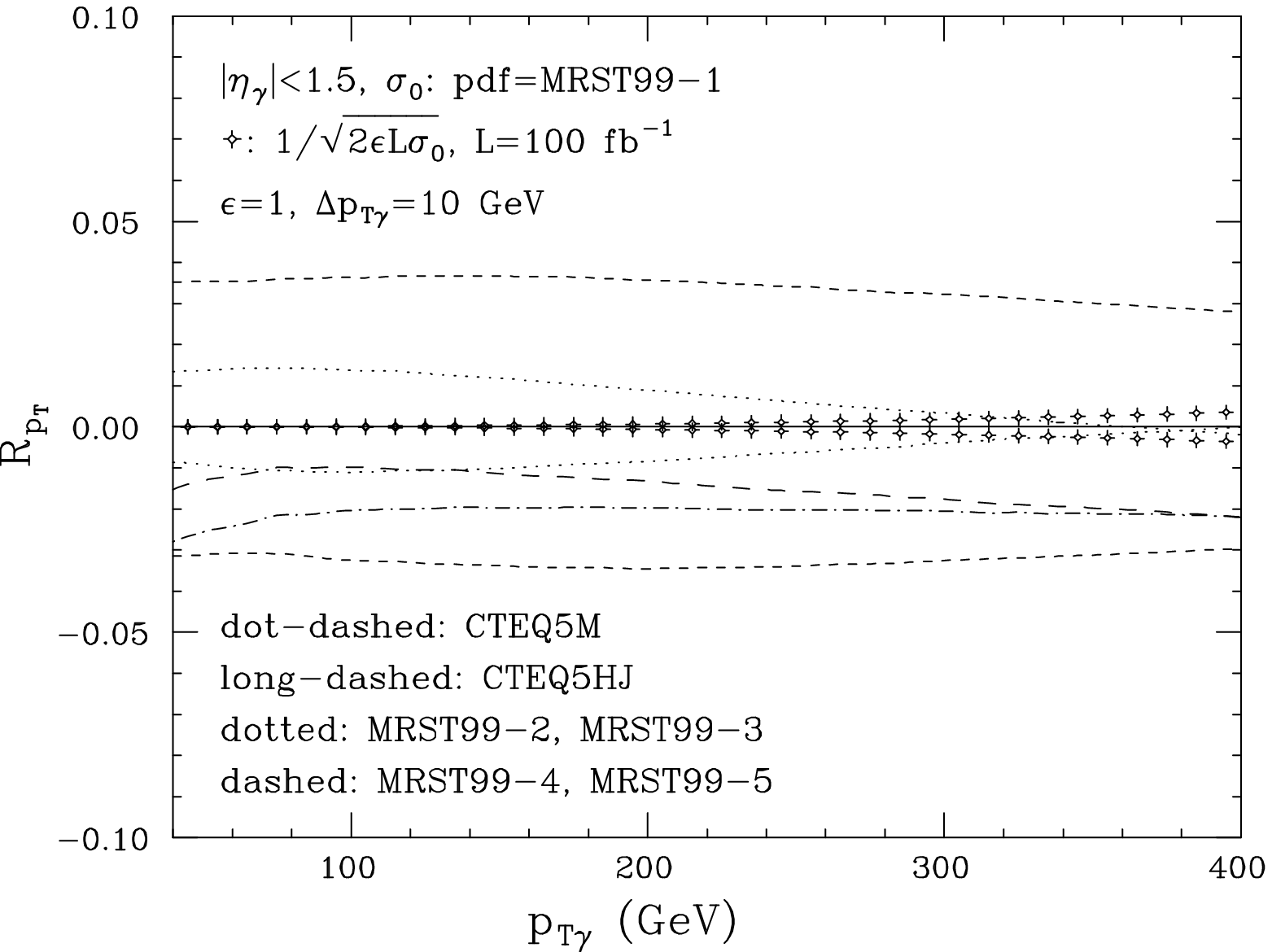,width=0.48\textwidth,clip=}
   \hfill
   \epsfig{figure=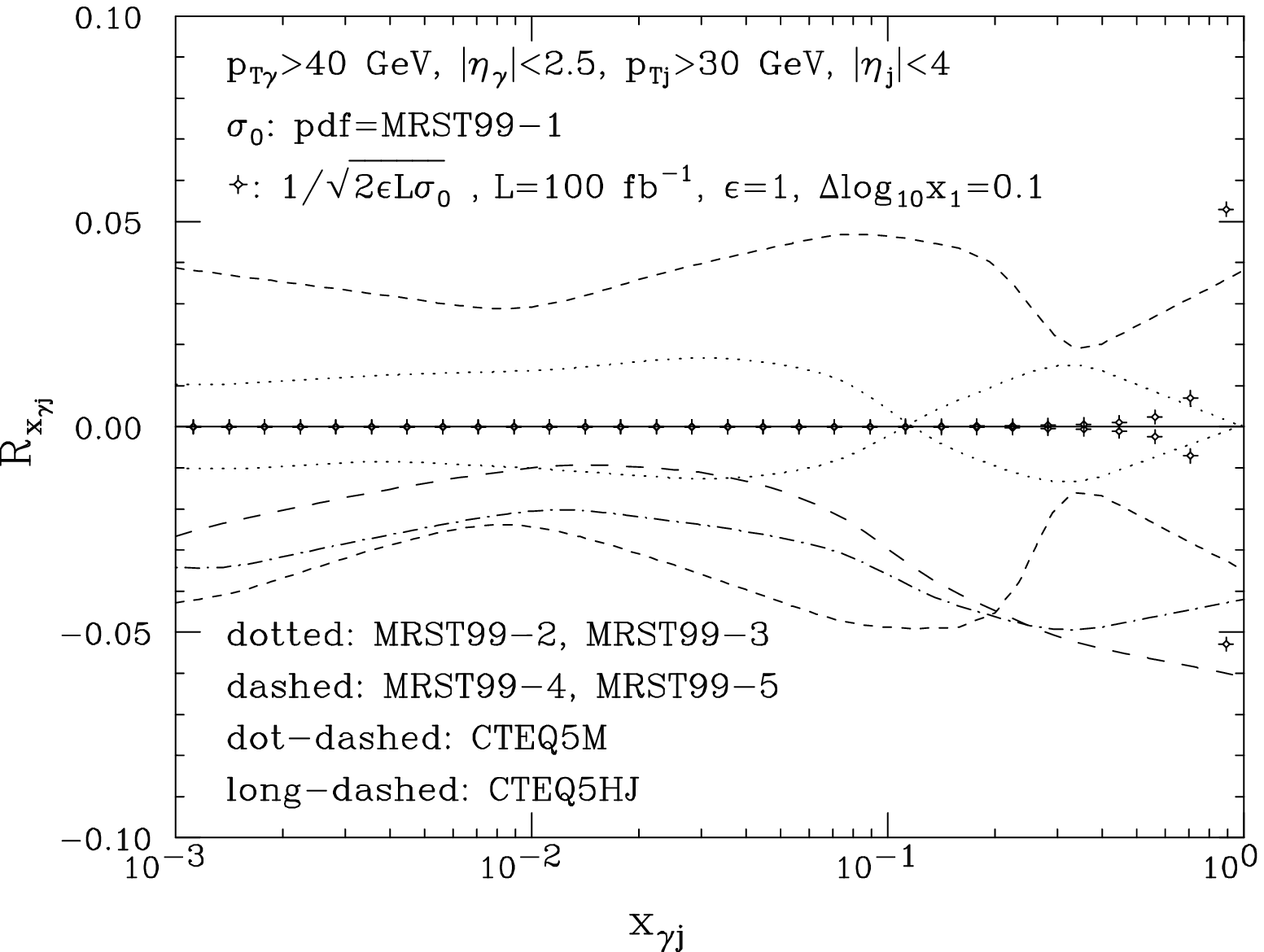,width=0.48\textwidth,clip=} }
\ccaption{}{ \label{fig:pdfdep}
Dependence of isolated-photon and isolated-photon-plus-jet cross
section upon parton densities, as a function of $\ptg$ and $x_{\gamma j}$.
}
\end{figure}

In fig.~\ref{fig:pdfdep} we present our predictions for ${\cal R}_x$.
In the left panel of the figure we have chosen $x=\ptg$, while in the
right panel we have $x=x_{\gamma j}$, where
\beq
x_{\gamma j}=\frac{\ptg\exp(\etag)+\ptj\exp(\etaj)}{\sqrt{S}}.
\eeq
In this equation $\sqrt{S}$ is the center-of-mass energy of the colliding
hadrons, and $\ptj$ and $\etaj$ are the transverse momentum and rapidity
of the hardest jet recoiling against the photon. In order to reconstruct
the jets, we adopted here a $k_T$-algorithm, namely that proposed in 
ref.~\cite{ESalg}, with $D=1$. Notice that $x_{\gamma j}$ exactly coincides 
at the leading order with the Bjorken-$x$ of the partons in one of the 
incoming hadrons; NLO corrections introduce only minor deviations.
For all the density sets considered, the dependence of ${\cal R}$
upon $\ptg$ is rather mild. The values in the low-$\ptg$ region
could also be inferred from table~\ref{tab:xsec}, since the cross
section is dominated by small $\ptg$'s. Analogously to what happens 
in the case of total rates, the sets MRST99-4 and MRST99-5 give rise
to extreme results for ${\cal R}_{\ptg}$, since the value of 
$\Lambda_{\sss QCD}$ is quite different 
from that of the default set. From the figure, it
is apparent that, by studying the transverse momentum spectrum, it
will not be easy to distinguish among the possible {\em shapes} of the
gluon density. On the other hand, it seems that, as far as the 
statistics is concerned, a distinction between any two sets
can be performed. Indeed, the symbols in the figure display
the quantity defined in eq.~(\ref{Rmin}), for ${\cal L}=100$~fb$^{-1}$,
$\Delta\ptg=10$~GeV and $\ep=1$. Of course, the latter value is not
realistic. However, a smaller value (leading to a larger 
$({\cal R})_{min}$), can easily be compensated by enlarging
$\Delta\ptg$ and by the fact that the total integrated luminosity 
is expected to be much larger than that adopted in fig.~\ref{fig:pdfdep}.

Turning to the right panel of fig.~\ref{fig:pdfdep}, we can see
a much more interesting situation. Actually, it can be shown that
the pattern displayed in the figure is rather faithfully reproduced
by plotting the analogous quantity, where one uses the gluon densities
instead of the cross sections. This does not come as a surprise.
First, $x_{\gamma j}$ is in an almost one-to-one correspondence
with the $x$ entering the densities. Secondly, photon production
is dominated by the gluon-quark channel, and therefore the cross
section has a linear dependence upon $g(x)$, which can be easily 
spotted. It does seem, therefore, to be rather advantageous to
look at more exclusive variables, like photon-jet correlations
(this is especially true if one considers the procedure of unfolding
the gluon density from the data: in the case of single-inclusive
variables, the unfolding requires a de-convolution, which is not
needed in the case of correlations). Of course, there is a price
to pay: the efficiency $\ep$ will be smaller in the case of
photon-jet correlations, with respect to the case of single-inclusive
photon observables, mainly because of the jet-tagging. However,
from the figure it appears that there should be no problem with 
statistics, except in the very large $x_{\gamma j}$ region.

Finally, we would like to comment on the fact that, for the case of
single-inclusive photon observables, we also computed the cross
section by isolating the photon according to definition A, using
\mbox{$\epc=2$~GeV$/\ptg$}. The two definitions return a $\ptg$
spectrum almost identical in shape, with definition B higher by a
factor of about 9\%. It is only at the smallest $\ptg$ values that we
considered, that definition B returns a slightly steeper spectrum.
The fact that such different definitions produce very similar cross
sections may be surprising. This happens because, prior to applying
the isolation condition, partons tend to be radiated close to the
photon; therefore, most of them are rejected when applying the
isolation, no matter of which type. This situation has already been
encountered in the production of photons at much smaller energies. The
reader can find a detailed discussion on this point in
ref.~\cite{fv99}.


\setcounter{figure}{0}
\setcounter{table}{0}
\setcounter{section}{0}
\setcounter{equation}{0}
\newpage


%
%
%




\begin{center}
\vspace*{1.2cm}
{\Large\sc \bf Direct photon pair production at colliders, \\
an irreducible background to Higgs boson searches at the LHC } \\
\vspace*{1.cm} 
{\sc T. Binoth, J.P. Guillet, V.A. Ilyin, 
E. Pilon and M. Werlen}
\vspace*{1.cm}
\end{center}

\setcounter{footnote}{0}



\begin{abstract}
Direct~\footnote{``Direct" or ``prompt" mean that these photons do not result
from the decay of $\pi^{0}$ and $\eta$.} photon pairs with large invariant
mass are the so-called irreducible background in the search for Higgs bosons
at the LHC in the channel $h \rightarrow \gamma \gamma$, in the mass range  
$80 - 140$ GeV$/c^{2}$. This huge background requires 
an understanding and quantitative evaluation. Photon pair production at the Tevatron
offers the opportunity to already test our understanding of this process.
In the same mass range, the production of a Higgs boson ($h rightarrow \gamma \gamma$) in association with a hard jet at the LHC is a promising channel, 
as the corresponding $\gamma \gamma +$ jet background may be under better
control. 
\end{abstract}

\section{Role and relevance of higher order corrections}\label{higher-orders}

Our theoretical understanding of direct photon pair production (as any hard
hadronic process, cf. \cite{catani2}) is based on the QCD improved parton
model, according to which long and short distance effects factorize from each
other. Short distance subprocesses are safely computed in perturbative QCD.
Long distance effects cannot be completely calculated from QCD at present,
although their scaling violations can be. Instead, they are extracted from experimental data
and encoded into non-perturbative quantities, such as the parton distribution
functions in incoming hadrons and, if necessary, inclusive fragmentation
functions of partons into observed outgoing particles, e.g. photons. Yet these
quantities are universal, i.e. independent of the hard subprocess:
schematically they can be measured in one process, then transported to predict
another one.

However, the border between short and long distance scales is arbitrary. The
separation requires the introduction of unphysical parameters; e.g. the
factorization scale $M^{2}$, and similarly the fragmentation scale $M_{f}^{2}$
in the case of a fragmentation process. In an ideal exact  calculation, the
dependence on these spurious parameters (as well as on the arbitrary
renormalization scale $\mu^{2}$) would cancel between the short and long
distance parts. In an actual expansion in powers of  $\alpha_{s}$ truncated at
some finite order, this cancellation is only  partial; it holds up to a term of
the lowest uncalculated order.  As the order of truncation increases,
theoretical estimates become  flatter and flatter over broader and broader
ranges of these spurious  scales. The uncertainty induced by this actual
dependence restricts the  accuracy and predictive character of QCD
calculations. In particular,  the result of a lowest order calculation is
plagued by a large monotonic dependence with respect to $M^{2}$, $M_{f}^{2}$
and $\mu^{2}$. It changes by a large factor (two or more) when these scales are
varied around the typical hard scale of the process; such a lowest order
estimate is {\it not at all quantitative}. This is the first reason  why any
tentatively accurate QCD calculation has to be carried out to at least
next-to-leading order (abbreviated below as NLO).

Another important motivation is that higher order corrections to some given
process may reveal new mechanisms, whose rates may be not necessarily
negligible compared to the leading order contribution. The production of
photons is a typical example of this phenomenon, as will be explained below.
This amounts to large higher order corrections, which affect substantially both
the magnitude and the shape of the distributions, not due to poor apparent
convergence of the first terms in the perturbative expansion, but for
physically understood reasons.

Finally, finite order calculations may not be accurate enough, as in the case
of infrared sensitive observables, i.e. observables controlled by multiple soft
gluon emission~\footnote{See for example  \cite{catani2,collins-soper,balacz}}.
Yet, in some less well-known cases of infrared sensitivity, they may reveal
perturbative instabilities or even divergences plaguing the calculation at any
further order  {\it inside} the physical spectrum \cite{catani-webber}. This is,
for  example, the case for the transverse momentum distribution of pairs of 
isolated photons, as will be discussed below. The calculation of higher  order
corrections is therefore the first step towards a deeper  understanding of what
happens in such cases.

\section{Mechanisms of production.}\label{mech}

Schematically, three possible mechanisms may produce prompt photon pairs with
large invariant mass: one (which may be called ``two direct") in which both
photons take part directly in the hard subprocess, another one (``one
fragmentation") in which one of the photons undergoes the hard subprocess
while the other results from the fragmentation of a hard parton (quark or
gluon), itself produced at large tranverse momentum, and yet another mechanism (``two
fragmentation") in which both photons result from such a fragmentation. This
schematical splitting into these three contributions emerges from a
factorization procedure sketched in what follows. Although this splitting
provides a convenient picture, one must however keep in mind that it is
arbitrary; none of these contributions can be measured separately. Only their
sum is physical.

\subsection{Direct vs. fragmentation mechanisms}

From a topological point of view, a photon produced from fragmentation is with a high 
probability accompanied by a jet of hadrons. From a technical point of view, the lowest
order of the ``one fragmentation" contribution emerges in the calculation of
higher order perturbative corrections to the ``two direct" contribution
given by the Born process $q \bar{q} \rightarrow \gamma \gamma$. Some of these
higher order corrections, such as $q g \rightarrow q\gamma \gamma$, are plagued
by final state collinear singularities associated with the collinear splitting
$q \rightarrow q \gamma$. The latter have to be factorized and absorbed into
quark and gluon fragmentation functions to a photon, $D_{\gamma / q \; or \;
g}(z,M_{f}^{2})$ defined at some fragmentation scale~\footnote{and in some given
factorization scheme. Here we use the $\overline{MS}$ scheme, \cite{bfg}.}
$M_{f}^{2}$. Analogously to the so-called anomalous component of the photon
structure function, a collinear logarithmic enhancement occurs, 
induced by the pointlike quark-photon coupling. To all orders in $\alpha_{s}$,
this phenomenon results in $D_{\gamma / q \; or \; g}(z,M_{f}^{2})$  behaving
asymptotically~\footnote{i.e. when the fragmentation scale $M_{f}^{2}$ (chosen
of the order of the hard scale of the subprocess) is large compared to  any
typical hadronic scale $\sim 1$ GeV$^{2}$.} as $\alpha/\alpha_{s}(M_{f}^{2})$.
This compensates the one extra power of  $\alpha_{s}$ involved in the short
distance subprocess, so that fragmentation contributions are asymptotically of
the same order as the Born term, by power counting in $\alpha_{s}$. What is
more, given the high gluon density at LHC, the $g q$ (or $\bar{q}$)
initiated process involving one photon from fragmentation even dominates the
inclusive production rate in the range $80 - 140$ GeV.

In turn, higher order corrections to ``one fragmentation" reveal the
``two fragmentation" mechanism. Similarly, a collinear enhancement associated
with each photon fragmentation compensates two extra powers of $\alpha_{s}$ in
the short distance subprocess, so that the power counting in $\alpha_{s}$ is
here also asymptotically the same as for the Born and ``one fragmentation"
parts. Higher order corrections to both fragmentation contributions have in
principle to be computed in order to provide a consistent NLO study. This has
been done in \cite{bgpw}. The actual quantitative significance of these
contributions is discussed in sect. \ref{phenomenology}.

\subsection{The box contribution}\label{box}

Beyond this, the gluon-gluon fusion contribution  $g g \rightarrow \gamma
\gamma$, of the ``two direct" type, cannot be neglected. Indeed, although it is
an ${\cal O}(\alpha^{2} \alpha_{s}^{2})$ i.e. next-to-next-to-leading order
contribution, the suppression due to higher powers of $\alpha_{s}$ is
compensated  by the large gluon luminosity at colliders. This is especially
true at LHC in the relevant range for Higgs search, where the so-called box
contribution has the same magnitude as the  Born term, roughly 50 to 80 \%.

Moreover it is the lowest order of a new mechanism, whose spurious scale
dependences are thus monotonic, and only higher corrections to it would reduce
the sensitivity with respect to spurious scale dependences. Finally, this
lowest order is a $2 \rightarrow 2$ process which yields only back to back
photons in the direction transverse to the beam axis; it gives no contribution
to the tail of  the transverse momentum distribution of photon pairs. An
evaluation of the distortion of the transverse momentum distribution of photon
pairs due to the process of gluon-gluon fusion requires the computation of at
least the next order correction \cite{balacz}. The sum ``two direct + box" will
be refered to as the ``direct" contribution.

\section{Isolation}

At TeV colliders, the fragmentation contributions are far from negligible. In
particular, the ``one fragmentation" component  dominates the inclusive
production of  photon pairs in the lower range of the invariant mass spectrum;
in the  range of interest for Higgs search at LHC it happens to be 2 to 5
times larger than the ``two direct" contribution, depending on choice of scales
~\footnote {As already mentioned in the beginning of sect. \ref{mech},
this statement is strongly fragmentation scale dependent.} \cite{bgpw}.
A NLO evaluation of the fragmentation contribution is thus necessary 
to have a tentatively reliable prediction.

Actually, collider experiments do not measure {\it inclusive} photon pairs.
Isolation cuts~\footnote{More about isolation issues in processes involving
direct photons is discussed in \cite{frixione}.} are imposed experimentally to
drastically reduce the gigantic background coming from decays of $\pi^{0}$ and
$\eta$ mesons. Schematically, a candidate-photon is considered isolated if, in some
given cone in azimuthal angle and rapidity about the photon defined by 
\begin{equation}\label{crit1}
(\phi - \phi_{\gamma})^{2} + (y - y_{\gamma})^{2}  \leq  R^{2},
\end{equation}
the deposited hadronic transverse energy $E_{T}^{had}$ is smaller than some 
maximal amount $E_{T}^{max}$, 
\begin{equation}\label{crit2}
E_{T}^{had} \leq E_{T}^{max}
\end{equation}
$R$ and $E_{T}^{max}$ being fixed by the experiments. Such isolation cuts
affect also the production rate of direct photon pairs, especially the
``single-" and ``two fragmentation" contributions, whose topologies are similar
to the one of the background. Those are severely reduced when $E_{T}^{max}$ is
chosen to be very small compared to the transverse
momenta of the photons.

Yet a NLO evaluation of fragmentation contributions is still relevant for 
various reasons. First, the actual isolation cuts used by collider experiments
may be quite more complicated than the schematical criterion given by eqns.
(\ref{crit1},\ref{crit2}). Higher order partonic calculations are not designed
to account for such criteria accurately, contrary to Monte-Carlo event
generators such as {\tt PYTHIA} \cite{PHYTIA} or {\tt HERWIG} \cite{HERWIG4}. Since these
Monte-Carlos and NLO partonic calculations are based on different QCD
approximations, it is worthwhile to compare these two approaches whenever
possible, as for the inclusive production rate, as well as with rather simple
isolation cuts such as the one of eqns. (\ref{crit1},\ref{crit2}). Secondly,
the above cone-type isolation criterion induces infrared sensitivity {\it
inside} the physical spectrum for observables such as the $q_{T}$ spectrum of
photon pairs. This effect appears at the NLO - and every higher order - in the
``one fragmentation" component, as will be shown in the next section.


\begin{figure}[tp]
\begin{center}
\epsfxsize=12cm
\epsfysize=12cm
\mbox{\epsfbox{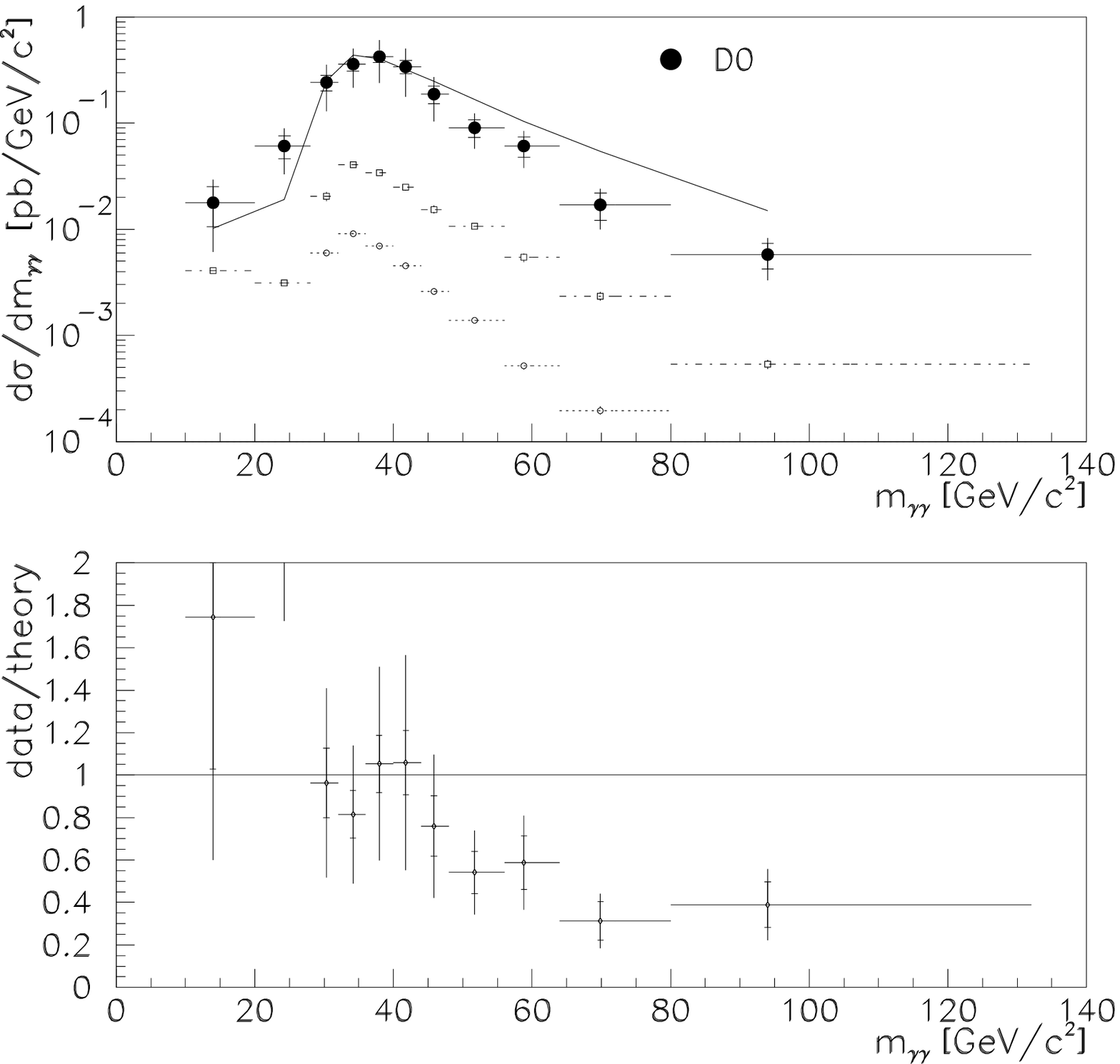}}
\end{center}
\caption{
 \small \sf Diphoton differential cross section $d\sigma/dm_{\gamma \gamma}$ 
 vs. $m_{\gamma\gamma}$, the mass of the photon pair, at the Tevatron,
 $\sqrt{S}=1.8$ TeV.
 Preliminary data points (statistical errors and systematics in quadrature)
 from the D0 collaboration~\protect\cite{d0} are compared to the theoretical
 predictions; the full NLO prediction is shown as the solid line
 while open squares (open circles) represent the single 
 (double) fragmentation contribution.
 }
\label{Fig:d0_mgg}
\end{figure}


\begin{figure}[tp]
\begin{center}
\epsfxsize=12cm
\epsfysize=12cm
\mbox{\epsfbox{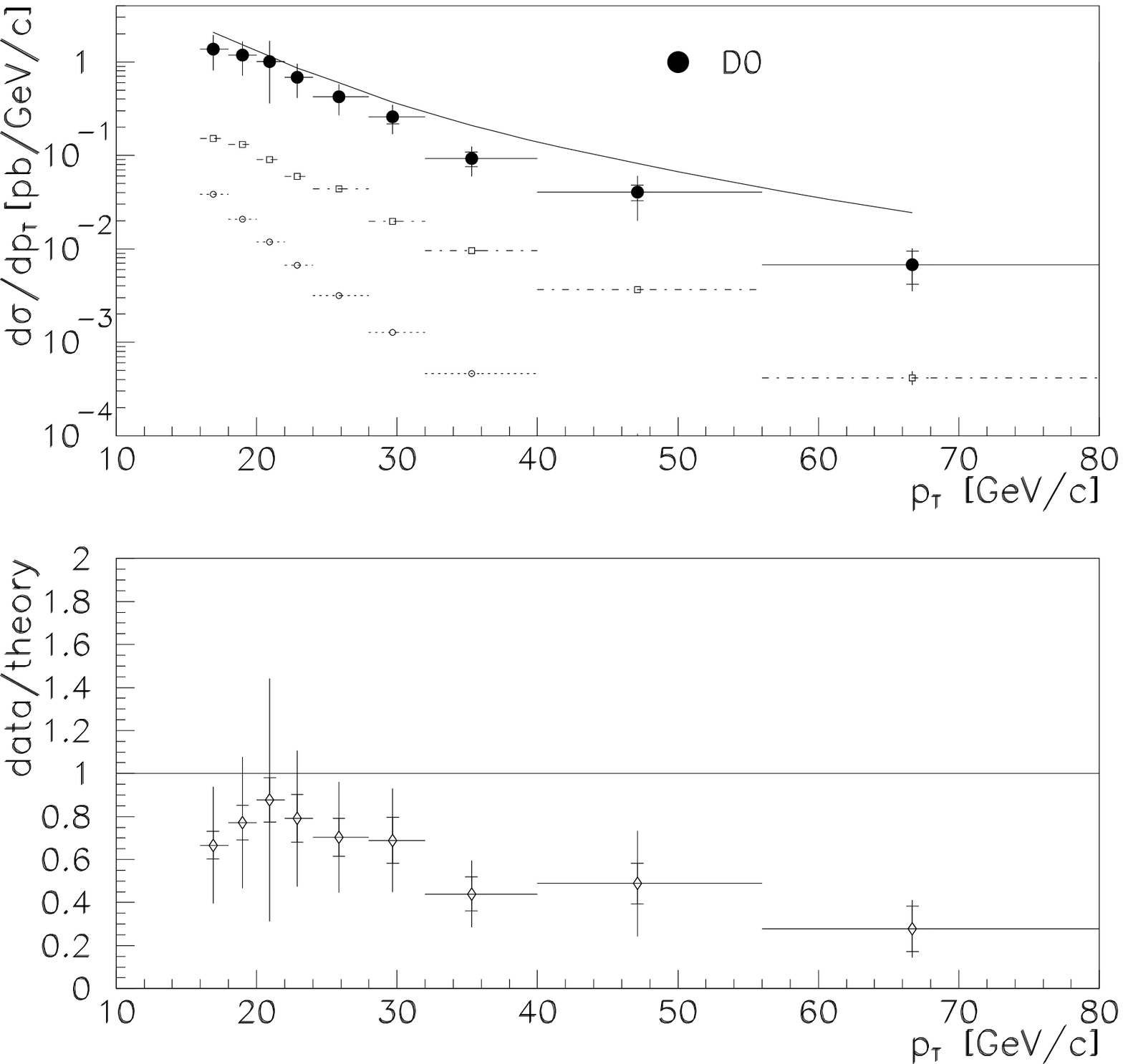}}
\end{center}
\caption{
 \small \sf Diphoton differential cross section $d\sigma/dp_T$ vs $p_T$, 
 the transverse energy of each photon, at Tevatron, $\sqrt{S}=1.8$ TeV.
  Preliminary data points (statistical errors and systematics in quadrature) 
 from the D0 collaboration~\protect\cite{d0} are compared to the theoretical
 predictions; the full NLO prediction is shown as
 the solid line while open squares (open circles) represent the single 
 (double) fragmentation contribution. The ratio data/(full NLO theory)
 is shown below.
 }
\label{Fig:d0_det}
\end{figure}


\begin{figure}[tp]
\begin{center}
\epsfxsize=12cm
\epsfysize=12cm
\mbox{\epsfbox{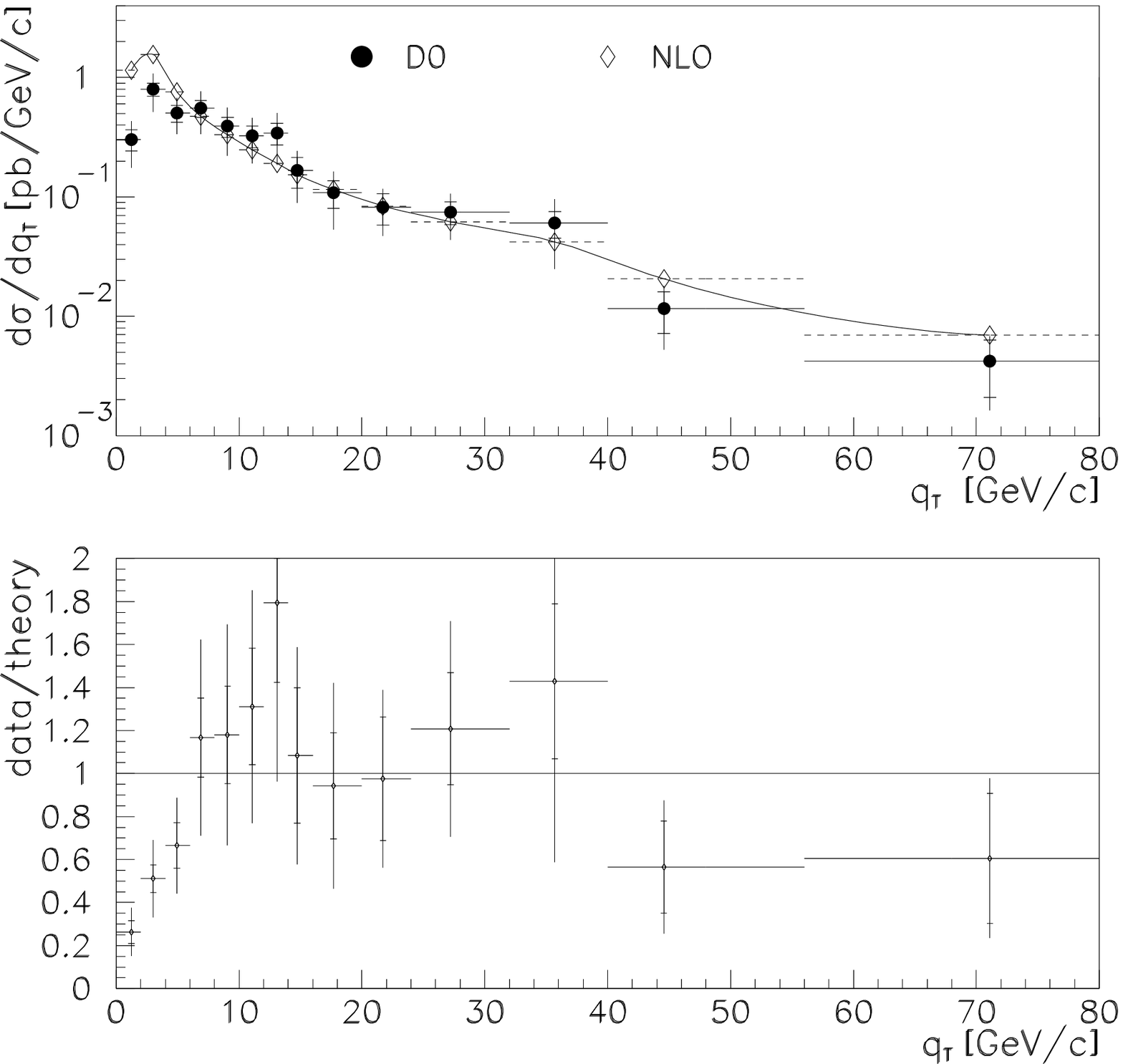}}
\end{center}
\caption{
\small \sf Diphoton differential cross section $d\sigma/dq_T$ vs. $q_T$, the
 transverse momentum of the photon pair, at the Tevatron, $\sqrt{S}=1.8$ TeV.
 Preliminary data points (statistical errors and systematics in quadrature) 
 from the D0 collaboration~\protect\cite{d0} are compared to the theoretical
 predictions; the full NLO prediction is shown as the solid line.
 }
\label{Fig:d0_ptpair}
\end{figure}


\begin{figure}[tp]
\begin{center}
\epsfxsize=12cm
\epsfysize=12cm
\mbox{\epsfbox{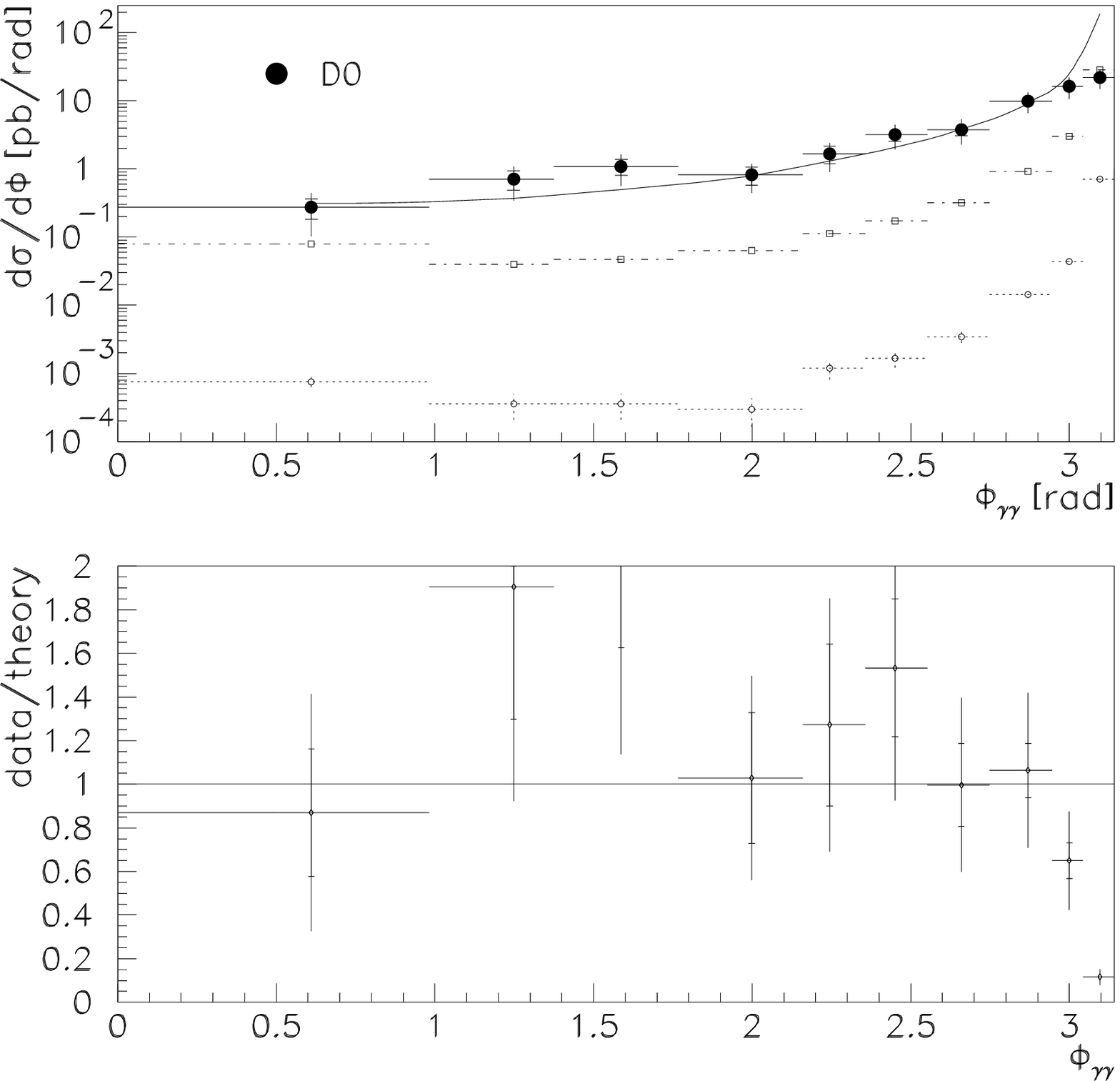}}
\end{center}
\caption{
 \small \sf Diphoton differential cross section $d\sigma/d \phi_{\gamma\gamma}$  
 vs. $\phi_{\gamma\gamma}$, the azimuthal angle between the two 
 photons, at the Tevatron, $\sqrt{S}=1.8$ TeV.
 Preliminary data points (statistical errors and systematics in quadrature) 
 from the D0 collaboration~\protect\cite{d0} are compared to the theoretical
 predictions; the full NLO prediction is shown as the solid line.
 The direct contribution is shown as the dashed line 
 while open squares (open circles) represent the single 
 (double) fragmentation contribution.
 }
\label{Fig:d0_dphi}
\end{figure} 

\section{Phenomenology}\label{phenomenology}

In earlier works on di-photon production \cite{abfs_aachen}, only the ``two
direct" contribution was calculated at NLO, while the fragmentation
contribution included only the lowest order ``one fragmentation"
part~\footnote{The ``box" contribution was included too.}. Moreover, these works
were not implemented in a form suited to compute observables such as the invariant
mass relevant for Higgs search, nor flexible enough to accomodate experimental
selection cuts. A further refinement \cite{owens3} has implemented the same
approximation in a more flexible approach combining analytical and Monte-Carlo
integration techniques, thus allowing the computation of several observables
within the same calculation, and the possibility to account for
selection/isolation cuts. Two recent developments are presented in this
workshop (see also C. Balazs' contribution). 

We have implemented the ``two direct", ``single-" and ``double fragmentation"
contributions at NLO accuracy, together with the box gluon-gluon contribution,
into a general purpose computer program of ``partonic event generator" type
({\it DIPHOX}) presented in detail in \cite{bgpw}. The results which we present
here are derived from this analysis.

\subsection{Comparison with Tevatron data}

The NLO results agree with the preliminary D0 data \cite{d0}  reasonably,  as
seen in Figs. (\ref{Fig:d0_mgg}-\ref{Fig:d0_dphi}), except for the  tails of
each photon's transverse momentum distribution  $d \sigma / d E_{T}$ and the
invariant mass distribution of pairs  $d \sigma / d M_{\gamma \gamma}$ at large
$E_{T}$ and  $M_{\gamma \gamma}$ respectively, where the three highest data
points  are affected by correlated systematic uncertainties due to background 
evaluation in both cases. However, more instructive conclusions will be drawn
after a finalized understanding of the systematics, and even more so after the
Tevatron Run II with the statistics improved by a factor of 20.

\subsection{Estimates for LHC}

\begin{figure}[tp]
\begin{center}
\epsfxsize=12cm
\epsfysize=12cm
\mbox{\epsfbox{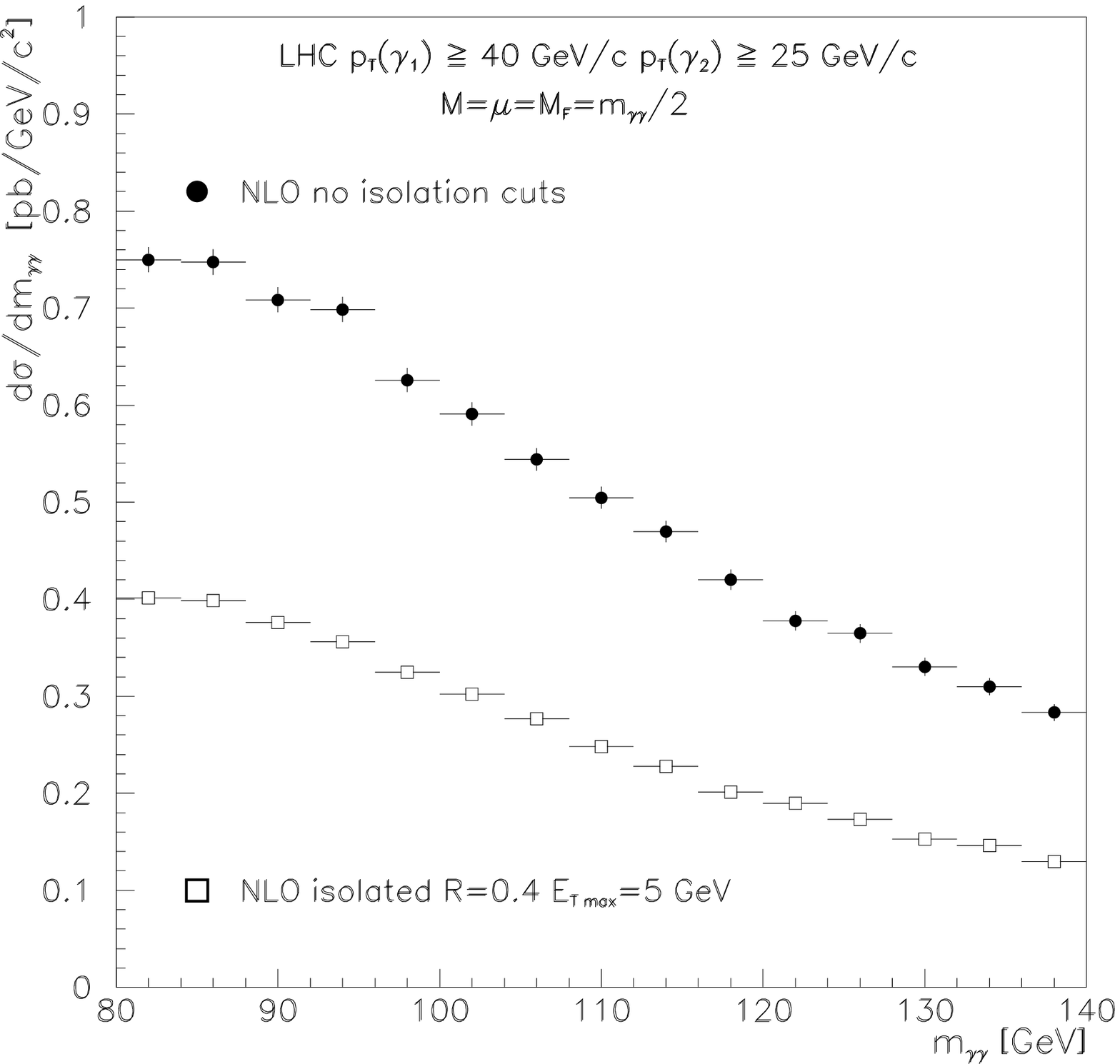}}
\end{center}
\caption{\small \sf 
 Diphoton differential cross section $d\sigma/d m_{\gamma \gamma}$ vs.
 $m_{\gamma \gamma}$ at the LHC, $\sqrt{S}=14$ TeV, without and 
 with isolation criterion $E_{T max}=5$ GeV in $R=0.4$. 
 Same kinematic cuts as in Fig. (\ref{Fig:lhc_mgg}). 
 The scale choice is $M=M_f=\mu=m_{\gamma \gamma}/2$.%
}
\label{Fig:lhc_comp_cuts}
\end{figure}

We now give some theoretical estimates in the domain relevant for Higgs
search at the LHC, for the invariant mass distribution cf. 
Fig.~(\ref{Fig:lhc_comp_cuts}) with and without isolation.
One has to keep in mind that the theoretical uncertainties are still
large. Firstly these results are still plagued by rather large scale 
uncertainties, as discussed below. Secondly, for a given scale choice, 
they may still underestimate the actual background to Higgs search. 

\subsection{Critical examination of various theoretical issues}

\subsubsection{Scale uncertainties}

As mentioned above all results depend on three unphysical scales.
Varying these between $M_{\gamma \gamma}^{2}/4$ and  $4 M_{\gamma
\gamma}^{2}$ along the first diagonal  $\mu^{2} = M^{2} = M_{f}^{2}$, the NLO
results for the invariant mass distribution appear surprisingly stable, since
they  change by about 5\% only. Alternatively, anti-diagonal variations of 
$\mu^{2}$ and $M^{2} = M_{f}^{2}$ in the same interval about the  central value
$M_{\gamma \gamma}^{2}$ lead to a variation still rather large (up to 20 \%). 
This is because variations with respect to $\mu^{2}$ and $M^{2}$ act in
opposite ways. When $\mu^{2}$ is increased, $\alpha_{s}(\mu^{2})$ and hence the NLO
corrections decrease; on the other hand the relevant values of the momentum
fraction of incoming partons  are small, $\sim {\cal O}(10^{-3}$ to
$10^{-2})$, so that the gluon and sea quark distribution functions increase when
$M^{2}$ is increased. Scale changes with respect to $\mu^{2}$ and $M^{2}$ thus
nearly cancel again each other along the first diagonal but add up in the other
case. Actually, the stability along the first diagonal is accidental at this
order~\footnote{In processes for which the lowest order involves some power of
$\alpha_{s}$, an explicit $\mu^{2}$ dependence appears in next-to-leading order
correction, which partially compensates the $\mu^{2}$ dependence in
$\alpha_{s}(\mu^{2})$. Unlike this, in the two direct component which dominates
the cross section when a drastic isolation is required, the lowest order
involves no $\alpha_{s}$. The explicit $\mu^{2}$ dependence would thus appear
only at  ${\cal O}(\alpha_{s}^{2})$, i.e. at next-to-next-to leading order.  At
next to leading order, the $\mu^{2}$ dependence occurs only through  the
monotonous decrease of the $\alpha_{s}(\mu^{2})$ weighting the first
correction: there is no partial cancellation of $\mu^{2}$ dependence.  The
mechanism is more complicated in the fragmentation components, and the
situation becomes mixed up between all components when the severity of
isolation is reduced.}. These observations hold separately for the box
contribution.

In conclusion, the $\mu^{2}$, $M^{2}$ dependences are thus not completely under
control yet. On the other hand,  accounting for the NLO corrections to the
fragmentation components provides stability with respect to $M_{f}^{2}$
variations about orthodox choices of the fragmentation scale. 

\subsubsection{Quantitative importance of fragmentation contributions}

For orthodox choices of the fragmentation scale, $M_{f}^{2}$ of order
$M_{\gamma \gamma}^{2}$, the ``single  fragmentation" contribution is small at
Tevatron, given the stringent isolation cuts used, and the ``two fragmentation"
one is even smaller. For example, as can be seen in Figs.
(\ref{Fig:d0_mgg},\ref{Fig:d0_det}) the ``one fragmentation" contribution is
about one order of magnitude less than the ``two direct" one. It may still have
a small visible effect, as in the tail of the azimutal angle distribution $d
\sigma / d \phi_{\gamma \gamma}$ ($\phi_{\gamma \gamma}$ being the azimutal
angle between the two photons of a pair) in the low $\phi_{\gamma \gamma}$
range, cf. Fig. (\ref{Fig:d0_dphi}). 

\begin{figure}[tp]
\begin{center}
\epsfxsize=12cm
\epsfysize=12cm
\mbox{\epsfbox{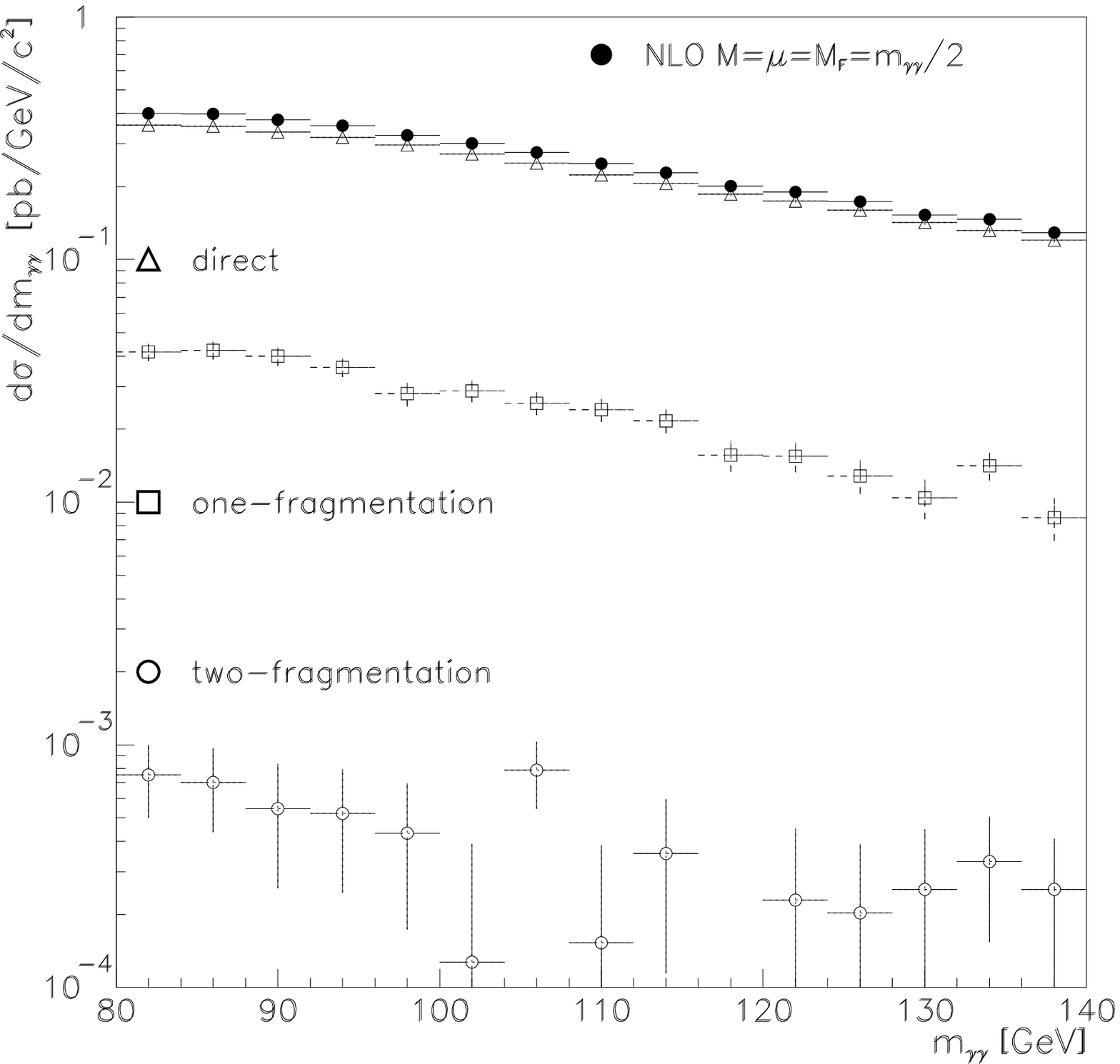}}
\end{center}
\caption{
 \small \sf Splitting of the diphoton differential cross 
 section $d\sigma/dm_{\gamma \gamma}$ at the LHC,
 $\sqrt{S}=14$ TeV with isolation criterion $E_{T max}=5$ GeV in $R=0.4$, 
 into the ``direct",``one
 fragmentation" and ``two fragmentation" components, shown for the scale 
 choice $\mu = M = M_{f}=m_{\gamma\gamma}/2$. 
 The following kinematic cuts are applied:
 $p_{T}(\gamma_{1}) \geq 40$ GeV, $p_{T}(\gamma_{2}) \geq 25$ GeV, 
 $|y(\gamma_{1,2})| \leq 2.5$.
 }
\label{Fig:lhc_mgg}
\end{figure}

The situation is the same for LHC predictions, see Fig. (\ref{Fig:lhc_mgg}).
Contributions from fragmentation are drastically reduced when very stringent
cuts are imposed, e.g. $E_{T}^{max} = 2.5$ GeV in $R = 0.4$. However, in
practice  such isolation cuts will be nearly saturated by underlying events: 
their veto on the hard event itself is thus even more severe, allowing almost
no transverse energy leakage from the hard process inside the cone.  This may be
experimentally most suitable. However, requiring that {\it  no transverse
energy} be deposited in a cone of fixed size about a photon is {\it not}
infrared safe order by order in perturbation. It means that finite but very
stringent isolation cuts imposed in fixed order partonic calculations would
lead to unreliable results. When less severe isolation cuts are used, the
``one-", and to a lesser extend, ``two fragmentation" components are
subdominant but not negligible.


\subsubsection{Infrared sensitive distributions}

Being based on a finite order calculation, our computer code is not
suited for the study of observables controled by multiple soft gluon
emission \cite{catani2,collins-soper}. Among those, one may distinguish 
the following examples, most of which would require an improved account 
of soft gluon effects.\\

\noindent
{\bf The transverse momentum distribution of pairs 
$d \sigma / d q_{T}$ near $q_{T} = 0$}\\
The problematics of the ``two direct" contribution is similar to the 
well-known Drell-Yan process, see \cite{balacz}. On the other hand, the 
fragmentation contributions do not diverge order by order when 
$q_{T} \rightarrow 0$. Indeed, in the ``one fragmentation" case,
\begin{eqnarray}
\mbox{parton 1} + \mbox{parton 2} & \rightarrow & 
     \gamma_{1} + \mbox{parton 3}  \label{hard1}  \\
\mbox{parton 3} & \rightarrow &  \gamma_{2} + X \label{hard2} 
\end{eqnarray}
the NLO contribution to the hard subprocess (\ref{hard1}) 
yields a double logarithm
\begin{equation}
\sim \alpha_{s} \ln^{2} \| {\mbox {\bf p}}_{T}(\gamma_{1}) + 
{\mbox {\bf p}}_{T}(\mbox{parton}\, 3) \| 
\end{equation}
when ${\mbox {\bf p}}_{T}(\gamma_{1}) + 
{\mbox {\bf p}}_{T}(\mbox{parton} \, 3) \rightarrow {\boldmath 0}$. 
However the extra convolution associated with the fragmentation
(\ref{hard2}) involves an integral over 
$z_{2} = p_{T}(\gamma_{2})/p_{T}(\mbox{parton} \, 3)$ which 
smears out this integrable singularity. The ``two fragmentation" 
contribution involves two such convolutions, and hence one more smearing.\\

\noindent
{\bf The azimuthal angle distribution $d \sigma / d \phi_{\gamma \gamma}$ near 
$\phi_{\gamma \gamma} = \pi$} \\
This case differs from the previous one for two reasons.
Firstly, not only the ``two direct" contribution diverges order by 
order when $\phi_{\gamma \gamma} \rightarrow \pi$, but also both ``single-" and 
``double-fragmentation" contributions do, as can be seen in Fig.
(\ref{Fig:d0_dphi}). Moreover, in both fragmentation cases, 
soft gluons may couple to both initial- and final-state hard emitters.
Indeed, consider the example of the ``one fragmentation case", cf.
eqns. (\ref{hard1}). Selecting $\phi_{\gamma \gamma} \rightarrow \pi$ 
emphasizes $\phi(\mbox{parton} \, 3) - \phi(\gamma_{1}) \rightarrow \pi$, 
so that all the emitted partons besides parton 3 have to be collinear 
to either of the incoming or outgoing particles, and/or soft, which 
yields double logarithms
\begin{equation}
\sim \alpha_{s} \ln^{2} \left[ \pi - 
\left( \phi(\mbox{parton} \, 3) - \phi(\gamma_{1}) \right) \right]
\end{equation}
associated with each of the hard partons 1,2,3 - plus single
logarithms as well. For the observable $d \sigma / d \phi_{\gamma \gamma}$ near
$\phi_{\gamma \gamma} = \pi$, the integral involved in the convolution  of the
hard subprocess with the fragmentation functions does not smear these
logarithmic divergences, since the fragmentation variable  
$z_{2} = p_{T}(\gamma_{2})/p_{T}(\mbox{parton} \,3)$ is 
decoupled from the azimutal variable 
$\phi(\mbox{parton} \, 3)$ equal to $\phi(\gamma_{2})$ in the 
soft and collinear limits. A similar explanation holds for the ``double
fragmentation component". An analagous problem affects the  $q_{T}$
distribution of a pair photon $+$ jet at low $q_{T}$.\\

\noindent
{\bf The azimuthal angle distribution $d \sigma / d \phi_{\gamma \gamma}$ near 
$\phi_{\gamma \gamma} = 0$}\\ 
Both fragmentation contributions to $d \sigma / d \phi_{\gamma \gamma}$ diverge 
also order by order when $\phi_{\gamma \gamma} \rightarrow 0$. Here also soft
gluons may couple to both initial- and final-state hard emitters.
Actually, given the large invariant mass of the pairs, the 
vicinity of $\phi_{\gamma \gamma} = 0$ is never probed, so that an improved
treatment of soft gluon effects is not needed in this case. Yet,
as a consequence, an increase of the ``single-fragmentation"
contribution can be seen~\footnote{A similar behaviour occurs also for
the ``two fragmentation" contribution, which is however too tiny to
have any significant effect.} in the lower range of the $\phi_{\gamma \gamma}$
spectrum, cf. Fig. (\ref{Fig:d0_dphi}). \\

\noindent
{\bf An infrared divergence for $d \sigma / d q_{T}$ {\it inside} 
the physical spectrum}\\
Besides the well-known issue at $q_{T} = 0$, another infrared
sensitive point appears in the $q_{T}$ spectrum due to isolation, at
the critical value $q_{T} = E_{T}^{max}$. Indeed, at lowest
order the ``one fragmentation" component is not smooth, it instead 
behaves as a step function \cite{bbmry},
\begin{equation}
\left( \frac{d \sigma}{ d q_{T}} \right) ^{``single \;fragm", (LO)} 
\propto \Theta \left( E_{T}^{max} - q_{T} \right) 
\end{equation}
Consequently, in agreement with the general study of
\cite{catani-webber}, the NLO - and every higher order - correction has 
a double logarithmic divergence at the critical point. Such 
singularities are very sensitive to the kinematical constraints and the 
observable considered. The present case has a double logarithm 
{\it below} the critical point,
\begin{equation}
\left(  \frac{d \sigma}{d q_{T}} \right) ^{``single \;fragm", (NLO)} 
\propto - \alpha_{s}
\ln^{2} \left( E_{T}^{max} - q_{T} \right)  
\Theta  \left( E_{T}^{max} - q_{T} \right) + \cdots
\end{equation}

\begin{figure}[tp]
\begin{center}
\epsfxsize=12cm
\epsfysize=12cm
\mbox{\epsfbox{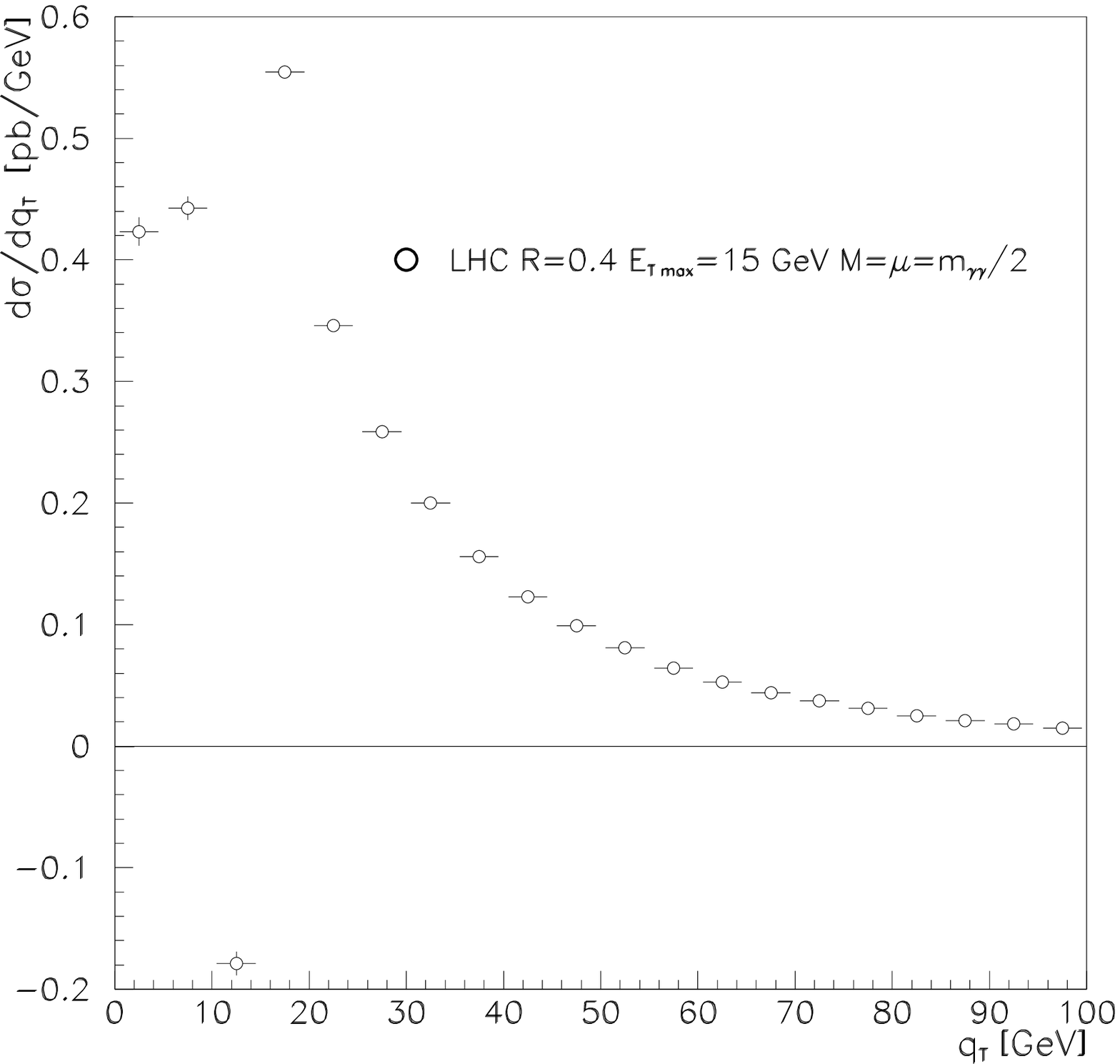}}
\end{center}
\caption{
\small \sf Diphoton differential cross section $d\sigma/dq_T$ at LHC, $\sqrt{S}=14$
 TeV, with isolation criterion $E_{T max}=15$ GeV in $R=0.4$.
 }
\label{Fig:lhc_r04_e15_qt}
\end{figure}

The infrared sensitive behaviour at this critical point can be infered
on Fig. (\ref{Fig:lhc_r04_e15_qt}), where a rather large value for 
$E_{T}^{max}$ is used in order to split this critical point from 
the small $q_{T}$ region. 


This effect is not visible on the theoretical prediction of Fig. 
(\ref{Fig:d0_ptpair}) because of the low value $E_{T}^{max} = 4$ GeV used by
the D0 collaboration. The two infrared sensitive regions ($q_{T} \rightarrow 0$
and  $q_{T} \sim E_{T}^{max}$) are not separated enough, and the bin smearing
averages over both effects. The phenomenon is thus camouflaged. An all order
summation of this soft gluon effect has to be  carried out to restore a
sensible shape to the $q_{T}$ distribution. A similar observation would be made
about the $q_{T}$ distribution of a pair \{photon $+$ jet\}. 

\section{Perspectives}

Future improvements of the theoretical understanding of di-photon
production will require various inputs. Next-to-next to leading order
corrections would hopefully stabilize the scale dependences and 
correct the normalisations. 
Another important quantitative improvement would be the higher order 
corrections to the box contribution, which is within reach thanks to some 
recent achievements. Progresses in this direction are reported on elsewhere in 
these proceedings \cite{delduca}. 

This accounting of multiple soft gluon effects, already implemented in
\cite{balacz} for the ``two direct" contribution in absence of isolation cuts,
is needed in the {\it DIPHOX}-based study to provide correct distributions in
the infrared sensitive regions. This affects  the $q_{T}$ distribution at low
$q_{T}$. It also concerns the $\phi_{\gamma \gamma}$ distribution when 
$\phi_{\gamma \gamma} \rightarrow \pi$, which has been less studied, and for
which the soft gluon effects in the fragmentation cases are more involved than
in the ``two direct" case \cite{cmmo,sterman}. It is also important for the
$q_{T}$ distribution in the vicinity of the critical point $q_{T} =
E_{T}^{max}$, induced by the fixed cone type isolation criterion.

The understanding of the background to the Higgs search is quantitatively
not yet on the same footing as for the signal. Hence, accounting for the
higher order correction to the signal in numerical simulations might be 
instructive, but the results should be considered with care in order to 
avoid  statements that are too optimistic.

\section{Higgs search in association with a hard jet}

In order to overcome the present insufficient control of higher order
corrections in inclusive production, it has been suggested \cite{Hjet1} to study
the associated production of $h(\to \gamma \gamma) +$ jet, for which both signal $S$
and background $B$ are lower (but still at the level of hundred signal events
at low luminosity). The lowest order estimate has shown that the $S/B$ ratio is
improved critically (up to $1/2-1/3$) with the same level of the significance
$S/\sqrt{B}$. Furthermore, higher order corrections to the background have
been shown recently \cite{kdf} to be under better control than in the inclusive
(i.e. unassociated) case. 

\subsection{Background: associated vs. inclusive}

Indeed, in the inclusive case, the magnitude of the NNLO box contribution is
comparable to the LO cross section essentially because the latter is initiated
by $q \bar{q}$, whereas the former involves $g g$. The $g g$ luminosity, much
larger than the $q \bar{q}$ one, compensates numerically the extra $\alpha_s^2$
factor of the box. On the contrary, in the channel $\gamma \gamma$ + jet, the
LO cross-section is dominated by a $q g$ initiated subprocess. The $q g$
luminosity is sizeably larger than the $q \bar{q}$ one, so that the
corresponding NNLO remains small compared to the LO result \cite{kdf}. Thus, 
expecting that the subprocess $g g \rightarrow \gamma \gamma  g$ gives the main
NNLO correction, a quantitative description of the background with an accuracy
better than 20\% could be achieved already at NLO in the $\gamma \gamma$+ jet
channel for a high  $p_{T}$ ($\geq 30$ GeV) jet. All the helicity amplitudes
needed for the implementation of the (``direct" contribution to the) background
to NLO accuracy~\footnote{We remind the reader unfamiliar with the LO, NLO,
NNLO, etc, terminology that this terminology does not refer to the absolute
power of $\alpha_s$ involved, but instead to the relative power with respect to
the Born term of the process considered. Hence, NLO corrections to $\gamma
\gamma+$ jet are also part of the NNLO corrections to $\gamma \gamma$ inclusive
(converse not true, cf. box).}are now available 
\cite{Bern:1995fz,Signer:1995nk,DelDuca:1999pa}.

\subsection{Signal vs. background}

The origin of this improvement of the $S/B$ ratio is the following. At
LO, the signal of associated $h(\to \gamma \gamma) + jet$ production has 
basically a 2-body kinematics, due to the extremely small Higgs width (a few
MeV). On the contrary, the LO background contributions $q {\bar q}\to\gamma
\gamma g$ and $gq \to \gamma + \gamma + q$  (as well as the NNLO $g g \to
\gamma \gamma g$ one) have 3-body kinematics. This is in contrast to the
inclusive channel where both signal and background LO subprocesses have 2-body
kinematics. This circumstance opens the room for more refined cuts to suppress
the background more efficiently  \cite{Hjet1}. The contributions $q {\bar q}
\to \gamma \gamma g$ and $g g \to \gamma \gamma g$ are suppressed down to 40\%
of the signal (to be compared with $S/B \sim 1/7$ for the  inclusive channel).
Unfortunately, as found in \cite{Hjet1}, the contribution of  the other
subprocess $g q \to \gamma \gamma q$, which dominates, yields the  overall
ratio $S/B \sim 1/2 - 1/3$.

The cuts which allow this efficient suppression of the irreducible background
at LO are based on the differences in the shapes of the angular distributions
in the partonic c.m.s.. Due to helicity and total angular momentum
conservation, the S-wave does not contribute to the dominant signal
subprocess $g g \rightarrow H g$. On the contrary, all angular momentum states
contribute to the subprocesses $g q  \rightarrow \gamma \gamma q$ and $q
\bar{q} \rightarrow \gamma \gamma g$. Therefore, the signal has a more
suppressed threshold behaviour compared to the background. The $S/B$ ratio can
thus be improved by increasing the partonic c.m.s. energy $\sqrt{\hat s}$ far
beyond threshold, and a cut $\sqrt{\hat s}>300$ GeV has been found to give
the best S/B ratio for the LHC. Actually, the effect can not be fully explained 
by the threshold behavior only, since that would result in a uniform suppression
factor. It was shown in \cite{Hjet1,Abdullin:1998er} (see Figs. 5 and 6
there) that the dependences of the background and the signal on the c.m.s.
angular variables are quite different; therefore, the strong $\hat s$ cut
affects them with different suppression factors (see
\cite{Hjet1,Abdullin:1998er} for more details). This effect can be exploited to
enhance the significance $S/\sqrt{B}$  at the same level as $S/B$. If the cut
$\cos({\vartheta^*)({j \gamma})}<-0.87$  on the jet-photon in  the partonic
c.m.s. is applied  for $\sqrt{\hat s}<300$ GeV and  combined with the cut
$\sqrt{\hat s}>300$  GeV, the change on $S/B$ is rather small, while the
significance is improved by a factor $\sim$ 1.3. The same effect can be
observed with the cut on the jet angle in the partonic c.m.s.
$(\vartheta^*({j})$, cf. Fig. 5 of \cite{Hjet1,Abdullin:1998er}, but 
one should notice that the two variables, $\vartheta^*({j \gamma})$ and
$\vartheta^*({j})$, are correlated. Therefore, it is desirable to perform a
multi-variable optimization of the event selection. Notice that the present
discussion is based on a LO analysis, and concerns only what was defined above
as the ``direct" component of the  irreducible background. One now has to
understand how this works at NLO. 

Other, reducible, sources of background are potentially dangerous. The 
above-defined  ``one  fragmentation" component to the so-called irreducible
background, and the reducible background coming from misidentification of jet
events were treated on a similar footing in the LO analysis of 
\cite{Hjet1,Abdullin:1998er} as a {\it de facto} reducible background. 
In \cite{Hjet1,Abdullin:1998er}, a rough analysis found that this
reducible background is less than 20\% of  the irreducible one after cuts are
imposed.  The misidentification rate is given mainly by the subprocesses $g q
\rightarrow \gamma g q$, $g g \rightarrow \gamma q \bar{q}$ and $q
q'\rightarrow \gamma q(g) q'(g)$, when the final state parton produces an
energetic isolated photon but other products of the hadronization escape the
detection as a jet. There, a $\gamma(\pi^0)/jet$ rejection factor equal to 2500
for a jet misidentified as a photon and 5000 for a well separated
$\gamma(\pi^0)$ production by a jet were used. No additional $\pi^0$ rejection
algorithms were applied. Furthermore, this reducible background is expected to
be suppressed even more strongly than the irreducible background of ``direct"
type when a cut on $\sqrt{\hat s}$ is applied.

In summary, the associated channel $H(\rightarrow \gamma \gamma) +$ jet with
jet transverse energy $E_T>30$ GeV and rapidity $|\eta|<4.5$ (thus
involving forward hadronic calorimeters) opens a promising possibility for
discovering the Higgs boson with a mass of 100-140 GeV at LHC even at low 
luminosity. However, to perform a quantitative analysis, the NLO calculations, 
namely of the background, have to be completed, and included in a more realistic
final state analysis.

\section{Acknowledgements}

T. B. is a EU fellow supported by the EU Fourth Programme ``Training and
Mobility of Researchers", Network ``Quantum Chromodynamics and the Deep
Structure of Elementary Particles", contract FMRX-CT98-0194 (DG12 - MIHT).
V.A. I. acknowledges support by the CERN-INTAS grant 377 and RFBR-DFG
grant 99-02-04011.
LAPTH is a {\it Unit\'e Mixte de Recherche (UMR 5108) associ\'ee au CNRS
et \`a l'Universit\'e de Savoie}.


\setcounter{figure}{0}
\setcounter{table}{0}
\setcounter{section}{0}
\setcounter{equation}{0}
\newpage

\end{document}